\documentclass[11pt,a4paper]{article} 
\pdfoutput=1 

\usepackage{jcappub} 

\usepackage[T1]{fontenc} 

\usepackage{amsmath,amsfonts,amssymb}

\usepackage{graphicx}
\usepackage{fixltx2e}
\usepackage{natbib}
\usepackage{epstopdf}
\usepackage{epsf,color}
\usepackage{ifthen}
\usepackage{relsize}
\usepackage{placeins}
\usepackage{morefloats}

\usepackage{mathrsfs}

\bibpunct{(}{)}{;}{a}{}{,}

\def\reff@jnl#1{{\rm#1\/}}

\def\aj{\reff@jnl{AJ}}                  
\def\araa{\reff@jnl{ARA\&A}}            
\def\apj{\reff@jnl{ApJ}}                
\def\apjl{\reff@jnl{ApJ}}               
\def\apjs{\reff@jnl{ApJS}}              
\def\ao{\reff@jnl{Appl.Optics}}         
\def\apss{\reff@jnl{Ap\&SS}}            
\def\aap{\reff@jnl{A\&A}}               
\def\aapr{\reff@jnl{A\&A~Rev.}}         
\def\aaps{\reff@jnl{A\&AS}}             
\def\azh{\reff@jnl{AZh}}                        
\def\baas{\reff@jnl{BAAS}}              
\def\jcap{\reff@jnl{JCAP}}              
\def\jrasc{\reff@jnl{JRASC}}            
\def\memras{\reff@jnl{MmRAS}}           
\def\mnras{\reff@jnl{MNRAS}}            
\def\pra{\reff@jnl{Phys. Rev. A}}         
\def\prb{\reff@jnl{Phys. Rev. B}}         
\def\prc{\reff@jnl{Phys. Rev. C}}         
\def\prd{\reff@jnl{Phys. Rev. D}}         
\def\prl{\reff@jnl{Phys. Rev. Lett}}      
\def\physrep{\reff@jnl{Phys. Rep.}}      
\def\pasp{\reff@jnl{PASP}}              
\def\pasj{\reff@jnl{PASJ}}              
\def\qjras{\reff@jnl{QJRAS}}            
\def\skytel{\reff@jnl{S\&T}}            
\def\solphys{\reff@jnl{Solar~Phys.}}    
\def\sovast{\reff@jnl{Soviet~Ast.}}     
 \def\ssr{\reff@jnl{Space~Sci.Rev.}}    
\def\zap{\reff@jnl{ZAp}}                
\def\nat{\reff@jnl{Nature}}             
\def\procspie{\reff@jnl{Proceedings of the SPIE}}             

\def\ben{\begin{enumerate}}
\def\een{\end{enumerate}}
\def\bi{\begin{itemize}}
\def\ei{\end{itemize}}
\def\be{\begin{equation}}
\def\ee{\end{equation}}
\def\bea{\begin{eqnarray}}
\def\eea{\end{eqnarray}}
\def\ba{\begin{align}}
\def\ea{\end{align}}

\def\bdd{\boldsymbol{d }}

\def\bdm{\boldsymbol{m }}
\def\bdn{\boldsymbol{n }}
\def\bds{\boldsymbol{s }}

\def\mA{\boldsymbol{\rm A}}

\def\mC{\boldsymbol{\rm C}}

\def\mS{\boldsymbol{\rm S}}

\def\mN{\boldsymbol{\rm N}}

\makeatletter
\newcommand{\ion}[2]{#1$\;$\textsc{\rmfamily\@roman{#2}}\relax}
\makeatother
\newcommand{\thedate}{\today}

\newcommand{\hii}{\ion{H}{2}}

\newcommand{\tA}{{\mA}}

\newcommand{\tN}{{\mN}}

\newcommand{\tC}{{\mC}}

\newcommand{\tS}{{\mS}}

\def\mX{\boldsymbol{\rm X}}

\newcommand{\tX}{{\mX}}
\def\trace{\mathrm{tr}}
\def\inv{^{-1}}

\newcommand{\healpix}{{\tt HEALPix}}
\newcommand{\nside}{\ensuremath{N_{\mathrm{side}}}}
\newcommand{\wg}[6]
{{
\left(
\begin{array}{lcr} #1 & #2 & #3 \\ #4 & #5 & #6 \end{array}
\right)
}}

\def\setsymbol#1#2{\expandafter\def\csname #1\endcsname{#2}}
\def\getsymbol#1{\csname #1\endcsname}

\def\Planck{\textit{Planck}}
\def\core{\textit{CORE}}

\newcommand\ltsima{$\; \buildrel < \over \sim \;$}
\newcommand\simlt{\lower.5ex\hbox{\ltsima}}
\newcommand\gtsima{$\; \buildrel > \over \sim \;$}
\newcommand\simgt{\lower.5ex\hbox{\gtsima}}
\newcommand\simprop{\lower.5ex\hbox{$\; \buildrel \propto \over \sim \;$}}

\title{Exploring Cosmic Origins with CORE: $B$-mode Component Separation}

\author[1]{M.~Remazeilles,}
\affiliation[1]{Jodrell Bank Centre for Astrophysics, Alan Turing
  Building, School of Physics and Astronomy, The University of
  Manchester, Oxford Road, Manchester, M13 9PL, U.K.}

\author[2,3]{A.~J.~Banday,}
 \affiliation[2]{Universit\'{e} de Toulouse, UPS-OMP, IRAP, F-31028
   Toulouse cedex 4, France }
 \affiliation[3]{CNRS, IRAP, 9 Av. colonel Roche, BP 44346, F-31028
   Toulouse cedex 4, France }

\author[4,5]{C.~Baccigalupi,}
 \affiliation[4]{SISSA, Via Bonomea 265, 34136, Trieste, Italy}
 \affiliation[5]{INFN, Via Valerio 2, I - 34127 Trieste, Italy} 

\author[6,4]{S.~Basak,}
\affiliation[6]{ Department of Physics, Amrita School of Arts \&
  Sciences, Amritapuri, Amrita Vishwa Vidyapeetham, Amrita University,
  Kerala 690525, India }

\author[1]{A.~Bonaldi,}

\author[7]{G.~De Zotti,}
 \affiliation[7]{INAF-Osservatorio Astronomico di Padova, Vicolo
   dell'Osservatorio 5, I-35122 Padova, Italy}

\author[8]{J.~Delabrouille,}
 \affiliation[8]{APC, AstroParticule et Cosmologie, Universit\'e Paris
   Diderot, CNRS/IN2P3, CEA/Irfu, Observatoire de Paris Sorbonne Paris
   Cit\'e, 10, rue Alice Domon et Leonie Duquet, 75205 Paris Cedex 13,
   France}

\author[1]{C.~Dickinson,}

\author[9]{H.~K.~Eriksen,}
 \affiliation[9]{Institute of Theoretical Astrophysics, University of
   Oslo, PO Box 1029, Blindern, NO-0315 Oslo, Norway}

\author[10]{J.~Errard,}
 \affiliation[10]{Institut Lagrange, LPNHE, Place Jussieu 4, 75005
   Paris, France.}

% ...

\emailAdd{mathieu.remazeilles@manchester.ac.uk}

\abstract{
We demonstrate that, for the baseline design of the \core\ satellite mission, the polarized foregrounds can be controlled at
  the level required to allow the detection of the primordial cosmic microwave background (CMB) $B$-mode polarization with the desired accuracy at both reionization and recombination scales,  for tensor-to-scalar ratio values of  ${r\gtrsim 5\times 10^{-3}}$. 
We consider detailed sky simulations based on state-of-the-art CMB
  observations that consist of CMB polarization with $\tau=0.055$ and
  tensor-to-scalar values ranging from $r=10^{-2}$ to $10^{-3}$, 
Galactic synchrotron, and thermal dust polarization with
  variable spectral indices over the sky, polarized anomalous
  microwave emission, polarized infrared and radio sources, and
  gravitational lensing effects. Using both parametric and blind approaches, we perform full component separation and likelihood analysis of the simulations, allowing us to quantify both uncertainties and biases on the reconstructed primordial $B$-modes. 
Under the assumption of perfect control of lensing effects, \core\ would measure an unbiased estimate of $r=\left(5 \pm 0.4\right)\times 10^{-3}$ after foreground cleaning.
  In the presence of both gravitational lensing effects and astrophysical foregrounds, the significance of the detection is lowered, with \core\ achieving a $4\sigma$-measurement of $r=5\times 10^{-3}$ after foreground cleaning and $60$\% delensing. For lower tensor-to-scalar ratios ($r=10^{-3}$) the overall uncertainty on $r$ is dominated by foreground residuals, not by the 40\% residual of lensing cosmic variance. Moreover, the residual contribution of unprocessed polarized point-sources can be the dominant foreground contamination to primordial B-modes at this $r$ level, even on relatively large angular scales, $\ell \sim 50$.
Finally, we report two sources of potential bias for the detection of the primordial $B$-modes by future CMB experiments: (i) the use of incorrect foreground models, e.g. a modelling error of $\Delta\beta_s = 0.02$ on the synchrotron spectral indices may result in an excess in the recovered reionization peak corresponding to an effective $\Delta r > 10^{-3}$; (ii) the average of the foreground line-of-sight spectral indices by the combined effects of pixelization and beam convolution, which adds an effective curvature to the foreground spectral energy distribution and may cause spectral degeneracies with the CMB in the frequency range probed by the experiment.
}

\keywords{Cosmology: observations --- methods: data analysis --- Polarization --- cosmic background radiation --- diffuse radiation --- inflation}

\begin{document}

\thedate
\maketitle
\flushbottom


\section {Introduction}
\label{sec:introduction}

The standard model of cosmology is based on the inflationary paradigm
\citep{Starobinsky1980,Guth1981,Linde1982,Linde1983,Albrecht1982}, yet
direct observational evidence of an inflationary epoch remains
elusive.  Measurements of the cosmic microwave background (CMB)
provide the cleanest experimental approach to address this issue, in
particular since primordial gravitational waves, generated during an
inflationary phase in the early Universe, induce a specific signature
in its polarization properties.

The CMB polarization signal can be divided into even and odd parity
$E$- and $B$-modes \citep{Kamionkowski1997}. The former are generated
by both scalar and tensor perturbations, the latter by tensor modes
only.  $E$-mode polarization has been detected at high significance,
as shown in studies of stacked fields centred on temperature hot and
cold spots around which radial and tangential polarization patterns
can be observed \citep{Komatsu2011,Planck_iands_2015}.  However,
primordial $B$-modes, arising only from tensor perturbations which are
intrinsically weaker than the $E$-mode generating scalar
perturbations, are yet to be discovered.

We quantify constraints on $B$-modes in terms of the ratio, $r$, of
the tensor fluctuations (gravitational waves) to scalar (density) fluctuations, evaluated at
a given spatial wavenumber. The $B$-mode power spectrum has a peak at
the horizon scale at recombination ($\ell \sim 90$) with an amplitude
proportional to this value. Reionisation then introduces an additional
peak at low-$\ell$ ($\ell \sim 10$) with an amplitude that depends on
the optical depth of the Universe, $\tau$. Recent results from
\Planck\ \citep{Planck_lowl_2016} have determined a value for $\tau$
of $0.055\pm 0.009$, a decrease from
the WMAP9 result of $0.089\pm 0.014$ \citep{Hinshaw2013} and the
previous \Planck\ result of $0.078\pm 0.019$
\citep{Planck_params_2015} obtained when combining a low-$\ell$
likelihood based on the 70\,GHz polarization data and a high-$\ell$
temperature-based likelihood.  This will have some implications for
the possibility of detection of the primordial $B$-modes. Since the primordial $B$-mode power spectrum scales as $r\times\tau^2$ at the reionization scales $\ell \sim 10$, the current $15$\% uncertainty on $\tau=0.055$ translates into a 30\% uncertainty, and a possible shift, on the amplitude of the reionization bump of $B$-modes, and hence on $r$.
Nevertheless, the determination of the $B$-mode power spectrum will
provide a powerful probe of the physics of inflation. 
The current upper limit on the tensor-to-scalar ratio from the BICEP2
and Keck Array experiments \citep{BICEP_Keck2016} is $r \sim 0.07$.

Current measurements of the Galactic foreground emission
\citep{Planck_Dust_EB_2016,Krachmalnicoff2016,Choi2015} imply that
primordial $B$-modes will be sub-dominant relative to foregrounds on all angular scales and over all observational frequencies in the microwave regime.  The
detection of $B$-modes must, therefore, be regarded as a component
separation problem.  This issue has been addressed previously in the
literature in the context of dedicated $B$-mode satellite experiments
\citep{Armitage-Caplan2012,Baccigalupi2004,Betoule2009,Bonaldi2011,Dunkley2009,Katayama2011,Remazeilles2016,Errard2016,Stompor2016,hervias_2017}.
However, some of the conclusions are open to question due, in some
cases, to the assumption of simplified foreground emission properties
or the adoption of a higher $\tau$ value.

In this paper, one of a series of publications dedicated to the
preparation of a future post-\Planck\ CMB space mission, the Cosmic
Origins Explorer 
\citep[\core,][]{ECO_mission}, we focus on evaluating the accuracy with
which \core\ can measure $r$ in the presence of foregrounds. Closely
related papers include the companion papers on inflation
\citep{ECO_inflation} and on cosmological parameters
\citep{ECO_parameters2016}. Other obstacles to the observation of
$B$-modes, due to instrumental noise and parasitic systematic signal
contributions, or the effects of gravitational lensing, are addressed
in detail in \cite{ECO_systematics} and \cite{ECO_lensing},
respectively.  Nevertheless, we do consider relevant issues related to
gravitational lensing, which mixes the $E$ and $B$ polarization modes
and creates a lensed $B$-mode spectrum peaking at $\ell \approx 1000$,
in the context of determining $r$. Additionally, given the utilisation
of the polarization $E$-modes for delensing purposes, we also briefly
present the corresponding spectra, and infer the quality of the
$E$-mode reconstruction by fitting the $\tau$ parameter.

Component separation will be the most critical step for measuring the
primordial CMB $B$-mode signal at a level of $r\sim 10^{-3}$. 
A common approach to estimating the constraints on $r$ that a
given experiment might achieve is via the Fisher forecasting formalism. However, the
predicted uncertainties are usually optimistic, and do not capture
potential biases in the recovery of the tensor-to-scalar ratio. 
Therefore we directly perform full component separation analysis on
simulated \core\ sky maps for several values of $r$, and include
challenging simulations of foreground emission including contributions
from synchrotron, dust, anomalous microwave emission (AME) and radio
and infrared point sources. We then adopt an approach close in spirit
to the analysis of actual real-world data. Specifically, various
component separation approaches are applied to the simulated data, and
Galactic diffuse and point source masks are inferred directly from the
analysis, before evaluating $r$ via a likelihood method. 
This paper can be regarded as a follow-up to the comprehensive
tretament in \cite{Leach2008} of component separation issues for
intensity observations, but focussed instead on polarization.  

The paper is organised as follows.  In Sect.~\ref{sec:foreground_sky},
we provide an overview of the important foregrounds that must be
addressed when searching for primordial $B$-modes. In
Sect.~\ref{sec:simulated_sky}, we produce and describe challenging sky
simulations for \core. In Sect.~\ref{sec:methods}, we perform a full
component separation analysis for \core, as it would be for real data
analysis, on the sky simulations: point-source detection and
pre-processing; component separation with parametric, blind and
semi-blind methods; and likelihood estimation of the tensor-to-scalar
ratio. Note that details of the component separation methods are given
in Appendix~\ref{sec:methods_appendix}.  Then, in
Sect.~\ref{sec:compsep_results}, we present a hybrid likelihood
analysis of simulations with $r=10^{-3}$, explicitly combining results
from multiple component separation approaches.
Section~\ref{sec:discussion} highlights several important issues to be
addressed in order to improve component separation approaches to
future data sets.  We conclude in Sect.~\ref{sec:conclusions},
including a comparison to a forecasting approach for $r$ described in
more detail in Appendix~\ref{sec:fisher}.

\section{Complexity of foregrounds}
\label{sec:foreground_sky}

It has by now been established that the primordial $B$-mode CMB signal
cannot be measured without correction for foreground emission. Here we
provide a synthesis of the current understanding about the nature of
such foregrounds. Since lensing induced $B$-modes are an effective
foreground to the primordial $B$-mode signal, we also include a brief
overview of their nature.

\subsection{Diffuse Galactic emission}
Our picture of the Galactic emission components in the microwave
frequency range largely originates in the WMAP and \Planck\
observations of the microwave sky from 23 to 857\,GHz. The total
intensity sky maps are consistent with an overall picture of the
Galactic foreground that comprises four components \citep[for a recent
review, see e.g.][]{Delabrouille2007}: synchrotron emission from
relativistic cosmic ray electrons, free-free (thermal bremstrahlung)
emission in the diffuse ionised medium, thermal (vibrational) emission
from dust heated by the interstellar radiation field, and finally an
anomalous microwave emission (AME) component strongly correlated
spatially with the thermal dust emission but that exhibits a rising
spectrum towards lower frequencies. The latter has been associated
with rotational modes of excitation of small dust grains (so-called
'spinning dust'). A time-variable contribution on large angular scales
from interplanetary dust (zodiacal light emission) has also been
detected by \Planck\ \citep{Planck_zodi}, which may lead to
systematic leakage from temperature to polarization at a level
that might be non-negligible for very sensitive $B$-mode experiment,
depending on the specifics of the scanning strategy.

In contrast to the situation for intensity where the foreground
emission dominates over only 20\% of the sky, the polarized flux at
20\,GHz exceeds the level of CMB polarization over the full sky, and
reveals the presence of large coherent emission features.  Analysis of
the WMAP and \Planck\ data has demonstrated that the polarized
Galactic emission is well-described by a simple two component model of
the interstellar medium comprising synchrotron radiation and thermal
dust emission. However, this picture is likely to become more complex
as the sky is measured with increasing accuracy.

Synchrotron emission is produced by cosmic-ray electrons spiralling in
the Galactic magnetic field. The measured synchrotron emission is
dependent on the density of the relativistic electrons along a given
line-of-sight, and approximately to the square of the plane-of-sky
magnetic field component, and can be strongly polarized in the
direction perpendicular to the Galactic magnetic field. It constitutes
by far the most important component of the polarized foreground at low
frequencies ($< 50$\,GHz). In detail, the observed polarized emission
is seen to arise mainly in a narrow Galactic plane and well-defined
filamentary structures -- the loops and spurs well-known in total
intensity measurements -- that can extend over 100 degrees across the
sky and be polarized at a level of $\sim$40\%
\citep[see][]{Vidal2015}. However, away from these features, the
polarization fraction remains relatively low, corresponding to values
of less than $\sim$15\% at high latitudes.

The synchrotron spectral energy distribution (SED) is typically
modelled as a power law, often with some form of spectral curvature
that is relevant for observations at microwave frequencies
\citep{Kogut2007}. There is no precise determination of the spatial
variation of the synchrotron spectral index, either in intensity or
polarization. In the former case, this is, in part, due to the
difficulty of separating the emission from free-free, AME and CMB,
which may dominate the integrated emission in the 20$-$100\,GHz range.
It is also a consequence of the fact that the fidelity of current
low-frequency data \citep[e.g.][]{Haslam1982,Reich2001} is not as good
as for CMB data in general, although new measurements from experiments
such as C-BASS \citep{Irfan2015} should improve this situation.  This
uncertainty will have obvious implications for our attempts to model
the diffuse emission in this paper. A further important observation is
that the polarized synchrotron and dust contributions are
spatially correlated \citep{Page2007, Planck_PIP_XXII} on large
angular scales.

Polarized dust emission results from non-spherical grains that adopt a
preferential orientation with the Galactic magnetic field and then
emit thermal radiation along their longest axis. This will be
perpendicular to the Galactic magnetic field and so the observed
thermal dust emission is polarized in the same direction as the
synchrotron emission. Studies of the Planck data have yielded a wealth
of new knowledge about the nature of this emission over the entire
sky.

\citet{Planck_PIP_XIX} have shown that over a very small fraction of
the sky the polarization fraction may reach $\sim$25\%, although
values are typically smaller ($\sim$12\% at high latitudes) yet
strongly variable and scale-dependent.  This can be interpreted in
terms of the structure of the turbulent part of the line-of-sight
magnetic field and an associated depolarization effect.

There is no single theoretical emission law for dust, which is
composed of many different populations of particles of matter.
However, on average, an SED can be fitted to the observational data,
generally in the form of a modified blackbody spectrum.
\citet{Planck_PIP_XXII} have determined the mean SED of dust emission
in both intensity and polarization from WMAP and \Planck\ data that is
spatially correlated with the \Planck\ 353~GHz emission. This is well
fitted by a mean dust temperature of 19.6\,K, and an opacity spectral index of
$1.59\pm 0.02$ for polarization, slightly lower than that measured for
total intensity. This modest evidence for a different frequency
dependence in intensity and polarization may be connected to the
variation of alignment efficiencies for various types of dust grains.
It should also be noted that \citet{Meisner2015, FDS1999} have
demonstrated that a two-component dust model, with independent
spectral indices and dust temperatures for the cold and hot
components, provides a marginally better fit in intensity when
combining the \Planck\ and DIRBE data.

There is no precise estimate to date of the spatial variation of the
polarized dust spectral index. However, \citet{Planck_PIP_L} provides
evidence for significant variations of the dust polarization SED at
high Galactic latitude, larger than those measured for dust intensity.
Moreover, \cite{Planck_PIP_L} further demonstrates that the polarized
dust emission may decorrelate across frequencies, because of the
co-addition of different dust component spectra along the
line-of-sight, in which case the use of  fixed spectral indices
across frequencies might be inadequate. 
Such variations can lead to an erroneous detection of primordial
$B$-modes if not properly taken into account in any component
separation analysis.

Although the synchrotron and thermal dust emission are clearly the
dominant contributors to the diffuse polarized Galactic foreground
emission, uncertainties in the current data may still allow other
components, more evident in intensity measurements, to contribute at
fainter levels.

If AME is solely due to spinning dust particles \citep{Draine1998},
then we expect it to have a very low polarization percentage,
$\lesssim$1\%, with a specific level determined by the alignment
efficiency of small grains in the interstellar magnetic field, and a
polarization fraction that decreases with increasing frequency.
Recent theoretical work \citep{Draine2016} predicts that dissipative
processes suppress the alignment of small grains contributing to the
AME through rotational emission, such that negligible polarization
($\sim 10^{-6}$) will be observed at frequencies above 10~GHz,
although such low levels still need to be confirmed empirically.
However, AME might arise from other physical mechanisms. For example,
\citet{Draine2013, Hoang2016} have suggested that part of the observed
AME emission may be due to magnetic dipolar emission, which would lead
to a contribution with polarization perpendicular to that of thermal
dust.  
Recent measurements place upper limits on the AME polarization at the few per cent level in individual clouds \citep{Dickinson2011, Lopez-Caraballo2011, Rubino-Martin2012}. Moreover, \cite{Planck2015_XXV_lowf} have determined a 2$\sigma$ upper limit of 1.6\% in the Perseus region, although setting such limits in other areas of the sky was hindered by significant synchrotron contamination. More recently, \cite{Genova-santos2017} obtained the most stringent upper limit on AME polarization of $< 0.22\%$ from the W43 molecular complex. \cite{Macellari2011} have analysed the AME polarization in diffuse regions of the sky, obtaining an upper limit of 5\% for the diffuse AME polarization. 
Although it appears to be low, it is difficult to infer the true intrinsic AME polarization due to various potential depolarization effects such as averaging of polarization along the line-of-sight or within the telescope beam.
Nevertheless, while the level at which the AME is polarized appears to be low, a failure to account for it could bias the measurement of $r$ in $B$-mode searches \citep{Remazeilles2016}.

Conversely, the contribution from free-free emission at high latitudes
and away from bright \hii\ regions is negligible in polarization \citep{Macellari2011},
as expected theoretically from the randomness of Coulomb interactions in \hii\ regions.

\begin{figure}[htbp]
\centering
\includegraphics[width=0.5\textwidth]{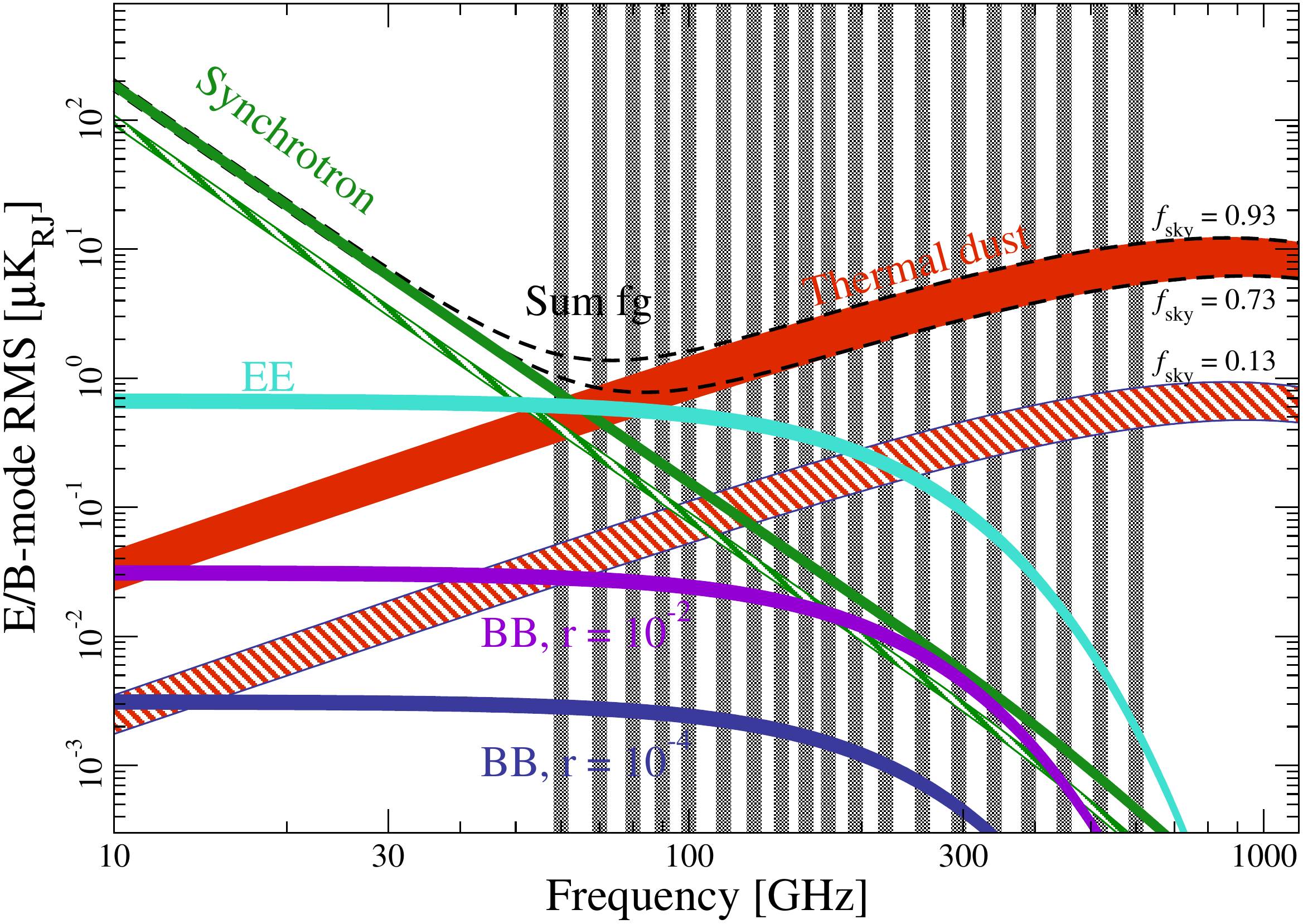}
\caption{Brightness temperature spectra of diffuse polarized
  foregrounds, based on \cite{Planck2015_X} computed on $40'$ angular
  scales, compared to the $E$- and $B$-mode CMB polarization
  spectra. Even for the quietest regions constituting $\sim 10$\,\% of
  the sky, the polarized foreground emission (the green and red lines
  for synchroton and dust respectively) dominate the primordial CMB
  $B$-mode signal (indicated by the purple line for $r=10^{-2}$ and
  the blue line for $r=10^{-4}$) by a few orders of magnitude over the
  entire frequency range covered by \core\ (denoted by grey vertical
  bands).}
  \label{Fig:sed}
\end{figure}

In summary, diffuse Galactic foregrounds are potentially less complex
for polarization studies because only a subset of the Galactic
foreground emissions seen in intensity are significantly
polarized. However, the component separation problem for $B$-mode
polarization is more challenging because the CMB $B$-mode signal is
itself intrinsically weak, especially if $r\sim10^{-3}$ and
$\tau=0.055$, compared to the foreground minimum at $\sim 70$\,GHz
(Fig.~\ref{Fig:sed}), so that a more precise understanding of the
polarized foreground properties is required. It is worth noting that,
contrary to expectations, an experiment with restricted low frequency
coverage, e.g., without detector bands $< 150$\,GHz, still cannot
avoid synchrotron contamination to $B$-modes: Fig.~\ref{Fig:sed}
demonstrates that at frequencies in excess of 200\,GHz the synchrotron
foreground has a similar spectral shape and amplitude to the
primordial CMB $B$-mode signal for $r\lesssim 10^{-2}$. This may
prevent multi-frequency component separation methods from
disentangling the CMB and synchrotron $B$-modes in the absence of
low-frequency observations acting as lever arms. Therefore, wide
frequency coverage is essential to allow the accurate measurement of
the primordial CMB $B$-modes, as will be provided by a CMB $B$-mode
satellite mission like \core.

\subsection{Point sources}
\label{subsec:xtragal}

The two significant contributors at mm and sub-mm wavelengths are
radio sources and dusty star-forming galaxies. Our understanding of
both populations in the \core\ spectral range has greatly improved in
recent years, primarily thanks to \textit{Planck}'s all sky surveys
\cite{PCCS2_2015} and to the much deeper surveys over limited sky
areas carried out by the \textit{Herschel} satellite and by
ground-based facilities such as the South Pole Telescope
\cite[SPT;][]{Mocanu2013}, the Atacama Cosmology Telescope
\cite[ACT;][]{Marsden2014} and SCUBA-2 on the James Clerk Maxwell
Telescope \cite[JCMT;][]{Geach2017}.

The dominant radio source populations in the \core\ frequency range are
the compact flat- and inverted- spectrum ones, primarily blazars (BL
Lac objects and flat-spectrum radio quasars). Observations with the
Australia Telescope Compact Array (ATCA) in the frequency range
between 4.5 and 40\,GHz of 3 complete samples of such sources (for a
total of 464 objects), carried out almost simultaneously with the
first two \textit{Planck} surveys, have shown that the spectra of most
objects steepen above $\simeq 30\,$GHz, consistent with synchrotron
emission becoming optically thin \cite{Massardi2016}. The median
high-frequency ($\nu \ge 70\,$GHz) slope was found to be in the range
$0.6 \simlt \alpha \simlt 0.7$ ($S_\nu \propto \nu^{-\alpha}$).
However, individual sources show a broad variety of spectral shapes:
flat, steep, upturning, peaked, inverted, downturning; \cite[see
also][]{PlanckCollaborationXIVextreme_radio_sour2011,
  PlanckCollaboration2011radioSED,
  PlanckCollaboration2016radio_spectra}. This complexity greatly
complicates the removal of the point source contamination from CMB
maps.

Extended, steep-spectrum radio sources are minor contributors at mm
and sub-mm wavelengths. Nevertheless WMAP and \textit{Planck} surveys
have detected a few tens of these sources \citep{LopezCaniego2007,
  Massardi2009, Gold2011, PlanckCollaborationXIIIstat_prop2011,
  PCCS2_2015}. A small fraction of them were resolved by
\textit{Planck}, in spite of its large beam and will also be resolved
by \core\, complicating their removal from the CMB maps.

The local population of dusty star-forming galaxies was characterized
by the InfraRed Astronomy Satellite \cite[IRAS;][]{Neugebauer1984}.
IRAS detected, in addition to relatively quiescent galaxies like the
Milky Way, Luminous InfraRed Galaxies (LIRGs) with star-formation
rates of tens to hundred $M_\odot$/yr and infrared luminosities in the
range with $10^{11}\,L_\odot < L_{\rm IR} < 10^{12}\,L_\odot$ and
UltraLuminous Infrared Galaxies (ULIRGs) with $L_{\rm IR} >
10^{12}\,L_\odot$ and up to $\ge 10^{13}\,L_\odot$, and star-formation
rates of up to thousands $M_\odot$/yr.
The dust emission of these galaxies is reasonably well described by a
grey-body spectrum, which peaks at rest-frame wavelengths $\sim
100\,\mu$m. At mm and sub-mm wavelengths such a spectrum is approximated
by $S(\nu)\propto \nu^{2+\beta}$ where $\beta$ is the dust emissivity
index, which typically takes values in the range $1.5\le \beta\le 2$.

As mentioned above, the \textit{average} frequency spectra of the two
populations are widely different: the radio emission declines with
increasing frequency while the dust emission steeply increases. This
makes the crossover frequency between radio and dust emission
components only weakly dependent on their relative intensities.
Moreover, dust temperatures tend to be higher for distant high
luminosity sources, partially compensating for the effect of redshift.
As a consequence there is an abrupt change in the populations of
bright sources above and below $\sim 1\,$mm: radio sources dominate at
longer wavelengths, while in the sub-mm region dusty galaxies take
over.

The wavelength at which the contribution of extragalactic sources is
minimum is therefore shorter than that of minimum Galactic emission
($\simeq 5\,$mm, see Fig.~\ref{Fig:sed}). The power spectra of extragalactic
sources are also very different than those of Galactic foregrounds.
The Galactic dust power spectrum scales approximately as
$C_\ell\propto \ell^{-2.7}$  or $\ell^{-2.8}$ for $\ell>110$
\citep{PlanckCollaborationXXX2014} and the Galactic synchrotron power
spectrum is similarly steep \citep{LaPorta2008}. The point source power
spectrum is much flatter. It is the sum of two components: Poisson
fluctuations with $C_\ell=\hbox{constant}$ and clustering.
However, the contribution of clustering to the angular power spectrum of radio
sources is strongly diluted by the broadness of their luminosity
function which mixes up, at any flux density level, sources
distributed over a broad redshift range. As a consequence, the power
spectrum is dominated by the Poisson term.

On the contrary, the power spectrum of dusty galaxies making up the
Cosmic Infrared Background (CIB) is dominated by clustering for $\ell
\simlt 2000$ \citep{PlanckCollaborationXXX2014,DeZotti2015}, while the
Poisson contribution takes over on smaller scales (higher multipoles).
Although the clustering power spectrum deviates from a simple power
law, a reasonably good approximation is  $C_\ell\propto \ell^{-1.2}$
\citep{PlanckCollaborationXXX2014}. The flatter point source power
spectra compared to diffuse Galactic emissions imply that
extragalactic sources are the main contaminants of CMB maps on small
angular scales. This happens already for $\ell \simgt 200$ for $\nu
\simlt 100\,$GHz, where the dominant population are radio sources and
for $\ell \simgt 1000$--2000 at higher frequencies, where dusty
galaxies dominate.

The most extensive study of the polarization properties of
extragalactic radio sources at high radio frequencies was carried out
in \cite{Massardi2013}. These authors obtained polarization data
for 180 extragalactic sources extracted from the Australia Telescope
20-GHz (AT20G) survey catalogue and observed with the Australia
Telescope Compact Array (ATCA) during a dedicated, high-sensitivity
run ($\sigma_{\rm p}\simeq 1\,$mJy). Complementing their data with
polarization information for seven extended sources from the 9--yr
Wilkinson Microwave Anisotropy Probe (WMAP) co-added maps at 23\,GHz,
they obtained a roughly $99\%$ complete sample of extragalactic
sources brighter than $S_{20\rm GHz} = 500\,$mJy at the selection
epoch. The distribution of polarization degrees was found to be well
described by a log-normal function with mean of $2.14\%$ and
dispersion of $0.90\%$. Higher frequency surveys indicate that the
distribution does not change appreciably, at least up to $\sim
40\,$GHz (cf. \cite{Battye2011} and \cite{Galluzzi2016}). 
The log-normal distribution of the polarization fractions of the radio sources with a mean of $\sim 3$\% has now been confirmed up to $353$\,GHz by \cite{Bonavera2017}.

In the case of star-forming galaxies, the polarized emission in the
\core\ frequency range is dominated by dust at wavelengths $\simlt
3\,$mm. At longer wavelengths the synchrotron emission takes over; but
at these wavelengths the extragalactic sky is dominated by radio sources, also in polarization. 

Polarization properties of dusty galaxies as a whole at (sub-)mm
wavelengths are almost completely unexplored. The only available
information has come from  SCUPOL, the polarimeter for SCUBA on the
James Clerk Maxwell Telescope, that has provided polarization
measurements at $850\,\mu$m for only two galaxies, M\,82
\citep{GreavesHolland2002} and M\,87 \citep{Matthews2009}. However the
global polarization degree has been published only for M\,82 and is
$\Pi = 0.4\%$. Integrating the \textit{Planck} dust polarization maps
over a $20^\circ$ wide band centred on the Galactic plane, 
\cite{DeZotti2016} found an average value of the Stokes $Q$
parameter of about 2.7\%. We may then expect a similar value for
spiral galaxies seen edge-on. For a galaxy seen with an inclination
angle $\theta$ the polarization degree is reduced by a factor
$\cos(\theta)$. If all galaxies are about as polarized as ours, the
average polarization fraction for unresolved galaxies, averaged over
all possible orientations, should be about half of 2.7\%, i.e.~around
1.4\%.

\subsection{Lensing $B$-modes}

Large-scale structures induce gravitational lensing in the CMB which
mixes the $E$ and $B$ polarization modes
\citep{Blanchard1987,Cole1989,Bernardeau1997,Zaldarriaga1998,2001Benabed,Challinor2005}. The
lensing $B$-mode power spectrum approximates that of white noise on
large angular scales, peaks at $\ell \sim 1000$, and for
$r \sim 0.07$, the current upper limit from the BICEP2 and Keck Array
experiments \citep{BICEP_Keck2016}, its amplitude is always larger
than the primordial signal for scales smaller than the reionisation
bump. Such a signal therefore acts as an effective foreground in the
search for primordial $B$-modes. The recovery of the $B$-mode
polarization may be attempted through a process called \lq
delensing'. This requires an unlensed estimate of the $E$-mode signal
and of the lensing potential \citep{Hu2002,Hirata2003}. The latter can
be derived from the CMB itself \citep{Carron2017}, or from alternative
measures of large-scale structure, e.g., the CIB \citep{Simard2015,
  Sherwin2015}. In this context, it is worth noting that
\citet{Larsen2016} have recently provided the first demonstration on
\Planck\ temperature data of CIB delensing, supporting its utility for
lensing removal from high precision $B$-mode measurements.

In a companion paper \citep[][in prep.]{ECO_lensing} it is shown that, for a
\core-like experiment, 60\% of the lensing effect will be removed. We
describe in a later section our approach to this signal and its
treatment, and assess its impact on component separation and derived
results on $r$.


\section{Sky simulations}
\label{sec:simulated_sky}

We produce detailed simulations of the polarized emission of the sky
by using a modified version of the publicly released Planck Sky Model
\citep[{\sc PSM} version 1.7.8,][]{Delabrouille2013}.\footnote{In our
  modified version of the {\sc PSM} version 1.7.8, we have added the
  options to generate polarization for spinning dust and polarization
  for thermal dust with a modified blackbody spectrum as parametrized
  by the \Planck\ {\tt GNILC} dust model
  \citep{Planck_PIP_XLVIII}. These additional models will be included
  in future releases of the {\sc PSM}.}  The simulation is more
challenging than has generally been considered in the literature to
date.

The sky simulation consists of: (i) CMB $E$- and $B$-mode polarization
with a low optical depth to reionization, $\tau=0.055$, and
tensor-to-scalar ratios spanning the range $r=10^{-2}$ down to $10^{-3}$,
including or not gravitational lensing effects;
(ii) polarized synchrotron radiation with a power-law spectrum and
variable spectral index over the sky; (iii) polarized thermal dust
radiation with a modified blackbody spectrum and variable spectral
index and temperature over the sky; (iv) polarized anomalous microwave
emission (AME); and (v) infrared and radio polarized
point-sources. 
The main characteristics of these components are summarized in
Table~\ref{tab:sky}. We analyse a set of 5 simulations, spanning
different values of the tensor-to-scalar ratio and different amounts
of gravitational lensing effects. The specific content of each
simulation is given in Table~\ref{tab:sims}. 
All of the simulated maps are provided in {\tt
  HEALPix} format \citep{Gorski2005},\footnote{\url{http://healpix.sourceforge.net}}
 with a pixel size defined by the $N_{\rm side}$ parameter, here set
 to a value of $2048$. A lower resolution set of simulations generated
 directly at  $N_{\rm side} = 16$ are also provided for the {\tt Commander}
 analysis (see Sects.~\ref{subsec:parametric}  and \ref{subsec:degrade}).

\begin{table}[htbp]
\small
  \centering
  \begin{tabular}{lll}
\hline
Component & Emission law & Template \\
\hline
\hline
CMB  & Blackbody derivative & $r=10^{-2}$ (simulation \#1)\\
  &  & $r=5\times 10^{-3}$ (simulation \#2)\\
 &  & $r=10^{-3}$ (simulation \#3)\\
  &  & $r=10^{-3}$, with lensing (simulation \#4)\\
 &  & $r=10^{-3}$, with 40\% lensing (simulation \#5)\\
\hline
Synchrotron & Power-law $\nu~^{\beta_s}$  & WMAP $23$\,GHz polarization maps \\
& Non-uniform $\langle\beta_s\rangle=-3$ & \citep{Miville-Deschenes2008} \\
\hline
Thermal dust & Modified blackbody & \Planck\ {\tt GNILC} $353$\,GHz map\\
& $\nu~^{\beta_d}B_\nu(T_d)$  & \citep{Planck_PIP_XLVIII}  \\
& Non-uniform $\langle\beta_d\rangle=1.6$ & $[Q_\nu,U_\nu]=f_dg_dI_\nu^{\tt GNILC}[\cos\left(2\gamma_d\right),\sin\left(2\gamma_d\right)]$  \\
&Non-uniform $\langle T_d\rangle=19.4$\,K & $f_d =$ 15\%, $\langle f_dg_d \rangle =$ 5\% \\
& & $g_d$ and $\gamma_d$ coherent with synchrotron polarization\\
\hline
AME & Cold Neutral Medium  & Thermal dust map rescaled by 0.91 K/K\\
& & at $23$\, GHz. Same polarization angles as \\
& & thermal dust. Uniform 1\% polarization fraction.\\
\hline
Point-sources & Four power-laws   & Radio source surveys at $4.85$, $1.4$, $0.843$\,GHz  \\
& & 2.7\% to 4.8\%  mean polarization fraction \\
& & \\
              & Modified blackbodies+free-free &  IRAS ultra-compact \hii\ regions \\
& & \\
              &  Modified blackbodies   & IRAS infrared sources \\
& &  1\% mean polarization fraction \\
\hline 
  \end{tabular}
 \caption{Summary of simulated sky components.}
  \label{tab:sky}
\end{table}

\begin{table}[htbp]
\small
  \centering
  \begin{tabular}{llccccc}
\hline
  & CMB & dust & synchrotron & AME & sources & lensing \\
\hline
\hline
Simulation \#1 & $r=10^{-2}$ & $\checkmark$ & $\checkmark$ & - & - & - \\
Simulation \#2 & $r=5 \times 10^{-3}$ & $\checkmark$ & $\checkmark$ &- &- &- \\
Simulation \#2-bis & $r=2.5 \times 10^{-3}$ & $\checkmark$ & $\checkmark$ &- &- &- \\
Simulation \#3 & $r=10^{-3}$ & $\checkmark$ & $\checkmark$ &- &- &- \\
Simulation \#4 & $r=10^{-3}$ & $\checkmark$ & $\checkmark$ & $\checkmark$ & $\checkmark$ & $\checkmark$ \\
Simulation \#5 & $r=10^{-3}$ & $\checkmark$ & $\checkmark$ & $\checkmark$ & $\checkmark$ &  (40\%)\\
\hline 
  \end{tabular}
 \caption{Set of simulations. Checkmarks indicate which components are included.}
  \label{tab:sims}
\end{table}

\subsection{CMB}

By using the Boltzmann solver {\tt CAMB} \citep{Lewis2000}, we
generate both lensed and unlensed $E$- and $B$-mode CMB angular power
spectra from a $\Lambda$CDM+$r$ cosmology,  an optical depth to
reionization, $\tau=0.055$, motivated by the latest \Planck\ results
\citep{Planck_lowl_2016} and a tensor-to-scalar ratio varying from
$r=10^{-2}$ down to $r=10^{-3}$, which is the ambitious detection goal for
the \core\ space mission. The other $\Lambda$CDM cosmological
parameters are set to the \Planck\ best-fit values from
\cite{planck2015_overview}. The CMB $B$-mode angular power spectrum,
$C_\ell^{BB}$, generated by {\tt CAMB} therefore is the combination of
a pure tensor power spectrum and a lens-induced power spectrum:
\bea
C_\ell^{BB} = \,C_\ell^{tensor}(r\,,\,\tau=0.055)\, +A_{lens}\,
C_\ell^{lensing}, 
\eea 
where $r$ is either set to $10^{-2}$, $5\times 10^{-3}$, $2.5\times 10^{-3}$, or $10^{-3}$, and
$A_{lens}$ is either set to 0 (unlensed), 1 (no delensing), or 0.4
(60\% delensing). The delensed case is idealized -- when analyzing
observations of the real sky, delensing would be applied
post-component separation \citep{Carron2017}, but such a treatment is beyond the scope of this work. Instead, we assume a scale-independent delensing
efficiency, and simply rescale the full lensing contribution. The
application of component separation methodologies to such sky
realizations does not affect the effectivness of the foreground
removal, and allows the generation of signal covariance matrices for the
likelihood analysis described later in Sect.~\ref{subsec:ilc}.

Gaussian random realizations of the CMB Stokes $Q$ and $U$
polarization components can then be simulated with the appropriate
$E$- and $B$-mode power spectra using the \healpix\ routine {\tt
  synfast}. 
The corresponding $E$- and $B$-mode maps are then generated from the 
spherical harmonic transforms of the $Q$ and $U$ components computed
by the \healpix\ {\tt anafast} routine, where the resulting $a_{\ell
  m}^E$, and $a_{\ell m}^B$ (pseudo)scalar coefficients are transformed
to full-sky maps using {\tt synfast}.
In this paper, we consider only a single CMB realization
per simulation. 

The lensed CMB $Q$ polarization map with $r=10^{-3}$, smoothed to two
degree resolution (FWHM) for illustrative purposes, is shown in the top left
panel of Fig.~\ref{Fig:sky}.
The CMB polarization $Q$ and $U$ maps are scaled across the \core\
frequency channels through the derivative of a blackbody spectrum that
is achromatic in thermodynamic temperature units. 
The component of
interest, i.e. the primordial CMB $B$-mode polarization map, is shown for
$r=10^{-3}$ and $\tau=0.055$ in the top left panel of
Fig.~\ref{Fig:sky-bmode}, while the lensed CMB $B$-mode polarization
map is shown in the top right panel of Fig.~\ref{Fig:sky-bmode}. 
Note that gravitational lensing effects add significant small-scale noise to
the anisotropies of the primordial CMB $B$-mode polarization, for
which the bulk of the cosmological signal is on the degree scale and larger.

\subsection{Synchrotron}

The Galactic synchrotron radiation is simulated by extrapolating the
\emph{WMAP} 23\,GHz polarization maps, $Q_{23\, GHz}$ and $U_{23\,
  GHz}$, to \core\ frequencies through a power-law frequency dependence
\bea
Q^{sync}_\nu &=& Q_{23\, GHz}\left({\nu\over 23\,{\rm GHz}}\right)^{\beta_s}, \cr
U^{sync}_\nu &=& U_{23\, GHz}\left({\nu\over 23\,{\rm GHz}}\right)^{\beta_s},
\eea
with an average spectral index, $\langle\beta_s\rangle = -3$, and including spatial
variations over the sky. The variable spectral index map was
estimated by fitting a power-law to the Haslam et al. 408\,MHz map and a
\emph{WMAP} 23\,GHz synchrotron map derived using polarization data \citep{Miville-Deschenes2008}. 
The Stokes $Q$ map of the synchrotron polarization
component at $23$\,GHz and the synchrotron spectral index map are shown 
in Fig.~\ref{Fig:sky}. The synchrotron $B$-mode map at $60$\,GHz is
shown in the bottom left panel of Fig.~\ref{Fig:sky-bmode}.

To date there is still no concensus in the literature as to an optimal estimate of the synchrotron spectral indices \citep[see, e.g.,][to justify the choice
of templates]{Dickinson2009}.  The characterization of the synchrotron
spectral indices is problematic due to three main reasons: the
difficulty in separating synchrotron from free-free emission (and AME
at higher frequencies) in intensity; uncertainties in modelling the
spectral shape, which is not well described by a single power-law over
the wide frequency range considered; and the quality of the
low-frequency 408\,MHz data arising from significant variations of the calibration with angular scale, that can result in artificial
variations in inferred spectral index maps. 

For our simulations we elected to use the spectral index map estimated in each pixel by \cite{Miville-Deschenes2008} from the \emph{WMAP} 23\,GHz polarization map and the Haslam et al. 408\,MHz intensity data, as currently implemented in the {\sc PSM} software. This template has a representative mean value of $-3$ which is close to typical values observed in the literature at CMB frequencies \citep{Davies1996,Planck2015_X,Kogut2007,Dickinson2009,Miville-Deschenes2008,Bennett2013}.
The exact choice of the spectral index template is not critical for the simulations, as long as a reliable, physically-motivated SED (here a power-law with a mean spectral index of $-3$) is used to scale the synchrotron emission across frequencies.

\subsection{Thermal dust}

In this study, we focus on the spectral variations in the dust
emission over the sky as the main complexity of the dust foreground,
and postpone the inclusion of other potentially important effects,
such as the frequency decorrelation\footnote{The spectral index of the
  Galactic dust emission varies along a given line-of-sight and with
  frequency due to the emission from multiple components, so that the
  actual spectrum is not a power-law.} noted in
Sect.~\ref{sec:foreground_sky}, for future investigations. We
therefore consider only a single modified blackbody dust component in
the simulations. However, we note that mismodelling the dust emission
in a parametric component separation method, e.g., by parametrising
the emission with a single modified blackbody when two are required to
fit the data accurately, can strongly bias the estimate of the
tensor-to-scalar ratio \citep{Remazeilles2016}.  The accurate
characterization of the spectral properties of dust is essential for
$B$-mode foreground studies and active research is being pursued in
this field.

\begin{figure}[htbp]
\centering
\includegraphics[width=0.5\textwidth]{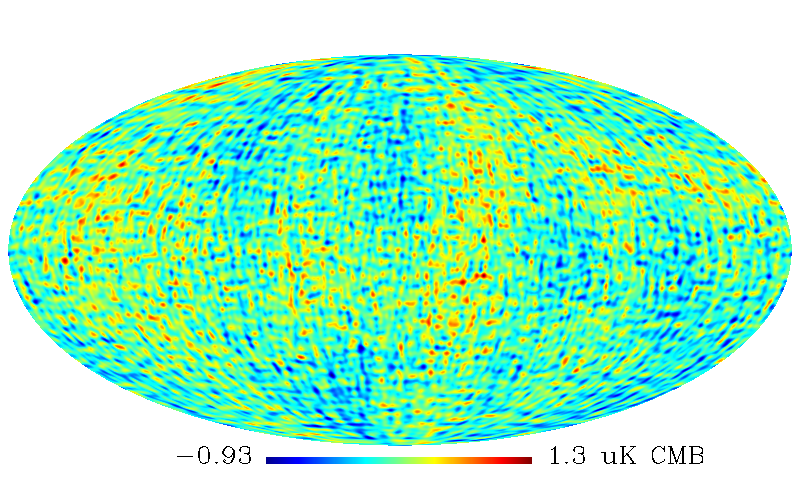}~
\includegraphics[width=0.5\textwidth]{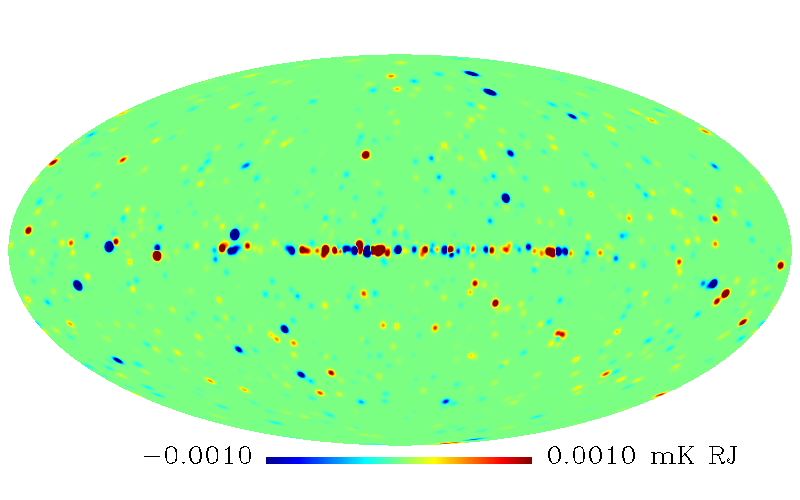} \\
\includegraphics[width=0.5\textwidth]{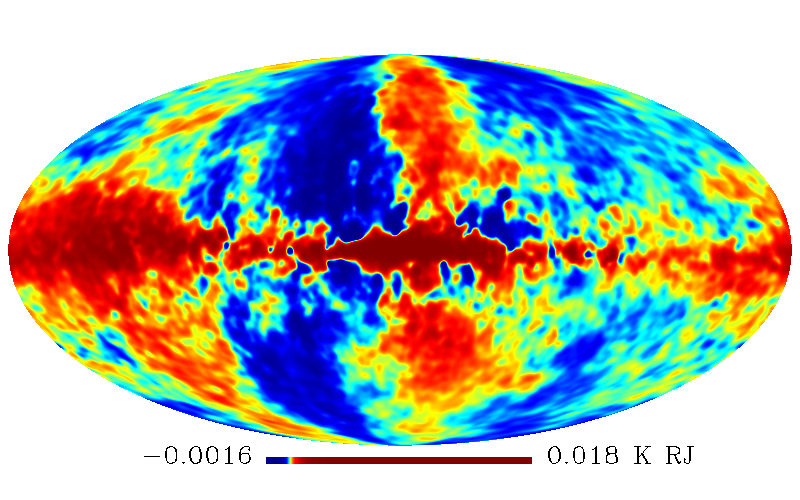}~
\includegraphics[width=0.5\textwidth]{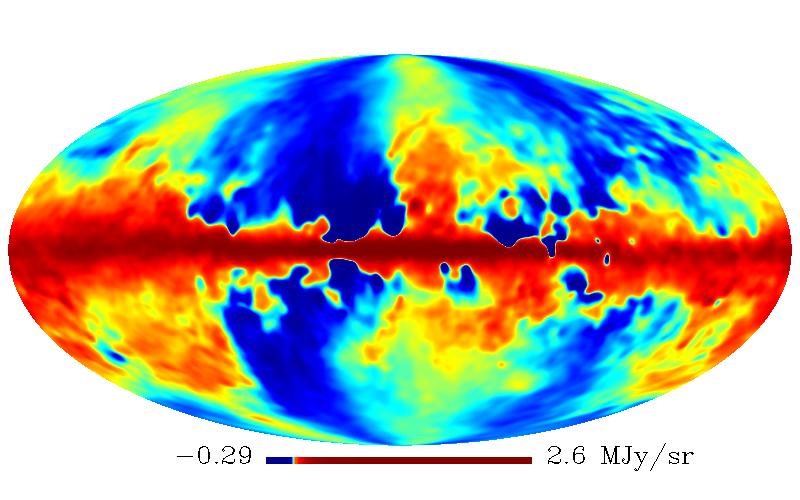}\\
\includegraphics[width=0.5\textwidth]{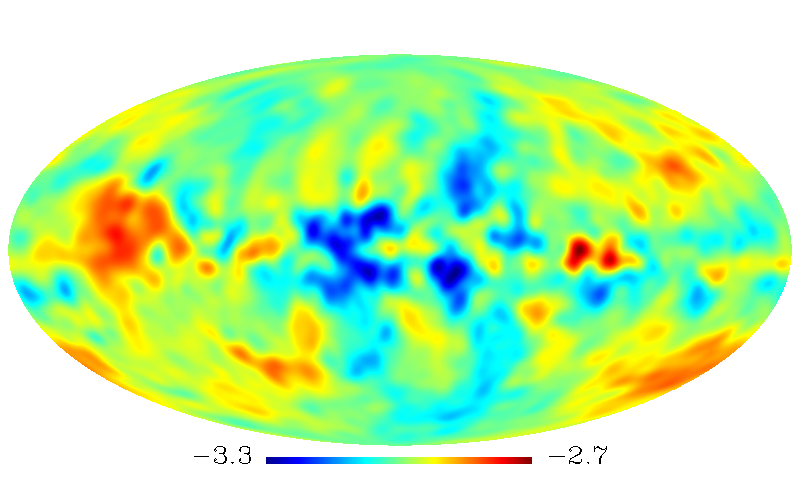}~
\includegraphics[width=0.5\textwidth]{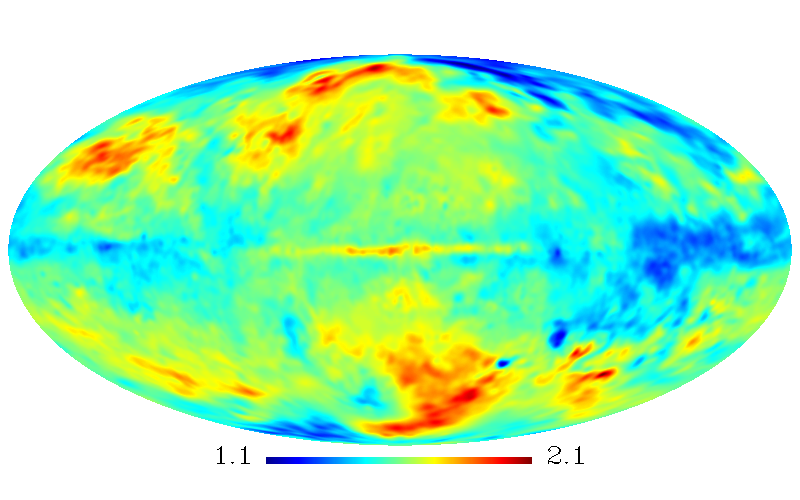}\\
\includegraphics[width=0.5\textwidth]{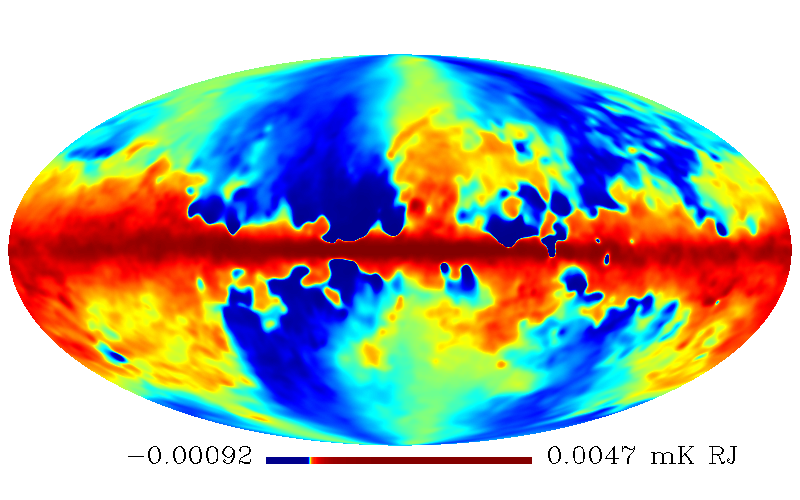}~
\includegraphics[width=0.5\textwidth]{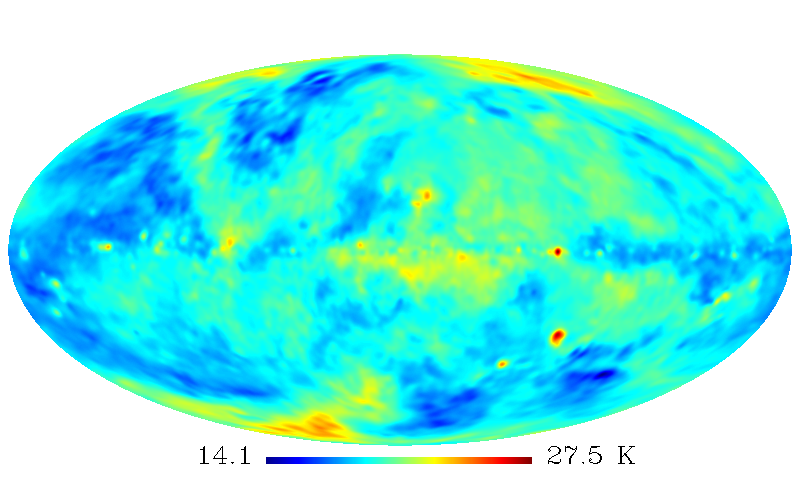}\\
\caption{Simulated sky components (smoothed to $2$ degrees for
  illustrative purposes).  \emph{First row}: lensed CMB $Q$ map with
  $r=10^{-3}$, $\tau=0.055$ (\emph{left}); point-source $Q$ map at
  $60$\,GHz (\emph{right}).  \emph{Second row}: synchrotron $Q$ map at
  $23$\,GHz (\emph{left}); thermal dust $Q$ map at $353$\,GHz
  (\emph{right}).  \emph{Third row}: synchrotron spectral index
  (\emph{left}); dust spectral index (\emph{right}).  \emph{Fourth
    row}: AME $Q$ map at $60$\,GHz (\emph{left}); dust temperature
  (\emph{right}).  Note that the synchrotron, thermal dust and AME $Q$
  maps are shown with histogram-equalized colour scales.}

  \label{Fig:sky}
\end{figure}

\begin{figure}[htbp]
\centering
\includegraphics[width=0.5\textwidth]{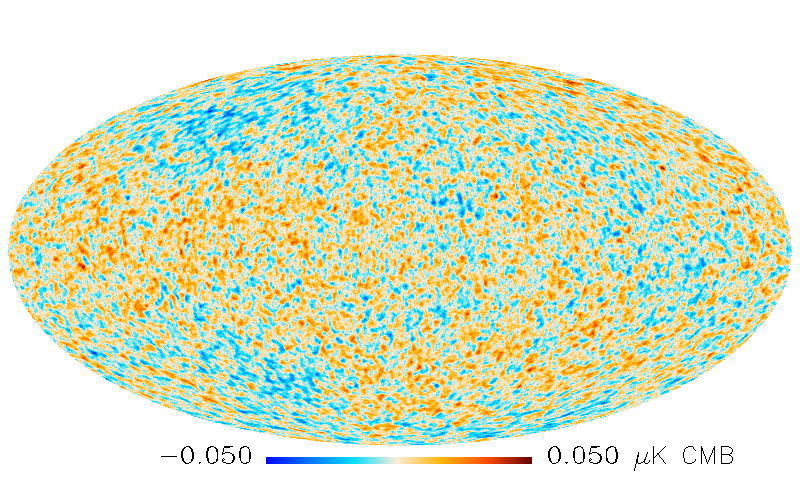}~
\includegraphics[width=0.5\textwidth]{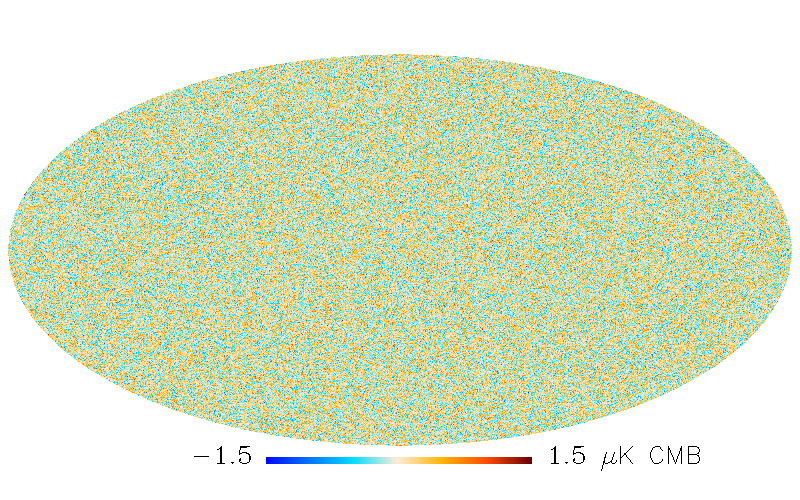} \\
\includegraphics[width=0.5\textwidth]{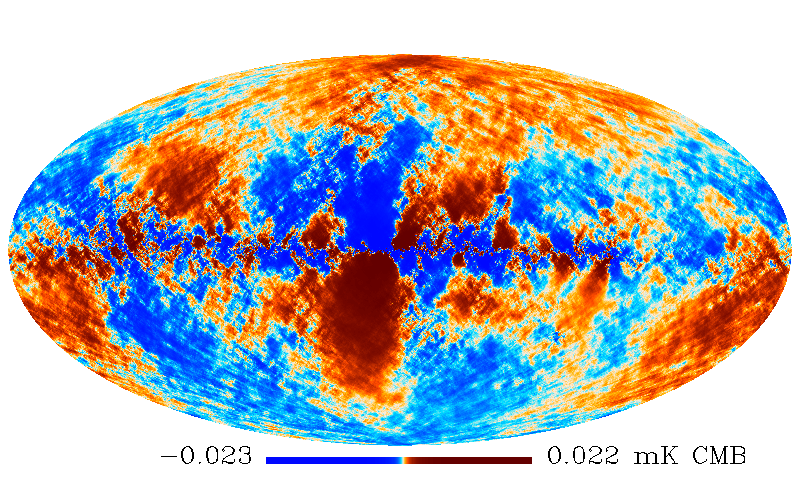}~
\includegraphics[width=0.5\textwidth]{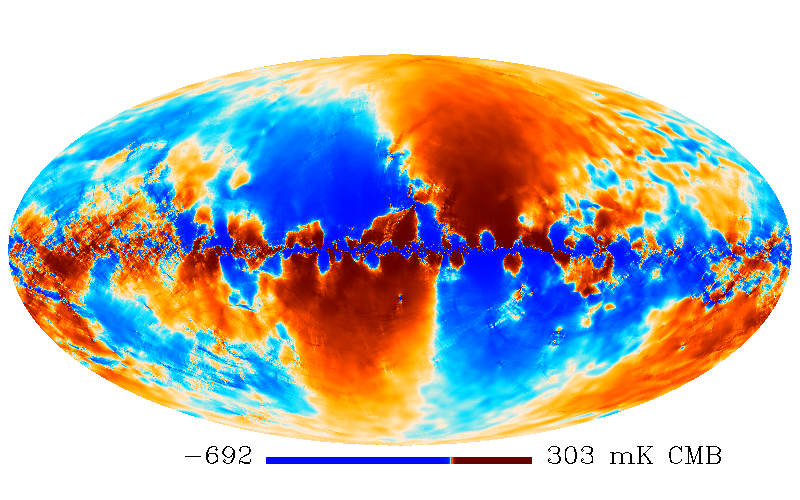}\\
\caption{$B$-mode polarization maps of simulated sky
  components. \emph{Top left}: primordial CMB signal for $r=10^{-3}$
  and $\tau=0.055$. \emph{Top right}: lensed CMB signal. \emph{Bottom
    left}: the Galactic synchrotron contribution at a reference
  frequency of 60\,GHz. \emph{Bottom right}: the Galactic dust
  contribution at 600\,GHz. Note that the synchrotron and dust maps
  are presented in histogram-equalized colour scales.}

  \label{Fig:sky-bmode}
\end{figure}

Here, the polarization maps of the Galactic thermal dust radiation are 
generated from the intensity map of the \Planck\ {\tt GNILC} 2016 dust
model \citep{Planck_PIP_XLVIII}, from which the CIB fluctuations have
been removed.
\bea
Q^{dust}_\nu &=& f_d\,g_d\,I^{\tt GNILC}_\nu\,\cos\left(2\gamma_d\right), \cr
U^{dust}_\nu &=& f_d\,g_d\,I^{\tt GNILC}_\nu\,\sin\left(2\gamma_d\right),
\eea
where the dust intensity map $I^{\tt GNILC}_\nu$ is scaled to the 
\core\ frequencies through a modified blackbody spectrum,
\bea
I^{\tt GNILC}_\nu = \tau_{353} \left({\nu\over 353\,{\rm GHz}}\right)^{\beta_d} B_{\nu}(T_d).
\eea
Here, $\tau_{353}$, $\beta_d$, and $T_d$ are respectively the \Planck\
{\tt GNILC} dust optical depth map at $353$\,GHz, the \Planck\ {\tt
  GNILC} dust emissivity map, and the \Planck\ {\tt GNILC} dust
temperature map that were derived in \cite{Planck_PIP_XLVIII}. The
dust emissivity and temperature are both variable over the sky with
average values, $\langle\beta_d\rangle=1.6$ and
$\langle T_d\rangle=19.4$\,K, respectively. The maps of dust
emissivity and dust temperature are shown in Fig.~\ref{Fig:sky}. 
The dust $B$-mode map at $600$\,GHz is shown in the bottom right panel
of Fig.~\ref{Fig:sky-bmode}.

We make the assumption that the dust polarization angle map,
$\gamma_d$, and the geometric depolarization map, $g_d$, due to the
specific magnetic field configuration \citep{Miville-Deschenes2008},
are coherent with those of the polarized synchrotron model. The dust
polarization fraction $f_d$ is set to 15\,\% on the sky, which, after
modulation with the geometric depolarization factor, gives an overall
polarization fraction fraction of $f_dg_d\sim 5$\,\% on
average\footnote{In fact, the dust polarization fraction adopted in
  the public version of the {\sc PSM} corresponds to about half of the
  currently accepted value. }, with spatial variations over the
sky. The Stokes $Q$ polarization map of thermal dust is shown in the
right panel of the second row of Fig.~\ref{Fig:sky}.

\FloatBarrier

\subsection{AME}

For sensitive CMB experiments, AME, even with a low polarization
fraction, may be a relevant low-frequency foreground to the primordial
CMB $B$-modes, especially for low values of the tensor-to-scalar ratio
\citep{Remazeilles2016}.  Therefore, we include in the sky simulation
an AME component with a uniform $\pi=1$\,\% polarization fraction over
the sky.
\bea
Q^{AME}_\nu &=& \pi\,I^{AME}_\nu\,\cos\left(2\gamma_d\right), \cr
U^{AME}_\nu &=& \pi\,I^{AME}_\nu\,\sin\left(2\gamma_d\right),\cr
I^{AME}_{23\,{\rm GHz}} &=& (0.91\,{\rm K/K}) I^{\tt GNILC}_{353\,{\rm GHz}},  
\eea
where the AME intensity map, $I^{AME}_\nu$, is the \Planck\ thermal
dust intensity map at 353\,GHz, $I^{\tt GNILC}_{353}$
\citep{Planck_PIP_XLVIII}, but rescaled by a factor $0.91$\,K/K at
$23$\,GHz using the correlation coefficient between AME and thermal
dust measured by \cite{Planck2015_XXV_lowf}, and extrapolated to
\core\ frequencies from the $23$\,GHz value by assuming a Cold Neutral
Medium (CNM) for modelling the emission law \citep{Ali-Haimoud2009,Draine1998}.
Because of the correlation between AME and thermal dust, we choose the
AME polarization angles, $\gamma_d$, to be identical to those of
thermal dust. The Stokes $Q$ polarization map for AME at $60$\,GHz is
shown in the bottom left panel of Fig.~\ref{Fig:sky}.

\subsection{Point-sources}

While polarized compact extragalactic sources are a negligible
foreground for CMB $B$-modes on very large angular scales near the
reionization peak ($\ell \lesssim 12$), they are expected to be the dominant foreground for $r=10^{-3}$ once delensing has been applied to the data, from the
recombination peak to smaller angular scales \citep[$\ell > 47$,][]{Curto2013}. Therefore, we include in
the sky simulation both radio and infrared extragalactic sources in
polarization.

Radio sources are taken from radio surveys at $4.85$, $1.4$, and
$0.843$\,GHz \citep{Delabrouille2013}, and extrapolated to \core\
frequencies assuming four kinds of power-laws for assigning the radio
sources to either a steep- or flat-spectrum class. The polarization
degree of the radio sources is randomly selected from the observed
sample of flat and steep radio sources \citep{Ricci2004}, so that the
polarization fraction of the radio sources in our PSM simulation
is 2.7\% on average for flat sources and 4.8\% on average for steep
sources. The $Q$ polarization map of strong radio sources at $60$\,GHz is shown in the top right panel of Fig.~\ref{Fig:sky}.

Strong and faint infrared sources are taken from the IRAS point-source
catalogue \citep{Beichman1988,Moshir1992} and extrapolated to \core\
frequencies by assuming modified blackbody spectra (see
\cite{Delabrouille2013} for more details). The polarization fraction
of the infrared sources is distributed around an average value of 1\%
through a chi-square distribution. We also include ultra compact \hii\
regions extracted from IRAS, which we extrapolate to \core\
frequencies by assuming both modified blackbody spectra and power-law
spectra due to free-free emission.

\subsection{\core\ instrumental specifications}
\label{subsec:coreplus}

The proposed \core\ space mission can observe the polarized sky
emission in 19 frequency bands ranging from $60$ to $600$\,GHz. The
goal is that this wide frequency coverage will provide lever arms that
allow non-trivial foregrounds at low and high frequencies to be
modelled adequately.  The large number of frequency channels is also
essential for component separation when facing multiple degrees of
freedom for foregrounds, e.g., decorrelation effects that may result
from multi-layer dust emission \citep{Planck_PIP_L}, spectral index
curvature, or as-yet undiscovered foregrounds.

The instrumental specifications of the \core\ space mission are
summarized in Table~\ref{tab:specs}.  The optical performance of
\core\ will allow high-resolution observations with FWHM $< 10'$ over
the primary CMB frequency channels and a few arcminute resolution at
high frequencies. The high resolution of the \core\ observations will
play an important role in the correction of the primordial CMB
$B$-modes for gravitational lensing effects.  \core\ will have
unprecedented sensitivity with detector noise levels of order
$\sim 5\,\mu$K.arcmin in polarization observations at the primary CMB
frequencies.

\begin{table}
 \centering
  \begin{tabular}{lll}
\hline
Frequency   & Beam  & $Q$ and $U$ noise RMS \\
$[\rm GHz]$   & $[\rm arcmin]$  & $[\rm \mu K.arcmin]$  \\
\hline \hline
60 & 17.87 & 10.6 \\
70 & 15.39 & 10.0 \\
80 & 13.52 & 9.6 \\
90 & 12.08 & 7.3 \\
100 & 10.92 & 7.1 \\
115 & 9.56 & 7.0 \\
130 & 8.51 & 5.5 \\
145 & 7.68 & 5.1 \\
160 & 7.01 & 5.2 \\
175 & 6.45 & 5.1  \\
195 & 5.84 & 4.9 \\
220 & 5.23 & 5.4 \\
255 & 4.57 & 7.9 \\
295 & 3.99 & 10.5 \\
340 & 3.49 (4.0) & 15.7 \\
390 & 3.06 (4.0) & 31.1 \\
450 & 2.65 (4.0) & 64.9 \\
520 & 2.29 (4.0) & 164.8 \\
600 & 1.98 (4.0) & 506.7 \\
\hline
   \end{tabular}
   \caption{Instrumental specifications for the \core\ mission. For the
     purposes of generating reasonably sized simulations at $N_{\rm side}=2048$ , the
     high-frequency observations ($\geq 340$\,GHz) have been simulated
     at $4'$ beam resolution instead of their native instrumental beam
     resolution.}
  \label{tab:specs}
\end{table}

We convolve the component maps of our sky simulation with Gaussian
beams with the FWHMs given in Table~\ref{tab:specs} for each frequency
channel. Note that we have limited the high-frequency observations
($\geq 340$\,GHz) to $4'$ beam resolution instead of the native
instrumental beam resolution in order to avoid an oversized data
set. This does not impact the results of this study in which we are
interested in detecting CMB $B$-modes on large angular scales. Sky
maps at each frequency are obtained by co-adding the beam-convolved
component maps.

For the purposes of this work it is sufficient to assume that the
noise is Gaussian and white, and uncorrelated between the $Q$ and $U$
Stokes parameters. We also ignore any variation of the noise
properties across the pixeized sky as would be introduced by a
realistic scanning strategy.  We do, however, consider the division of
the data into \lq half-mission' surveys. These are also idealized,
splitting the full-mission data into two equal parts.  We then
simulate two distinct realizations of white noise for the half-mission
$Q$ and $U$ maps at a given frequency by using the noise RMS values
listed in Table~\ref{tab:specs} multiplied by a factor of
$\sqrt{2}$. The two sets of noise realizations are then co-added to
the same sky realisation to generate two half-mission surveys which
have uncorrelated noise properties. The resulting two sets of
observation maps in the $19$ frequency bands are referred to as the
\core\ half-mission 1 (HM1) and half-mission survey 2 (HM2)
surveys. The corresponding full-mission survey (FM) is formed simply
by adding to the sky maps at each frequency the full-mission noise
maps, $n^{FM}$, that are related to the half-mission noise maps,
$n^{HM1}$ and $n^{HM2}$, by: \bea n^{FM} = {n^{HM1} + n^{HM2} \over
  2},\quad \mbox{with }\quad \langle n^{HM1}, n^{HM2}\rangle =0.  \eea

The component separation is performed on the full-mission simulation.
The appropriately cleaned half-mission simulations can be used to
compute the CMB power spectrum free from noise bias via cross-spectral
estimators.

A companion paper \citep{ECO_systematics} considers in detail more
realistic instrumental simulations, addressing topics including the
presence of 1/f noise, asymmetric beams, realistic scanning
strategies, temperature to polarization leakage, and bandpass
mismatch. A more comprehensive study of the \core\ mission and its
capabilities will need to apply component separation methods to such
simulations to assess the impact of these effects on the fidelity of
the component separation. Note that \cite{2010MNRAS.401.1602D} have
previously demonstrated the highly detrimental effect of calibration
errors on various classes of component separation algorithms.


\section{Component separation and likelihood analysis}
\label{sec:methods}

Several component separation approaches have been proposed in the
literature for cleaning foregrounds in CMB temperature and
polarization maps, although the focus has predominantly been on
intensity \citep{Leach2008}. For $B$-mode detection, several blind and
parametric methods have been proposed.

Blind methods
\citep[e.g.,][]{Tegmark1996,2003MNRAS.346.1089D,2004MNRAS.354...55B,delabrouille2009full,Betoule2009,Basak2012,Fernandez2016}
exploit minimal prior information on the foregrounds, therefore they
are not prone to systematic errors due to mis-modelling of the
foreground properties. One of the drawbacks of such approaches is that
foreground error estimation is difficult and must usually rely on
Monte-Carlo simulations or other bootstrapping techniques.

A complementary strategy is provided by parametric methods
\citep[e.g,][]{Eriksen2008,2009MNRAS.392..216S,ricciardi2010,Remazeilles2016},
that explicitly model the foreground properties by means of a set of
parameters that are fitted to the data. Such methods provide an easy
way to characterise and propagate foreground errors, but their
effectiveness depends on the consistency of the foreground model
adopted.

Which of these two strategies will yield the best results with future
experiments ultimately depends on the intrinsic complexity of the true
sky in polarization and our ability to model it. This, and the need
for cross-checks, continues to motivate the community to develop
multiple independent approaches.

In this work, we use the {\tt Commander} \citep{Eriksen2008}, and the
{\tt NILC} \citep{delabrouille2009full,Basak2013}, and {\tt SMICA} methods
\citep{2003MNRAS.346.1089D,Cardoso2008}, as representative of the parametric and
of the blind approaches, respectively. As detailed in
Appendix~\ref{subsec:commander_method} the {\tt Commander} code is
currently limited by computational resources to the analysis of
low-resolution maps and by the precision with which foregrounds can be
modelled. However, it is ideally suited for studies on large angular
scales and specifically the reionization peak ($2 \leq \ell \leq 47$).
Conversely, while {\tt NILC} (Appendix~\ref{subsec:nilc_method}) and
{\tt SMICA} (Appendix~\ref{subsec:smica_method}) can process full
resolution maps, their ability to clean the data from foregrounds at
low $\ell$ are limited by the minimum variance that can be reached
depending on the number of modes and available frequency channels. In
terms of $B$-mode detection, these methods more naturally target the
recombination bump ($\ell \sim 100$).  Since the two types of
approaches are complementary, they can be considered to form
low-multipole and high-multipoles analysis pipelines respectively.

The fidelity of the $B$-mode component separation is assessed by
evaluating the tensor-to-scalar ratio $r$ through an appropriate
cosmological likelihood function.  Since the $E$-mode spectrum is an
important input to delensing methods as applied to the $B$-mode
spectrum, we also quantify its accuracy by fitting the optical depth
to reionization, $\tau$ using a likelihood formalism. However, we fix
all other cosmological parameters to their input values, thus the
derived errors will not be representative of those determined from a
combined analysis of the spectra derived from both temperature and
polarization.

\subsection{Point-source detection and masking}

Point-sources are expected to constitute the dominant astrophysical foreground for
CMB $B$-modes on scales beyond the recombination peak \citep{Curto2013}. The detection
and subsequent masking or removal of such sources is therefore an
important pre-processing step to be applied to the data before the
application of component separation algorithms that predominantly
target diffuse Galactic foreground emission.

Two different procedures have been used to create the temperature and
polarization point-source masks. First, a blind source detection
pipeline is run on the 19 \core\ intensity maps, producing catalogues
of sources above a given signal-to-noise ratio in each of the
simulated maps. Then, a different pipeline is run on the polarization
$Q$ and $U$ maps using as input the positions of the sources detected
in intensity. In both cases, intensity and polarization, the single
frequency masks are built using a hole radius of about 1.25 times the
FWHM of each channel. All the intensity masks are then combined into a
single intensity union mask. Similarly, all the polarization masks are
combined together into a single polarization union mask. At the end of
the process we have 40 masks, 19 intensity masks, 19 polarization
masks and the two combined union masks.

\subsubsection{Intensity point-source analysis}

The detection of compact sources in the 19 simulated intensity \core\
frequency maps is attempted on a frequency-by-frequency basis by
searching for peaks above a given signal-to-noise ratio (SNR) in maps
that have been optimally filtered using the Mexican Hat Wavelet 2
\citep[hereafter MHW2][]{lopezcaniego2006,gonzaleznuevo2006}.  Such a
wavelet is employed as a filter because it simultaneously removes both
small scale noise fluctuations and large scale structures in the
vicinity of the compact sources, thereby improving the detection
process. The analysis is performed on flat patches of the sky,
generated from full sky maps in \healpix\ format. Specifically, a
given map is sub-divided into a sufficiently large number of
overlapping projected patches so as to effectively cover the full
sky. Each of these patches is then filtered with the MHW2 taking into
account the local statistics of the background in the vicinity of the
source in order to optimize the shape the filter. After filtering,
candidate sources are identified and repetitions from overlapping
regions removed, retaining those sources with the highest SNR. The
process is repeated twice, first, following this blind detection
procedure on the full sky, then, second, repeating the analysis with
the patches centred on each of the previously detected sources. In
this second iteration, the sources are characterized again and those
with a SNR > 5 are kept. As a result, a catalogue of sources is produced
with information about the position, amplitude and SNR of each
object. This procedure is identical to that used to produce the Planck
compact source catalogues~\citep{planck2013_pccs,planck2015_pccs}.

In order to produce the corresponding compact source masks, a simple
masking procedure has been followed, masking all the pixels within a
radius proportional to the beam FWHM of each simulated map. This
number depends on the amplitude of the source allowing for slightly
larger masked regions when the sources are very bright and slightly
smaller masked regions for weaker sources. Typically, the radius of
the masked holes is of the order of three times the beam ($3\sigma$),
where the beam is defined as $\sigma \approx FWHM/\sqrt{8\ln 2}$.

\subsubsection{Polarization point-source analysis}\label{subsubsec:source_pol}

\begin{figure}[htbp]
\begin{center}
\includegraphics[width=0.6\textwidth]{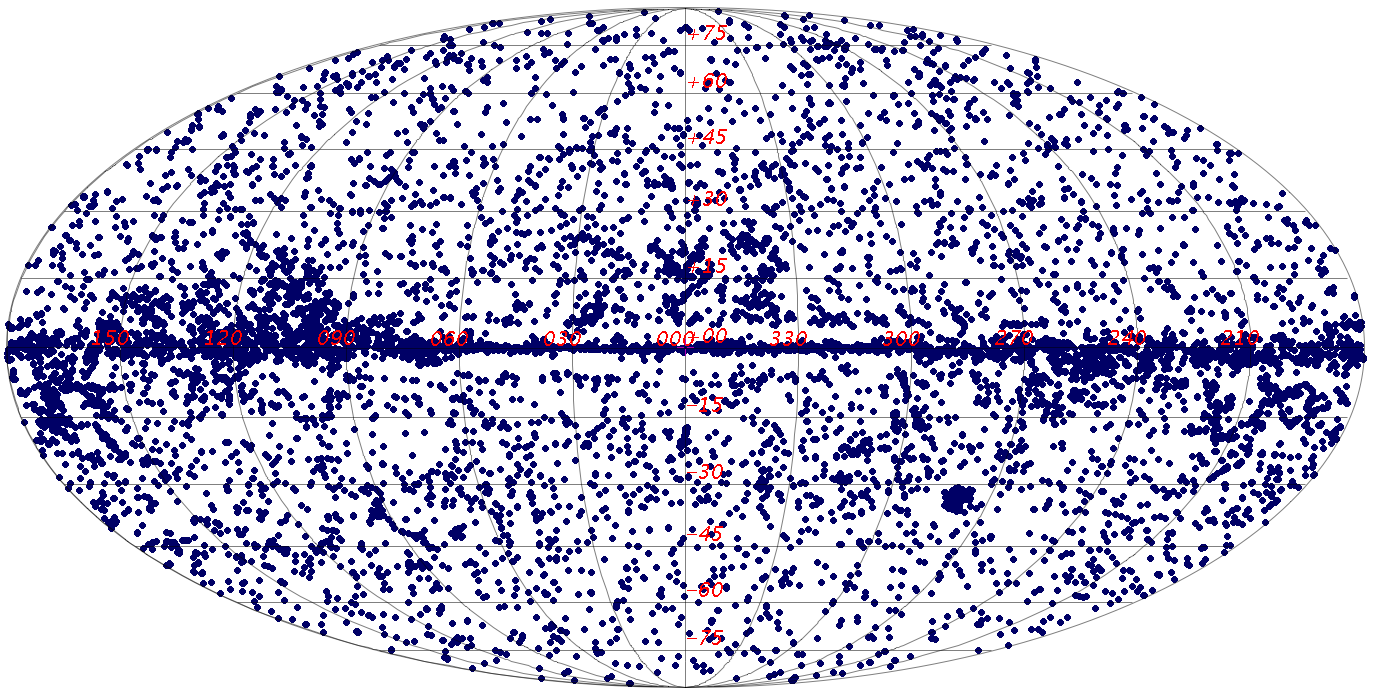}~
\end{center}
  \caption{Union of 60 to 600\,GHz polarization masks used in the
    analysis for mitigating the contamination from polarized compact
    sources. Individual polarized sources are detected in each
    frequency band of \core. 
} 
  \label{fig:union_ps_mask}
\end{figure}

The detection of compact sources in the \core\ polarization $Q$ and
$U$ maps is performed in a non-blind fashion by assessing the
significance of the polarized signal at the position of the sources
detected in the intensity maps.  The procedure is a two-step process
as proposed in \cite{lopez-caniego2009}. First, a maximum likelihood
filter is applied on the $Q$ and $U$ maps \citep{argueso2009}. Second, a
map of polarization intensity, $P=\sqrt{Q^2+U^2}$, is built from the
filtered maps of $Q$ and $U$, and the polarized flux density at the
position of the source is calculated. Then, analyzing the background
statistics of the $P$ map, a $99.90$,\% significance threshold is
calculated. If the flux density at the position of the source is above
this significance threshold, we consider that the signal is from the
source and not from the background. Figure~\ref{fig:union_ps_mask}
shows the union mask for the polarized compact sources that we have
detected in the nineteen \core\ frequency bands.

\subsubsection{Point-source pre-processing}

For some foreground cleaning methods, e.g. {\tt NILC} (see
Appendix~\ref{subsec:nilc_method}) and {\tt SMICA} (see
Appendix~\ref{subsec:smica_method}), the removal of polarized
point-sources from the sky maps prior to component separation is
required to optimize the reconstruction of the CMB $B$-mode power
spectrum. In the case of {\tt NILC}, the presence of any bright point-source
increases the value of the local frequency-frequency covariance matrix in
the neighbourhood of the source 
since the set of pixels from 
which the covariance is computed is more extended than the source
itself. The {\tt NILC} weights then ultimately adjust
themselves to suppress the point-source power as much as possible at
the expense of a lower control of the diffuse foreground signal around
the source. By pre-processing the point-sources in the \core\ sky
maps, the {\tt NILC} weights computed from the pre-processed maps 
readjust themselves to better suppress the contributions from diffuse foregrounds and
noise. In the case of {\tt SMICA}, the performance of foreground
cleaning relies on the correct assumption of the dimension of the
foreground subspace (i.e., the number of independent foreground degrees
of freedom or the rank of the foreground covariance
matrix). Appropriate masking of the diffuse Galactic foreground
emission must ensure that the dimension of the foregrounds is close to
constant over the sky, as assumed by  {\tt
  SMICA}. However, the presence of point-sources locally increases
the effective dimension of the foregrounds, which are then no longer
uniform over the sky. By
pre-processing the point-sources in the \core\ sky maps, we avoid
any local increase of the dimension of the foreground subspace,
therefore optimizing the foreground removal by {\tt SMICA}.

Many techniques have been proposed in the literature to remove
point-sources from sky maps, either relying on a direct fit
of the source profiles which are then subtracted, or on the restoration
of signal to the masked source pixels by extrapolating the background signal
determined from neighbouring pixels (often referred to as
'inpainting').  Different approaches have been developed for restoring
missing data, including sparse representation of the data in a wavelet
frame \citep[e.g.,][]{Abrial2008}, constrained Gaussian realisations
\citep[e.g.,][]{Bucher2012}, and minimum curvature spline surface
inpainting \citep[e.g.,][]{remazeilles2015}.

In this paper, a shortcut has been adopted to subtract the brightest
polarized sources from the \core\ sky maps. Specifically, the 
point-source component maps input to the {\sc PSM} simulation at each
\core\ frequency are masked using the source masks produced in
Sect.~\ref{subsubsec:source_pol}. These masked source
maps are then coadded to the other diffuse foreground component maps in
order to produce the simulated
\core\ sky maps. The resultant simulated skies contain no contribution
from those sources that have been detected in polarization
as would result from a perfect source subtraction approach, although a
background of undetected polarized sources remains.

\subsection{Bayesian parametric fitting at low multipoles using {\tt Commander}}
\label{subsec:parametric}

We have applied the {\tt Commander} algorithm \citep{Eriksen2008} to
the simulated multi-frequency \core\ $Q$ and $U$ sky maps described in
Sect.~\ref{sec:simulated_sky} in order to separate the individual
components of polarized emission. {\tt Commander} performs a
parametric fit to the data, on a pixel by pixel basis, in a Bayesian
framework by using Markov Chain Monte Carlo (MCMC) Gibbs sampling
\citep{Wandelt2004}. For details of the methodology, we refer the
reader to Sect.~\ref{subsec:commander_method}. Since the fit is
performed for each pixel and requires a large number of Gibbs samples
in the MCMC, the algorithm has a large computational cost. This
implies that the method is better suited for the analysis of large
angular scales and low-resolution maps, where the primordial CMB
$B$-mode signal is most significant. The {\tt Commander} component
separation method has previously been applied successfully to \Planck\
data to recover the CMB and foreground contributions at low multipoles
\citep{Planck2015_X}.

We follow the methodology described in \cite{Remazeilles2016} for the
analysis of $B$-modes. The simulated data, $\bdd(\nu,p)$, consist of
both Stokes $Q_\nu(p)$ and $U_\nu(p)$ polarization maps at \core\
frequencies, $\nu$, that were simulated directly at the \healpix\
$N_{\rm side}=16$ resolution. This avoids issues connected to the
downgrading of sky maps and its impact on foreground spectral
behaviour which has consequences for parametric component separation
algorithms. Specifically, using lower resolution simulations ensures
that any residuals in the data after component separation are not due
to a mismatch between the assumed spectral model and that induced by
the downgrading. We refer the reader to Sect.~\ref{subsec:degrade} for
a more detailed discussion.  Therefore, we use {\tt Commander} to
reconstruct the CMB $B$-mode power spectrum on scales corresponding to
reionization at low multipoles $2 \leq \ell \leq 47$.

\begin{figure}[htbp]
\centering
\includegraphics[width=0.6\textwidth]{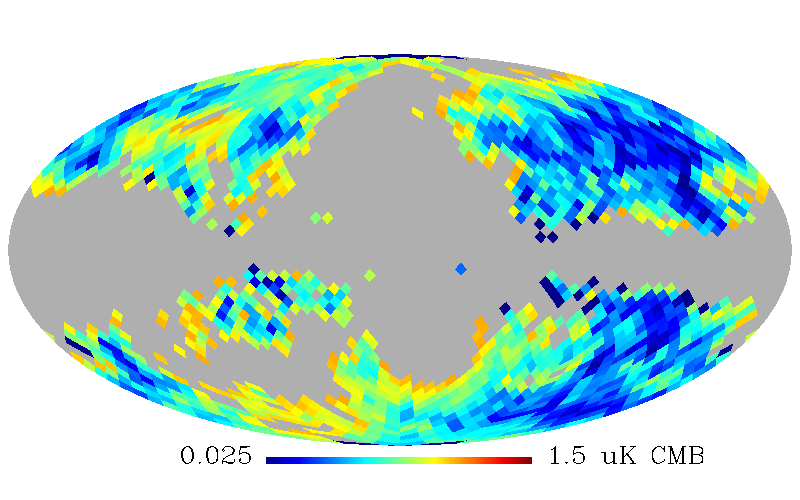}~
  \caption{Definition of the mask used in the {\tt Commander}
      analysis. The masked region (in grey) fits the morphology of the bright
    polarization intensity (in background colour) of the \core\
    $60$\,GHz (synchrotron tracer) and $600$\,GHz (dust tracer) sky
    maps extrapolated to $70$\,GHz through a power-law spectrum and a
    modified blackbody spectrum, respectively (see text). }
\label{Fig:comm_mask}
\end{figure}

In order to partially mitigate the impact of diffuse foreground
contamination on the Bayesian fit by {\tt Commander}, we have applied
a Galactic mask to the \core\ frequency maps so as to avoid the
brightest regions of the sky. The mask has been constructed as
follows. The polarization intensity, $P=\sqrt{Q^2+U^2}$, of the \core\
600\,GHz map, which is dominated by thermal dust emission, was
computed and extrapolated to $70$\, GHz (the foreground minimum)
through a modified blackbody spectrum with $\beta_d=1.6$ and
$T_d=19.4$\,K. A dust mask was then defined by thresholding the
extrapolated $P$ map with respect to a simulated CMB $P$ map at
$70$\,GHz. The regions of the sky where the extrapolated $P$ map
amplitude exceeds ten standard deviations of the simulated CMB $P$ map
signal at $70$\,GHz are masked.  Similarly, we have used the \core\
60\,GHz map, which is the map with most synchrotron emission, to
compute a corresponding $P$ map that is then extrapolated to $70$\,
GHz using a power-law spectrum with $\beta_s=-3$. A similar
thresholding procedure defines a synchrotron mask. The combination of
these two masks then mitigates the brightest synchrotron and dust
contamination. Figure~\ref{Fig:comm_mask} shows the Galactic mask
(grey), superimposed on the polarization intensity map of the combined
\core\ $60$ and $600$\,GHz data extrapolated to $70$\,GHz (background
colours). The mask leaves $f_{\rm sky}=51$\% of the sky usable. 
It should also be noted that, despite masking the sky maps, the CMB
map reconstructed by {\tt Commander} is still a full-sky map because
of the inference of the $C_\ell$ to the map and vice-versa through the
Gibbs sampling iteration scheme described by Eq.~\ref{eq:gibbs}. Of
course, the application of a mask is not mandatory for component
separation, and foreground parameters can be computed over the full
sky if required.

The {\tt Commander} component separation outputs consist of a set of
MCMC $C_\ell$ samples characterized by the posterior distribution
Eq.~(\ref{eq:cl_posterior}). As a consequence, the {\tt Commander} CMB
$C_\ell$ samples are particularly suited to a Blackwell-Rao type
estimation of the cosmological parameters $\tau$ and $r$. 
Note that all other cosmological parameters are set to the
\emph{Planck} 2015 best-fit values as used in the simulations.

Given that the theoretical CMB $E$-mode power spectrum scales as
\citep[e.g.,][]{Planck2015_XI}
\begin{equation}\label{eq:cl_ee_tau}
C_\ell^{th\, EE} (\tau) = \begin{cases} \left(\tau\over 0.05\right)^2\,C_\ell^{EE}(\tau=0.05) 									& 2 \leq \ell \leq 12 \\
																{e^{-2\tau}\over e^{-2 \times 0.05}} \,C_\ell^{EE}(\tau=0.05) & \ell > 12 
										\end{cases},
\end{equation}
then the posterior distribution of the optical depth to reionization, $\tau$,
can be computed with the Blackwell-Rao estimator using the MCMC Gibbs
samples, ${\widehat{C}}_\ell^{EE\, (i)}$, of the reconstructed CMB $E$-mode
power spectra:
\bea\label{eq:br_posterior_tau}
P\left(\tau\right) \approx {1\over N_G}\sum_{i=1}^{N_G} \mathcal{L}\left[{\widehat{C}}_\ell^{EE\, (i)}|C_\ell^{th\, EE}\left(\tau\right)\right],
\eea 
where the sum runs over $N_G$ Gibbs samples and the log-likelihood function reads as 
\bea\label{eq:br_likelihood}
-2\ln \mathcal{L}\left[{\widehat{C}}_\ell^{(i)}|C_\ell^{th}\right]  = \sum_{\ell} (2\ell+1)\left[ \ln\left({C_\ell^{th}\over{\widehat{C}}_\ell^{(i)}}\right) + {{\widehat{C}}_\ell^{(i)}\over C_\ell^{th}} - 1\right].
\eea

If we then consider the theoretical CMB $B$-mode power spectrum as the
combination of the tensor modes and the lensing modes,
\bea\label{eq:cl_bb_r}
C_\ell^{th\, BB} (r,A_{lens}) = {r\over 0.01}\,C_\ell^{tensor}(r=0.01)\, +\, A_{lens}\,C_\ell^{lensing}(r=0),
\eea
then the posterior distribution of the tensor-to-scalar ratio, $r$,
and the amplitude of the lensing $B$-modes, $A_{lens}$, can be computed
with the Blackwell-Rao estimator using the MCMC Gibbs samples, 
${\widehat{C}}_\ell^{BB\, (i)}$, of the reconstructed CMB $B$-mode power spectra.
\bea\label{eq:br_posterior}
P\left(r,A_{lens}\right) \approx {1\over N_G}\sum_{i=1}^{N_G} \mathcal{L}\left[{\widehat{C}}_\ell^{BB\,(i)}|C_\ell^{th\, BB}\left(r,A_{lens}\right)\right],
\eea 
where the log-likelihood function reads as in
Eq.~\ref{eq:br_likelihood}.
We specify the quantity
${\mathcal{D}_\ell \equiv \ell(\ell + 1)C_\ell/2\pi}$ for plotting
purposes.  We also define
${\Delta \mathcal{D}_\ell / \sigma_\ell=
  \left(\widehat{\mathcal{D}}_\ell -\mathcal{D}^{th}_\ell\right) /
  \sigma_\ell}$ as the difference between the estimated power
spectrum, $\widehat{\mathcal{D}}_\ell$, and that of the input
theoretical model sky, $\mathcal{D}^{th}_\ell$, normalized by the
error bar, $\sigma_\ell$, on $\widehat{\mathcal{D}}_\ell$.

Throughout the paper, the reconstructed angular power spectra are
binned over specific multipole ranges in order to minimize the correlations
between angular scales. We adopt the following weighting in the
binning procedure, 
\bea
 \label{eq:bin_weights}
 C_q = \sum_{\ell \in [\ell_{\min}(q),\ell_{\max}(q)]} {(2\ell+1)\over
   \sum_{\ell \in [\ell_{\min}(q),\ell_{\max}(q)]} (2\ell+1)}\, C_\ell.
 \eea Five multipole bins are defined for {\tt Commander}:
 $\ell \in [2,6]$, $[7,12]$, $[13,20]$, $[21,31]$, and $[32,47]$.

\begin{figure}[t]
\centering
\includegraphics[width=0.48\textwidth]{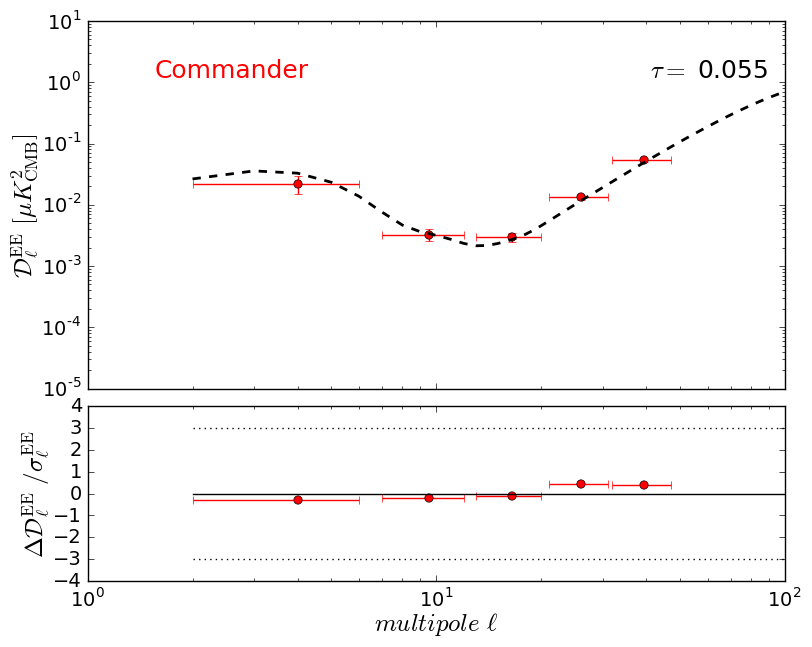}~
\includegraphics[width=0.52\textwidth]{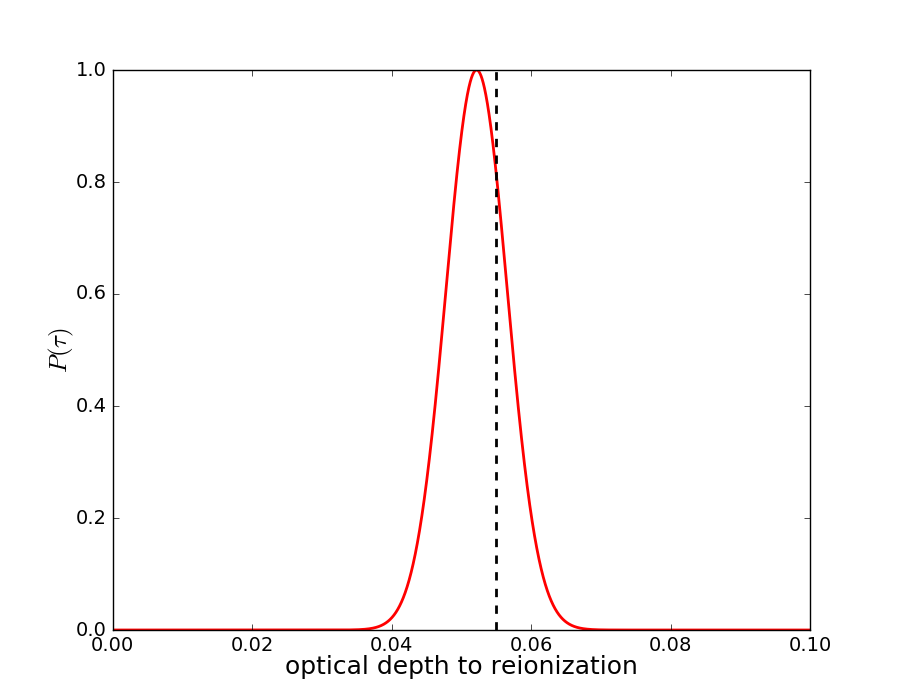}~\\
\caption{{\tt Commander} results for the CMB $E$-modes determined from
  a simulation with the CMB optical depth $\tau=0.055$, and
  synchrotron and dust foregrounds. \emph{Left panel}: CMB $E$-mode
  power spectrum reconstruction. The fiducial CMB $E$-mode power
  spectrum is indicated by the dashed black line, while the
  Blackwell-Rao power spectrum estimate is denoted by red points.  The
  horizontal dotted lines represent the 3$\sigma$ limits. \emph{Right
    panel}: Posterior distribution, $P(\tau)$, of the optical depth to
  reionization computed over the multipole range $2 \leq \ell \leq 12$.  }
\label{Fig:comm_tau}
\end{figure}

\begin{figure}[t]
\centering
\includegraphics[width=0.49\textwidth]{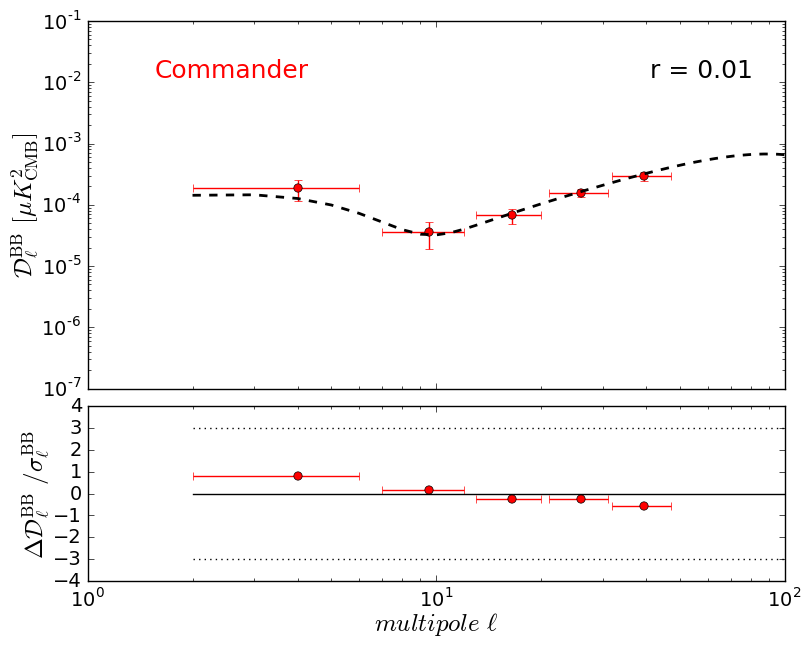}~
\includegraphics[width=0.51\textwidth]{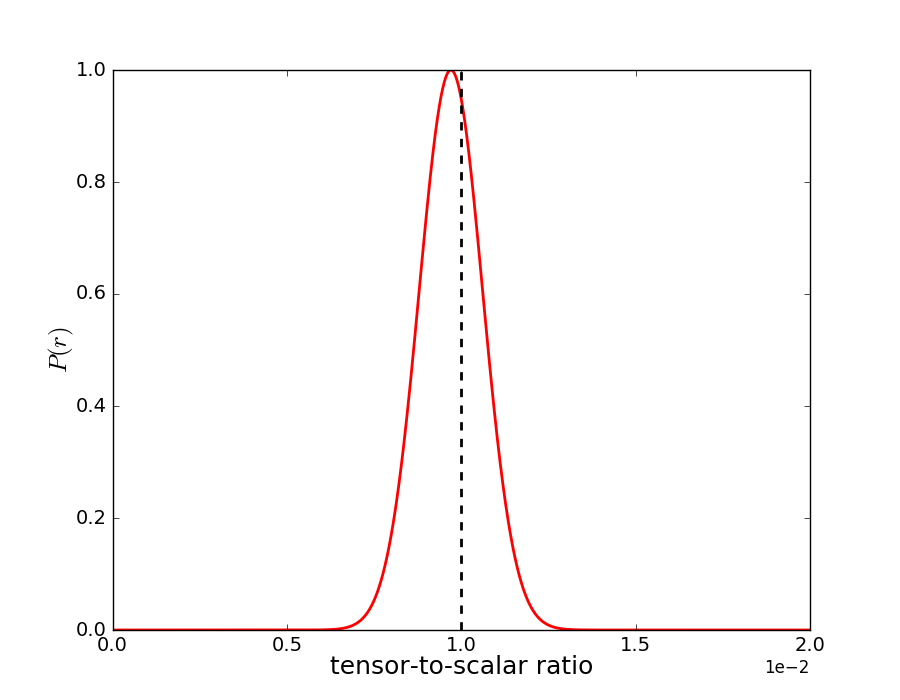}~\\
\includegraphics[width=0.49\textwidth]{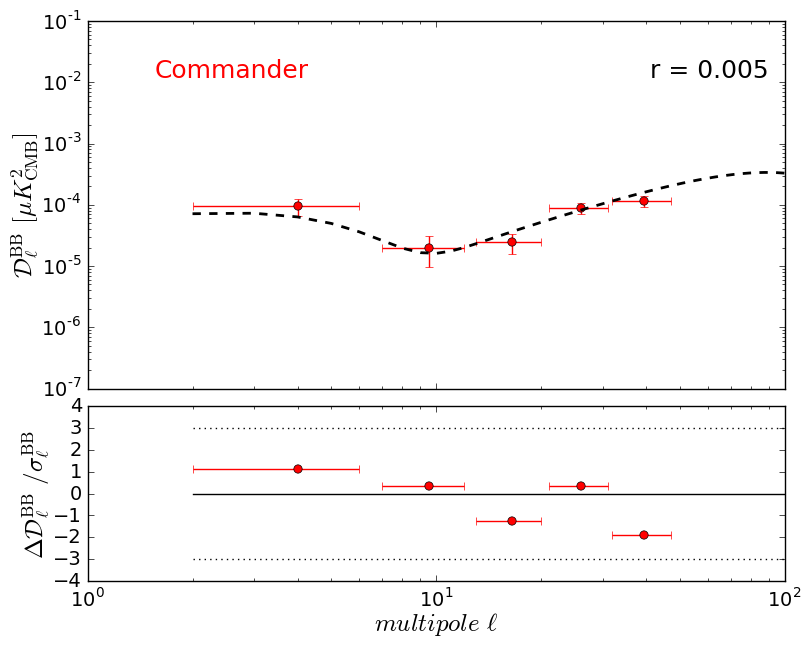}~
\includegraphics[width=0.51\textwidth]{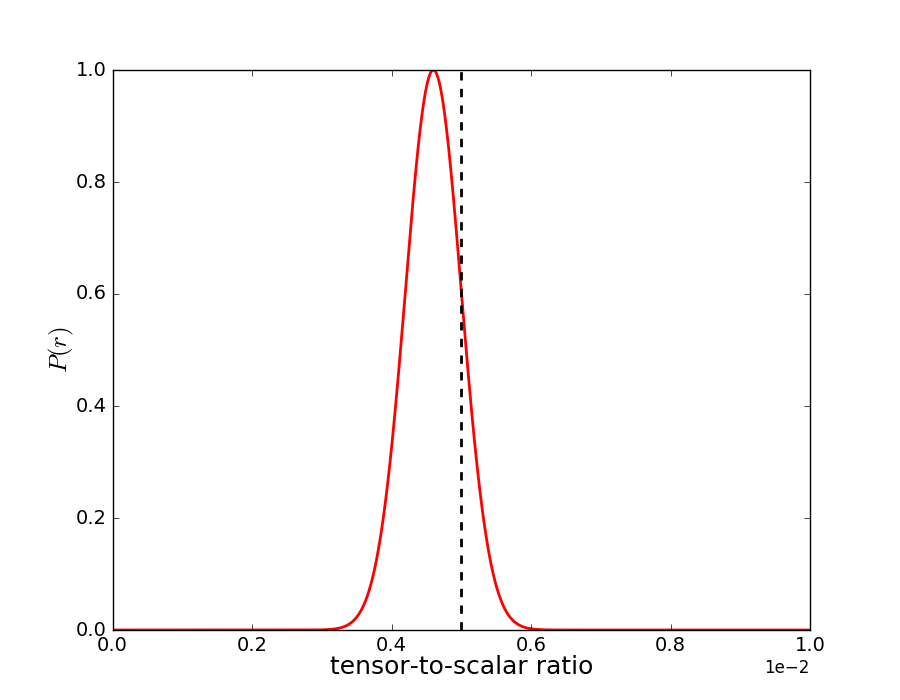}~\\
\includegraphics[width=0.49\textwidth]{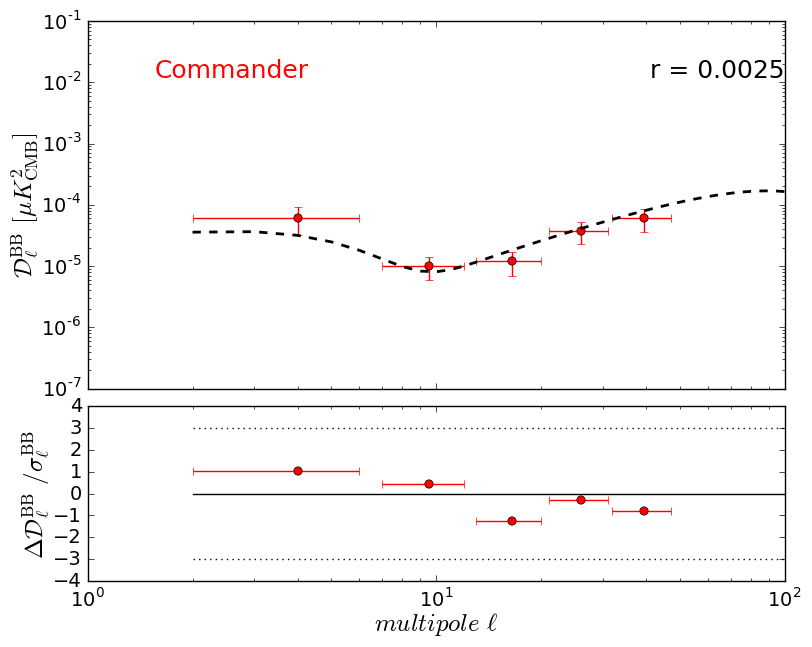}~
\includegraphics[width=0.51\textwidth]{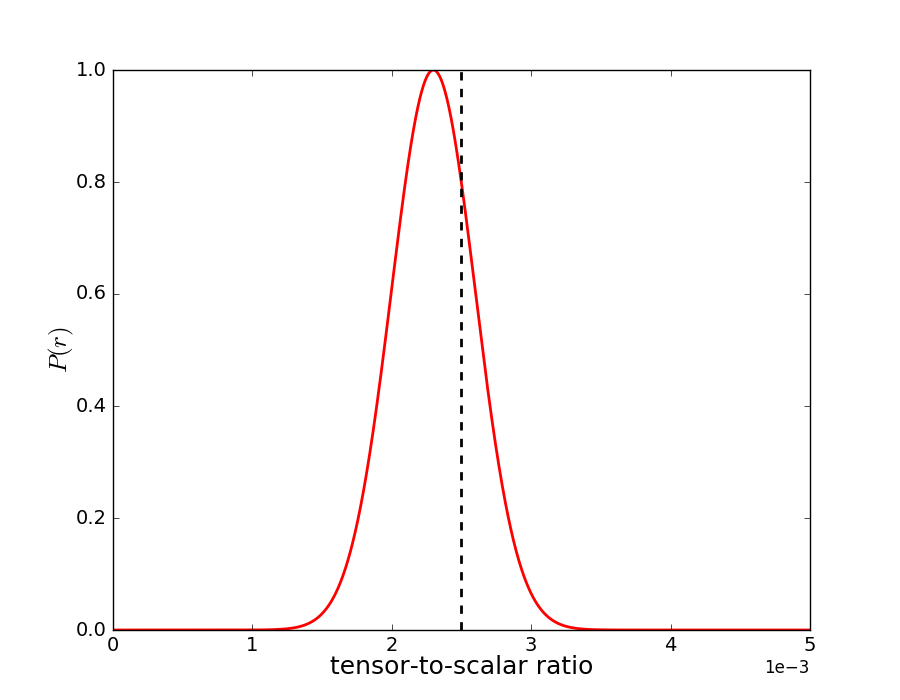}~\\
\caption{{\tt Commander} results on CMB $B$-modes for simulations \#1
  ($r=10^{-2}$, \emph{top}), \#2 ($r=5\times 10^{-3}$, \emph{middle}),
  and an additional simulation \#2-bis ($r=2.5\times 10^{-3}$, \emph{bottom}),
  including synchrotron and dust foregrounds. \emph{Left panels}: CMB
  $B$-mode power spectrum reconstruction. The fiducial primordial CMB
  $B$-mode power spectrum is denoted by a dashed black line, while the
  Blackwell-Rao power spectrum estimates are indicated by the red
  points.  The horizontal dotted lines represent the 3$\sigma$ limits.
  \emph{Right panels}: Posterior distribution, $P(r)$, of the
  tensor-to-scalar ratio computed over the multipole range $2 \leq \ell \leq 47$.}
\label{Fig:comm_r}
\end{figure}

In Fig.~\ref{Fig:comm_tau}, we first show the reconstruction of the
CMB $E$-mode power spectrum after component separation.  The
reconstruction of the CMB $E$-mode signal is easier than in the case
of the primordial CMB $B$-mode signal since the latter is sub-dominant
to foreground signals over all angular scales.  The accurate
reconstruction of the CMB $E$-modes fulfils several
functions: it serves as a validation criterion for the fidelity of the
component separation algorithm; E-modes can be used as an input to
delensing algorithms \citep[][in prep.]{ECO_lensing} operating via quadratic
estimators \citep{Hu2002,Hirata2003}; and finally it provides
important information on cosmological parameters, in particular the
optical depth to reionization \citep{ECO_parameters2016}.

By applying the Blackwell-Rao estimator
Eqs.~(\ref{eq:br_posterior_tau})-(\ref{eq:br_likelihood}) to the {\tt
  Commander} reconstructed $E$-mode power spectrum over the multipole
range $2 \leq \ell \leq 12$, we determine the posterior distribution
for the optical depth to reionization (shown in the right panel of
Fig.~\ref{Fig:comm_tau}), indicating an unbiased estimate of
$\tau=0.0522 \pm 0.0044$. This is a more than $11\sigma$ measurement
of $\tau$ by \core\ to be compared to the $6\sigma$ measurement from
the latest \Planck\ results \citep{Planck_lowl_2016}.

The left panels of Fig.~\ref{Fig:comm_r} present the reconstructed CMB
$B$-mode power spectra as determined by {\tt Commander} for different
values of the tensor-to-scalar ratio, in the absence of gravitational
lensing and when only synchrotron and dust foregrounds are
present. These correspond to simulations \#1 and \#2, as described in
Tables~\ref{tab:sky} and \ref{tab:sims}, plus an additional simulation \mbox{\#2-bis}
with $r=2.5\times 10^{-3}$.  In these three cases, {\tt Commander}
successfully recovers the primordial CMB $B$-mode signal from \core\
simulations over the multipole range $2 \leq \ell \leq 47$ and
provides unbiased measurements of the tensor-to-scalar ratio, finding
$r=\left(0.97\pm 0.09\right)\times 10^{-2}$,
$r=\left(4.6\pm 0.4\right)\times 10^{-3}$, and
$r=\left(2.3\pm 0.3\right)\times 10^{-3}$ respectively (as seen in the
right panels of Fig.~\ref{Fig:comm_r}).  These correspond to
detections of $r$ at a level exceeding $8\sigma$.

\begin{figure}[t]
\centering
\includegraphics[width=0.49\textwidth]{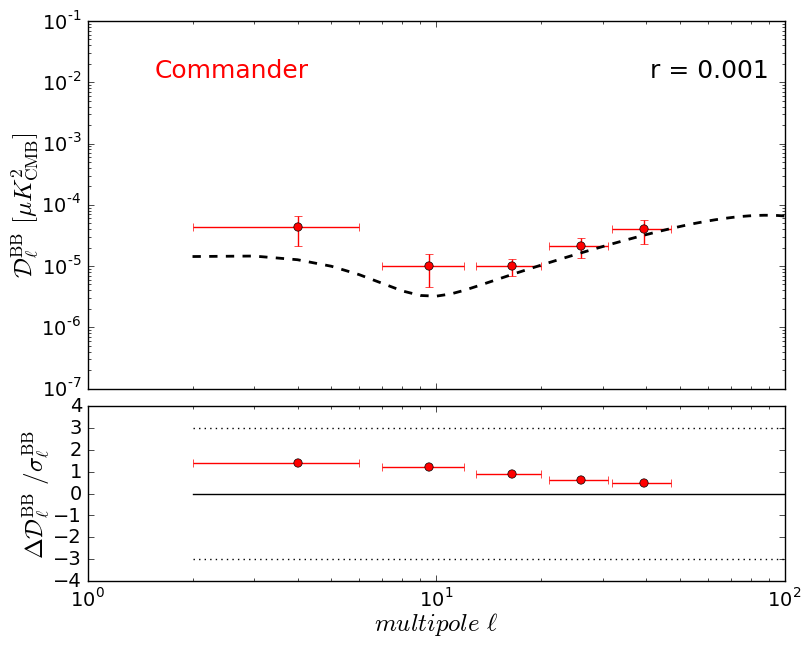}~
\includegraphics[width=0.51\textwidth]{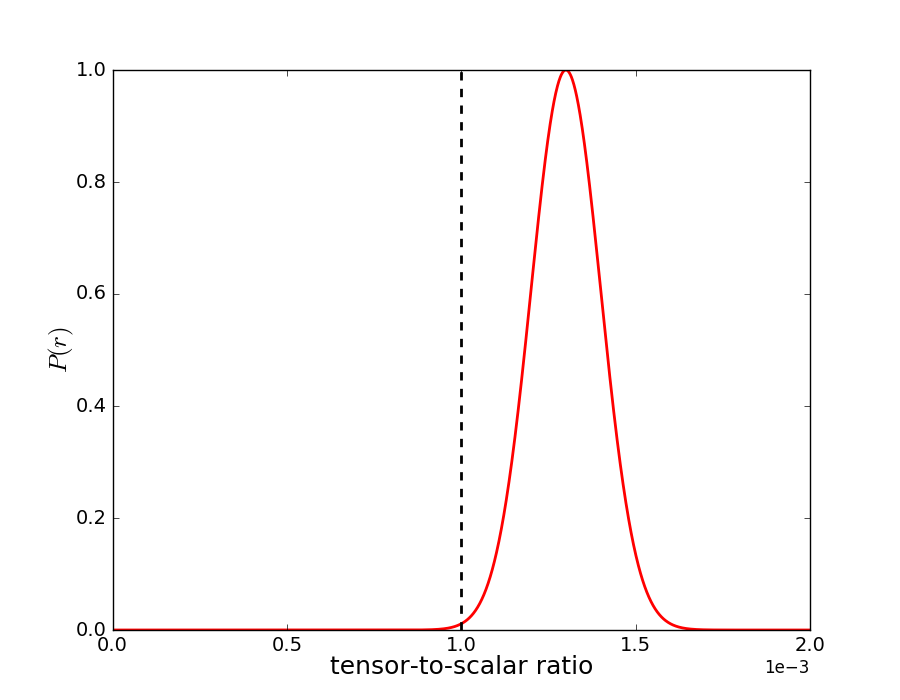}~\\
\includegraphics[width=0.49\textwidth]{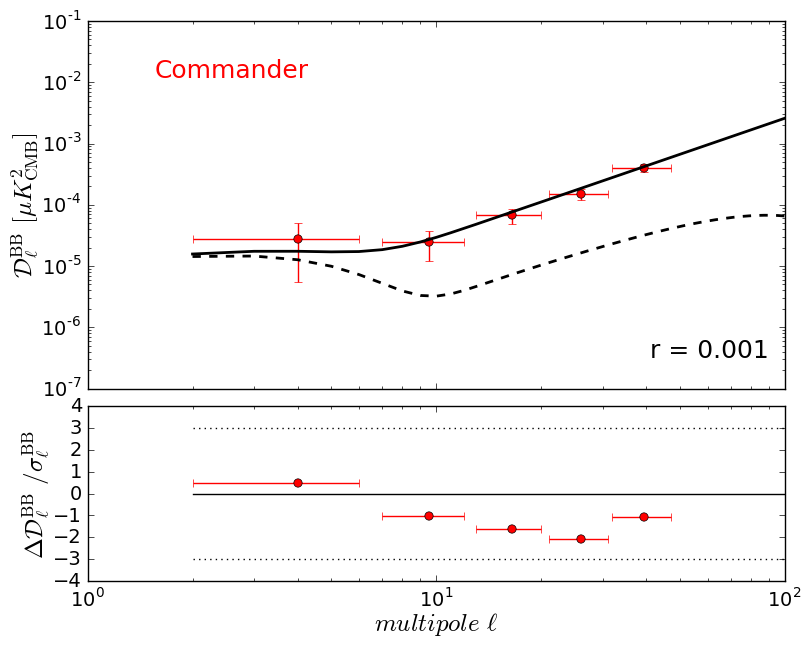}~
\includegraphics[width=0.51\textwidth]{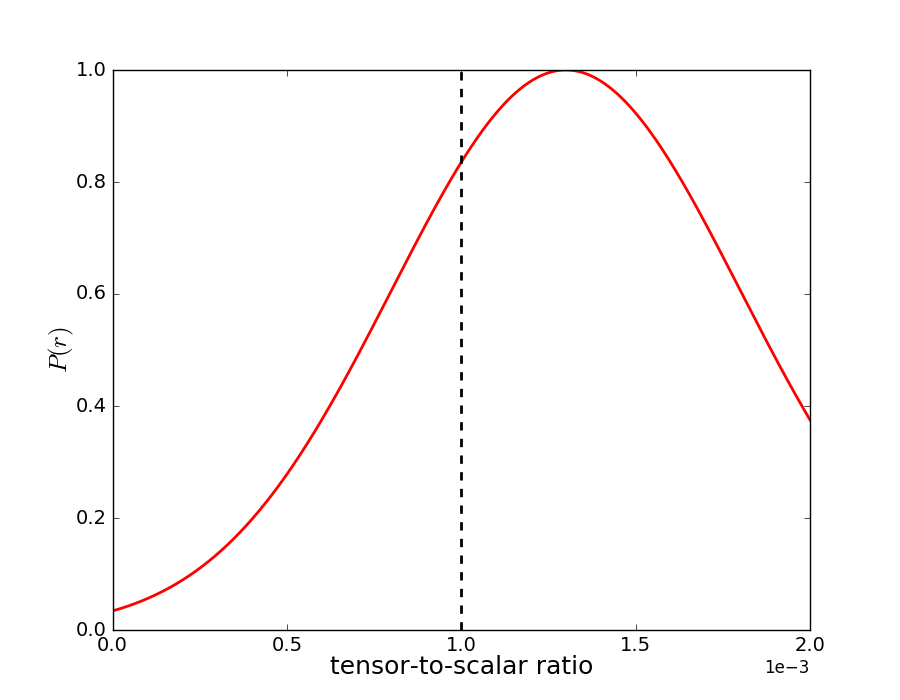}~\\
\caption{{\tt Commander} results on CMB $B$-modes for $r=10^{-3}$ in
  the absence of lensing (\emph{top}; simulation \#3, including dust
  and synchrotron foregrounds) and in the presence of lensing
  (\emph{bottom}; simulation \#4, including dust, synchrotron, AME,
  and point-source foregrounds).  \emph{Left panels}: CMB $B$-mode
  power spectrum reconstruction. The fiducial primordial CMB $B$-mode
  power spectrum is denoted by a dashed black line, while the solid
  black line shows the lensed CMB $B$-mode power spectrum. The
  Blackwell-Rao power spectrum estimates are indicated by the red
  points. The horizontal dotted lines represent the 3$\sigma$ limits.
  \emph{Right panels}: Posterior distribution, $P(r)$, of the
  tensor-to-scalar ratio  computed over the multipole range $2 \leq \ell \leq 47$.}
\label{Fig:comm_r10-3}
\end{figure}

When the tensor-to-scalar ratio falls to the $r=10^{-3}$
level, the reconstruction of the primordial CMB $B$-mode signal by
{\tt Commander} at low-$\ell$ becomes more challenging.
Figure~\ref{Fig:comm_r10-3} presents results for such a case in the
presence of dust and synchrotron foregrounds with variable spectral
indices, but in the absence of lensing (simulation \#3).  In
particular, the reconstructed CMB $B$-mode power spectrum (top left
panel) shows evidence of residual foreground contamination on scales
$\ell \leq 12$.  As a consequence, the Blackwell-Rao estimate of $r$
(top right panel) is biased by $3\sigma$. 
In the absence of lensing the uncertainty on $r$ after foreground cleaning is $\sigma(r=10^{-3})=10^{-4}$ (see Table~\ref{tab:final_results_table} in Sect.~\ref{sec:compsep_results}), therefore the sensitivity target of \core, $\sigma(r)=3\times 10^{-4}$, is achieved by {\tt Commander}, even when including the bias in the uncertainty.

\begin{figure}[t]
\centering
\includegraphics[width=0.5\textwidth]{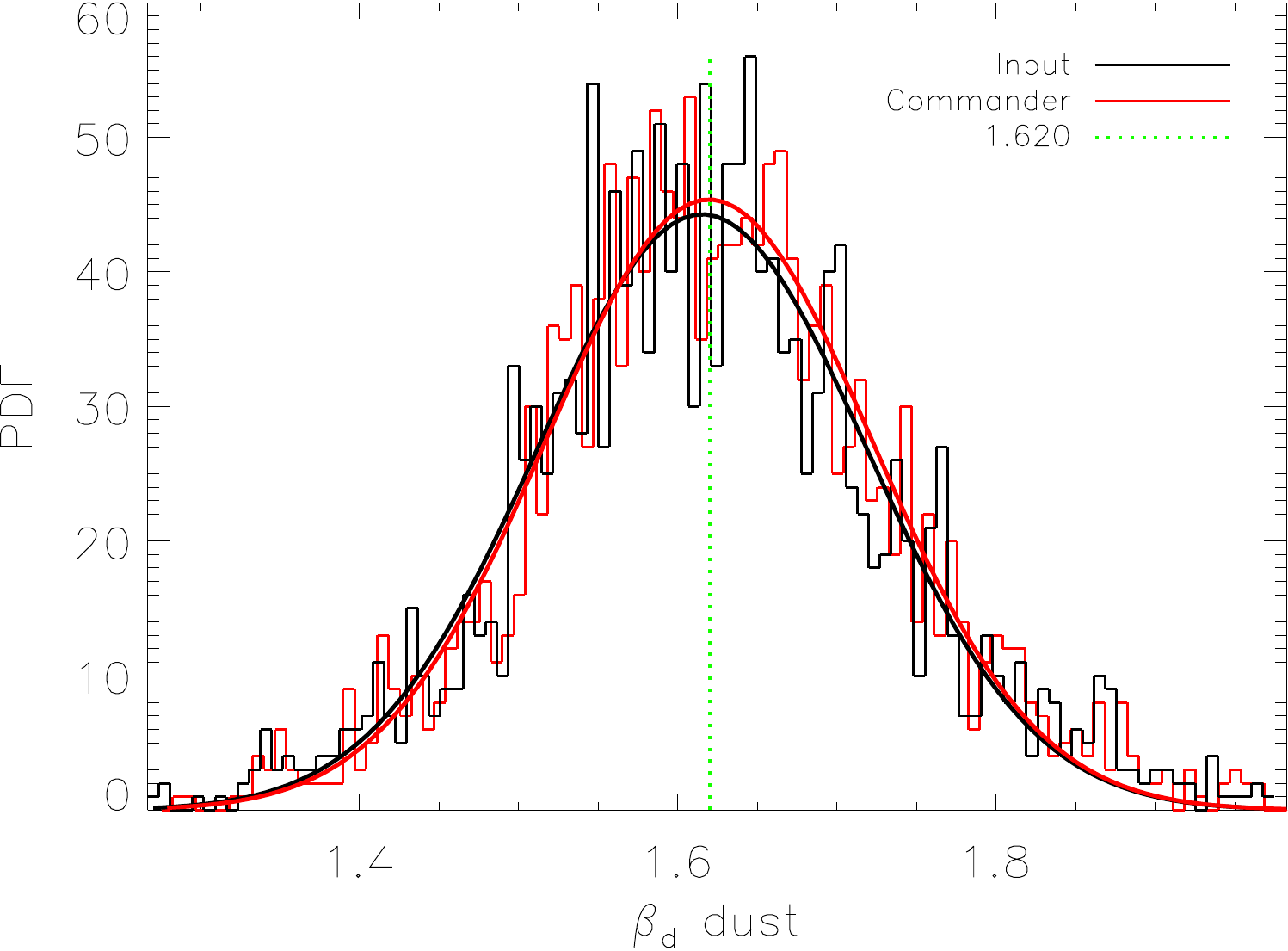}~
\includegraphics[width=0.5\textwidth]{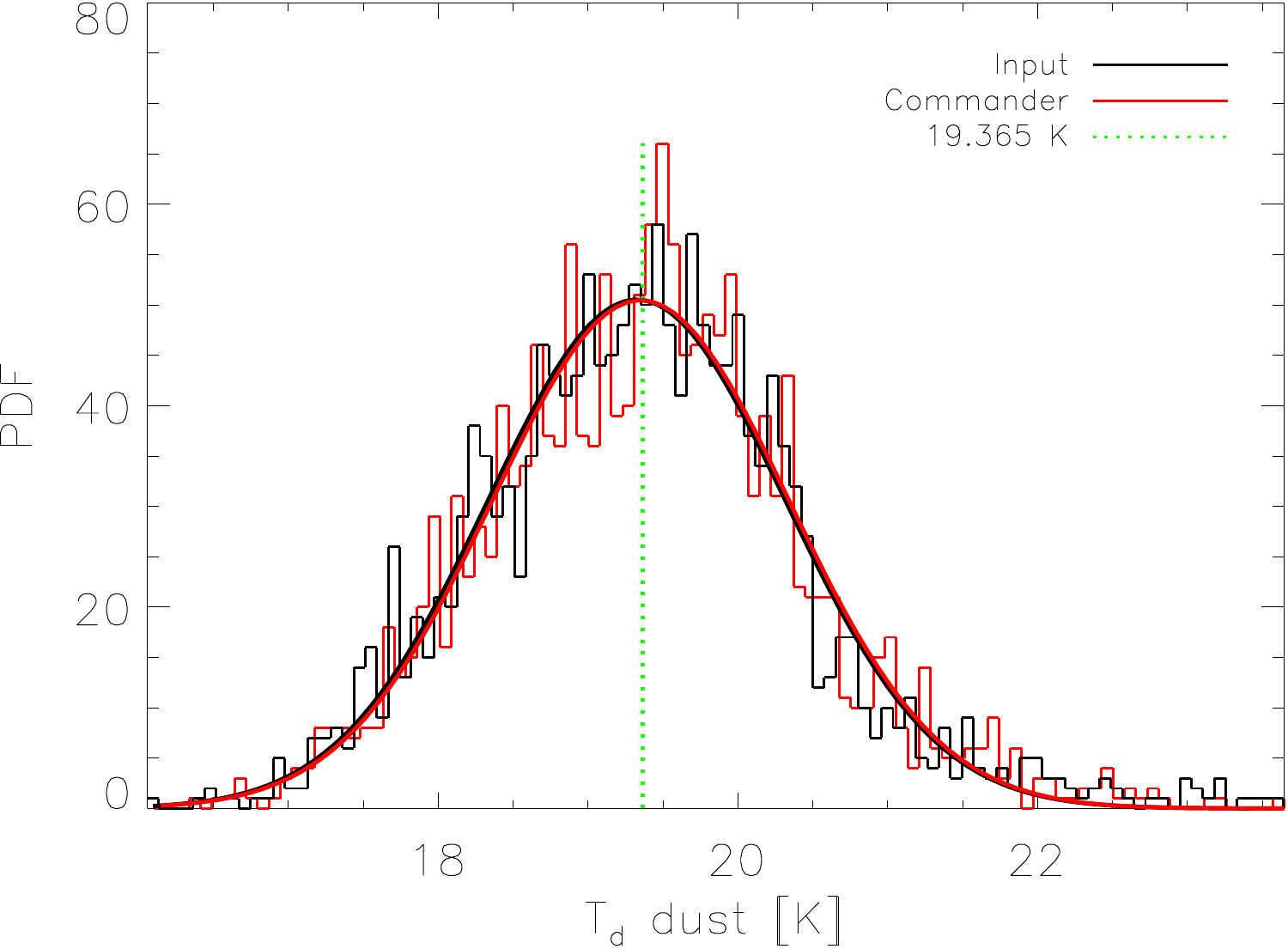}~\\
\includegraphics[width=0.5\textwidth]{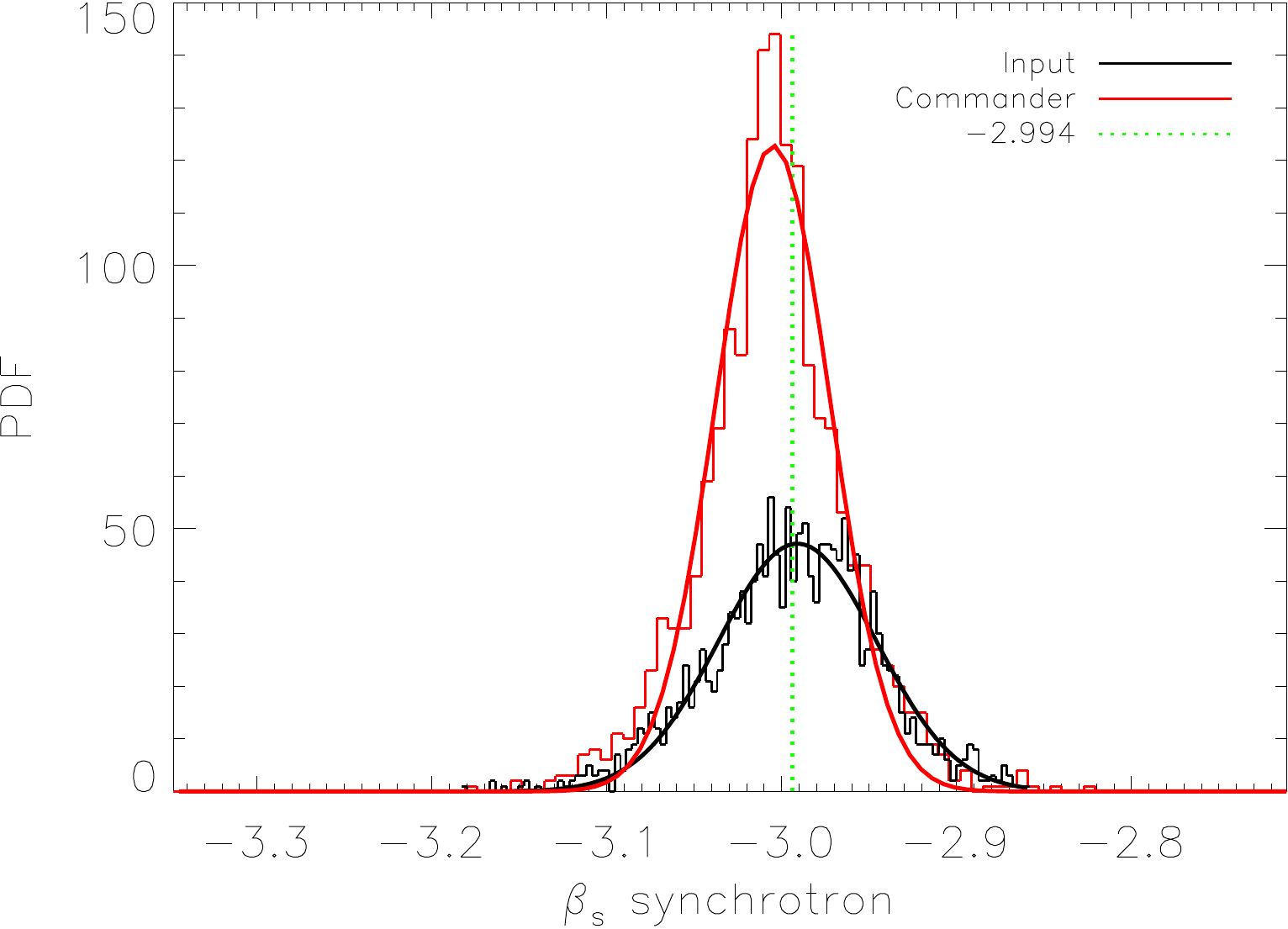}~
\includegraphics[width=0.5\textwidth]{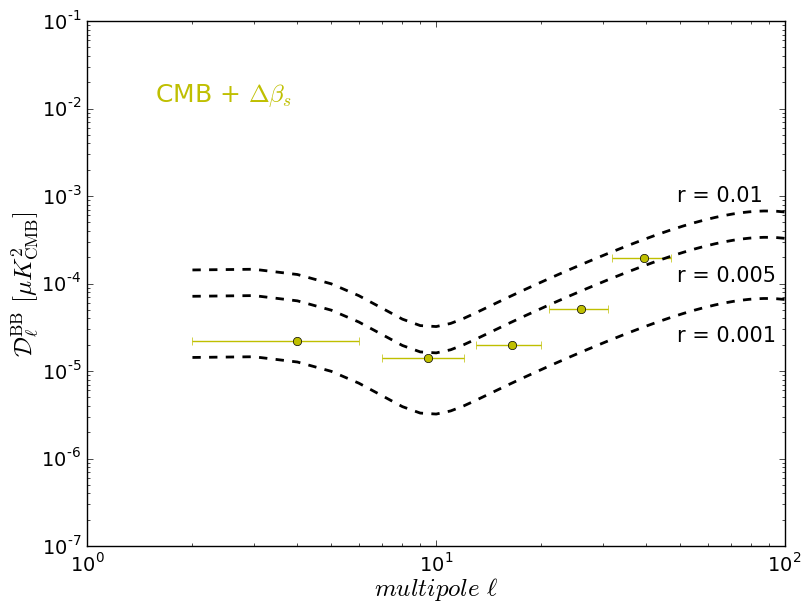}~\\
\caption{{\tt Commander} results on the estimation of the foreground spectral indices, and impact on CMB $B$-modes of slight inaccuracies in the recovered synchrotron spectral index. \emph{Top left:} Recovered distribution (red) of the dust spectral index with respect to the input distribution (black). \emph{Top right:} Recovered distribution (red) of the dust temperature with respect to the input distribution (black). \emph{Bottom left:} Recovered distribution (red) of the synchrotron spectral index with respect to the input distribution (black). Gaussian fits of these distributions are also plotted. \emph{Bottom right:} The slight discrepancies between the reconstructed and the input distributions of the synchrotron spectral index (bottom left panel) are sufficient to give an excess of $B$-mode power (yellow) at a magnitude larger than $r=10^{-3}$.}
\label{Fig:delta_beta_sync}
\end{figure}

In order to provide some intuition on the origin of the bias in the estimated primordial $B$-modes for $r=10^{-3}$, we plot in Fig.~\ref{Fig:delta_beta_sync} the recovered distributions of the foreground spectral indices by {\tt Commander} (red) with respect to the input distributions (black). While the dust spectral indices and dust temperatures are recovered with the desired accuracy, the recovered distribution of the synchrotron spectral index presents slight deviations, $\Delta\beta_s\sim 0.02$, with respect to the input distribution. Those small discrepancies might still be significant for $B$-modes at level of sensitivity of $r=10^{-3}$.
Scaling the synchrotron map at $145$\,GHz through either the recovered or the input spectral index distribution actually yields to a difference of $B$-mode power of order $\sim 2\times10^{-5}$\,$\mu {\rm K}^2$ on reionization scales, as shown in the bottom right panel of Fig.~\ref{Fig:delta_beta_sync}. Therefore, errors on the synchrotron spectral index of order $\Delta\beta_s\sim 0.02$ are sufficient to add an excess of $B$-mode power on top of the primordial signal at $r=10^{-3}$, but still below $r=5\times 10^{-3}$. Of course, the fit is multidimensional so that the exact shape of the excess power might be different, but this calculation at least provides the order of magnitude of those effects at reionization scales. Interestingly, the means and standard deviations of the recovered and the input distributions of the synchrotron spectral index are consistent, but the skewness and kurtosis are not because of the non-Gaussian nature of the distribution, which yields to imperfect recovery of the synchrotron spectral index over the sky. For accurate Bayesian fitting at the level of sensitivity of $r=10^{-3}$, we may think of using more informative priors on the foreground spectral indices instead of the Gaussian priors adopted in this analysis. This will be investigated in a future work. We also anticipate that ongoing low-frequency surveys, e.g. C-BASS \citep{Irfan2015}, will provide tighter constraints on the synchrotron spectral index at a precision smaller than $\Delta\beta_s\sim 0.02$, which will help the parametric component separation.
Paradoxically, using chi-square statistics to evaluate the agreement between the modelled total sky emission and the simulated data indicates that the fit is statistically adequate. However, the foreground residuals present in the reconstructed spectrum and the bias on $r$ indicate that the recovered CMB component is not perfectly fitted. As discussed in \cite{Remazeilles2016}, the absence of chi-square evidence for incorrect foreground modelling arises from a lack of frequency coverage. With more channels at frequencies below $60$\,GHz, any mismatch $\Delta\beta_s\sim 0.02$ between the input and modelled foregrounds would be indicated by the chi-square statistics, thus avoiding either biased or false detections of the tensor-to-scalar ratio. However, this assumes that the spectral index above 60\,GHz is the same, an assumption that is not guaranteed to hold at the required level of precision.

Residual $B$-mode foreground contamination at the reionization peak
could, in principle, be corrected for in the likelihood estimation of
$r$. After introducing a parametrized phenomelogical model with a
specific spectral shape for the residual foregrounds, then $r$ and the
foreground nuisance parameter, $A_{\rm FG}$ (the amplitude of the power
spectrum of residual foregrounds), could be fitted simultaneously to
the {\tt Commander} reconstructed CMB $B$-mode power spectrum.  We
defer an attempt to implement such a strategy to
Sect.~\ref{sec:compsep_results}. 

\begin{figure}[htbp]
\centering
\includegraphics[width=0.5\textwidth]{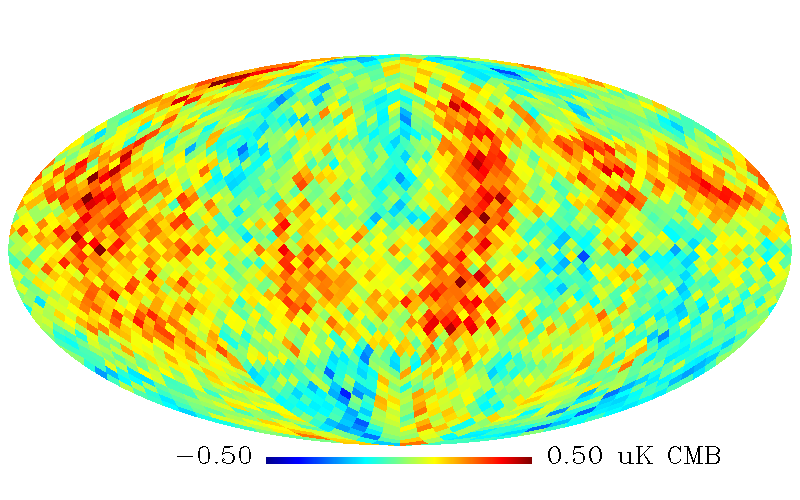}~
\includegraphics[width=0.5\textwidth]{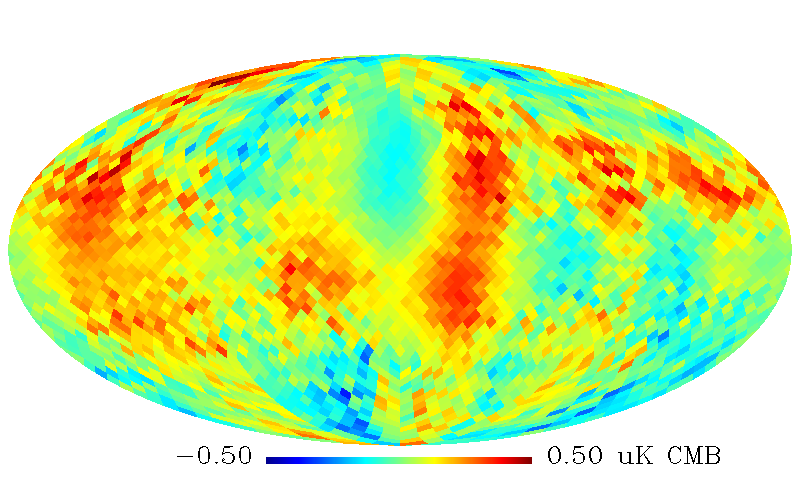}~\\
\includegraphics[width=0.5\textwidth]{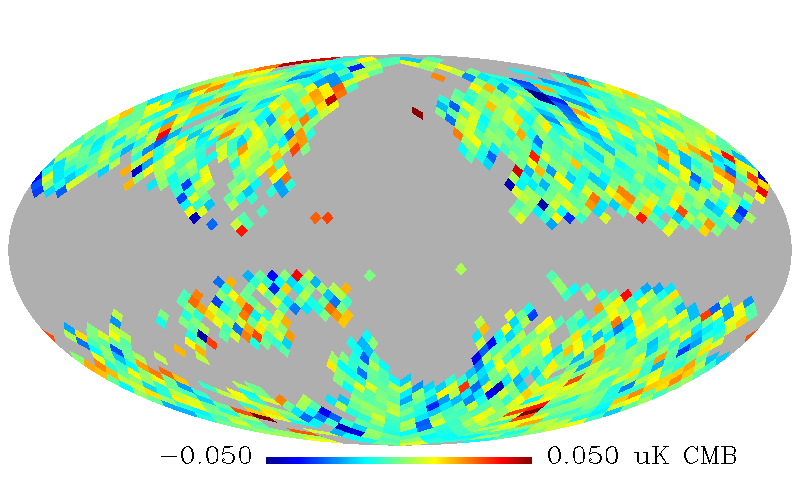}~
\includegraphics[width=0.5\textwidth]{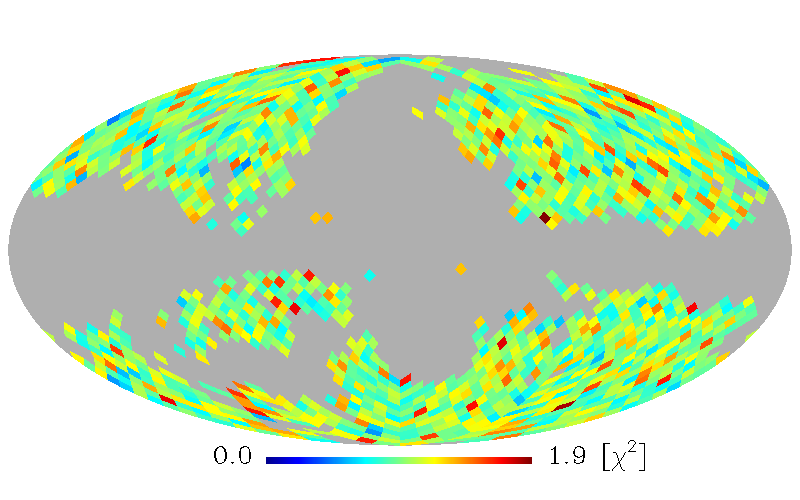}~\\
  \caption{{\tt Commander} CMB map reconstruction for simulation
      \#4. \emph{Top left}: input CMB $Q$
    map. \emph{Top right}:  {\tt Commander} CMB $Q$ map. \emph{Bottom
      left}: difference map between the {\tt Commander} and the input
    CMB maps. \emph{Bottom right}: {\tt Commander} $\chi^2$ map measuring
    the goodness-of-fit per pixel. All maps are at $N_{\rm side}=16$.}
\label{Fig:comm_cmb_map}
\end{figure}

\begin{figure}[htbp]
\centering
\includegraphics[width=0.5\textwidth]{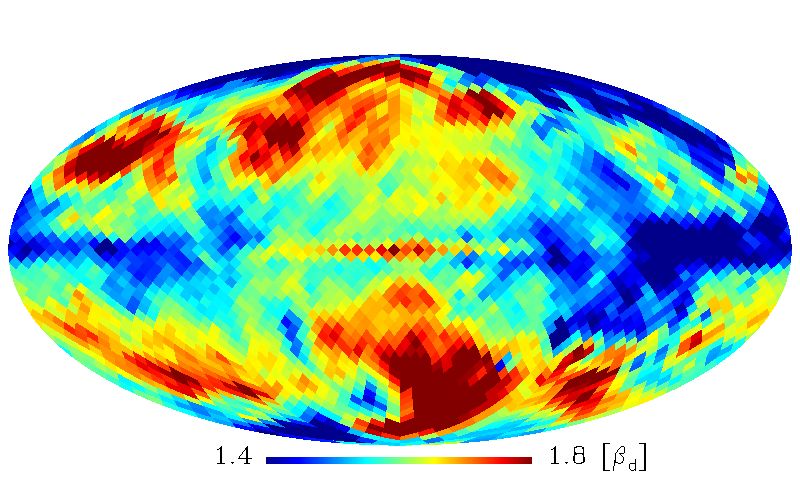}~
\includegraphics[width=0.5\textwidth]{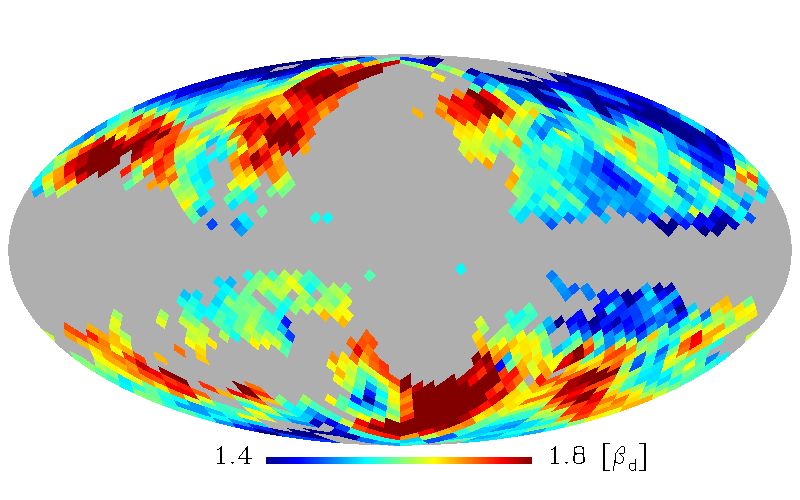}~\\
\includegraphics[width=0.5\textwidth]{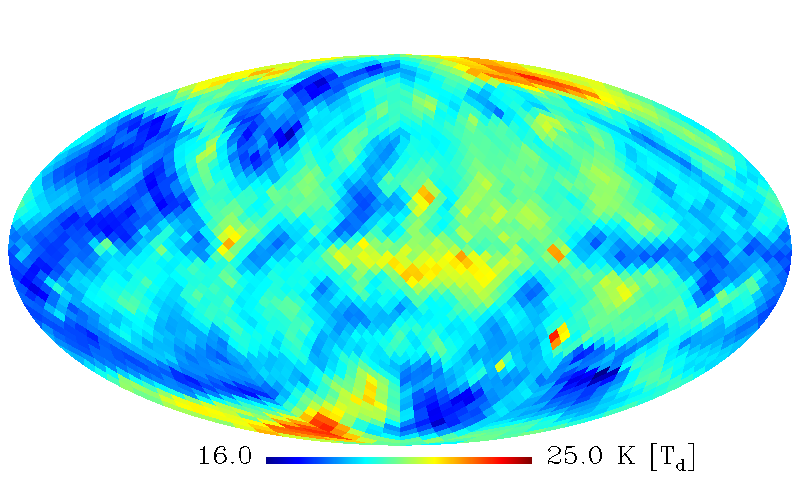}~
\includegraphics[width=0.5\textwidth]{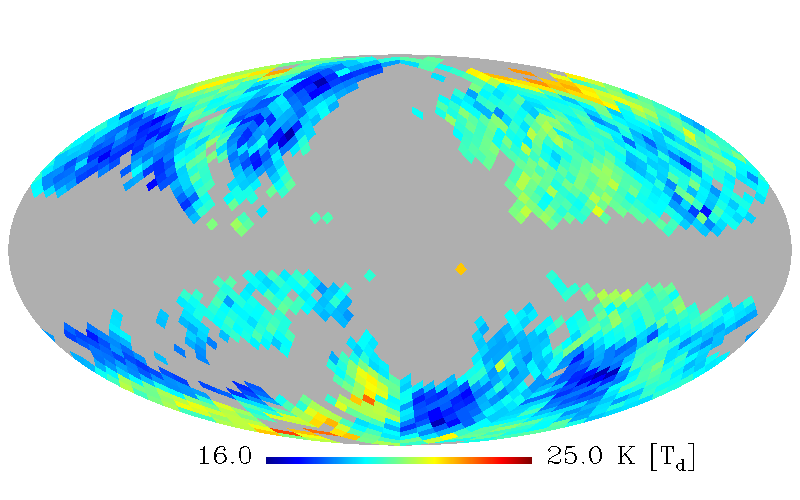}~\\
\includegraphics[width=0.5\textwidth]{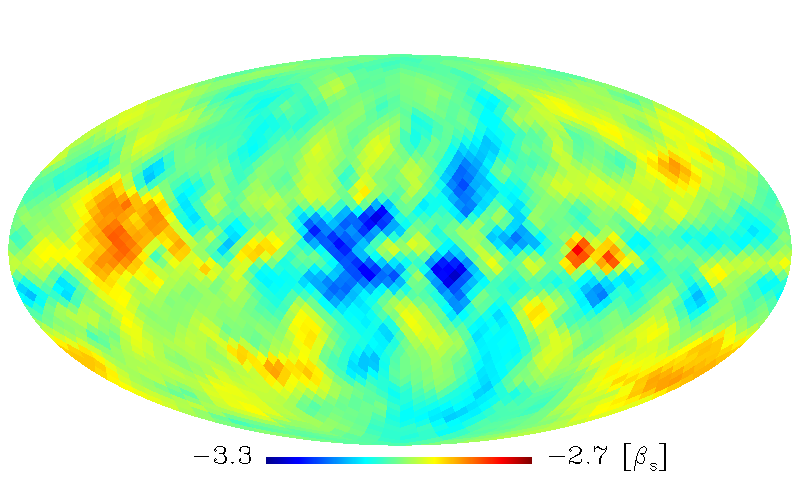}~
\includegraphics[width=0.5\textwidth]{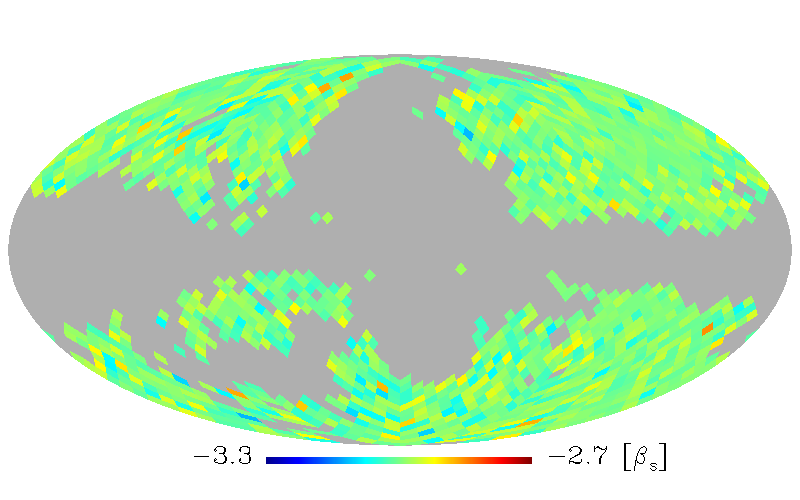}~\\
  \caption{{\tt Commander} foreground  reconstruction for simulation
    \#4. \emph{Top left}: input dust spectral index
    map. \emph{Top right}: {\tt Commander} dust spectral index
    map. \emph{Middle left}: input dust temperature map. \emph{Middle
      right}: {\tt Commander} dust temperature map. \emph{Bottom
      left}: input synchrotron spectral index map. \emph{Bottom
      right}: {\tt Commander} synchrotron spectral index map. All maps
    are at $N_{\rm side}=16$.}
\label{Fig:comm_foregrounds}
\end{figure}

The presence of gravitational lensing effects, which transform
$E$-modes into $B$-modes, is an additional complication for the
measurement of the primordial signal. In simulation \#4, we attempt to
analyse a simulation more closely matching reality by including both
the lensing $B$-mode contribution and additional foreground
contributions from AME, and point-sources. The results are shown in
the bottom panels of Fig.~\ref{Fig:comm_r10-3}.  Note that the
reconstruction of the \emph{lensed} CMB $B$-mode power spectrum is
unbiased even at the reionization peak.  We then apply the
Blackwell-Rao estimator
Eqs.~(\ref{eq:br_likelihood})-(\ref{eq:br_posterior}) to derive the
posterior distributions for $r$ by fixing $A_{lens}=1$.  Although
treating the lensing contribution effectively as a nuisance term does
allow the bias on $r$ to be accommodated, it does not correct for the
additional cosmic variance introduced by the lensing signal. The
effect of this is to reduce the significance of the measurement of $r$,
as indicated in the right hand panel of the
plot. Specifically, we obtain
$r=\left(1.3\pm 0.5\right)\times 10^{-3}$ when fixing the amplitude of the
lensing $B$-modes to $A_{lens}=1$.  Although the information on
$A_{lens}$ derived from $T$- and $E$-modes, together with the possible
lensing potential reconstruction from $E$-modes, would help this
analysis, the inclusion of additional modes at higher $\ell$ obtained
from appropriate component separation approaches such as {\tt NILC} or
{\tt SMICA} would clearly be beneficial. This is discussed in
Sect.~\ref{sec:compsep_results}. In addition, actual delensing of the
CMB map with the quadratic estimators proposed in 
\cite{Hu2002,Hirata2003} will reduce the cosmic variance on $r$
\citep[][in prep.]{ECO_lensing}.  With this in mind,
Sect.~\ref{sec:compsep_results} presents revised estimates of $r$
after imposing an effective $60$\% delensing of the data, in addition
to foreground cleaning (simulation \#5).

Of course, the {\tt Commander} results also include maps of the
reconstructed physical components.  Figure~\ref{Fig:comm_cmb_map}
compares the the input CMB $Q$ map (top left panel) with the
reconstructed CMB $Q$ map (top right panel), and presents their
difference in the bottom left panel.  The bottom right panel shows a
map of the {\tt Commander} $\chi^2$ statistic given by
Eq.~(\ref{eq:chi2}) divided by the number of input frequency channels,
which measures the mismatch between the fitted model and the data in
each pixel. In principle, the $\chi^2$ map can be useful for
constructing a posteriori an optimal mask to then be employed for
parametric fitting, by rejecting any pixels where the $\chi^2$ value
is too large. We have not adopted this approach here. 
We also note that the inspection of foreground residuals in the
reconstructed CMB $Q$ and $U$ maps is not strongly informative with
respect to $B$-modes, since they are dominated by the $E$-mode
contribution. However, it can highlight a significant failure in
component seperation, likely driven by incorrect assumptions in the
foreground models adopted by {\tt Commander}, also indicated by the
$\chi^2$ map.

In Fig~\ref{Fig:comm_foregrounds}, we show the reconstruction by 
{\tt Commander} of the foreground spectral indices for thermal dust and
synchrotron for simulation \#4. The inputs and {\tt Commander}
estimates can be compared between the left and right columns.  While
the dust spectral indices (top panels) are accurately recovered over
the (unmasked) sky, with a similar fidelity for the dust temperature
(middle panels), it is evident that the reconstructed synchrotron
spectral indices (bottom right panel) are noisy. 
\core\ does not have enough sensitivity to measure the synchrotron parameters in the $60-600$ GHz frequency range,
and would benefit from additional low-frequency information, either from
external measurements provided by other instruments such as C-BASS
\citep{Irfan2015} or QUIJOTE \citep{Rubino-Martin2012}, or by the
addition of a low-frequency channel below $60$\,GHz (see the
discussion in Sect.~\ref{subsec:improvements}).

\subsection{Needlet Internal Linear Combination at high multipoles}
\label{subsec:ilc}

\begin{figure}
\centering
\includegraphics[width=0.5\textwidth]{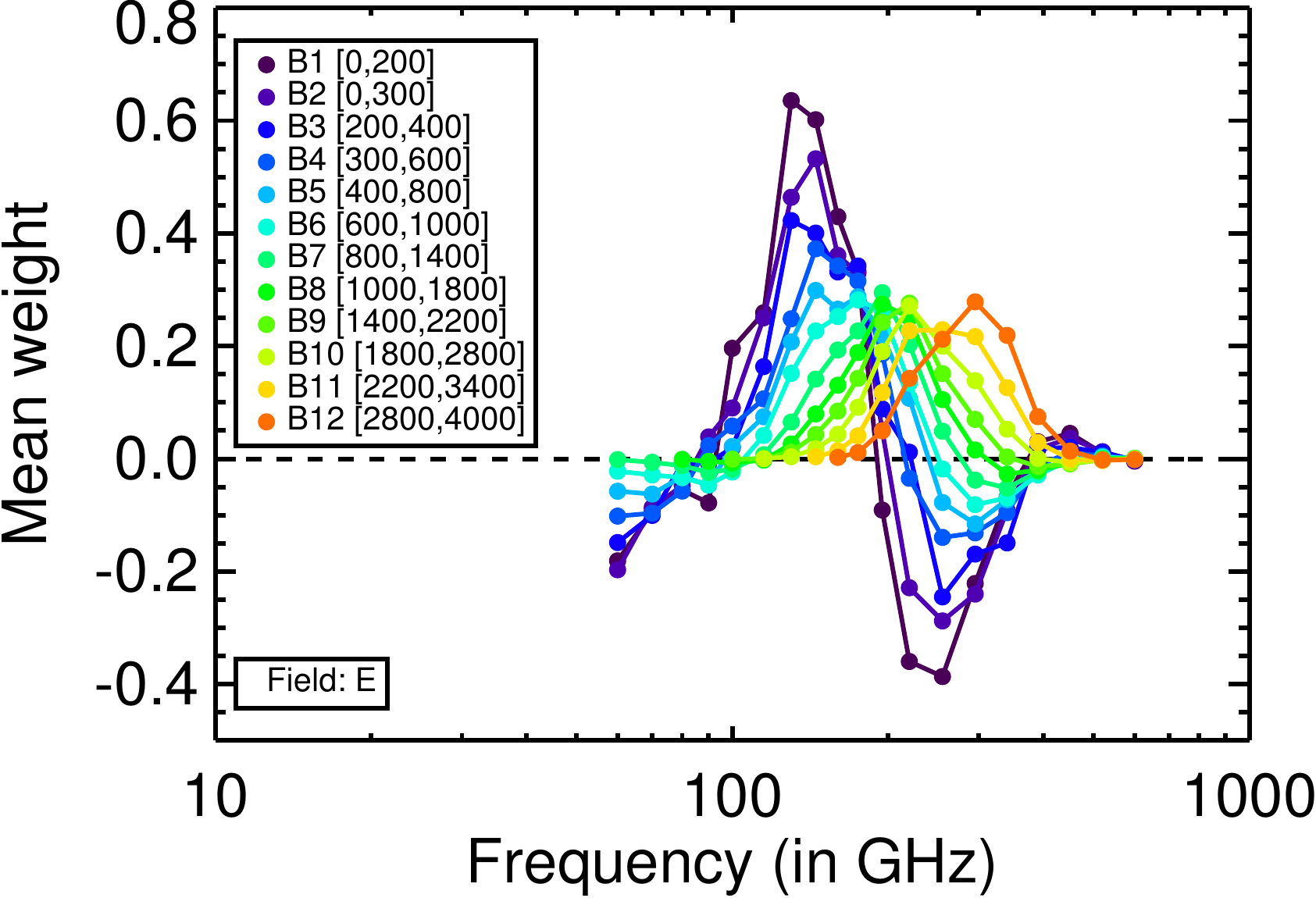}~
\includegraphics[width=0.5\textwidth]{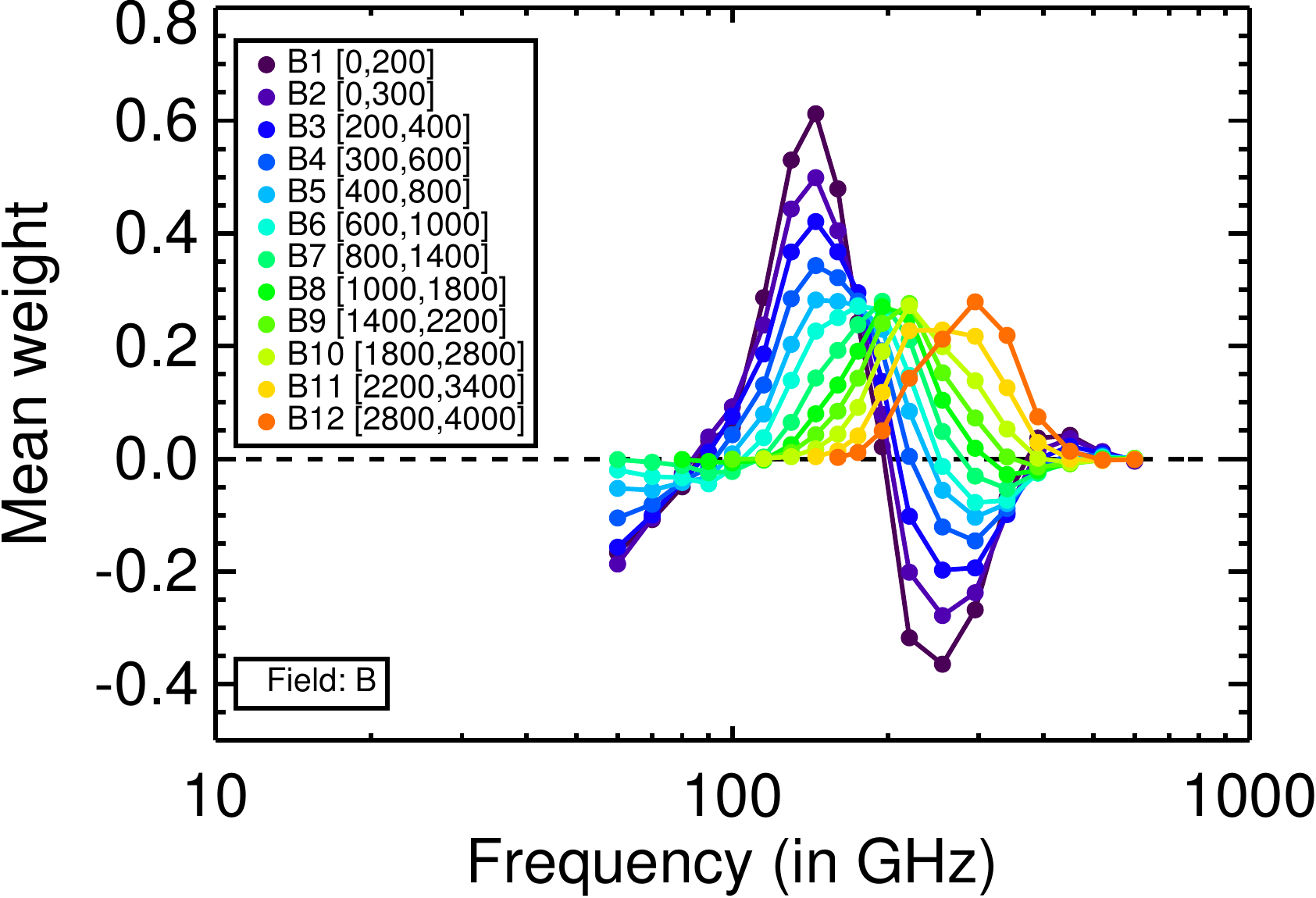}~
\caption{Full-sky average of the {\tt NILC} weights obtained for the
  complete simulation \#4 at different needlet scales and frequencies
  for the \core\ configuration. The left panel shows the weights for
  the $E$-modes and the right panel shows the weights for the
  $B$-modes. The legend denotes the needlet bands and associated
  multipole range.}
\label{fig:needlet-weight-core}
\end{figure}

The {\tt NILC} component separation method \citep{delabrouille2009full,Basak2012,
  Basak2013, planck2015_compsep} has been applied to the \core\
simulations to determine the power spectrum of the polarized CMB
signal at high-$\ell$. 
For details of the method, we refer the reader to
Sect.~\ref{subsec:nilc_method}.

\begin{figure}
\centering
\includegraphics[width=0.5\textwidth]{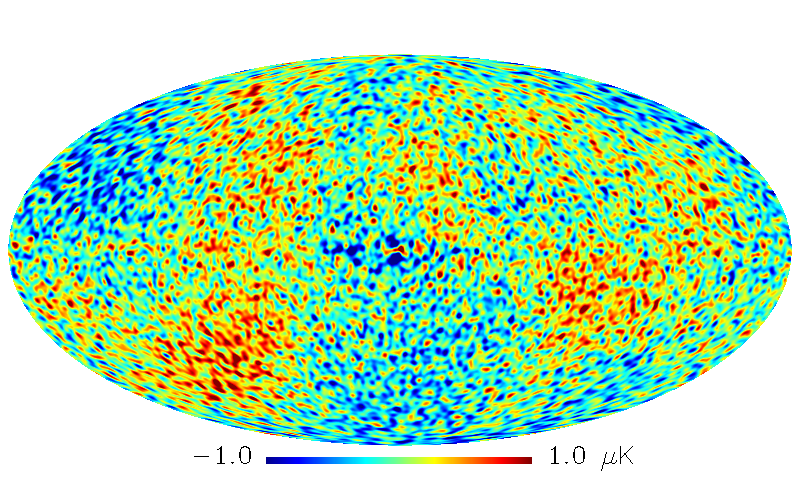}~
\includegraphics[width=0.5\textwidth]{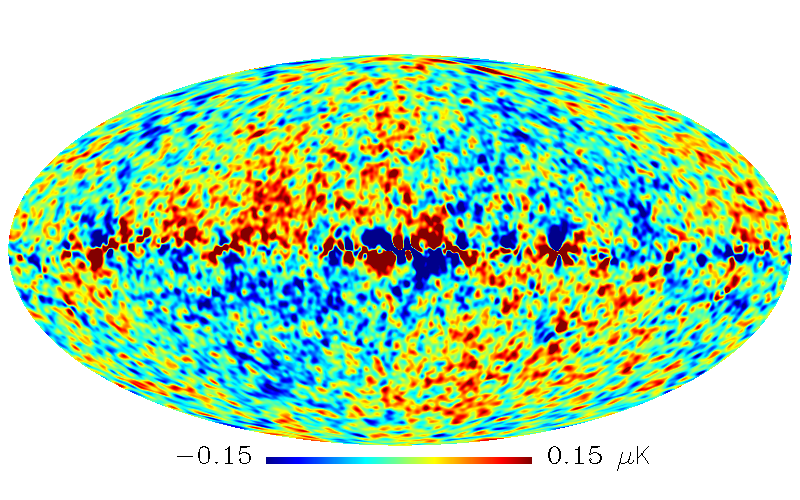}\\
\includegraphics[width=0.5\textwidth]{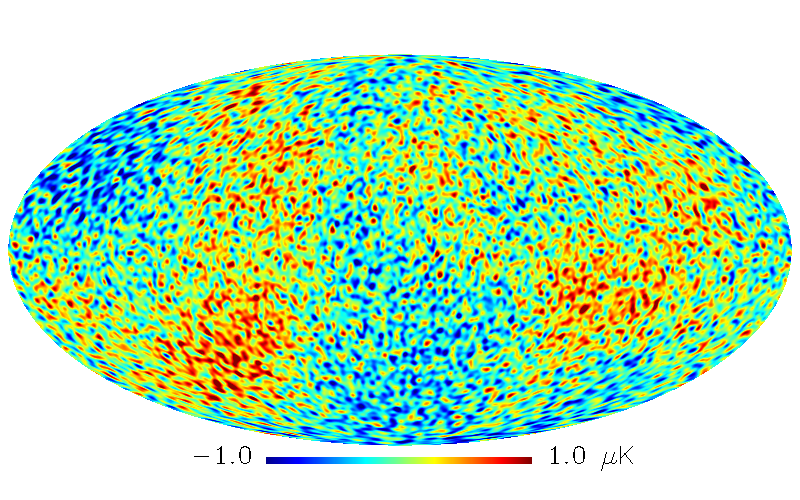}~
\includegraphics[width=0.5\textwidth]{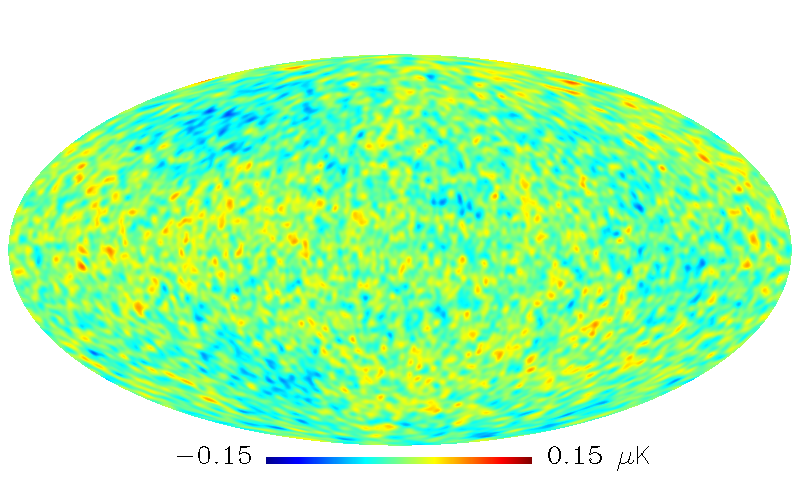}\\
\includegraphics[width=0.5\textwidth]{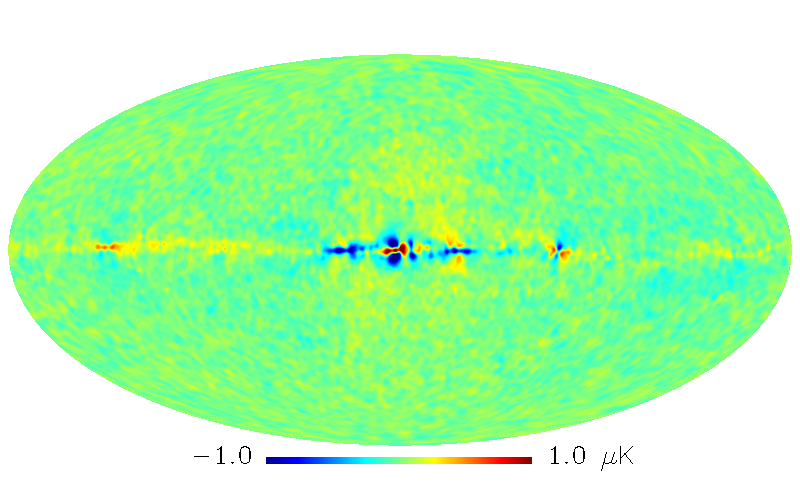}~
\includegraphics[width=0.5\textwidth]{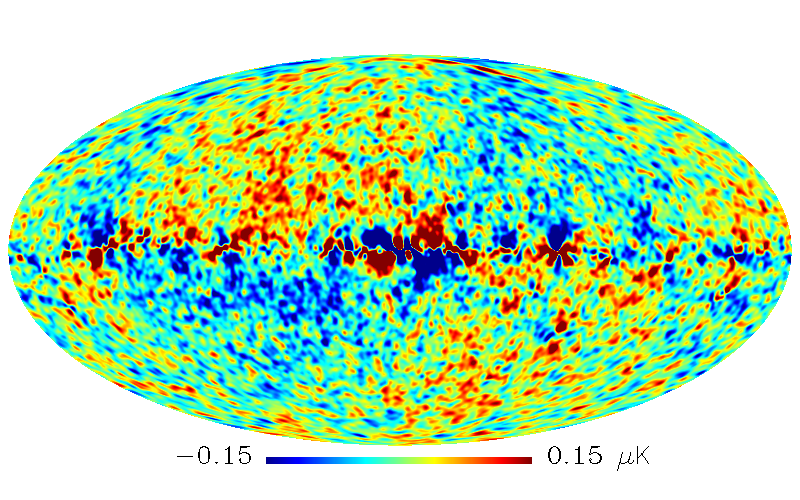}\\
\caption{{\tt NILC} reconstruction of the CMB $E$-mode (top left) and
  $B$-mode (top right) polarization maps from the \core\
  simulation \#1, at \nside\ $= 2048$ and $120'$ resolution. Middle panels show the input CMB $E$- and $B$-mode realisations of the simulation, at the same resolution. Bottom panels show the residuals in the {\tt NILC} maps with respect to the input CMB maps. 
}
\label{fig:recovered-ebcmb-core}
\end{figure}

\begin{figure}
\centering
\includegraphics[width=0.5\textwidth]{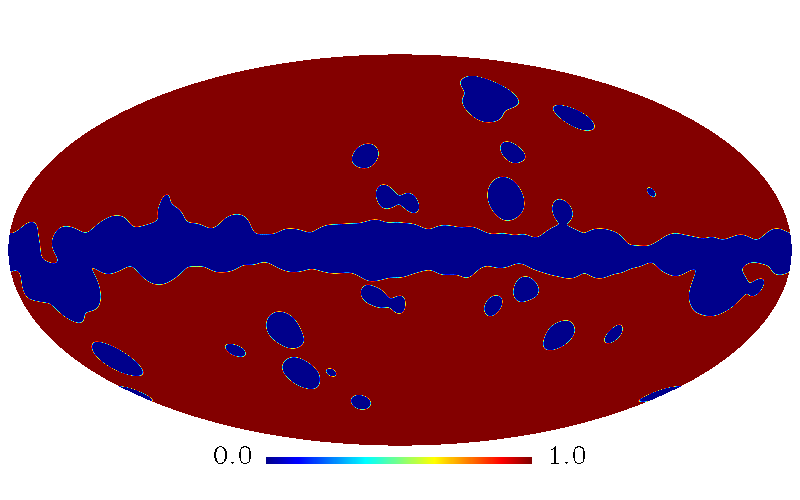}~
\includegraphics[width=0.5\textwidth]{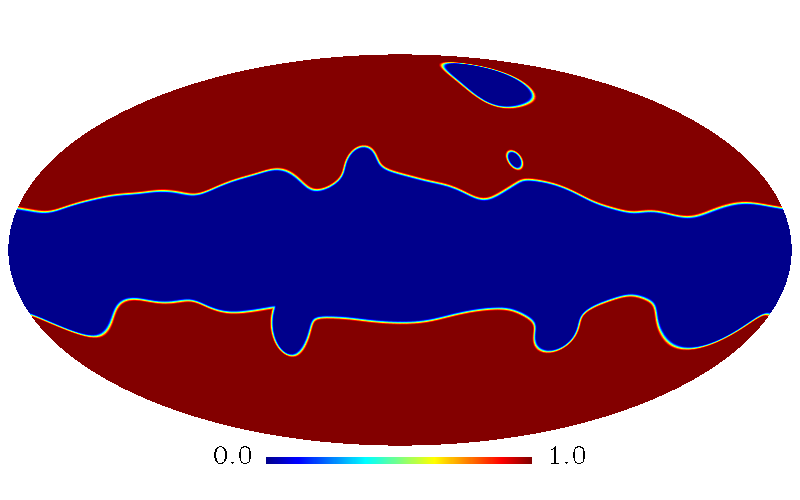}\\
\caption{Apodised masks used in the {\tt NILC} analysis to reduce the
  impact of residuals for foregrounds on the measurement of angular
  power spectra. The left panel shows the mask for $E$-modes
  ($f_{\rm sky}=0.80$) and the right panel shows the mask for
  $B$-modes ($f_{\rm sky}=0.60$).}
\label{fig:mask-core}
\end{figure}

The simulated sky maps are first convolved or de-convolved in harmonic space to 
put them all at the same angular resolution prior
to the application of the {\tt NILC} algorithm. Here we pick that of the smallest beam, 
i.e. $4'$ (see Table~\ref{tab:specs}).  Since the method as
currently implemented is applicable to scalar fields on the sphere,
sky maps of the $E$ and $B$ modes are constructed from the input
Stokes parameters, $Q$ and $U$, on the full sky (i.e., without
masking). The {\tt NILC} weights used to combine the multi-frequency
input data in order to estimate the CMB signal are then computed from
the full-mission $E$ an $B$ sky maps. The derived weights are also
applied to the half-mission maps, which are later used for both power
spectrum and noise estimation.

Figure~\ref{fig:needlet-weight-core} shows for the simulation \#4 the
full sky average of the {\tt NILC} weights for each frequency channel
and needlet band (as specified in Sect.~\ref{subsec:nilc_method}).
The overall features of the fullsky average of the {\tt NILC} weights at different frequencies and needlet bands do not notably change when AME, point-sources, and lensing effects are included or not in the simulation. The needlet weights are mostly determined by the galactic contamination, which dominates on large angular scales, and by the noise level, which dominates on small angular scales. It is apparent that most of the contributions to the reconstructed CMB polarization maps come from those channels with frequencies around $130$\,GHz and
$255$\,GHz. On large angular scales (the first needlet bands), the
$130$\,GHz channel gets a more significant weighting than the
$255$\,GHz channel. However, on small angular scales (the last needlet
bands), the situation is reversed due to the higher angular resolution
of the $255$\,GHz channel. Although the weights of the remainder of
the frequency channels are relatively low, they are important for
removing the Galactic foreground contamination.

However, as shown by Eq.~\ref{equ:ilc}, the reconstructed CMB $E$- and
$B$-mode maps cannot be completely free from contamination by residual
foregrounds and noise. Figure~\ref{fig:recovered-ebcmb-core}
clearly shows residual foreground contamination
along a narrow strip of the Galactic plane for E modes, while the B-mode reconstructed 
map is dominated by residuals of galactic contamination.  Therefore, for further
analysis, a set of conservative masks are derived from the
full-mission {\tt NILC} CMB maps.  These have been generated using the
following procedures.

For $E$-modes, the {\tt NILC} CMB map is filtered in harmonic space
through the Gaussian window
\begin{eqnarray}
f_{l}=\exp\left[-\frac{1}{2}\left(\frac{l-l_{centre}}{100}\right)^{2}\right],
\end{eqnarray} 
peaking at $l_{centre}=2275$ which corresponds to the multipole where
the sum of the power of the CMB and noise is minimum.  The resulting
map is then squared and smoothed with a Gaussian beam of
FWHM=$8^{\circ}$. The variance map obtained in this way is then
corrected for the noise contribution by subtracting a noise variance
map obtained using the same procedure as applied to the
half-difference of the {\tt NILC} half-mission maps. The $E$-mode
confidence mask shown in Fig.~\ref{fig:mask-core} (left panel) is then
obtained by thresholding the noise-corrected variance map.
The {\tt NILC} CMB $B$-mode map is dominated by
noise at almost all angular scales. Since most of the power of the CMB
$B$-modes is concentrated on large angular scales, the {\tt NILC} CMB
map is filtered through a Gaussian window function of FWHM=$30'$. The
resulting map is then squared and smoothed with a Gaussian beam of
FWHM=$20^{\circ}$. The resulting variance map is corrected for the
noise contribution as before. The confidence mask for $B$-modes shown
in the right panel of  Fig.~\ref{fig:mask-core} is then obtained by
thresholding the noise-corrected variance map.

For the computation of the angular power spectrum of the high
resolution {\tt NILC} CMB maps, we use a pseudo-$C_{\ell}$ estimator
\citep{Hivon2002,Chon2004,Szapudi2005}. This method is computationally
much faster than maximum likelihood and provides optimal results at
intermediate to high $\ell$s. In order to compute the covariance on
our measurement of angular power spectrum, we have followed the method
described in \cite{Tristram2005}. More details are given in
Appendix~\ref{subsec:nilc_method}.  The impact of instrumental noise
residuals on the measurement of angular power spectra is avoided by
evaluating cross-spectra between the {\tt NILC} half-mission
maps. Then, in order to reduce effects related to the sharp edges of
the masks on the measurement of the power spectra, the masks are
apodized through a cosine transition with length $1^{\circ}$ for $E$-modes and $2^{\circ}$ for
$B$-modes,
the apodization length for $B$-mode masks being larger because of the weakness of the signal compared to $E$-modes.
Given that the effective sky fraction of an apodized mask is 
$f_{\rm sky}=\sum_i w^2_i \Omega_i/(4\pi)$, where $w_i$ is the value of the mask in pixel
$i$ and $\Omega_i$ is the solid angle of the pixel, the resulting $E$-
and $B$-more masks have retained sky fractions of 80\% and 60\%
respectively. 
The computed pseudo-$C_{\ell}$ spectra are then used as input to a
likelihood-based estimation of the cosmological parameters.

\begin{figure}
\centering
\includegraphics[width=0.5\columnwidth]{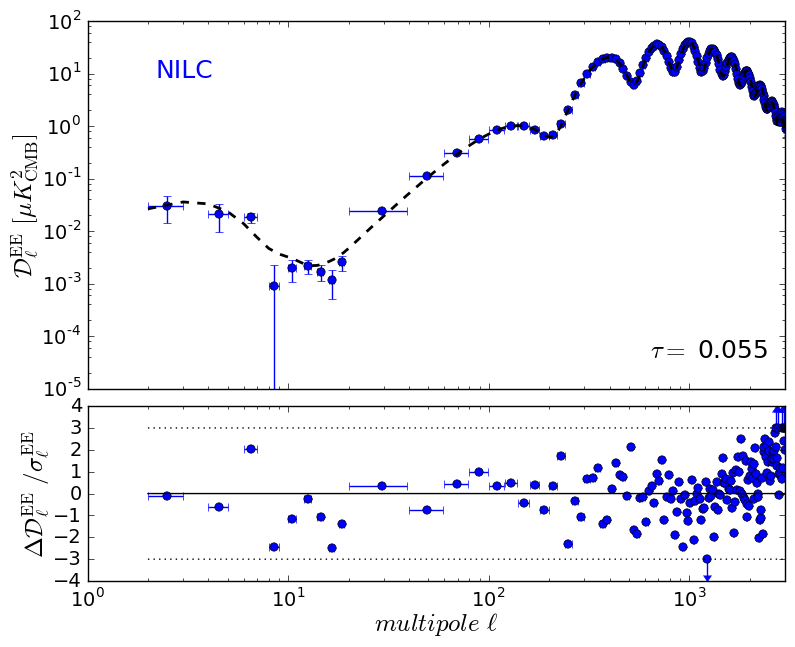}~
\includegraphics[width=0.5\columnwidth]{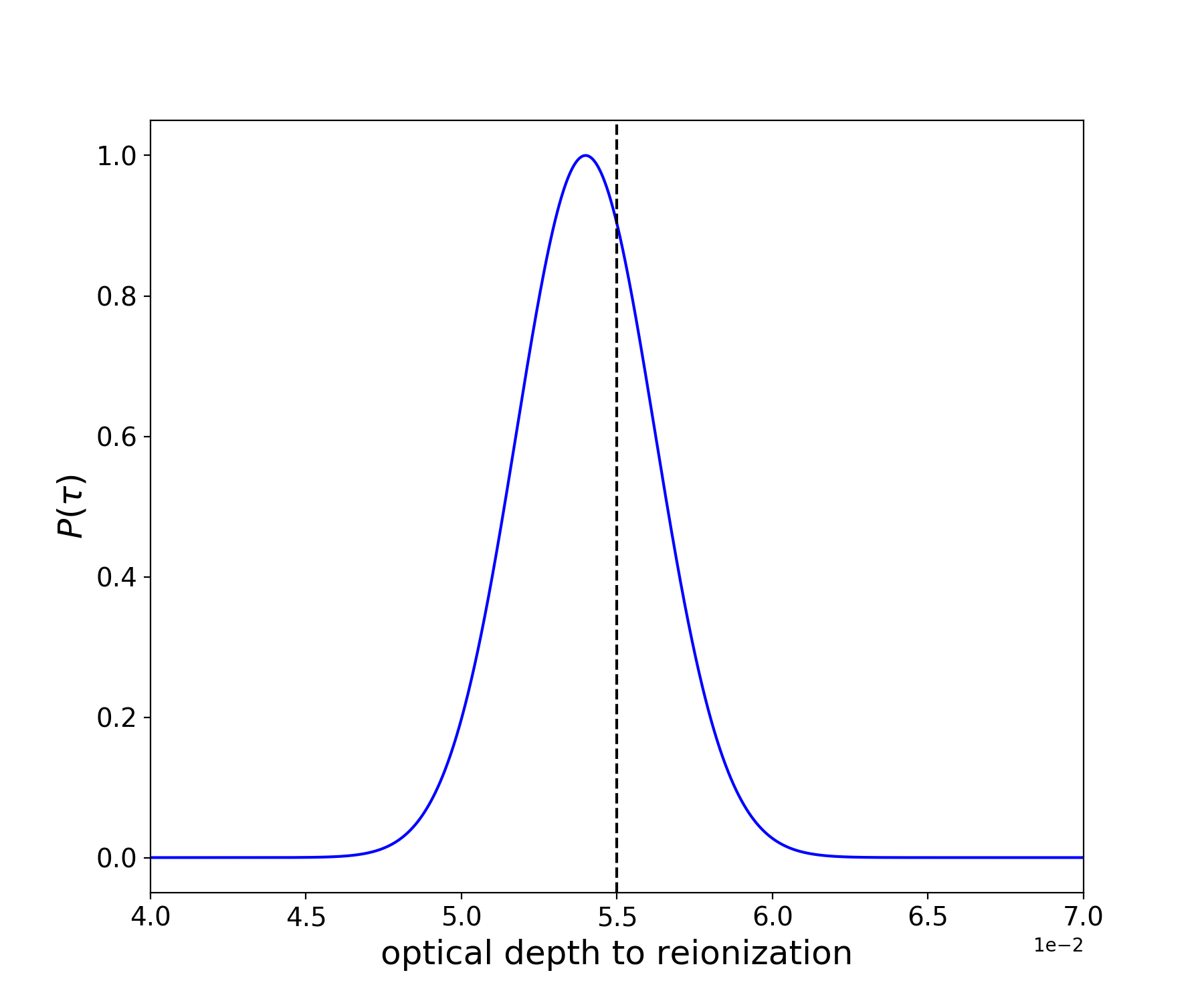}\\
\caption{{\tt NILC} results for the CMB $E$-modes determined from a
  simulation with the CMB optical depth $\tau=0.055$, and synchrotron
  and dust foregrounds. \emph{Left panel}: CMB $E$-mode power spectrum
  reconstruction.  The fiducial CMB $E$-mode power spectrum is
  indicated by the dashed black line while the power spectrum estimate
  is denoted by the dark blue points. The horizontal
    dotted lines show the 3$\sigma$ limits, while the vertical arrows
    signify outliers. \emph{Right
    panel}: Posterior distribution, $P(\tau)$, of the optical depth to
  reionization computed over the multipole range $20 \leq \ell \leq 359$.}
\label{fig:recovered-spectrum-core-ee}
\end{figure}

\begin{figure}
\centering
\includegraphics[width=0.5\columnwidth]{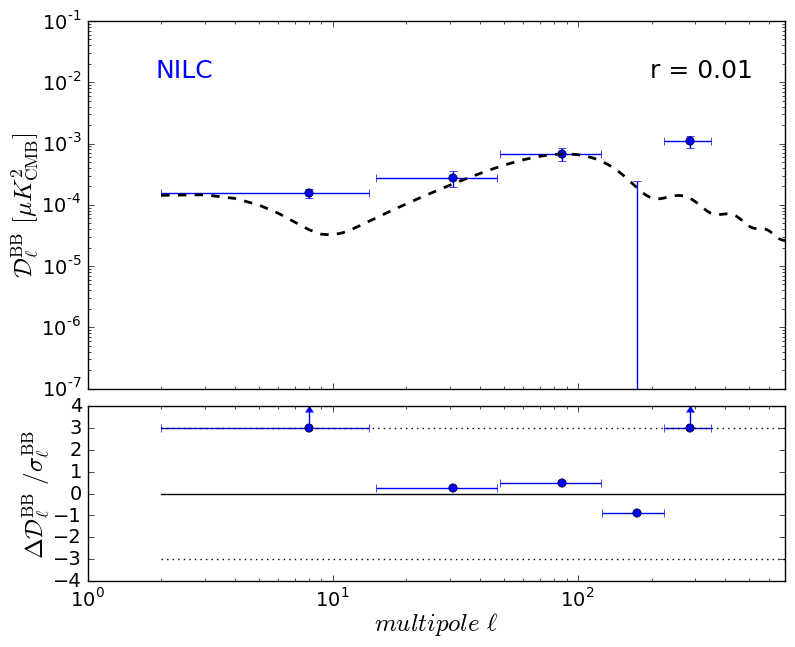}~
\includegraphics[width=0.5\columnwidth]{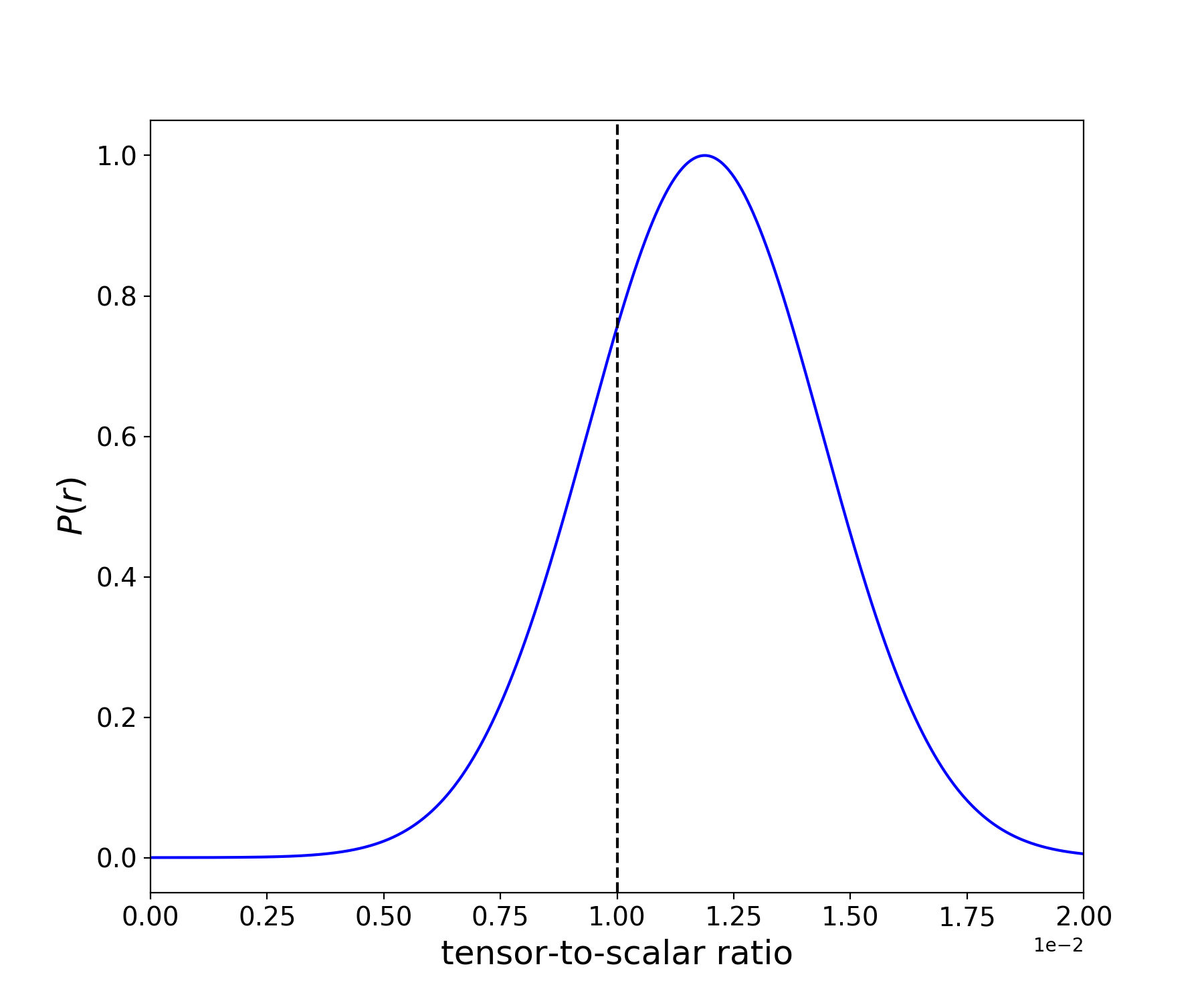} \\
\includegraphics[width=0.5\columnwidth]{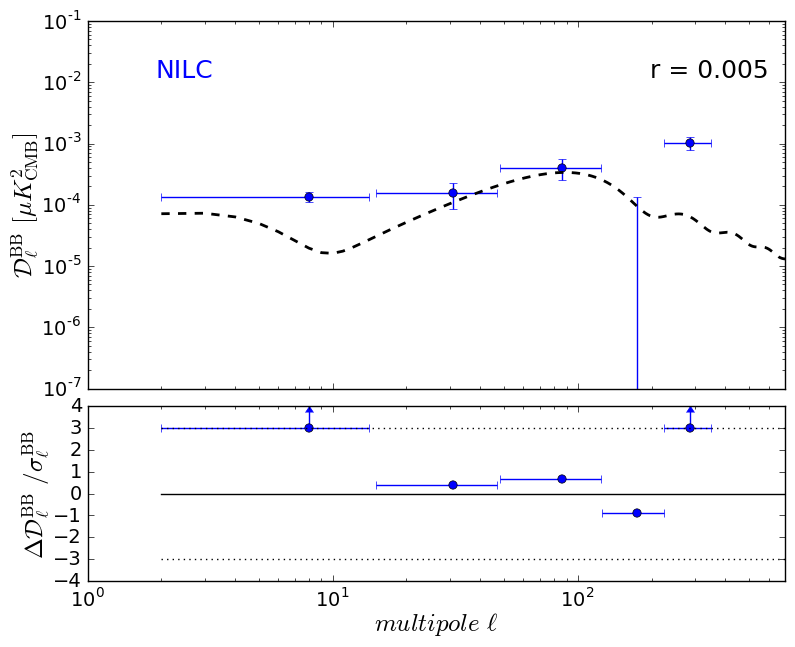}~
\includegraphics[width=0.5\columnwidth]{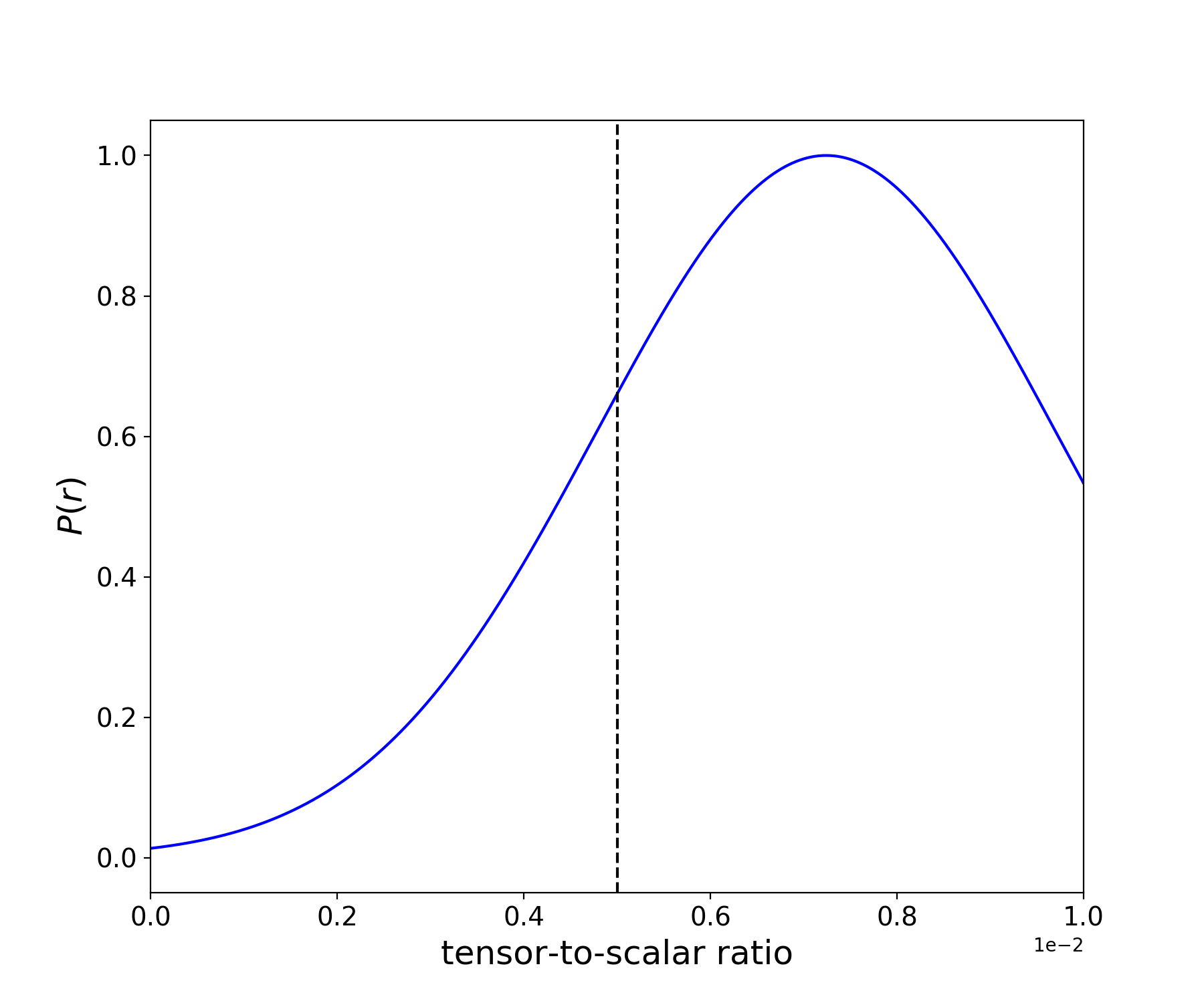}\\
\caption{{\tt NILC} results on CMB $B$-modes for simulations \#1
  ($r=10^{-2}$, \emph{top}), and \#2 ($r=5\times 10^{-3}$,
  \emph{bottom}), including synchrotron and dust
  foregrounds. \emph{Left panels}: CMB $B$-mode power spectrum
  reconstruction.  The fiducial primordial CMB $B$-mode power spectrum
  is denoted by a dashed black line, while the power spectrum
  estimates are indicated by the dark blue points. The
    horizontal dotted lines show the 3$\sigma$ limits, while the
    vertical arrows signify outliers.
  \emph{Right panels}: Posterior distribution, $P(r)$, of the
  tensor-to-scalar ratio calculated over the multipole  range $48 \leq \ell \leq 349$.}
\label{fig:recovered-spectrum-core-bb-larger-r}
\end{figure}

\begin{figure}
\centering
\includegraphics[width=0.5\columnwidth]{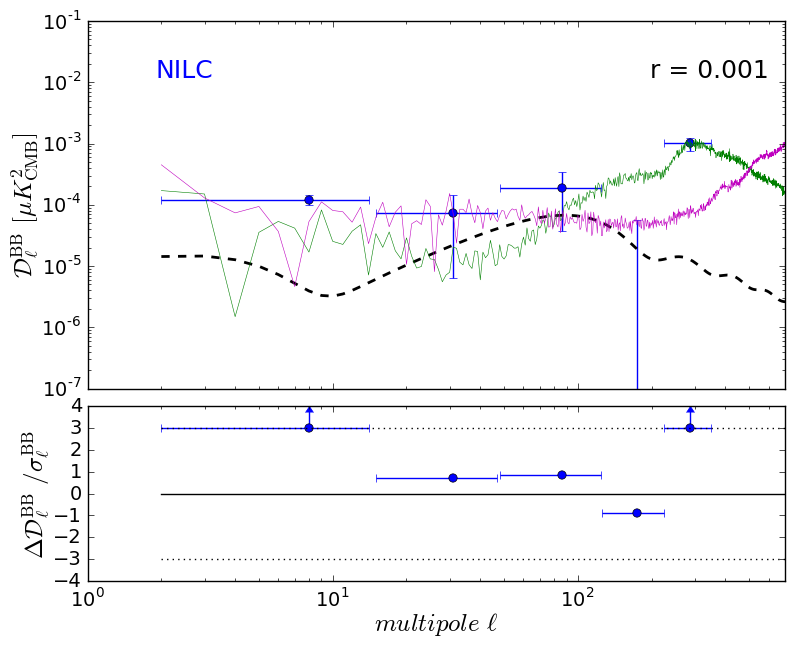}~
\includegraphics[width=0.5\columnwidth]{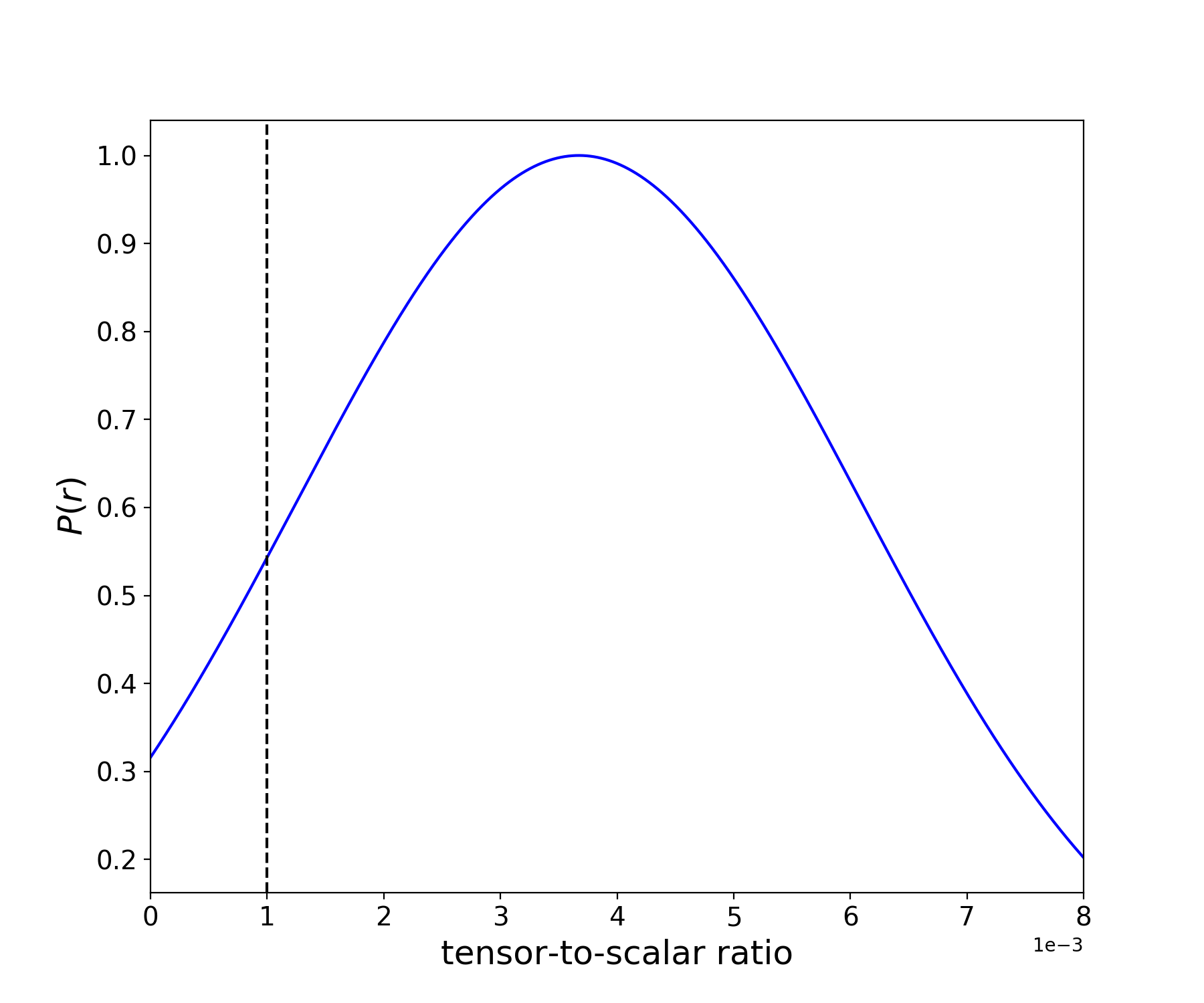}\\
\includegraphics[width=0.5\columnwidth]{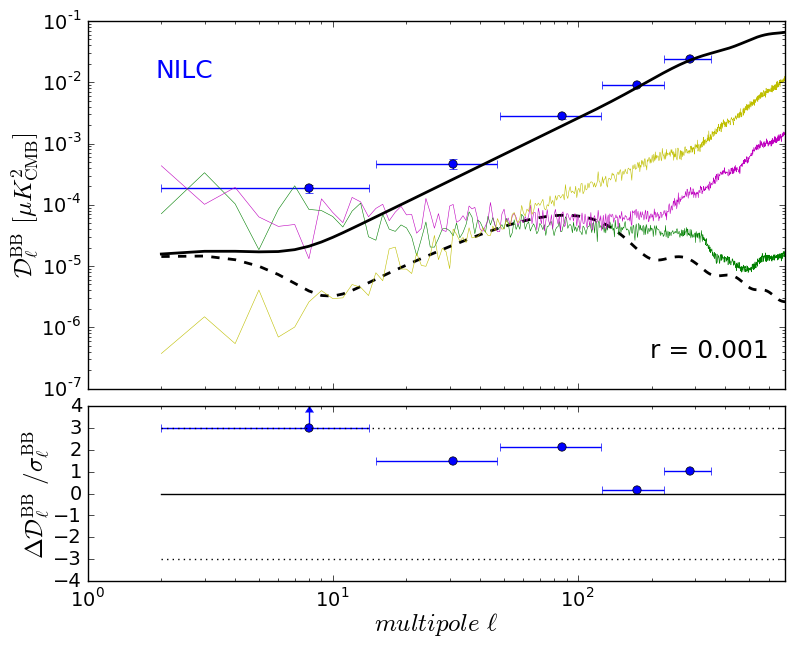}~
\includegraphics[width=0.5\columnwidth]{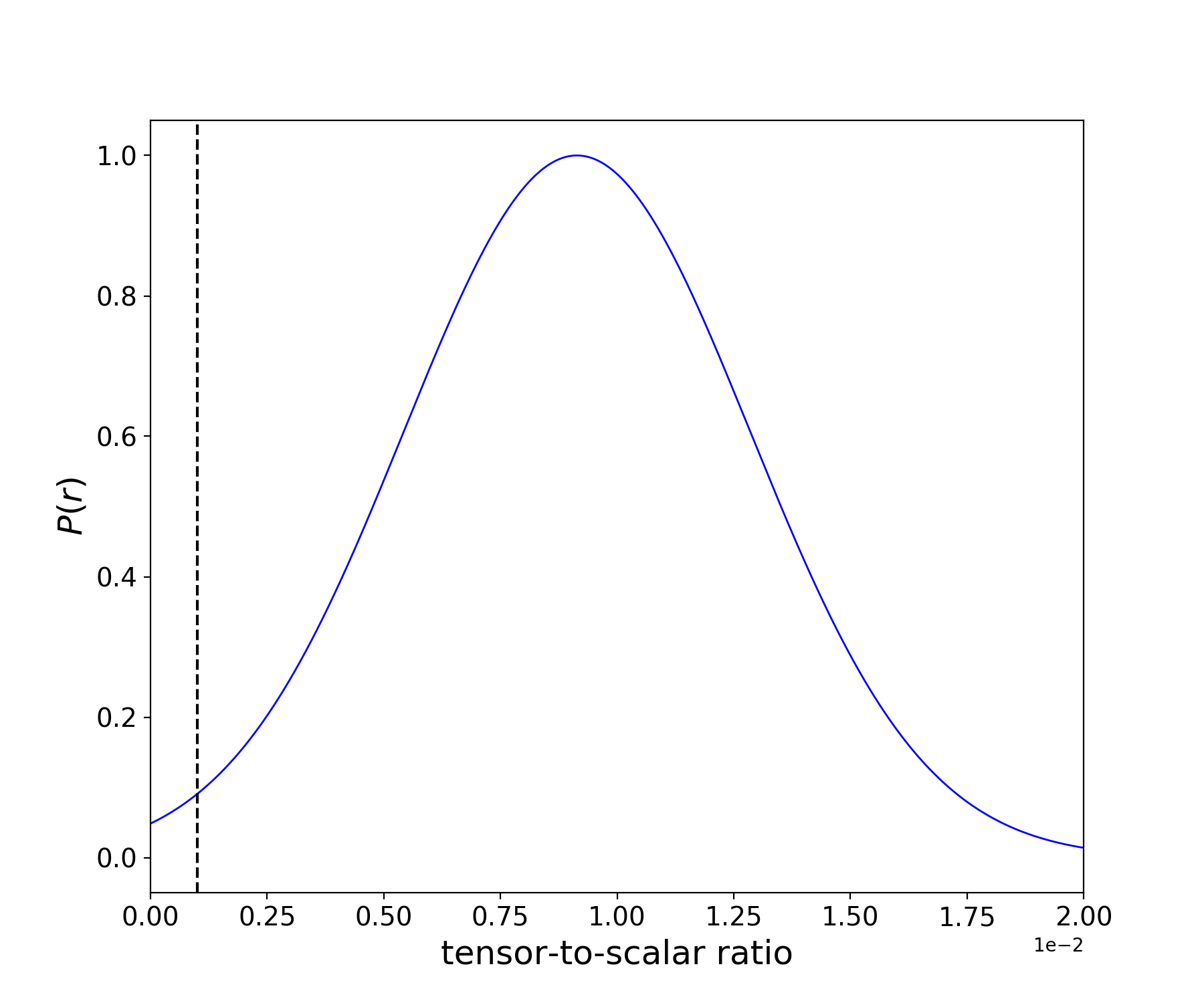}\\
\caption{{\tt NILC} results on CMB $B$-modes for $r=10^{-3}$ in the
  absence of lensing (\emph{top}; simulation \#3, including dust and
  synchrotron foregrounds) and in the presence of lensing
  (\emph{bottom}; simulation \#4, including dust, synchrotron, AME,
  and point-source foregrounds). \emph{Left panels}: CMB $B$-mode
  power spectrum reconstruction. The fiducial primordial CMB $B$-mode
  power spectrum is denoted by a dashed black line, while the solid
  black line shows the lensed CMB $B$-mode power spectrum. The power
  spectrum estimates are indicated by the dark blue points, while the foreground residuals are indicated by the coloured lines: synchrotron (green), dust (magenta), point-sources (yellow).
  The horizontal dotted lines show the 3$\sigma$ limits, while the
  vertical arrows signify outliers. \emph{Right panels}: Posterior
  distribution, $P(r)$, of the tensor-to-scalar ratio calculated over
  the multipole range $48 \leq \ell \leq 349$.}
\label{fig:recovered-spectrum-core-bb}
\end{figure}


Unlike in the case for {\tt Commander}, where a non-Gaussian
likelihood has been used via the Blackwell-Rao approximation to
calculate the posterior distribution of either the optical depth to
reionization, $\tau$, or the tensor-to-scalar ratio, $r$, here we
adopt a simple Gaussian likelihood.
Given an estimated and binned CMB power spectrum
$\widehat{C}^{XX}_{b}$, and a model CMB power spectrum
$C^{th\, XX}_{\ell}(p)$, which, in principle, can depend on any arbitrary
number of parameters $p$, we define the likelihood as 
\begin{eqnarray}
	\mathcal{L}(p) &=& \exp(-\frac{\chi^2(p)}{2}) \\
	\chi^2(p)	&=& \sum_{bb'} [\widehat{C}^{XX}_{b}-C^{th\, XX}_{b}(p)]\, \left(\widehat{\Xi}^{XX}\right)^{-1}_{bb'}\, [\widehat{C}^{XX}_{b'}-C^{th\, XX}_{b'}(p)] \text{,}
\end{eqnarray}
where $b,b'$ runs through the corresponding multipole bins,
$C^{th\, XX}_{b}(p)$ corresponds to the binned model power spectrum, and
$\left(\widehat{\Xi}^{XX}\right)^{-1}_{bb'}$ is the inverse of the
covariance matrix. 
The covariance matrix is computed for a fiducial
model, defined by the input tensor-to-scalar ratio.

When considering the likelihood for $\tau$, then $XX$=$EE$ and the
theoretical CMB $E$-mode power spectrum is given by
Eq.~\ref{eq:cl_ee_tau}.  In the case of $r$ and $A_{\rm lens}$, then
$XX$=$BB$ and the model $B$-mode power spectrum is defined as in
Eq.~\ref{eq:cl_bb_r}.  However, in either case, it may be necessary to
consider the addition of various nuisance terms to account for diffuse
foreground residuals at low-$\ell$ or point-source residuals at
high-$\ell$. These effects, $i$, will then be specified by a template
spectrum, $C_{\ell}^{BB,i}$, accompanied by the corresponding nuisance
amplitude, $A_i$, used to mitigate the effect \citep{hervias_2017}.

The $A_{\rm lens}$ parameter may not be well constrained by the
$B$-mode spectra alone. However, tight priors on it can be derived
from other CMB power spectra, together with the lensing power
spectrum. Indeed, the uncertainty on $A_{\rm lens}$ for \core\ has
been shown to be 0.013 from a joint analysis of the temperature and
$E$-mode spectra, improving to 0.012 when combined with the lensing
power spectrum, \citep[see][]{ECO_parameters2016}. This justifies the
choice below to fix $A_{\rm lens}$ at the fiducial value for a
$B$-mode only likelihood.

The left panel of Fig.~\ref{fig:recovered-spectrum-core-ee} shows the
$E$-mode angular power spectrum derived from the {\tt NILC} CMB
reconstruction. 
The reconstruction of the CMB $E$-modes is of good
  quality over a large range of angular scales up to $\ell=3000$, with
  only three outliers biased by more than $3\sigma$. The resulting
  estimate of the optical depth to reionization is
  $\tau=0.054 \pm 0.0022$, therefore a more than $24\sigma$
  measurement of $\tau$ by \core\ when accounting for multipoles
  $20 \leq \ell \leq 359$. When we add more multipoles to the likelihood, we measure unbiased $\tau$ estimates with higher significance, although in real-sky data analysis other $\Lambda CDM$ parameters should be jointly fitted with $\tau$ on small angular scales.
  Our $\tau$ estimates are just used as a validation criterion of the accurate CMB $E$-mode reconstruction, not as forecasts on the measurement of $\tau$ by \core. Detailed forecasts on the measurement of $\tau$ by \core\ are given in the accompanying paper on cosmological parameters \citep{ECO_parameters2016}, where it has been found that $\tau$ can be detected at $\sim 29 \sigma$ significance when using multipoles up to $\ell=3000$ and marginalizing over the other $\Lambda CDM$ cosmological parameters, but neglecting the foreground contamination. Using more multipoles and negelecting foregrounds obviously improves the significance of the detection, while marginalizing over other cosmological parameters decreases it.

As in the case of {\tt Commander}, the
accurate reconstruction of the CMB $E$-mode power spectrum serves as a
validation of the component separation method, while guaranteeing a
good proxy for potential delensing of CMB $B$-modes.

The left panels of Fig.~\ref{fig:recovered-spectrum-core-bb-larger-r}
present the reconstructed primordial CMB $B$-mode angular power
spectrum by {\tt NILC} for two values of the tensor-to-scalar ratio,
$r=10^{-2}$ (top) and $r=5\times 10^{-3}$ (bottom) in the absence of
gravitational lensing and when only synchrotron and dust foregrounds
are present. The binning of the reconstructed $B$-mode angular power
spectra again follows the weighting defined in
Eq.~\ref{eq:bin_weights}. The five multipole bins defined for {\tt
  NILC} are $\ell \in [2,14]$, $[15,47]$, $[48,124]$, $[125,224]$,
$[225,349]$.  While the reconstruction of the reionization peak of the
primordial CMB $B$-mode power spectrum is evidently biased by more
than $3\sigma$ in both cases, the recombination peak is accurately
recovered by {\tt NILC}.  The corresponding estimates of the
tensor-to-scalar ratio seen in
Fig.~\ref{fig:recovered-spectrum-core-bb-larger-r} (right panels) are
therefore based on the three {\tt NILC} multipole bins spanning the
range $48 \leq \ell \leq 349$.  In the two cases examined, we find
$r=\left(1.19\pm 0.25\right)\times 10^{-2}$ and
$r=\left(7.2\pm 2.4\right)\times 10^{-3}$ respectively.

When the tensor-to-scalar ratio falls to $r=10^{-3}$, the
reconstruction of primordial CMB $B$-modes becomes more
problematic. Figure~\ref{fig:recovered-spectrum-core-bb} shows the
$B$-mode power spectra derived from the {\tt NILC} CMB sky maps both
when lensing effects are either excluded (top) or included
(bottom). There is a significant bias on many angular scales, and
clear consequences for the estimated tensor-to-scalar ratio. {We have estimated the residual foreground contamination in the {\tt NILC} maps by applying the {\tt NILC} weights that were used for the reconstruction of the CMB $B$-mode map to the individual foreground components. As shown in Fig.~\ref{fig:recovered-spectrum-core-bb}, the residual Galactic foregrounds, thermal dust (solid magenta line) and synchrotron (solid green line), dominate the cosmological signal at $r=10^{-3}$ on large angular scales ($2\leq \ell \leq 50$), while residual extra-galactic point-sources (solid yellow line) are the main contaminant on the recombination peak and angular scales $\ell > 50$.

The origin of the bias is the fact that the ILC (and its needlet variant, the {\tt NILC}) is not
a power spectrum estimation tool, but a tool to make minimum variance maps of a component of interest.
The reconstructed map is contaminated by a mixture of foreground residuals and of noise. 
Residual foregrounds are seen in Fig.~\ref{fig:recovered-spectrum-core-bb} to be always larger than the 
primordial $B$ modes. Unlike the noise, which does not contribute to the estimated power spectrum as it is uncorrelated between the half-mission maps, 
foreground emission residuals bias the estimated power spectrum, with no natural prescription to de-bias. However, the level of this bias is quite informative as to what is the level of foreground residuals in the reconstructed maps.\\
Two ways to monitor such residuals with {\tt NILC} can be implemented.
One could modify the way that weights are computed, to reduce foreground contamination rather than noise in the maps, e.g. following the idea of the {\tt MILCA} algorithm \citep{2013A&A...558A.118H}. One can also at least avoid false detection of primordial B-modes by 
computing the power spectrum using variable masks, to check for a dependence of the measured power spectrum on the level of foreground emission.
If this test shows that there indeed is a dependence, it is a clear indication that the measurement is foreground-dominated.

\subsection{Spectral Matching Independent Component Analysis at high multipoles}
\label{subsec:smica}

As a third  independent component separation method, we apply the {\tt
  SMICA} algorithm \citep{2003MNRAS.346.1089D,Cardoso2008} to the
\core\ simulations. This method, based on matching the data to a
model in the spectral domain, allows the joint estimation of the spatial
power spectra of the components and of the noise, and of their mixing
coefficients. For details of the methodology, we refer the reader to
Sect.~\ref{subsec:smica_method}. In this paper, no attempt is made to
reconstruct maps corresponding to the derived best-fit spectra. 

Although the method works directly with $E$- and $B$-mode spectra, the
analysis utilises a mask constructed in a similar manner to that of
{\tt Commander} (Sect.~\ref{subsec:parametric}).  Polarization intensity
maps at 60 and 600~GHz are smoothed with a beam of 10 degrees FWHM and
extrapolated to the foreground minimum frequency of 70~GHz assuming
that the maps are dominated by synchrotron and thermal dust emission
respectively.  Each of the extrapolated maps is then thresholded with
respect to the smoothed 70~GHz simulated $P$ map to form two
independent masks, which are then combined. The resulting mask is then
apodized with an apodization length of 4 degrees, generating an
effective usable sky fraction of $f_{\mathrm{sky}}=0.40$, as shown in
Fig.~\ref{Fig:smica_mask}.

\begin{figure}[t]
\centering
\includegraphics[width=0.5\textwidth]{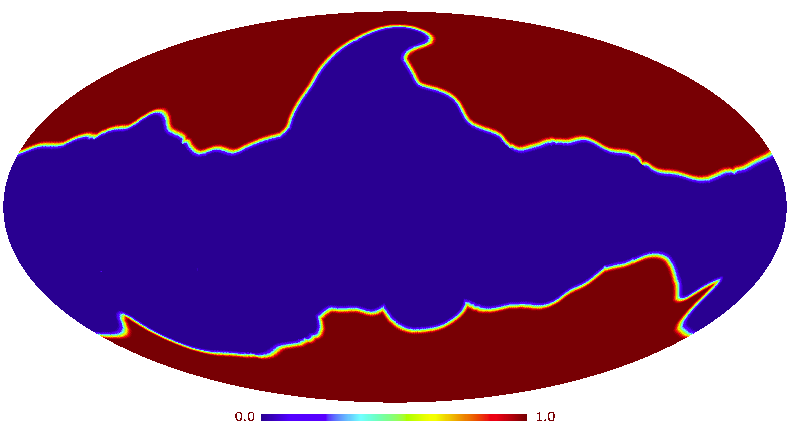}
\caption{Galactic mask used for the {\tt SMICA} analysis
  ($f_{\rm sky}=40$\%). The mask corresponds to a combination of masks
  built at the foreground minimum (70 GHz) starting from smoothed 60
  and 600 GHz maps extrapolated to that frequency.}
\label{Fig:smica_mask}
\end{figure}

\begin{figure}[t]
\centering
\includegraphics[width=0.5\textwidth]{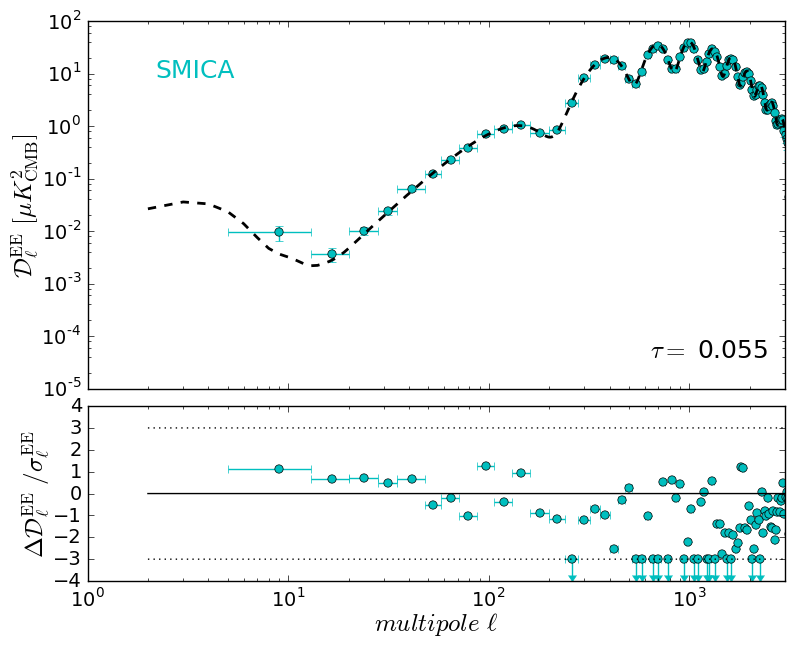}~
\includegraphics[width=0.5\textwidth]{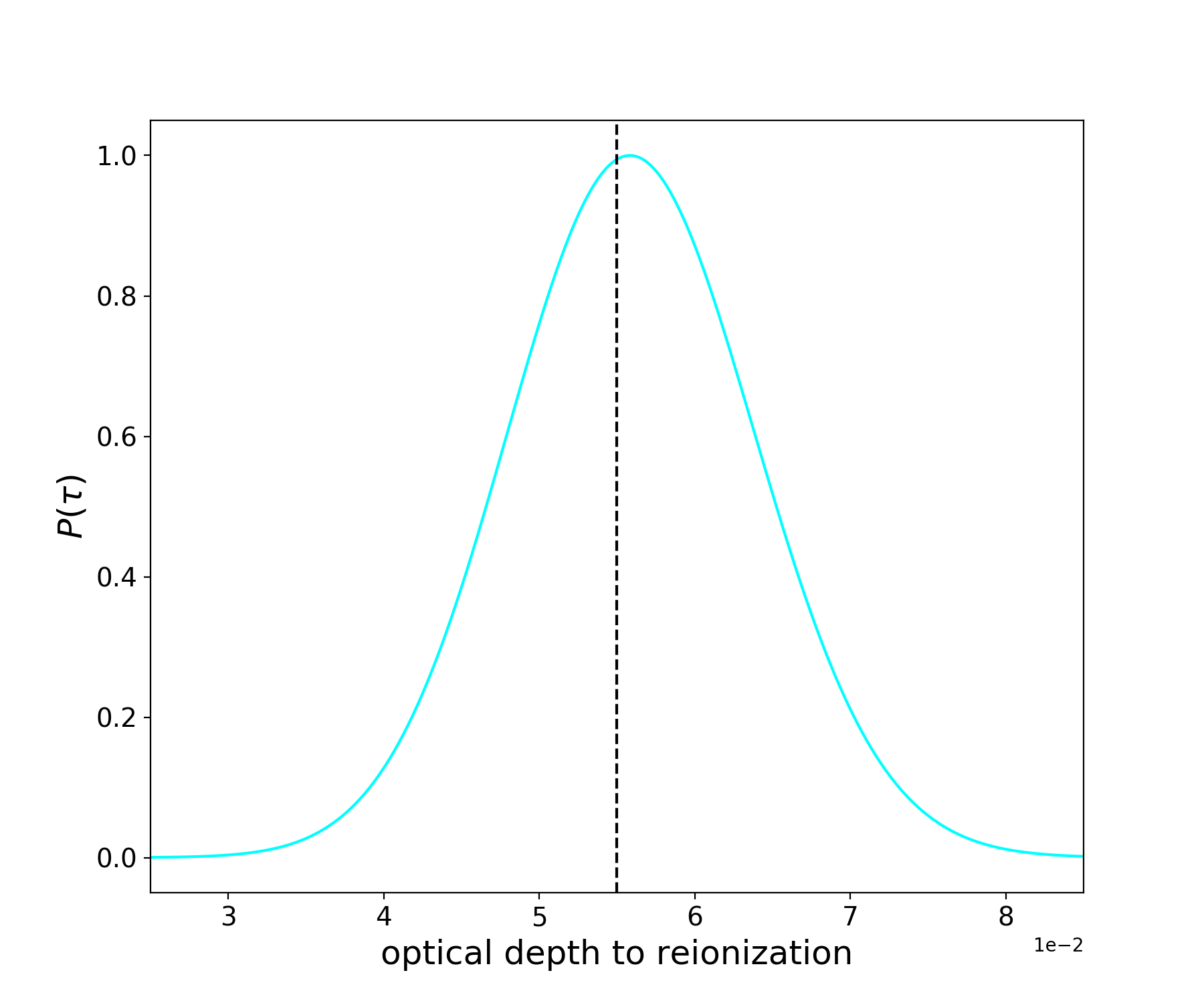}\\
\caption{{\tt SMICA} results on CMB $E$-modes determined from a
  simulation with the CMB optical depth $\tau=0.055$, and synchrotron
  and dust foregrounds. \emph{Left panel}: CMB $E$-mode power spectrum
  reconstruction. The fiducial CMB $E$-mode power spectrum is denoted
  by a dashed black line, while the power spectrum estimate is
  indicated by the light blue points. The horizontal dotted lines show
  the 3$\sigma$ limits, while the vertical arrows signify
  outliers.\emph{Right panel}: Posterior distribution, $P(\tau)$, of
  the optical depth to reionization computed over the multipole range
  $20 \leq \ell \leq 199$. }
\label{Fig:smica_ee}
\end{figure}

\begin{figure}[t]
\centering
\includegraphics[width=0.5\textwidth]{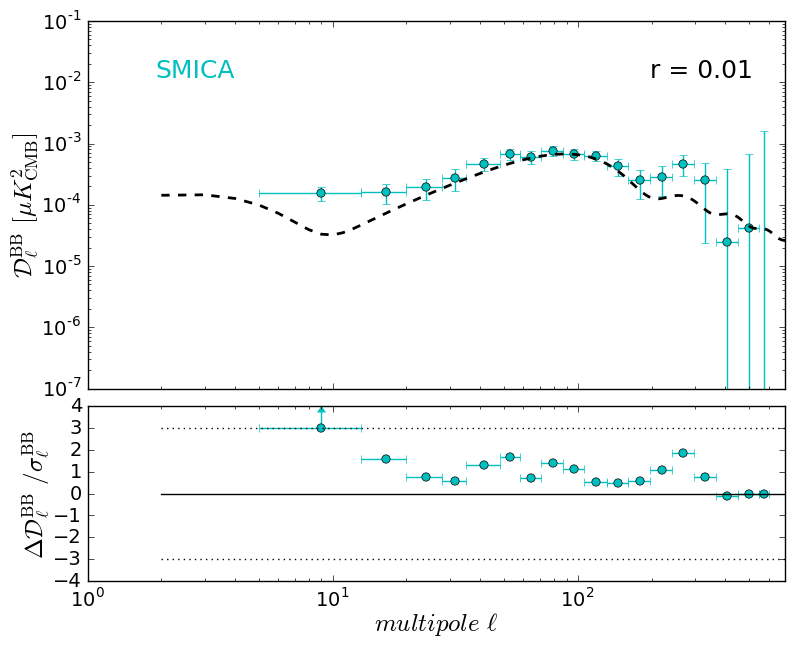}~
\includegraphics[width=0.5\textwidth]{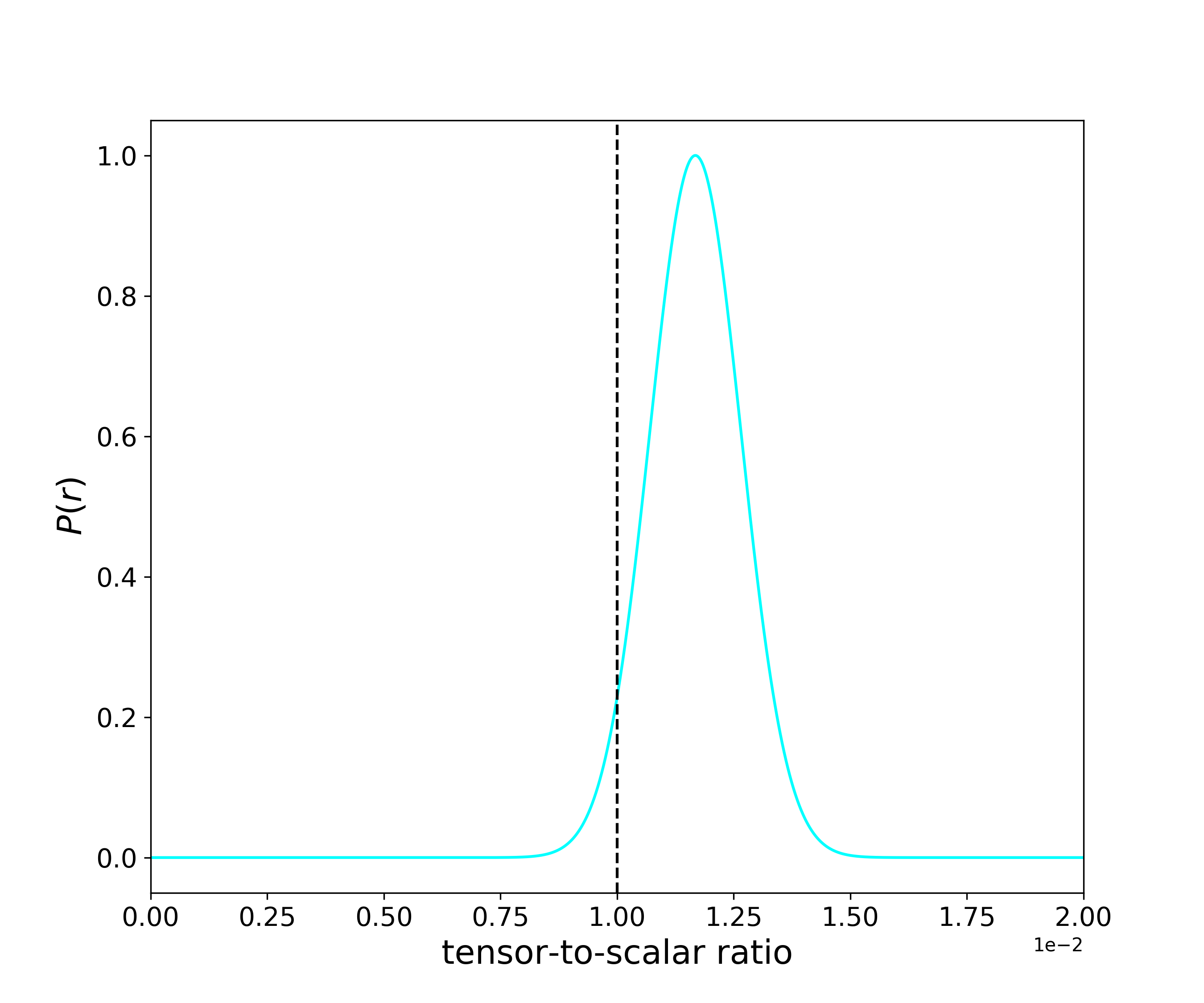}\\
\includegraphics[width=0.5\textwidth]{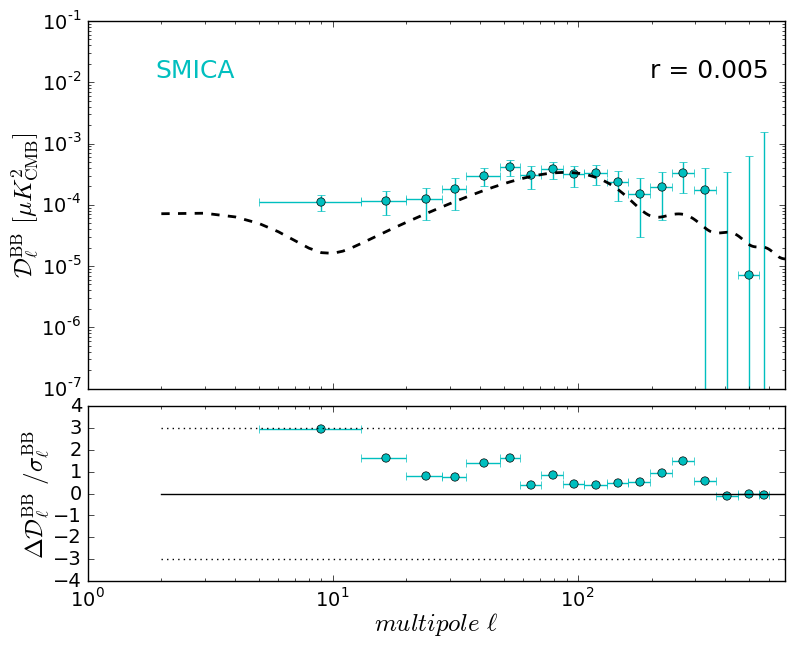}~
\includegraphics[width=0.5\textwidth]{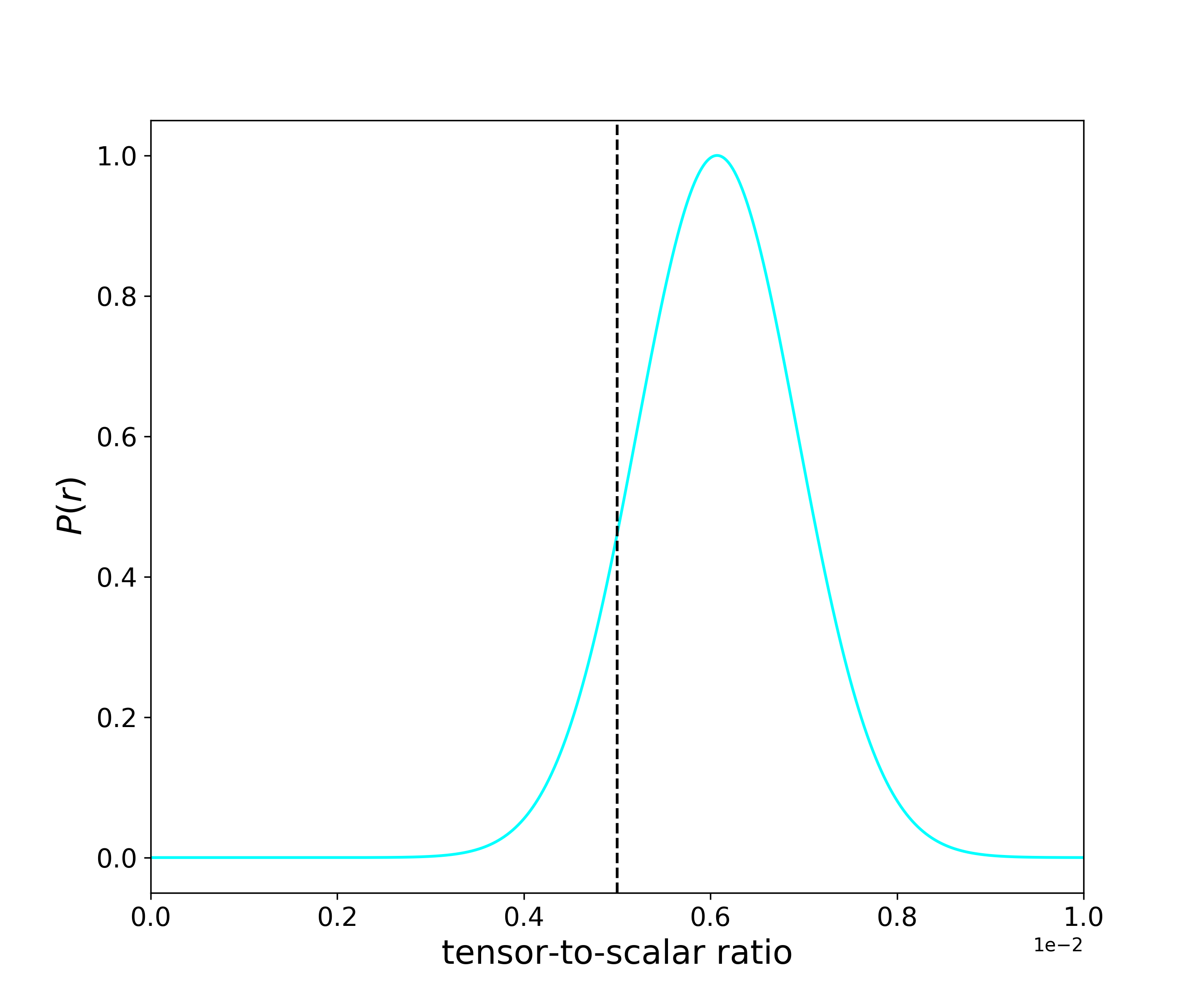}\\
\caption{{\tt SMICA} results on CMB $B$-modes for simulations \#1
  ($r=10^{-2}$, \emph{top}), and \#2 ($r=5\times 10^{-3}$,
  \emph{bottom}), including synchrotron and dust
  foregrounds. \emph{Left panels}: CMB $B$-mode power spectrum
  reconstruction.  The fiducial primordial CMB $B$-mode power spectrum
  is denoted by a dashed black line, while the power spectrum
  estimates are indicated by the light blue points. The horizontal
  dotted lines show the 3$\sigma$ limits, while the vertical arrows
  signify outliers. \emph{Right panels}: Posterior distribution,
  $P(r)$, of the tensor-to-scalar ratio computed over the multipole
  range $48 \leq \ell \leq 600$.}
\label{Fig:smica_r10-2}
\end{figure}

Figure~\ref{Fig:smica_ee} shows the reconstruction of the CMB $E$-mode
power spectrum by {\tt SMICA} for a simulation with optical depth to
reionization $\tau=0.055$, and dust and synchrotron foregrounds. The
recovered CMB $E$-mode power spectrum is consistent with the input
theory spectrum at the $2\sigma$ level over the majority of angular
scales up to $\ell=3000$, although there are quite a few outliers showing deviations exceeding $3\sigma$ on scales $\ell > 200$.  The 
reconstructed spectrum is an improvement on measurements by \textit{Planck}
\citep{Planck2015_XI} in terms of the absolute deviation from the
theory spectrum. However, the error-normalized deviations are more
significant since the uncertainties obtained with \core\ are smaller,
and indicate that further optimization in the {\tt SMICA} approach
should be explored.  The resulting estimate of the optical depth to
reionization is $\tau=0.0558 \pm 0.0082$, a measurement of $\tau$ by
\core\ at a level exceeding $6\sigma$ when including multipoles over
the range $20 \leq \ell \leq 199$ (and holding all other cosmological
parameters fixed at their input vaues). If we consider a multipole range with 
more bins, $20 \leq \ell \leq 600$, we measure $\tau = 0.077$. This result is
biased by the bins above $\ell \sim 500$.

The left panels of Fig.~\ref{Fig:smica_r10-2} show the reconstructed
primordial CMB $B$-mode power spectra determined by {\tt SMICA} from
simulations \#1 and \#2 for different values of the tensor-to-scalar
ratio, in the absence of gravitational lensing and when only
synchrotron and dust foregrounds are present.  The binning of the
reconstructed $B$-mode angular power spectra follows the weighting
scheme defined in Eq.~\ref{eq:bin_weights}.  The 18 multipole bins
defined for {\tt SMICA} are $\ell \in [5,12]$, $[13,19]$, $[20,27]$,
$[28,34]$, $[35,47]$, $[48,57]$, $[58,70]$, $[71,86]$, $[87,105]$,
$[106,130]$, $[131,159]$, $[160,196]$, $[197,241]$, $[242,296]$,
$[297,364]$, $[365,447]$, $[448,549]$, $[550,600]$.  In both cases,
the recombination peak, between $\ell\sim 60$ and $\ell\sim 200$, is
recovered by {\tt SMICA} within the $1\sigma$ error bound.  The
right panels in these figures show the estimated tensor-to-scalar
ratio obtained from the reconstructed CMB $B$-mode power spectra on
angular scales $48 \leq \ell \leq 600$. We recover values of
$r=\left(1.2\pm 0.1\right)\times 10^{-2}$ and
$\left(6.1\pm 0.9\right)\times 10^{-3}$ for the fiducial values of
$r=10^{-2}$ and $r=5\times 10^{-3}$ respectively. Using only those
multipole scales corresponding to the recombination peak and above,
these results are consistent with detections of the tensor-to-scalar
ratio at a level of $5\sigma$ or above.

However, as we have seen in previous sections, for a value of
$r=10^{-3}$, the reconstructed CMB $B$-mode power spectrum can show
evidence of foreground residuals and bias relative to the input theory
spectrum.  The top left panel of Fig.~\ref{Fig:smica_r10-3} presents
results for simulation \#3, and indicates that an excess of power is
seen on most angular scales for $\ell \leq 400$ due to foreground
residuals. The corresponding estimate of the tensor-to-scalar ratio
(top right panel), $r=\left(2.1\pm 0.8\right)\times 10^{-3}$, is not
formally biased, albeit more than $1\sigma$ high, but the error has
increased so that only an upper limit on $r$ can be claimed.

\begin{figure}[t]
\centering
\includegraphics[width=0.5\textwidth]{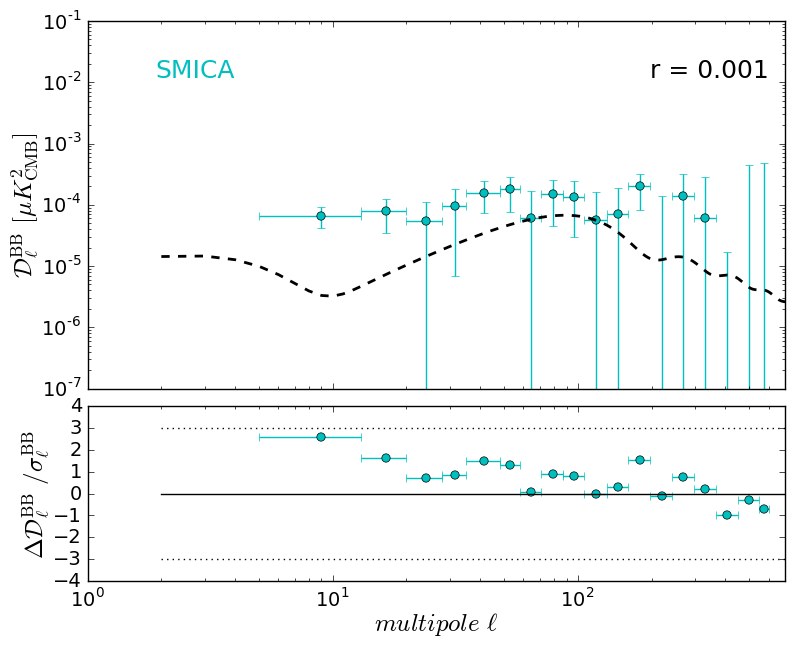}~
\includegraphics[width=0.5\textwidth]{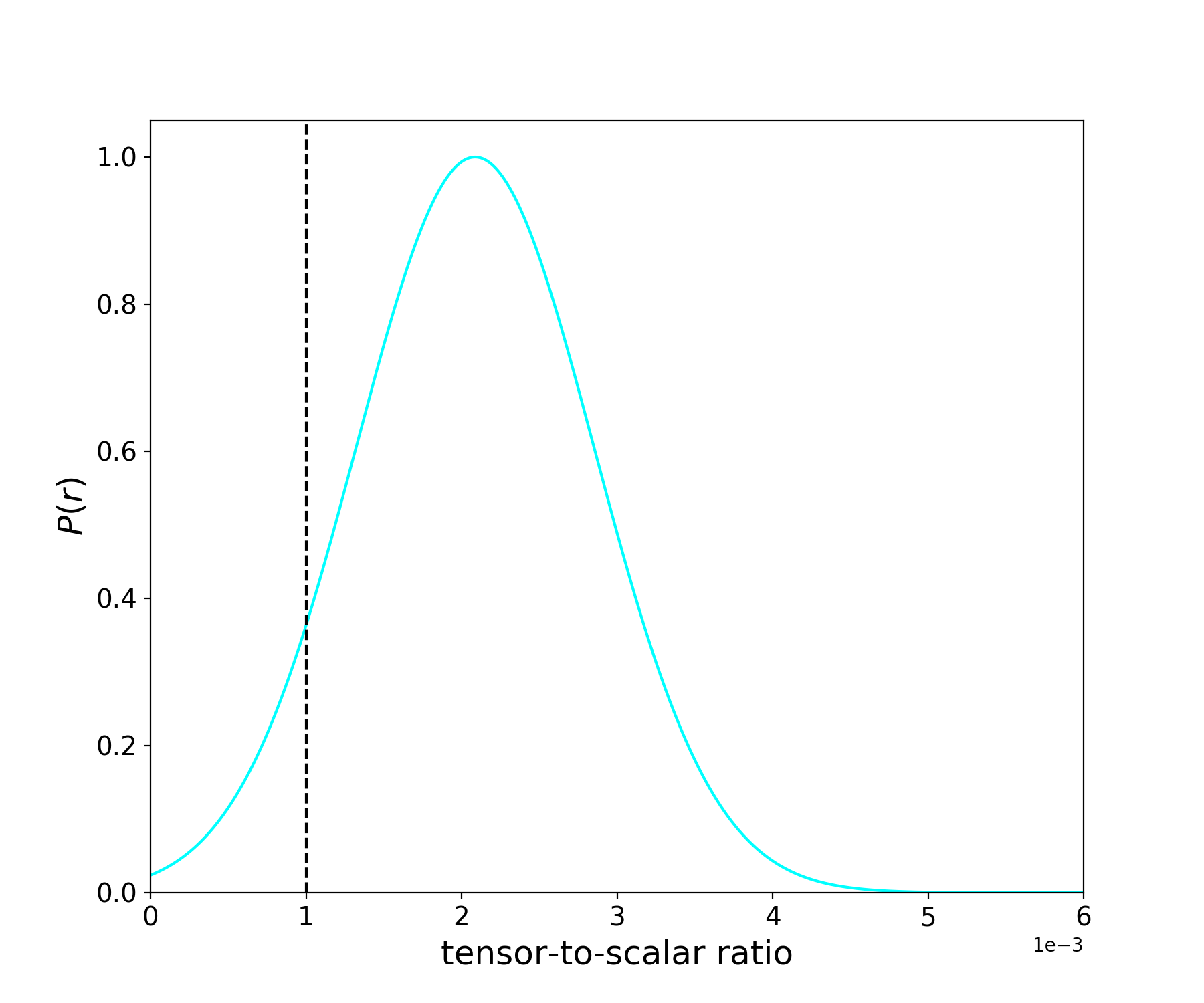}\\
\includegraphics[width=0.5\textwidth]{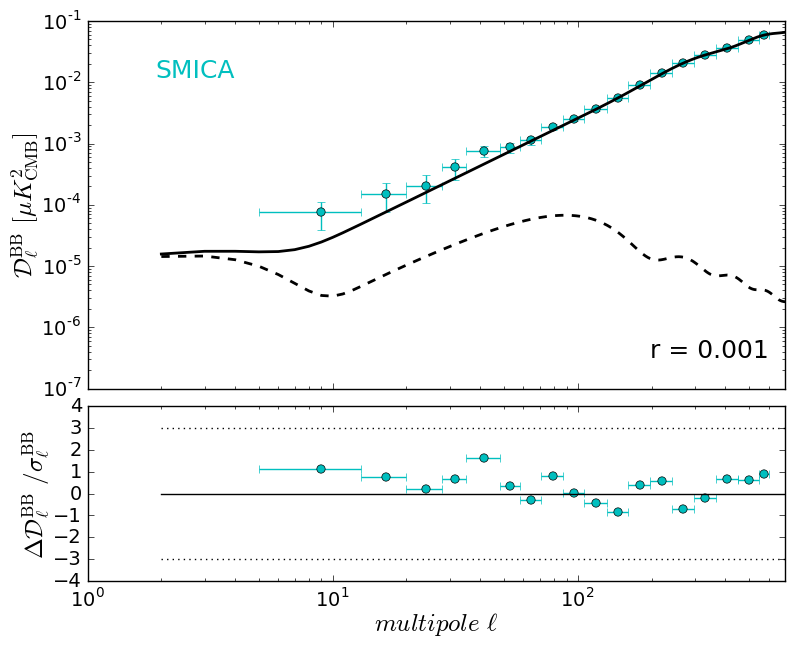}~
\includegraphics[width=0.5\textwidth]{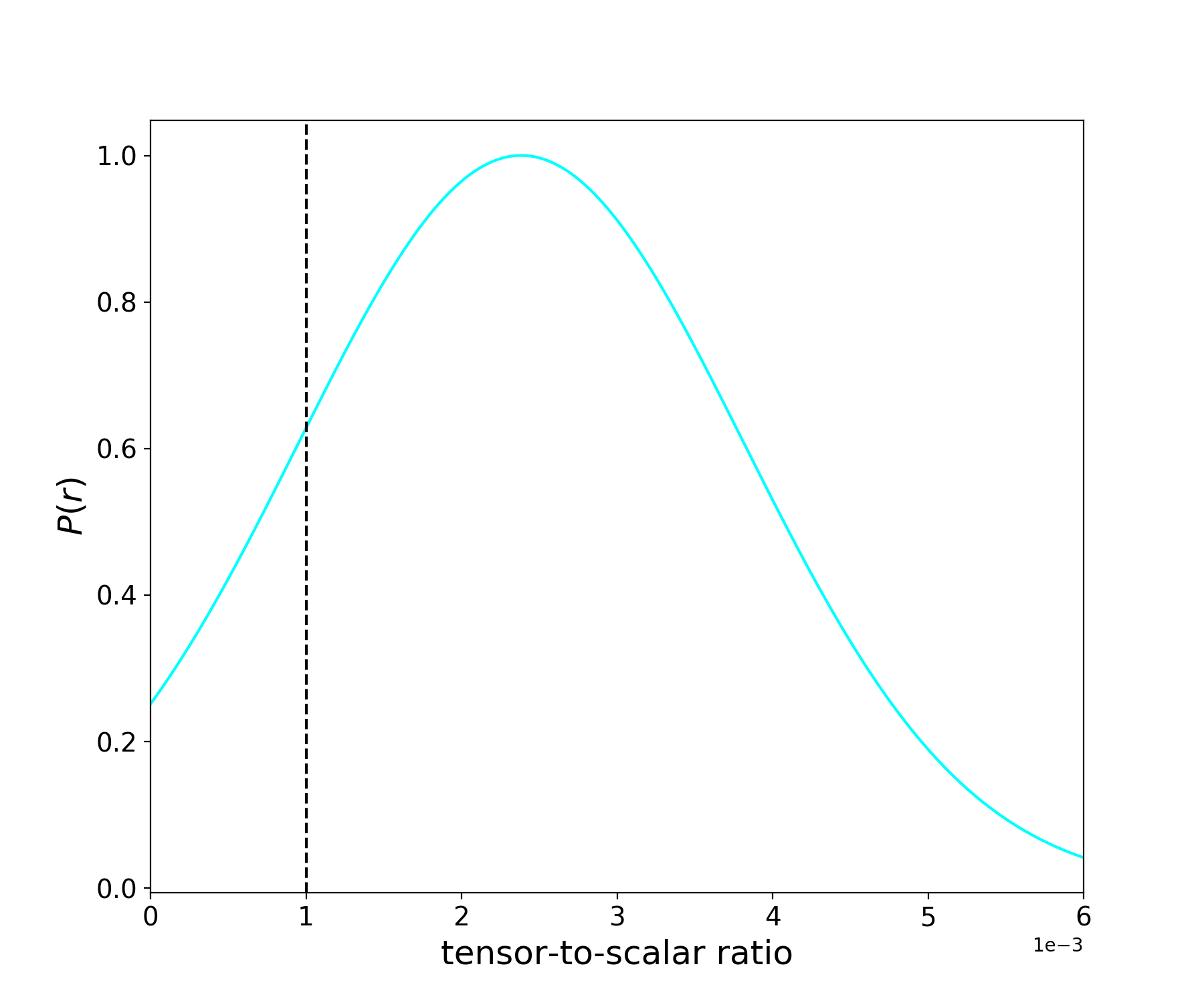}\\
\caption{{\tt SMICA} results on CMB $B$-modes for $r=10^{-3}$ in the
  absence of lensing (\emph{top}; simulation \#3, including dust and
  synchrotron foregrounds) and in the presence of lensing
  (\emph{bottom}; simulation \#4, including dust, synchrotron, AME,
  and point-source foregrounds). \emph{Left panels}: CMB $B$-mode
  power spectrum reconstruction. The fiducial primordial CMB $B$-mode
  power spectrum is denoted by a dashed black line, while the solid
  black line shows the lensed CMB $B$-mode power spectrum. The power
  spectrum estimates are indicated by the light blue points. The
  horizontal dotted lines show the 3$\sigma$ limits, while the
  vertical arrows signify outliers. \emph{Right panels}: Posterior
  distribution, $P(r)$, of the tensor-to-scalar ratio computed over
  the multipole range $48 \leq \ell \leq 600$. }
\label{Fig:smica_r10-3}
\end{figure}

In the presence of gravitational lensing effects (simulation \#4),
{\tt SMICA} performs an accurate reconstruction of the lensed CMB
$B$-mode power spectrum when $r=10^{-3}$ over a large multipole range
$20 < \ell< 600$, and is consistent with the fiducial lensed CMB $B$-mode
power spectrum within $1\sigma$ (see the bottom left panel of
Fig.~\ref{Fig:smica_r10-3}).  The tensor-to-scalar ratio $r=10^{-3}$
is recovered without significant bias (bottom right panel of
Fig.~\ref{Fig:smica_r10-3}), but with a large uncertainty, when either
fitting or fixing $A_{lens}$ in the likelihood, with an estimate of
$r=\left(2.4\pm 1.4\right)\times 10^{-3}$.
 
Currently, it appears that the measurement of the primordial CMB
$B$-mode power spectrum with {\tt SMICA} based on \core\ observations
can not be achieved at the required level of accuracy and precision 
when $r=10^{-3}$, yet is possible for $r=5\times 10^{-3}$ in the
absence of lensing effects.
Evidently, further optimization of the methodology is desirable.


\section{\core\ results on CMB $B$-mode measurements}
\label{sec:compsep_results}

\begin{figure}[htbp]
	\begin{center}
          \includegraphics[width=0.5\textwidth]{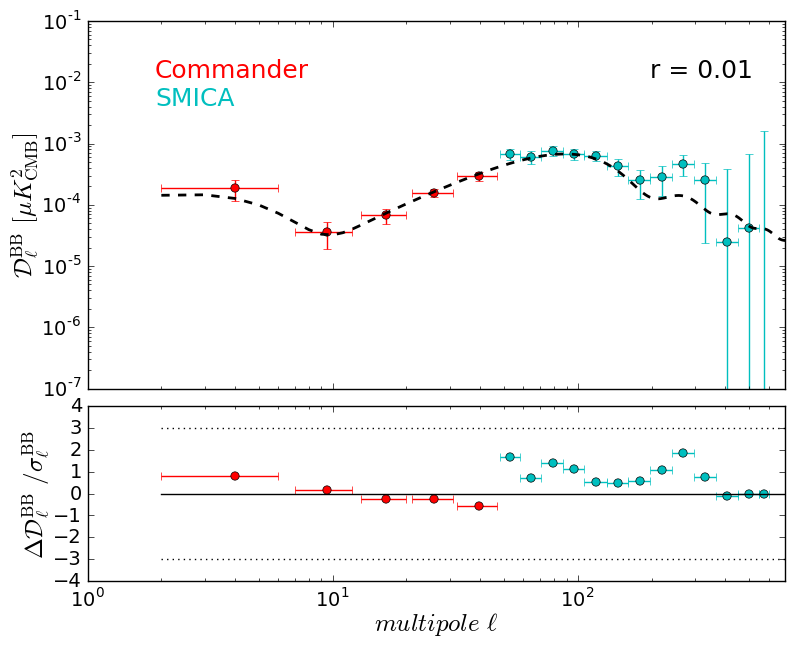}~
	  \includegraphics[width=0.5\textwidth]{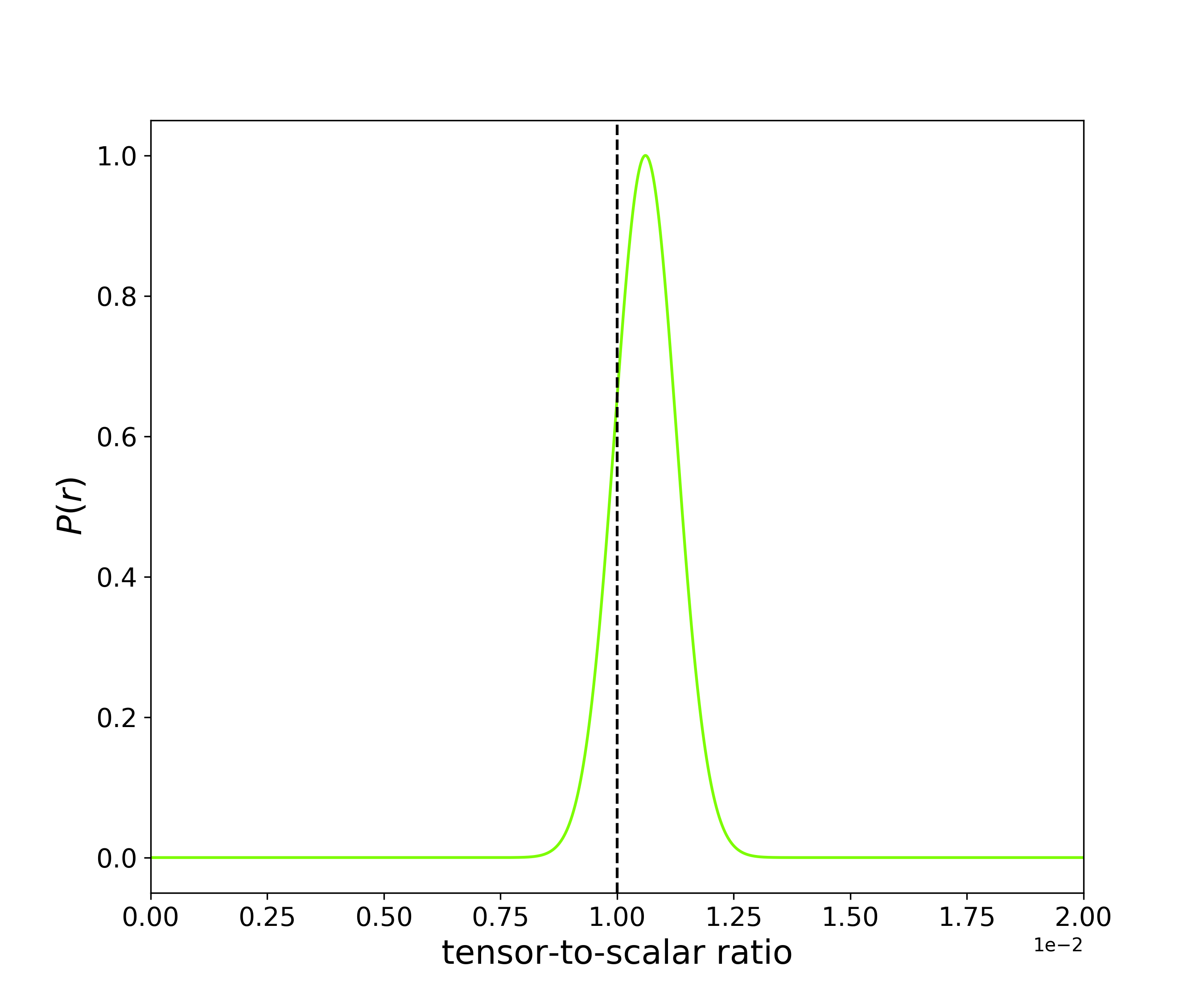}~\\
          \includegraphics[width=0.5\textwidth]{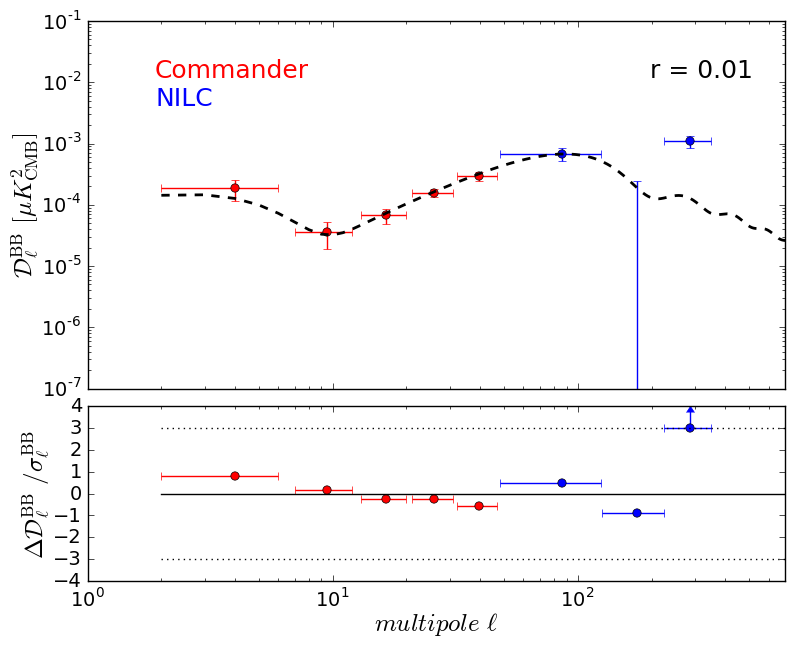}~
	  \includegraphics[width=0.5\textwidth]{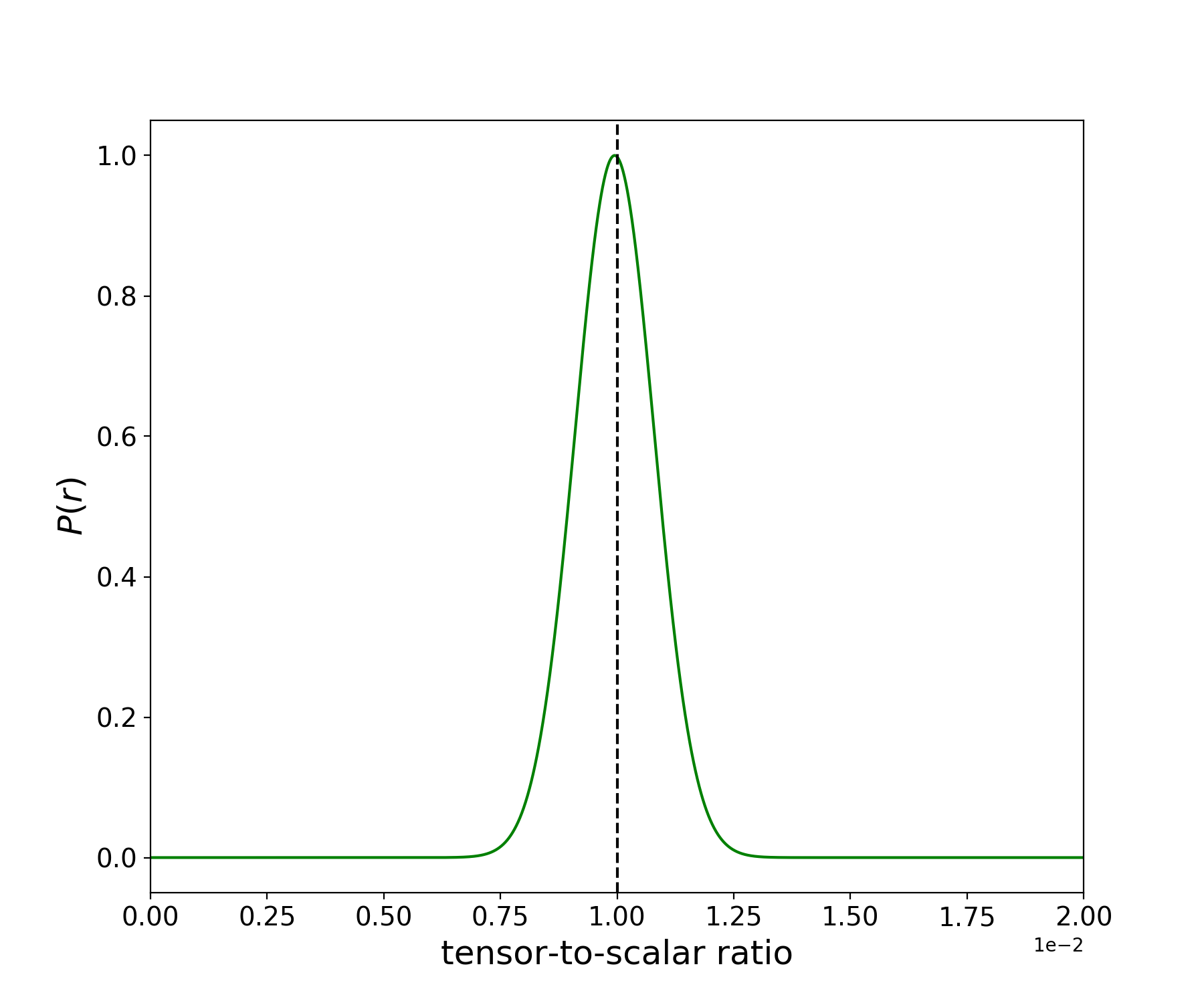}~\\
	\end{center}
        \caption{Joint results on simulation \#1: CMB with
          $r=10^{-2}$, and including synchrotron and dust
          foregrounds. \emph{Left panels}: CMB $B$-mode power spectrum
          reconstruction using the combined estimates from {\tt
            Commander} at low multipoles and either {\tt SMICA}
          (\emph{top}) or {\tt NILC} (\emph{bottom}) at intermediate
          multipoles. The fiducial primordial CMB $B$-mode power
          spectrum is denoted by a dashed black line. The horizontal
          dotted lines show the 3$\sigma$ limits, while the vertical
          arrows signify outliers.  \emph{Right panels}: posterior
          distribution, $P(r)$, of the tensor-to-scalar ratio derived
          from \emph{top} the joint {\tt Commander-SMICA} power
          spectrum calculated over the multipole range
          $2 \leq \ell \leq 600$, or \emph{bottom} the joint {\tt
            Commander-NILC} power spectrum for
          $2 \leq \ell \leq 349$.}
\label{Fig:r1ten-2}
\end{figure}

In this section, we attempt to determine improved estimates of the the
tensor-to-scalar ratio by adopting a hybrid likelihood approach,
following the work of \textit{Planck}
\citep{planck_2013_XV,Planck2015_XI} and \textit{WMAP}
\citep{WMAP_likelihood2009}. Specifically, we combine the
Blackwell-Rao estimate of the CMB $B$-mode power spectrum from {\tt
  Commander} on scales $2 \leq \ell \leq 47$ (including the reionisation peak) with a pseudo-$C_{\ell}$ estimate from either {\tt SMICA} or {\tt NILC} on scales spanning the recombination peak and
above ($\ell > 47$).  Specifically, we take the product of the
posterior on $r$ from \texttt{Commander}, obtained via the
Blackwell-Rao approximation and a non-Gaussian likelihood as presented
in Sect.~\ref{subsec:parametric}, with that from either \texttt{NILC}
or \texttt{SMICA}, obtained using the Gaussian likelihood formalism
described in Sect.~\ref{subsec:ilc}, to obtain the resulting joint
posterior.  \texttt{SMICA} and \texttt{NILC} provide their own
estimation of the covariance matrix (see e.g. Appendix
\ref{subsec:nilc_method}), to use in the Gaussian likelihood approach
for $\ell > 47$.

\begin{figure}[htbp]
\centering
    \includegraphics[width=0.5\textwidth]{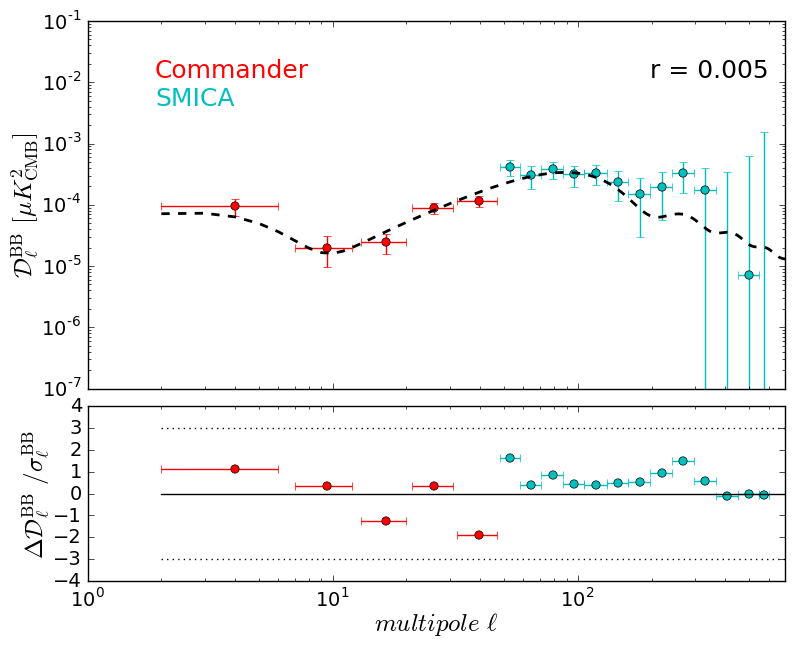}~
    \includegraphics[width=0.5\textwidth]{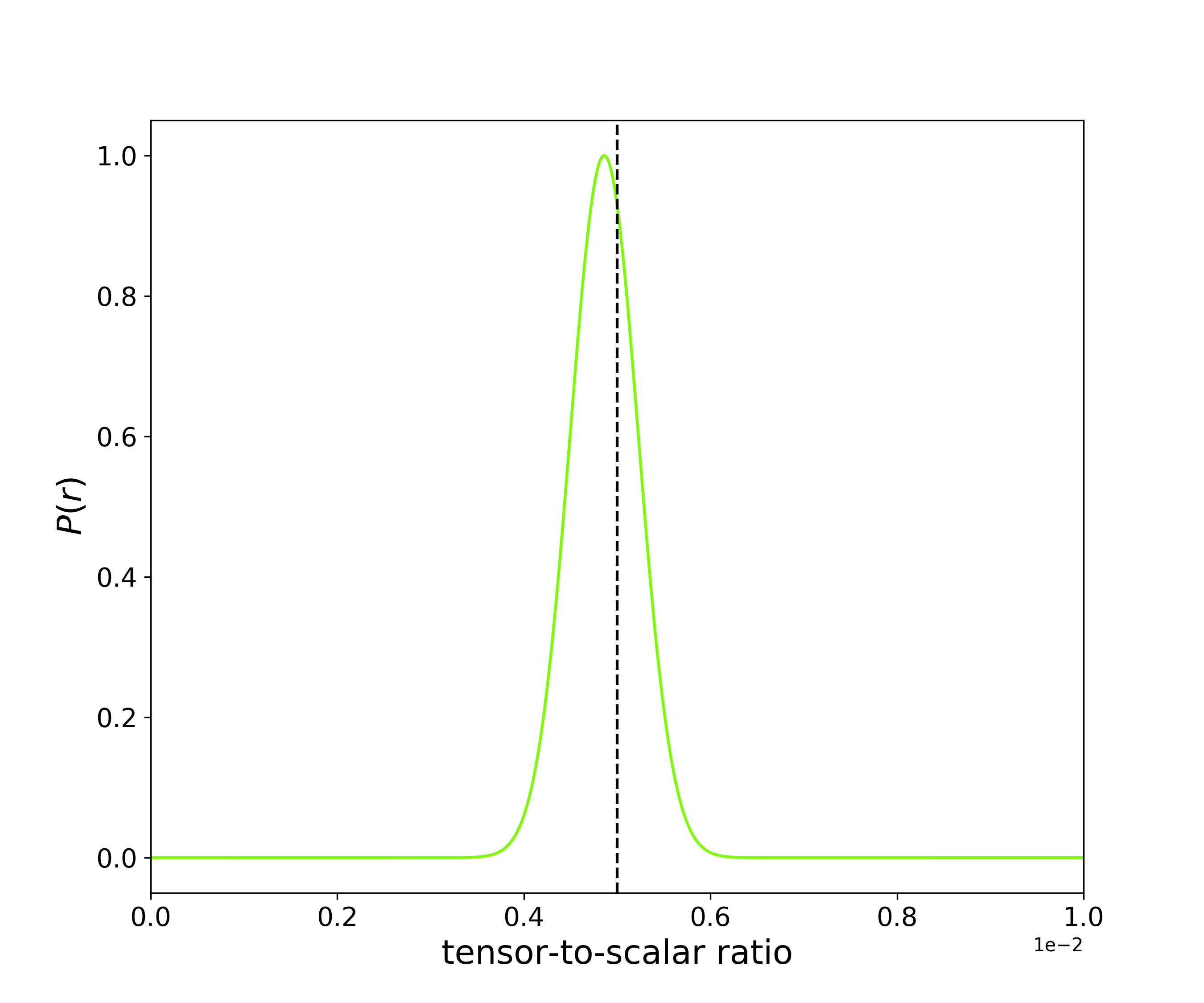}~\\
    \includegraphics[width=0.5\textwidth]{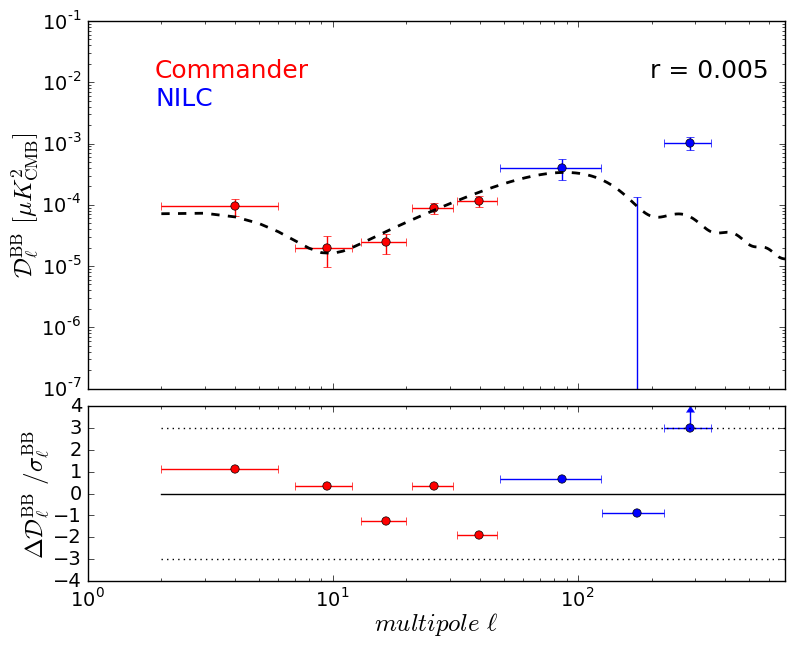}~
    \includegraphics[width=0.5\textwidth]{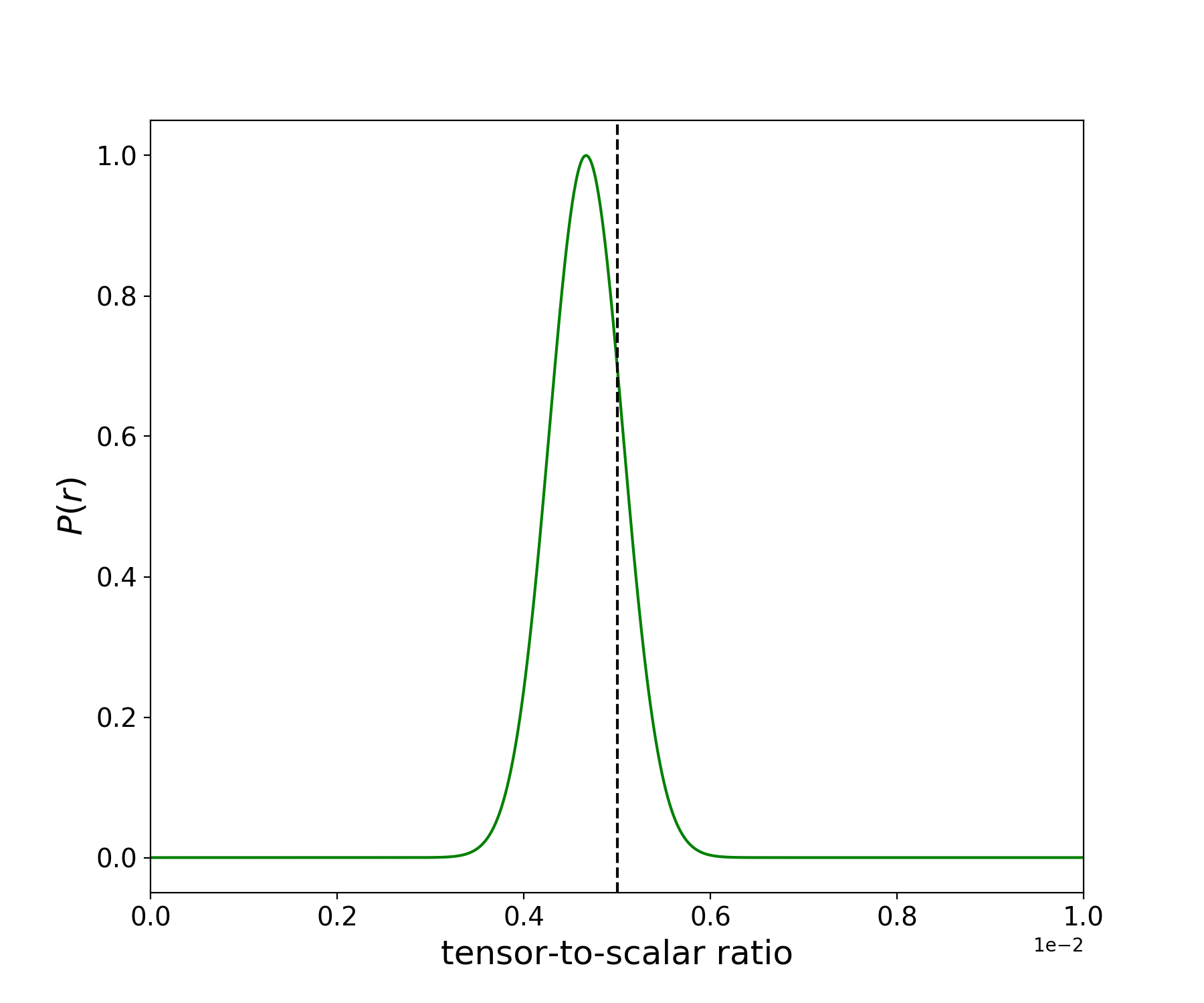}~\\
        \caption{Joint results on simulation \#2: CMB with
          $r=5\times 10^{-3}$, and including synchrotron and dust
          foregrounds. \emph{Left panels}: CMB $B$-mode power spectrum
          reconstruction using the combined estimates from {\tt
            Commander} at low multipoles and either {\tt SMICA}
          (\emph{top}) or {\tt NILC} (\emph{bottom}) at intermediate
          multipoles. The fiducial primordial CMB $B$-mode power
          spectrum is denoted by a dashed black line. The horizontal
          dotted lines show the 3$\sigma$ limits, while the vertical
          arrows signify outliers.  \emph{Right panels}: posterior
          distribution, $P(r)$, of the tensor-to-scalar ratio derived
          from \emph{top} the joint {\tt Commander-SMICA} power
          spectrum calculated over the multipole range
          $2 \leq \ell \leq 600$, or \emph{bottom} the joint {\tt
            Commander-NILC} power spectrum for
          $2 \leq \ell \leq 349$.}
\label{Fig:r5ten-3}
\end{figure}
\begin{figure}[htbp]
\centering
                \includegraphics[width=0.5\textwidth]{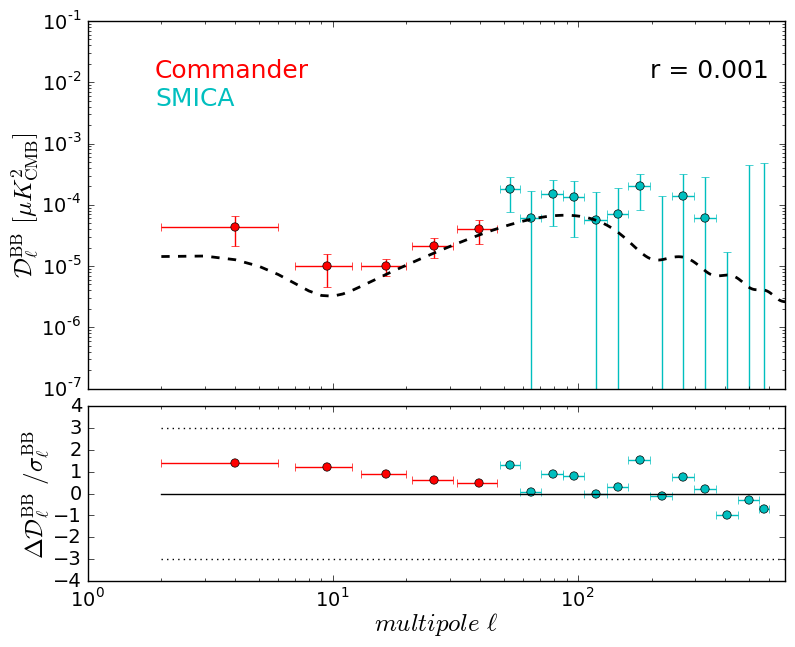}~
		\includegraphics[width=0.5\textwidth]{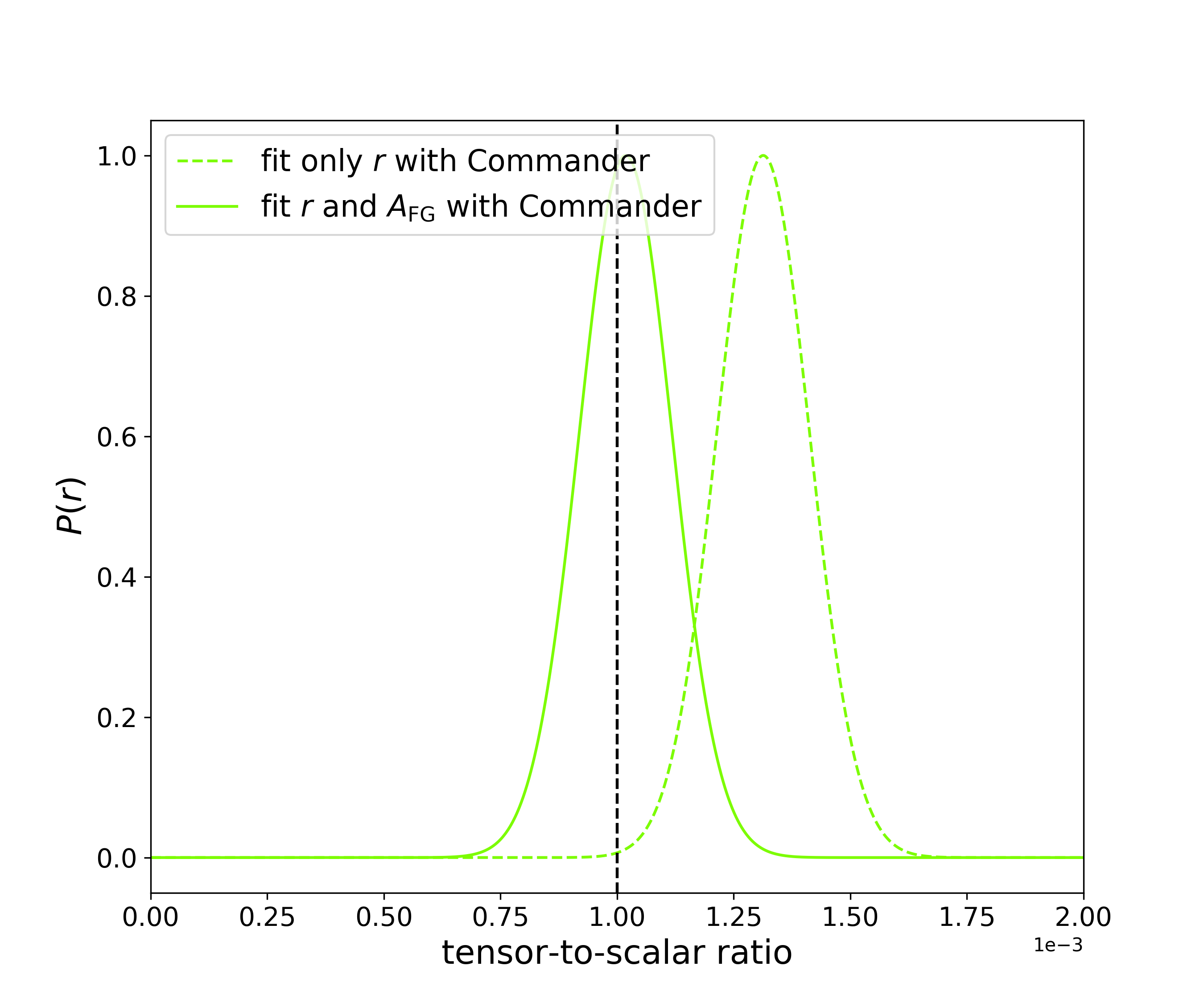}~\\
                \includegraphics[width=0.5\textwidth]{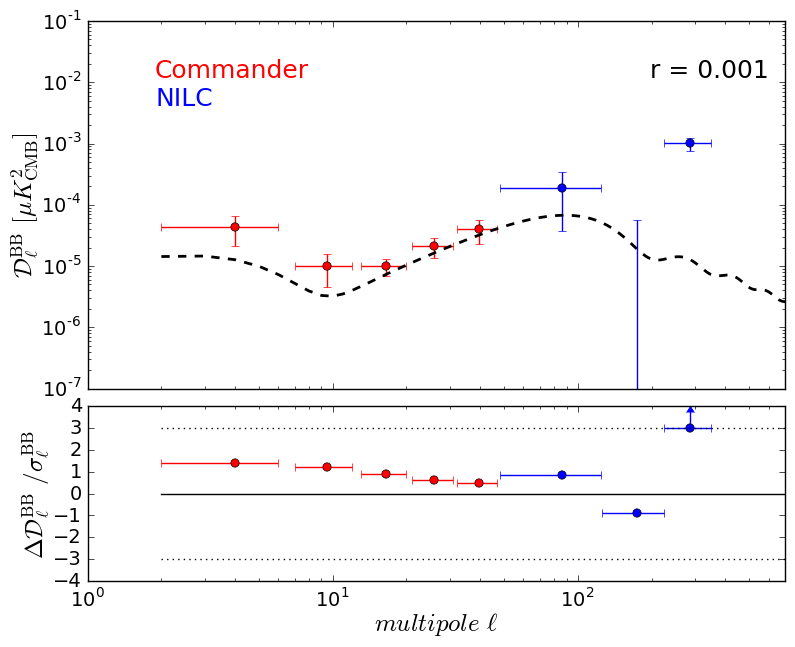}~
		\includegraphics[width=0.5\textwidth]{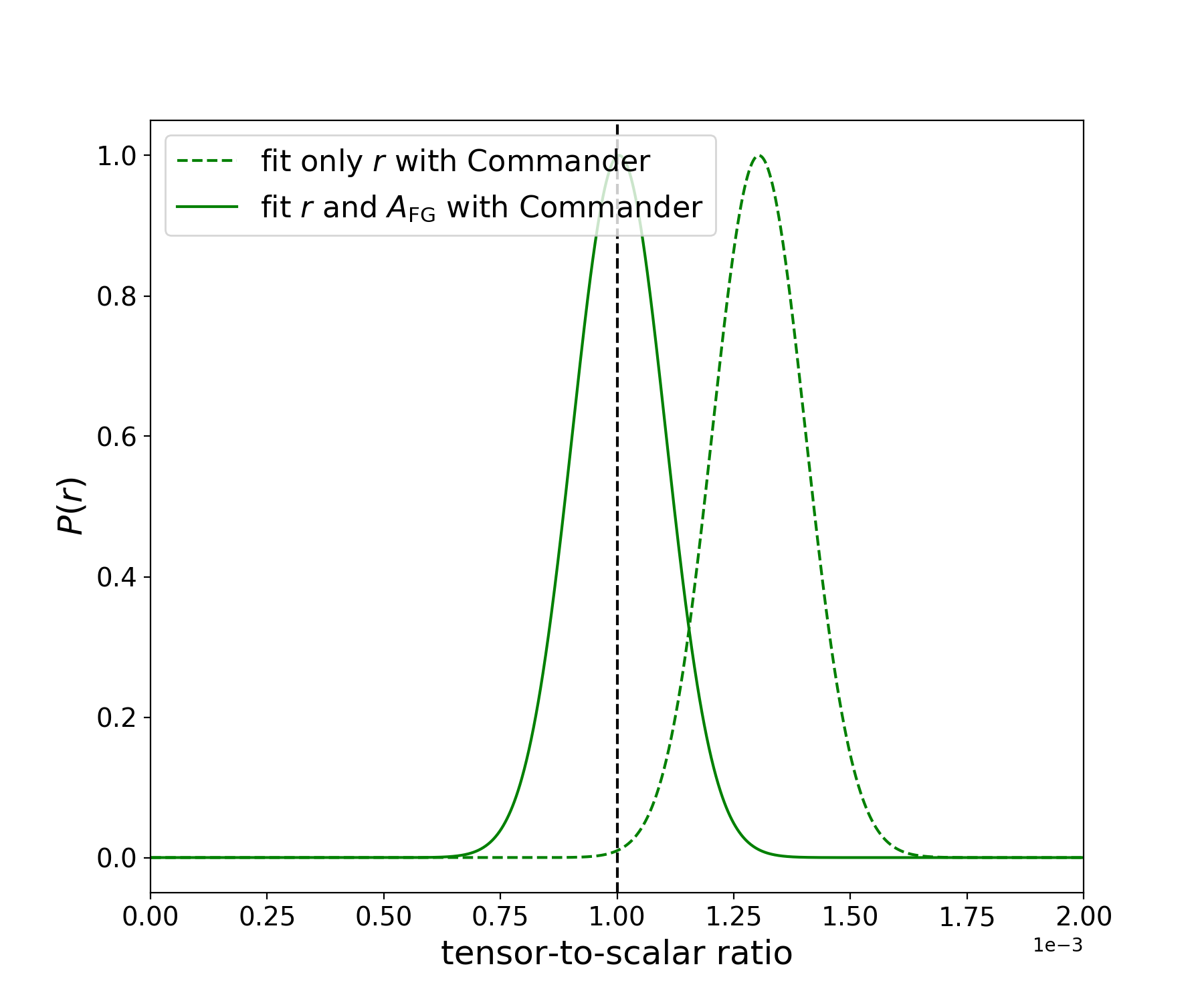}~\\
        \caption{Joint results on simulation \#3: CMB with
          $r=10^{-3}$, and including synchrotron and dust
          foregrounds. \emph{Left panels}: CMB $B$-mode power spectrum
          reconstruction using the combined estimates from {\tt
            Commander} at low multipoles and either {\tt SMICA}
          (\emph{top}) or {\tt NILC} (\emph{bottom}) at intermediate
          multipoles. The fiducial primordial CMB $B$-mode power
          spectrum is denoted by a dashed black line. The horizontal
          dotted lines show the 3$\sigma$ limits, while the vertical
          arrows signify outliers.  \emph{Right panels}: posterior
          distribution, $P(r)$, of the tensor-to-scalar ratio derived
          from \emph{top} the joint {\tt Commander-SMICA} power
          spectrum calculated over the multipole range
          $2 \leq \ell \leq 600$, or \emph{bottom} the joint {\tt
            Commander-NILC} power spectrum for
          $2 \leq \ell \leq 349$.}
  \label{Fig:joint}
\end{figure}

\begin{figure}[htbp]
\centering
		\includegraphics[width=0.5\textwidth]{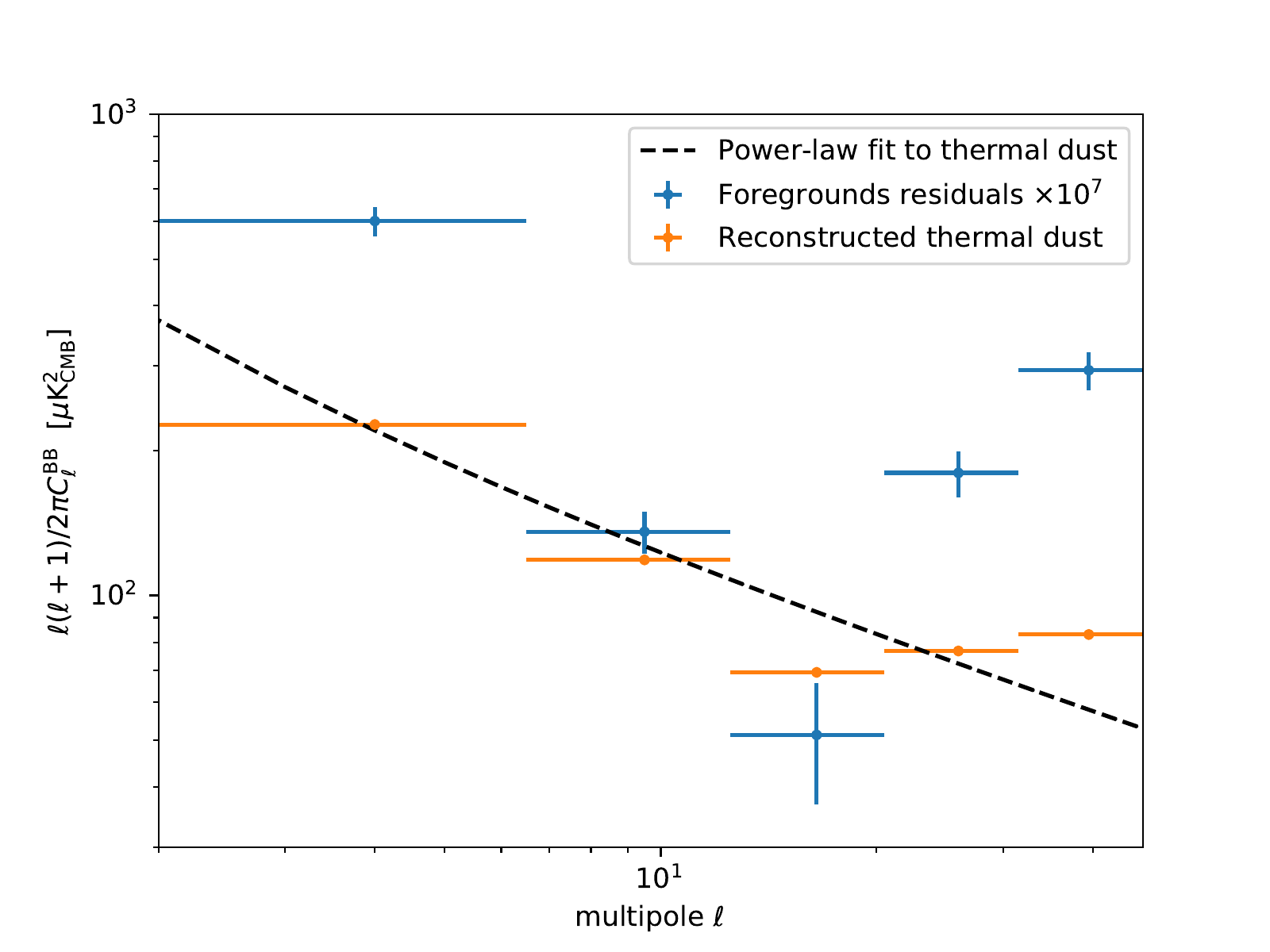}
  \caption{Procedure used to debias the tensor-to-scalar ratio
    estimate for simulation \#3 with $r=10^{-3}$. Specifically, 
    the amplitude of the foreground residuals as described by an assumed
    power-law spectrum is fitted on scales corresponding to the reionization peak.
    The dashed black
    line shows the power-law fit of the power spectrum of the {\tt
      Commander} reconstructed thermal dust map (orange dots). Blue
    dots indicate the power spectrum of the real foreground residuals that remain in
    the CMB $B$-mode map, multiplied by $10^7$ (to be visible on the same
    scale). These are calculated by differentiating the input and 
    reconstructed CMB maps. At the frequency range considered by \core\, most of the foreground residuals will be originated in dust.
\label{Fig:comm_r10-3_debiased}}
\end{figure}

\begin{figure}[htbp]
\centering
          \includegraphics[width=0.5\textwidth]{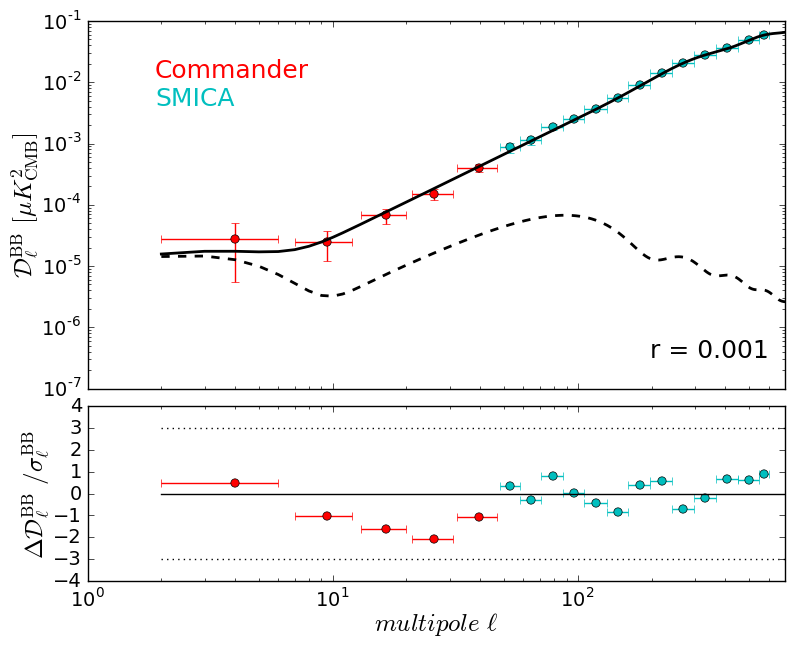}~
	  \includegraphics[width=0.5\textwidth]{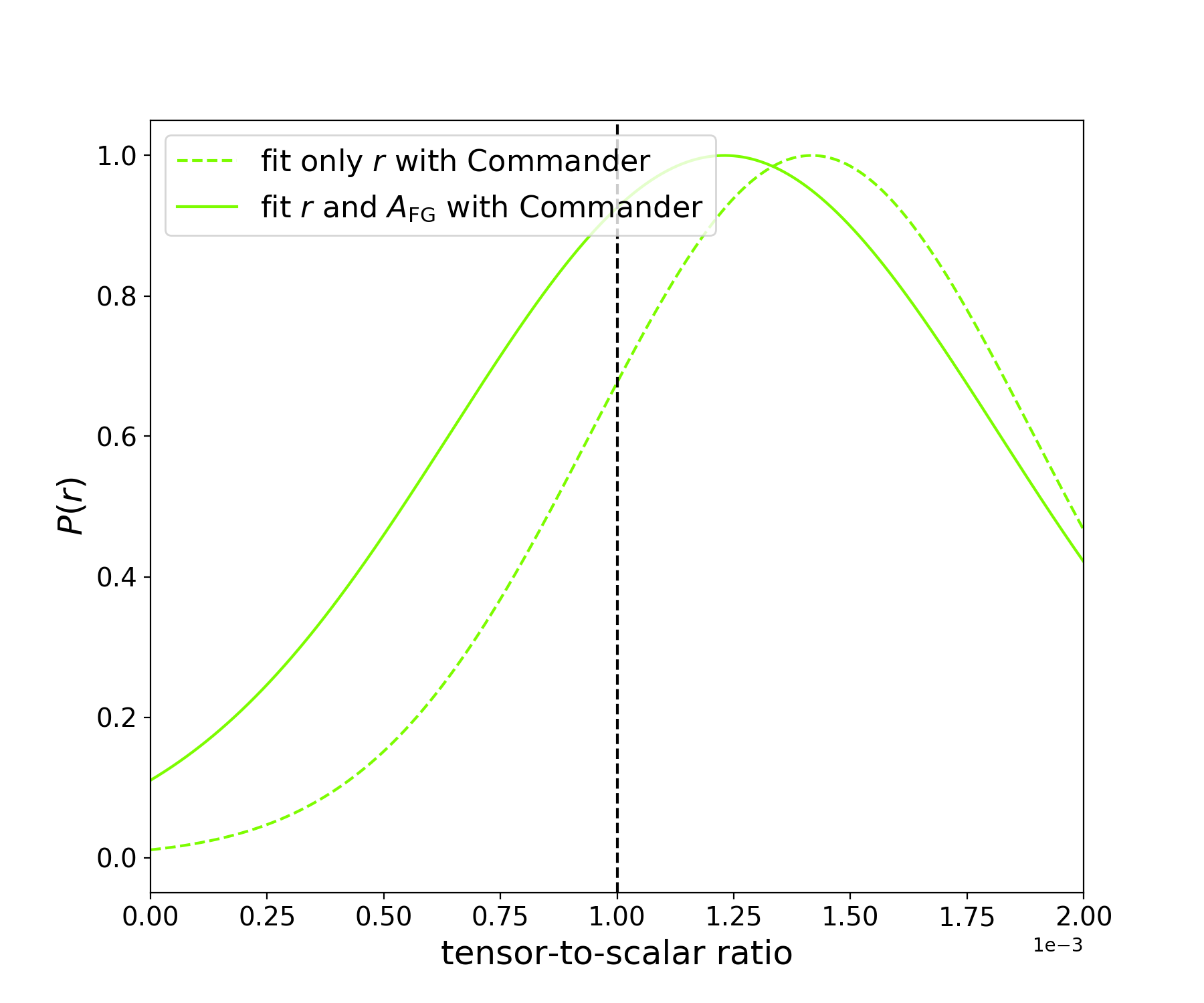}~\\
          \includegraphics[width=0.5\textwidth]{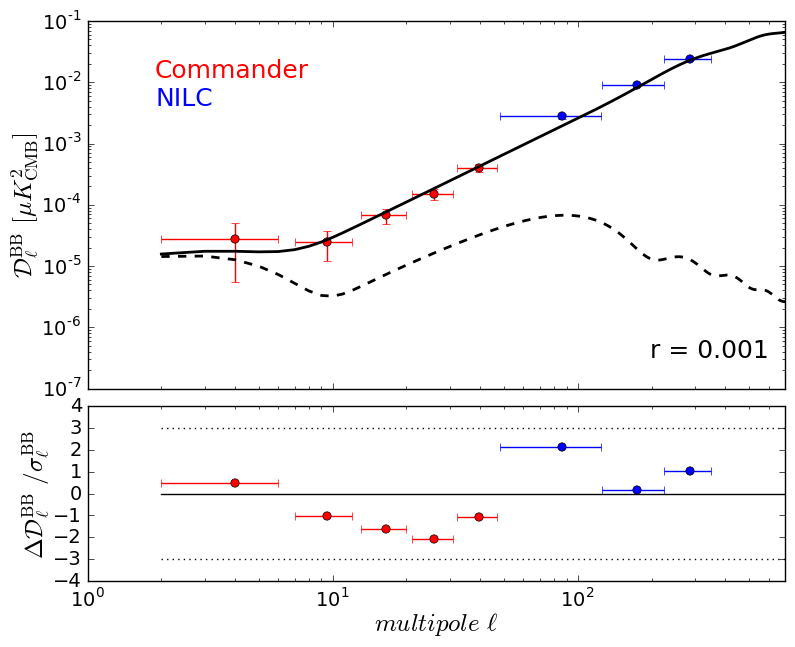}~
	  \includegraphics[width=0.5\textwidth]{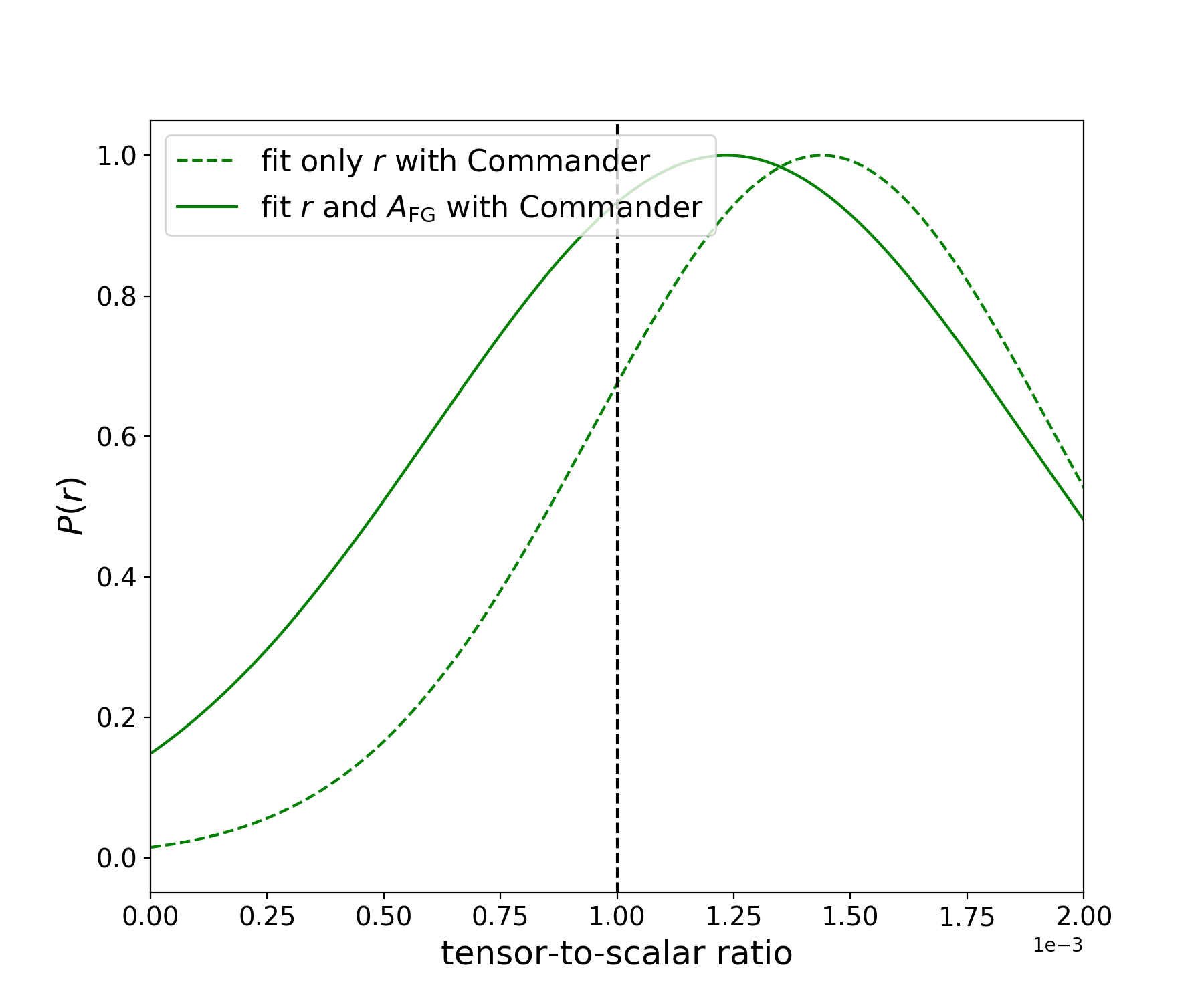}~\\
        \caption{Joint results on simulation \#4: CMB with $r=10^{-3}$
          and a lensing $B$-mode contribution, and including
          synchrotron, dust, AME, and point-source
          foregrounds. \emph{Left panels}: lensed $B$-mode power
          spectrum joint reconstruction using the combined estimates
          from {\tt Commander} at low multipoles and either {\tt
            SMICA} (\emph{top}) or {\tt NILC} (\emph{bottom}) at
          intermediate multipoles. The fiducial primordial CMB
          $B$-mode power spectrum is denoted by a dashed black line,
          while the solid black line shows the lensed CMB $B$-mode
          power spectrum. The horizontal dotted lines show the
          3$\sigma$ limits, while the vertical arrows signify
          outliers. \emph{Right panels}: posterior distribution,
          $P(r)$, of the tensor-to-scalar ratio derived from
          \emph{top} the joint {\tt Commander-SMICA} power spectrum
          calculated over the multipole range $2 \leq \ell \leq 600$,
          or \emph{bottom} the joint {\tt Commander-NILC} power
          spectrum for $2 \leq \ell \leq 349$.  }
\label{Fig:model18v6_joint}
\end{figure}

\begin{figure}[htbp]
\centering
\includegraphics[width=0.5\textwidth]{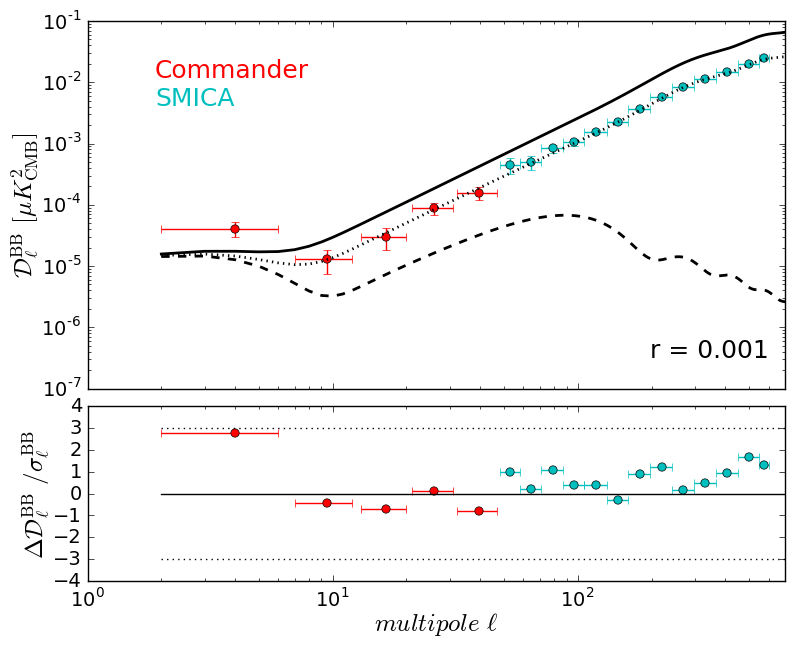}~
\includegraphics[width=0.5\textwidth]{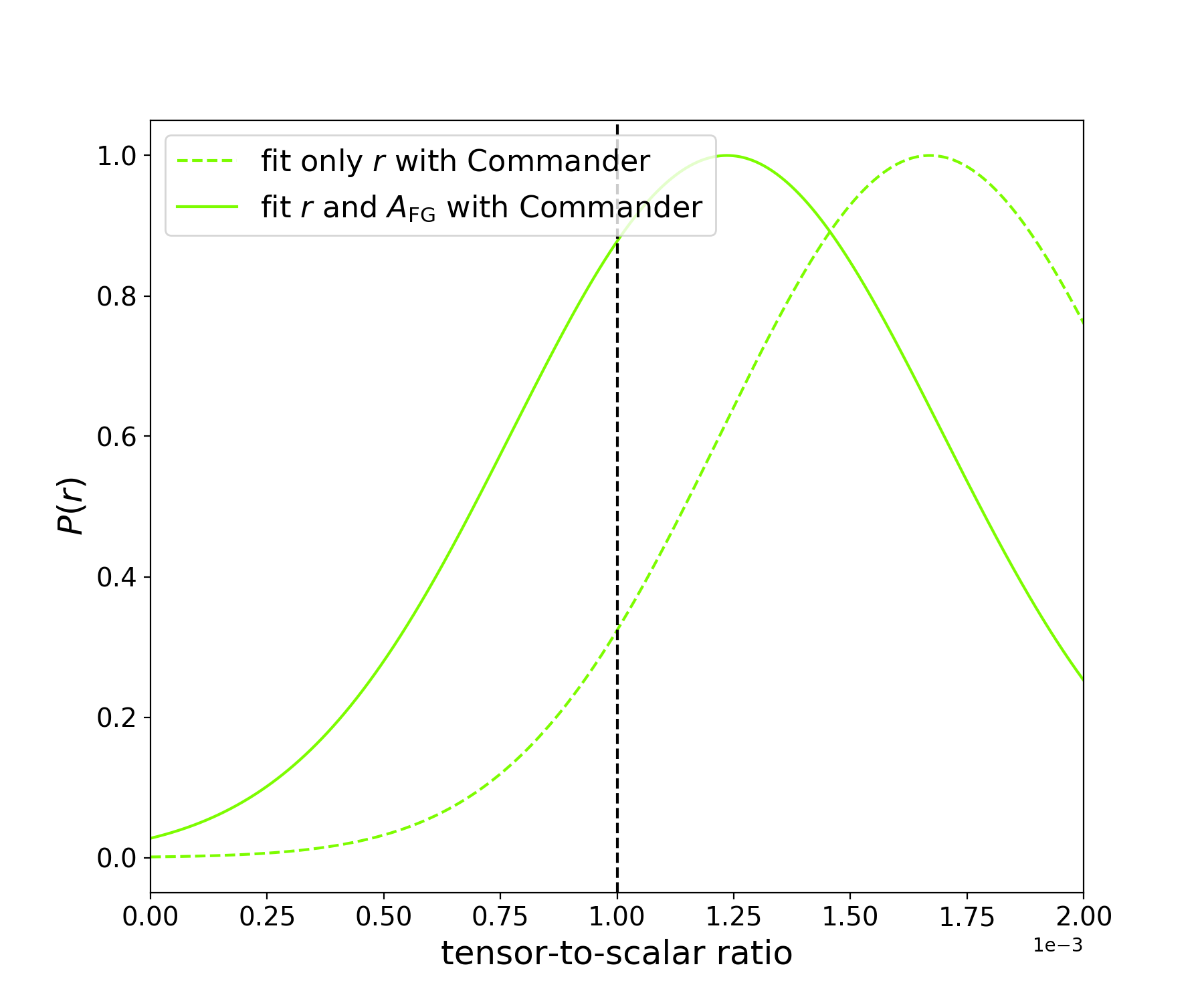}\\
\includegraphics[width=0.5\textwidth]{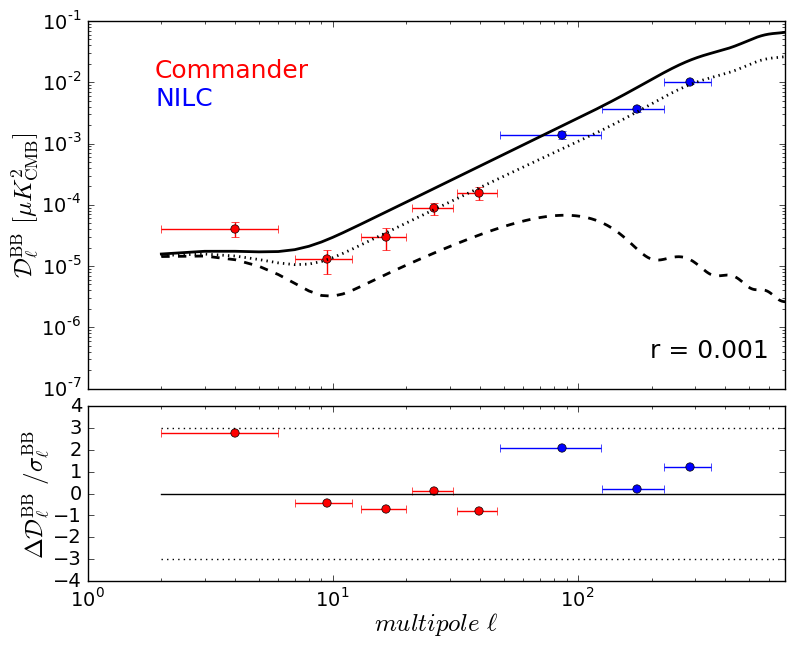}~
\includegraphics[width=0.5\textwidth]{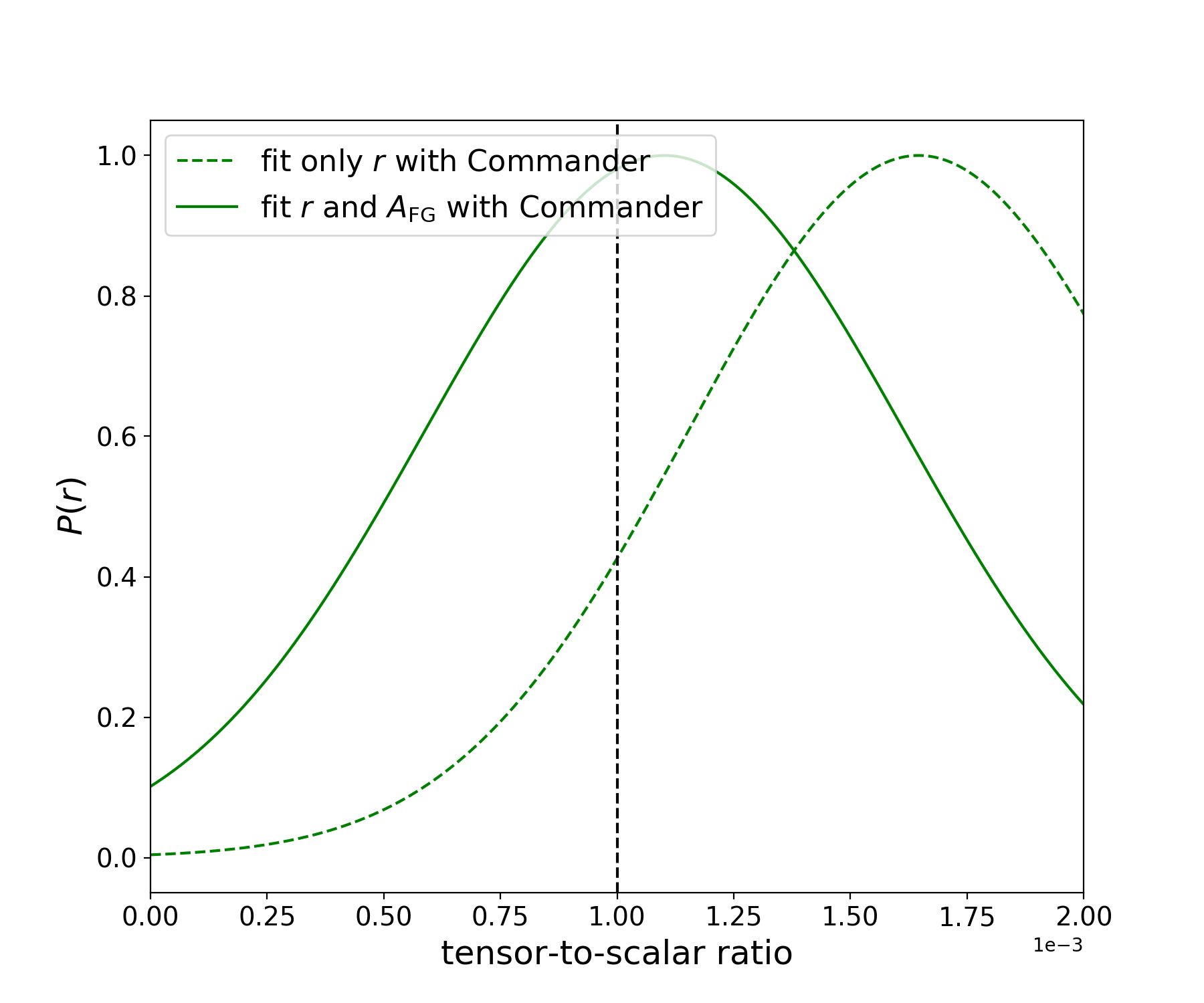}\\
\caption{Joint results on simulation \#5: CMB with $r=10^{-3}$, and a
  residual lensing contribution (mimicking a 60\% delensing capability
  of \core) and including synchrotron, dust, AME, and point-source
  foregrounds. \emph{Left panels}: partially delensed $B$-mode power
  spectrum joint reconstruction using the combined estimates from {\tt
    Commander} at low multipoles and either {\tt SMICA} (\emph{top})
  or {\tt NILC} (\emph{bottom}) at intermediate multipoles.  The
  fiducial primordial CMB $B$-mode power spectrum is denoted by a
  dashed black line, the solid black line shows the lensed CMB
  $B$-mode power spectrum, while the dotted black line shows the
  theoretical lensed CMB $B$-mode power spectrum after 60\%
  delensing. The horizontal dotted lines show the
    3$\sigma$ limits, while the vertical arrows signify outliers.
    \emph{Right panels}: posterior
  distribution, $P(r)$, of the tensor-to-scalar ratio derived from
  \emph{top} the joint  {\tt Commander-SMICA} power spectrum
          calculated over the multipole range $2 \leq \ell \leq 600$, or \emph{bottom} the joint {\tt
    Commander-NILC} power spectrum for $2 \leq \ell \leq 349$.  }

    \label{Fig:model18v6_lens40_joint} 
\end{figure}

\subsection{Component separation results on $\boldsymbol{r=10^{-2}}$ and $\boldsymbol{r=5\times 10^{-3}}$}\label{subsec:sim1-2}

Figures~\ref{Fig:r1ten-2} and \ref{Fig:r5ten-3} present the results of
the joint reconstruction of the primordial $B$-mode power spectra
(left panels) and the joint likelihood fits (right panels) to
simulations \#1 and \#2. For these values of the tensor-to-scalar
ratio, both the reionization and the recombination peaks of the
primordial $B$-modes are accurately recovered by \core\ between
$\ell=2$ and $\ell=200$, with neither of the the combined power
spectra deviating from the input theory spectrum by more than
$2\sigma$.

The right panels of Figs.~\ref{Fig:r1ten-2} and \ref{Fig:r5ten-3}
indicate that observations of a CMB sky with $r=10^{-2}$ by the
current \core\ configuration yield estimates of the
tensor-to-scalar ratio of $r=\left(1.06 \pm 0.07\right)\times 10^{-2}$
and $r=\left(1.00 \pm 0.09\right)\times 10^{-2}$ by
the hybrid {\tt Commander}-{\tt SMICA} or {\tt Commander}-{\tt NILC}
analyses, respectively. For an input value of $r=5\times 10^{-3}$, the corresponding estimates
are $r=\left(4.9 \pm 0.4\right)\times 10^{-3}$ and $r=\left(4.7 \pm 0.4\right)\times
10^{-3}$, respectively.
These results correspond to high significance detections of the
tensor-to-scalar ratio, at a level exceeding $10\sigma$ for unlensed
realisations of the CMB sky after correcting for foregrounds
consisting of synchrotron and thermal dust emission which have
spectral variations over the sky.

\subsection{Component separation results on $\boldsymbol{r=10^{-3}}$}\label{subsec:sim4}

We next consider the critical case where the tensor-to-scalar ratio is
of order $r=10^{-3}$, initially excluding lensing effects (simulation
\#3). Figure~\ref{Fig:joint} summarizes the results. As we have seen
previously in Sect.~\ref{subsec:parametric}, there is clear evidence
of residual foregrounds over the reionization scales probed by {\tt
  Commander}. As a consequence, the recovered estimate of the
tensor-to-scalar ratio for both combinations of component separation
methods is $r=\left(1.3 \pm 0.1\right) \times 10^{-3}$.  However, in
order to mitigate the potential bias due to the foreground residuals,
we implement a strategy outlined in Sect.~\ref{subsec:parametric}, and
fit a parametric nuisance model.  The actual shape of the power
spectrum of foreground residuals is not known a priori, and may differ
from that of the input foregrounds. This is particularly plausible for
those component separation methods which combine the frequency maps in
harmonic space, such as {\tt NILC}, which computes different weights
over different multipole ranges and regions, depending on the relative
ratio between foregrounds and noise contamination. 
A conservative first approximation, then, is to
assume a power-law shape for the power spectrum of the foreground
residuals at low multipoles, and consider that this, up to an
arbitrary amplitude to be fitted, can be described by the {\tt
  Commander} thermal dust map estimate, presented in
Fig.~\ref{Fig:comm_r10-3_debiased}. For the frequency coverage
considered in the \core\ simulations, thermal dust emission dominates,
and therefore most of the foreground residuals will originate therein.
The best-fitting spectral index of the power-law is determined by a
direct fit to the $B$-mode power spectrum of this thermal dust map,
calculated with a Quadratic Maximum Likelihood (QML) estimator. The
Blackwell-Rao likelihood is then modified in order to fit
simultaneously for both $r$ and the amplitude of the foreground residuals,
$A_{\rm FG}$, over the {\tt Commander} multipole range. This results
in an effective debiasing of the tensor-to-scalar ratio, as shown in
the right panels of Fig.~\ref{Fig:joint}.

Indeed, in this case, we find that the joint estimates of the
tensor-to-scalar ratio estimated from {\tt Commander} and either {\tt
  SMICA} or {\tt NILC} are $r=\left(1.02 \pm 0.10\right)\times 10^{-3}$ and
$r=\left(1.00 \pm 0.10\right)\times 10^{-3}$ respectively,
corresponding to $10\sigma$ detections of cosmological $B$-modes.

\subsection{Component separation results on $\boldsymbol{r=10^{-3}}$ with lensing and delensing}\label{subsec:sim5}

Finally, we consider challenging simulations where both gravitational
lensing effects and complex foregrounds comprising synchrotron,
thermal dust emission, AME, strong and faint point sources have been
included. Figure.~\ref{Fig:model18v6_joint} presents the results for
simulation \#4.  In order to account for the lensing contribution
which will otherwise bias the estimate of $r$, we fix $A_{\rm lens}=1$
and fit for $r$ only. The posterior distribution of the
tensor-to-scalar ratio from the joint $B$-mode power spectra analyses
is shown in the right panels of Fig.~\ref{Fig:model18v6_joint}. The
measured value of $r$ is $\left(1.4 \pm 0.47\right)\times 10^{-3}$ and
$\left(1.4 \pm 0.49\right)\times 10^{-3}$ for the {\tt
  Commander-SMICA} and {\tt Commander-NILC} joint likelihoods,
respectively.  As is the case for simulation \#3, foreground residuals
on the reionization scales are a source of bias at low-$\ell$.  If the
nuisance parameter, $A_{\rm FG}$, is again fitted in the likelihood,
the estimated tensor-to-scalar ratios are
$\left(1.2 \pm 0.6\right)\times 10^{-3}$ and
$\left(1.2 \pm 0.6\right)\times 10^{-3}$ for {\tt Commander-SMICA} and
{\tt Commander-NILC}, respectively. Note that that the presence of the
lensing signal leads to increased uncertainty in the measurement due
to the related cosmic variance contribution, and reduces the
measurements of $r$ effectively to upper limits.

Consequently, we consider the impact of delensing on the determination
of the tensor-to-scalar ratio. In principle, \core\ should allow the
reduction of the lensing contribution to the measured $B$-mode power
spectrum by a factor typically of order 60\% \citep[][in prep.]{ECO_lensing}. In
this paper, we adopt a shortcut to mimic such a delensing
analysis. Specifically, we analyse simulation \#5, which is identical
to simulation \#4 except that only 40\% of the pure $B$-mode lensing
signal is included in the CMB map. In order to properly account for
residuals from delensing in the likelihood estimation of the
tensor-to-scalar ratio, the $A_{lens}$ parameter is fixed to a value
of $0.40$ while fitting for $r$ only.
Figure~\ref{Fig:model18v6_lens40_joint} presents the results of the
analyses, indicating best-fit values for the tensor-to-scalar ratio of
$r=\left(1.70 \pm 0.45\right)\times 10^{-3}$ and
$r=\left(1.60 \pm 0.49\right)\times 10^{-3}$ for the joint likelihood
with {\tt Commander-SMICA} and {\tt Commander-NILC}, respectively.  As
before, fitting for foreground residuals at low-$\ell$ improves the
consistency with the input value of $r$, resulting in estimates of
$r=\left(1.20 \pm 0.46\right)\times 10^{-3}$ and
$r=\left(1.10 \pm 0.51\right)\times 10^{-3}$.  Note that there is no
clear improvement in the error term, $\sigma(r=10^{-3})$, due to the
delensing procedure. This is because the uncertainty is still
dominated by foreground residuals. We discuss this important issue
further in Sect.~\ref{subsec:delens}.

\begin{table}[htbp]
\footnotesize
\centering 
\begin{tabular}{c l l l l l} \hline\hline
 {\bf Simulation \#1} & \multicolumn{4}{l}{\bf  $\boldsymbol{r=10^{-2}}$,  dust,  synchrotron}\\ \hline\hline
            & $\ell_{\rm min}$ & $\ell_{\rm max}$ & $r\ [10^{-3}]$ & $\sigma(r)\ [10^{-3}]$ & $\vert r - r_{in}\vert / \sigma(r)$ \\ [1ex]\hline
{\tt Commander} 			& 2 	& 47 		& $9.7$ 		& $0.9$  & $0.3$	\\[1ex] \hline
{\tt NILC} 				& 48 	& 349  	& $11.9$ 	& $2.5$  & $0.8$	\\[1ex] \hline
{\tt SMICA} 				& 48 	& 600 	& $11.7$		& $1.0$  & $1.7$	\\[1ex] \hline
{\tt Commander} $+$ {\tt NILC} 	& 2 	& 349 	& $10.0$		& $0.9$	& $0.0$  \\[1ex] \hline
{\tt Commander} $+$ {\tt SMICA} 	& 2 	& 600 	& $10.6$ 	& $0.7$  & $0.9$	\\[1ex] \hline
& & & & \\\hline\hline
{\bf Simulation \#2} & \multicolumn{4}{l}{\bf  $\boldsymbol{r=5\times 10^{-3}}$,  dust,  synchrotron}\\ \hline\hline
     & $\ell_{\rm min}$ & $\ell_{\rm max}$ & $r\ [10^{-3}]$ &
                                                              $\sigma(r)\ {10^{-3}}$ & $\vert r - r_{in}\vert / \sigma(r)$ \\ [1ex]\hline
{\tt Commander} 			& 2 		& 47 	& $4.6$ 	& $0.4$ & $1.0$	\\[1ex] \hline
{\tt NILC} 				& 48 	& 349 	& $7.2$	& $2.5$	& $0.9$\\[1ex] \hline
{\tt SMICA} 				& 48 	& 600	& $6.1$	& $0.9$ & $1.2$	\\[1ex] \hline
{\tt Commander} $+$ {\tt NILC} 	& 2 	& 349       & $4.7$ 	& $0.4$ & $0.7$ 	\\[1ex] \hline
{\tt Commander} $+$ {\tt SMICA} 	& 2 	& 600 	& $4.9$ 	& $0.4$  & $0.2$	\\[1ex] \hline
& & & & \\\hline\hline
{\bf Simulation \#3} & \multicolumn{4}{l}{\bf  $\boldsymbol{r=10^{-3}}$,  dust,  synchrotron}\\ \hline\hline
 & $\ell_{\rm min}$ & $\ell_{\rm max}$ & $r\ [10^{-3}]$ & $\sigma(r)\ [10^{-3}]$ & $\vert r - r_{in}\vert / \sigma(r)$\\ [1ex]\hline
{\tt Commander} & 2   & 47     & $1.3 (1.0)$ & $0.1 (0.1)$ & $3.0 (0.0)$ \\[1ex] \hline
{\tt NILC} 	   & 48 & 349 	& $3.7$   	& $2.2$ & $1.2$ \\[1ex] \hline
{\tt SMICA}          & 48 & 600 	& $2.1$		& $0.8$ & $1.4$	\\[1ex] \hline
{\tt Commander} $+$ {\tt NILC}  & 2 & 349 & $1.3 (1.0)$ & $0.1 (0.1)$ & $3.0 (0.0)$	\\[1ex] \hline
{\tt Commander} $+$ {\tt SMICA} & 2 & 600 & $1.3 (1.0)$
                                                  & $0.1 (0.1)$ & $3.0 (0.0)$	\\[1ex] \hline
& & & & \\\hline\hline
{\bf Simulation \#4} & \multicolumn{4}{l}{\bf  $\boldsymbol{r=10^{-3}}$,  dust,  synchrotron, }\\
                                  & \multicolumn{4}{l}{\bf AME, sources, lensing}\\ \hline\hline
    & $\ell_{\rm min}$ & $\ell_{\rm max}$ & $r\ [10^{-3}]$ & $\sigma(r)\ [10^{-3}]$ & $\vert r - r_{in}\vert / \sigma(r)$ \\ [1ex]\hline
{\tt Commander} & 2 & 47 & $1.3 (1.0)$ & $0.5 (0.6)$ & $0.6 (0.0)$	\\ [1ex]\hline
{\tt NILC}         & 48 & 349 & $9.1$ & $3.7$ & $2.2$			\\[1ex]\hline
{\tt SMICA} 	& 48 	& 600	& $2.4$			& $1.4$  & $1.0$			\\[1ex]\hline
{\tt Commander} $+$ {\tt NILC} & 2 & 349 & $1.4 (1.2)$ & $0.5 (0.6)$
                                                                   & $0.8 (0.3)$	\\[1ex] \hline
{\tt Commander} $+$ {\tt SMICA} & 2 & 600 & $1.4 (1.2)$ & $0.5 (0.6)$
                                                                   & $0.8 (0.3)$	 \\[1ex] \hline
& & & & \\\hline\hline
{\bf Simulation \#5} & \multicolumn{4}{l}{\bf  $\boldsymbol{r=10^{-3}}$,  dust,  synchrotron, }\\
                                 & \multicolumn{4}{l}{\bf AME,  sources, 40\% lensing}\\ \hline\hline
    & $\ell_{\rm min}$ & $\ell_{\rm max}$ & $r\ [10^{-3}]$ & $\sigma(r)\ [10^{-3}]$ & $\vert r - r_{in}\vert / \sigma(r)$ \\ [1ex]\hline
{\tt Commander} & 2 & 47 & $1.5 (0.9)$ & $0.5 (0.5)$ & $1.0 (0.2)$	\\ [1ex]\hline
{\tt NILC} & 48 & 349 & $8.5$ & $3.4$  & $2.2$	\\[1ex]\hline
{\tt SMICA} & 48 & 600 & $2.3$ & $1.0$  & $1.3$	\\[1ex]\hline
{\tt Commander} $+$ {\tt NILC} & 2 & 349 & $1.6 (1.1)$ & $0.5 (0.5)$ &
                                                                       $1.2 (0.2)$  \\[1ex] \hline
{\tt Commander} $+$ {\tt SMICA} & 2 & 600 & $1.7 (1.2)$ & $0.4 (0.5)$
                                                                   & $1.7 (0.4)$	 \\[1ex] \hline
\end{tabular} 
\caption{Summary table of the tensor-to-scalar ratio estimates derived
  from simulated observations of the sky by
  \core. \emph{First column}: component separation
    methods. \emph{Second column}: minimum multipole used in the
    likelihood estimation. \emph{Third column}: maximum multipole used
    in the likelihood estimation. \emph{Fourth column}: recovered
    value of the tensor-to-scalar ratio. \emph{Fifth column}:
    uncertainty on $r$. \emph{Sixth column}: bias, defined as the
    difference between the recovered and input values of $r$. For the cases where
    $r=10^{-3}$, the values in parenthesis correspond to
    estimates after correction for residual foregrounds in the
    likelihood (see text).}
\label{tab:final_results_table} 
\end{table}
\FloatBarrier


\section{Discussion}
\label{sec:discussion}

\subsection{Concerning B-mode delensing versus foreground cleaning}
\label{subsec:delens}

It is apparent from Table~\ref{tab:final_results_table} that the
overall uncertainty on $r=10^{-3}$ is not reduced much after de-lensing, either with {\tt Commander} alone, 
or with {\tt Commander} in combination with {\tt SMICA} or {\tt NILC}. The uncertainty on the tensor-to-scalar ratio
remains at a level $\sigma(r=10^{-3})\sim 0.5\times 10^{-3}$
whether $60$\% of the lensing is subtracted (simulation
\#5) or not (simulation \#4).
This means that the error on $r$ is not dominated by residual lensing.
It is not dominated by noise in CMB channels either, since noise is lower than the lensing contamination. 
Hence, the dominant source of error can be either foreground uncertainty, propagating to the final power 
spectrum error, or cosmic variance, or a combination of both. 
Cosmic variance can dominate the error if most of the sensitivity to
$B$ modes comes from the reionization bump.

We also note, however, that when de-lensing is performed, the error on $r$ 
from the {\tt SMICA} fit alone decreases substantially, from $1.4\times 10^{-3}$ for
simulation \#4 to $1.0\times 10^{-3}$ for simulation \#5. As our implementation of 
{\tt SMICA} in this paper constrains $r$
in the range $48 \leq \ell \leq 600$, this decrease shows that de-lensing has an 
impact on the detectability of the recombination bump. The overall error on $r$, however, is
not impacted much because the lowest harmonic modes, measured by {\tt Commander} in this paper, 
dominate the final sensitivity in this simulation.

To clarify further the impact of delensing on $\sigma(r)$, we
apply {\tt Commander} to additional specific simulations, as follows.
In the case where there are no foregrounds at all in the observed maps, 
a run of {\tt Commander} finds that $\sigma(r=10^{-3})$ decreases from
$0.3\times 10^{-3}$ without delensing to $0.2\times 10^{-3}$ after 
$60$\% delensing of the CMB $B$-mode signal (left panel of
Fig.~\ref{Fig:delensing_vs_foregrounds}, corresponding to analyses
where no foregrounds are included). 
There also is a notable difference in $\sigma(r=10^{-3})$
between the cases with and without foreground emission.
It hence appears that when foreground emission is included, the global error budget is
dominated by foreground-induced uncertainties.

\begin{figure}[t]
\centering
\includegraphics[width=0.49\textwidth]{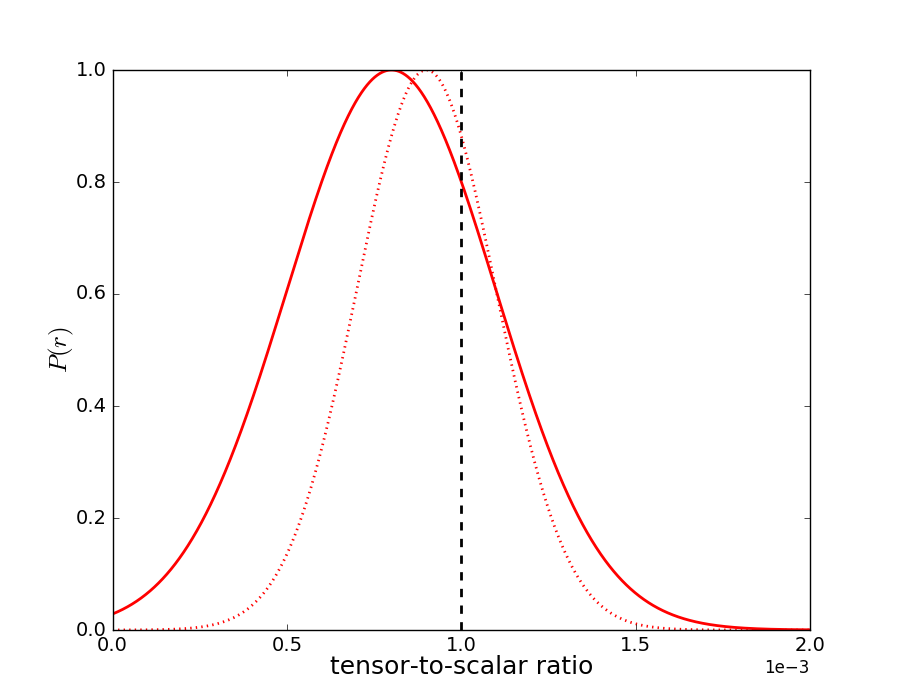}~
\includegraphics[width=0.51\textwidth]{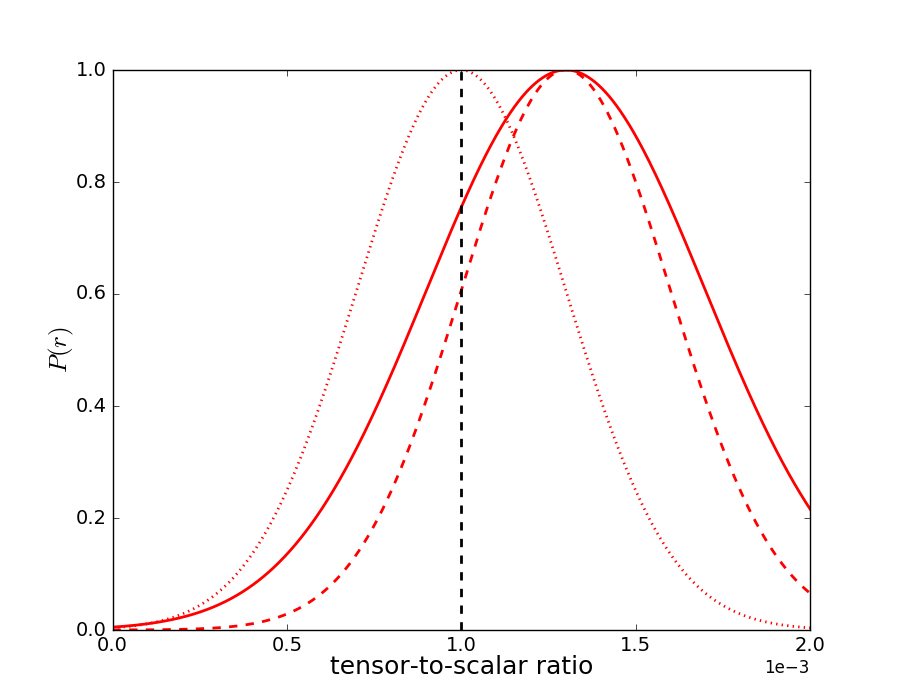}~\\
\caption{{\tt Commander} results on $r=10^{-3}$ after 60\% delensing
  as a function of the complexity of the foregrounds. \emph{Left
    panel}: posterior distribution $P(r)$ in the absence of
  foregrounds when either no delensing (\emph{solid red}) or 60\%
  delensing (\emph{dotted red}) is performed. \emph{Right panel}:
  posterior distribution $P(r)$ after 60\% delensing and foreground
  cleaning for \emph{constrained versions} of the simulation \#5: (i)
  in the absence of point-sources (\emph{solid red}), (ii) in the
  absence of point-sources, with synchrotron $\beta_s$ fixed rather
  than fitted (\emph{dashed red}), and (iii) in the absence of
  point-sources, with synchrotron $\beta_s$ and dust $T_d$ both fixed
  rather than fitted (\emph{dotted red}). See text for best-fit
  values.}
\label{Fig:delensing_vs_foregrounds}
\end{figure}

The right panel of Fig.~\ref{Fig:delensing_vs_foregrounds} presents
results from the analysis of three simplified versions of simulation
\#5. Firstly, we consider a simulated sky including $40$\% lensing
residuals but with no point-source contamination.  In this case
(denoted by the solid red line), the uncertainty on $r$ is reduced
to a level $\sigma(r=10^{-3})=0.4\times 10^{-3}$ (20\% lower than the 
case with point sources). When {\tt Commander} has to deal only 
with galactic foregrounds, it performs better than when the additional 
confusion due to polarized sources is included.

If in addition an idealized improvement in modelling the synchrotron foreground emission
is imposed on this simulation by fixing the synchrotron spectral indices to their
input values rather than fitting them, then the uncertainty on $r$ is reduced to the level
$\sigma(r=10^{-3})=0.3\times 10^{-3}$. 
Finally, when the
dust temperatures are fixed to their input values instead of fitting them, then in addition to reducing the uncertainty
to the level of ${\sigma(r=10^{-3})=0.3\times 10^{-3}}$ we also
suppress the shift  on the best fit value of $r$ previously seen in analyses of the
reionization scales.  

Hence,  when {\tt Commander} is given information about the real
frequency scaling in each pixel, the code can interpolate better
to CMB frequencies the foreground templates observed at 60\,GHz and 600\,GHz.
The final performance of the mission is as good as if there was no foreground contamination at all.

We note that combining the {\tt Commander} low-$\ell$ solution
with information from higher multipoles determined either by {\tt
NILC} or {\tt SMICA} does not seem to help much in reducing the uncertainty on $r$,
as compared with what is achieved by {\tt Commander} alone.
This is attributable to the fact that the two blind methods, as implemented in this work,
provide information in a range of $\ell$ where the primordial B-modes are
subdominant anyway. But one must keep in mind that the simulations analyzed here
have been tuned to match the {\tt Commander} model assumptions. The impact of
this is discussed in the next subsection. 

Possibly, further optimization of the methods can help to better detect the 
recombination peak, using as a prior improved understanding 
of the foregrounds (including
the point-source contribution at high-$\ell$). Detecting the recombination bump
in addition to the reionization bump is important to confirm the inflationary origin of any 
detected B-mode excess, as witnessed by the bias in {\tt Commander} data points at low $\ell$.
De-lensing is an important tool to achieve this.

\subsection{Concerning pixelization-related effects on foreground parametrization}
\label{subsec:degrade}

The emission laws of thermal dust and synchrotron are described, at
least in part, by power-law spectral indices that vary over the
sky. Even if we consider that, for any given line-of-sight, an
effective spectral index can be defined (ignoring the fact that it may
include emission with varying spectral properties along the
line-of-sight itself), maps from any CMB experiment are pixelized at a
finite spatial resolution and observed through a finite optical beam
response.  Thus a given pixel (and beam) encompasses many
line-of-sights with different spectral indices.  The coaddition of the
different power-laws with different spectral indices within the finite
pixel or beam solid angle can perturb the effective emission law
away from the idealized power-law dependence, and result in an
effective, unphysical, curvature of the spectral indices. This can
represent a difficulty for parametric component separation
methods faced with the challenge of correctly modelling foreground
contamination to measure the primordial CMB $B$-modes. A related
discussion on these issues for intensity can be found in
\cite{Chluba2017}.

\begin{figure}[htbp]
\centering
\includegraphics[width=0.5\textwidth]{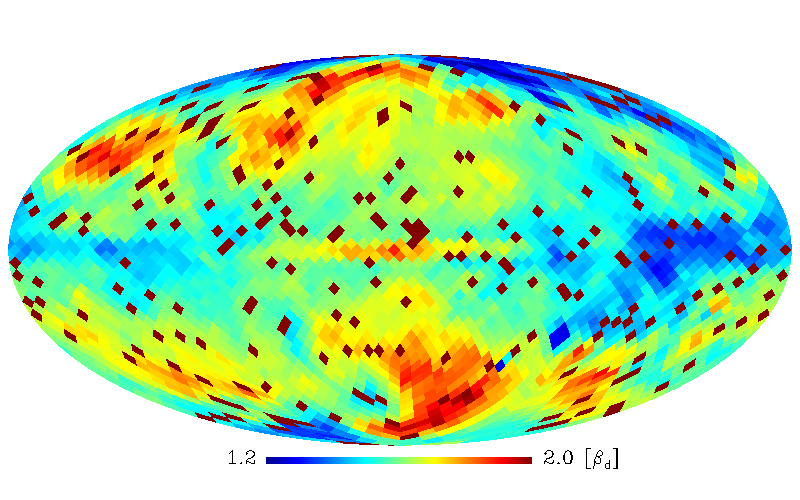}~
\includegraphics[width=0.5\textwidth]{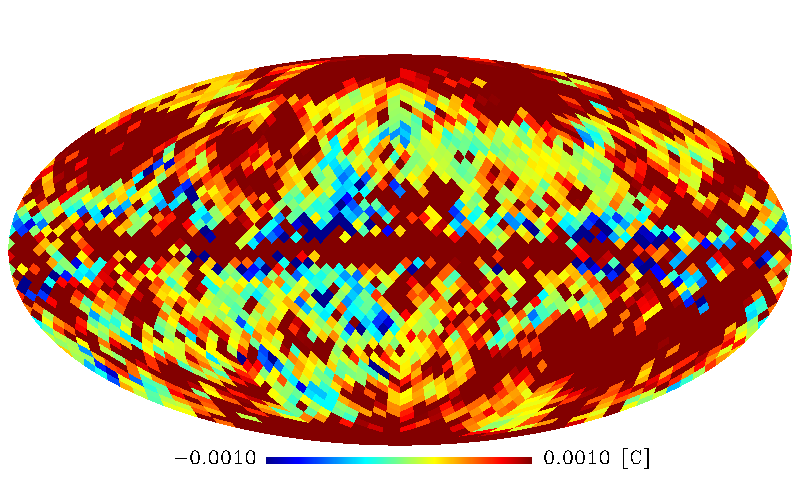}~\\
\includegraphics[width=0.5\textwidth]{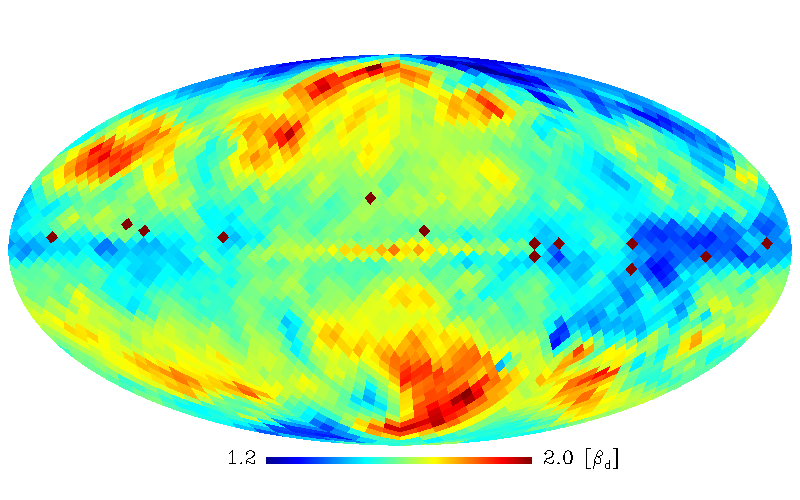}~
\includegraphics[width=0.5\textwidth]{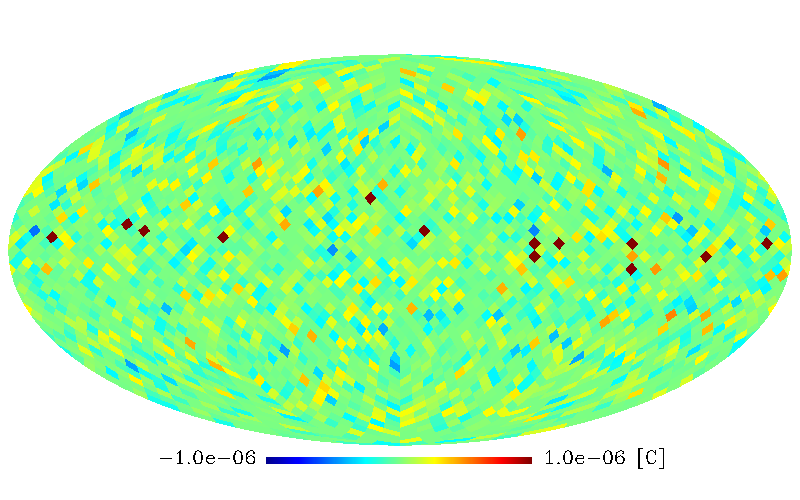}~\\
  \caption{Least-squares fit of dust spectral index, $\beta_d$,
      and curvature, $C$, on simulations with zero curvature:
      degrading the resolution of sky maps generates an effective
      curvature. \emph{Top panels}: the result of the fit of
    $\beta_d$ (left) and $C$ (right), obtained from $N_{\rm
      side}=2048$ simulated  frequency maps that have been degraded to
    $N_{\rm side}=16$, shows a non-zero (unphysical)
    curvature. \emph{Bottom panels}: same fit but obtained from direct
    $N_{\rm side}=16$ simulated frequency maps, in which case the
    curvature is consistent with zero.
}
\label{Fig:spurious_curvature}
\end{figure}

In particular, we have experienced this issue when utilising the
Bayesian parametric fitting method {\tt Commander} to fit the spectral
properties of foreground emission in low-resolution maps, at
${N_{\rm side}=16}$, which have been derived from simulated
$N_{\rm side}=2048$ data sets, within which the foreground spectral
dependencies vary from high resolution pixel-to-pixel.

We demonstrate the effect as follows. 
In Fig.~\ref{Fig:spurious_curvature}, we explicitly fit both a
spectral index, $\beta_d$, and a curvature term, $C$, i.e.,
\bea
\left({\nu\over \nu_{0}}\right)^{\beta_d + C \log \left({\nu\over \nu_{0}}\right)} 
\eea
to simulated \core\ dust frequency maps in which the simulated
emission law is a perfect power-law with zero curvature ($C=0$) in the
original simulation at $N_{\rm side}=2048$. We adopt a simple
chi-square fitting procedure to illustrate the issue. The fit is
either performed on the $N_{\rm side}=2048$ frequency maps degraded to
$N_{\rm side}=16$ (top panels of Fig.~\ref{Fig:spurious_curvature}) or
on direct $N_{\rm side}=16$ simulations with one spectral index per
low resolution pixel (bottom panels of
Fig.~\ref{Fig:spurious_curvature}). Clearly, degrading the pixelized
sky maps generates a non-zero unphysical curvature in the dust
spectral index (top right panel). Although the effective curvature of
the dust spectral index caused by degrading the pixelized sky maps is
low ($ C \sim 0.05$ at maximum), omitting this systematic curvature in
the fitted model can have a non-negligible
impact on low $r=10^{-3}$ $B$-mode signal detection, and this can
potentially bias the detection of CMB $B$-modes.

This issue is not specific to degrading sky maps to
lower pixel size. It is more general and due to the finite-sized
pixelization of the data and is a consequence of the foreground
spectral indices varying from one line-of-sight to another with a
characteristic scale smaller than the adopted pixel size. The scales
over which the effect may be important are those related to the detailed
physics of the interstellar medium, and the connection in relation to
polarized dust and synchrotron emission with the magnetic field
properties of the Galaxy.

\subsection{Concerning point-source processing for $B$-mode studies}
\label{subsec:pps}

The impact of point source contamination on $B$-mode analyses is an
issue that requires careful appraisal for both current and future CMB
experiments that aim to detect the tensor-to-scalar ratio at levels of
$10^{-3}$ and below.  In general, the compact source contribution does
not impact the large angular scales (near the reionization peak), but
can play an important role on intermediate and small angular scales
where the lensing-induced $B$-mode signal is present \citep{Curto2013}. 
Moreover, the consequences of point-source contamination of the $E$-mode 
signal to be used in delensing needs to be assessed.  In fact, a careful
investigation of the power spectrum of radio sources in polarization
was carried out by \cite{TucciToffolatti2012}
concluding that radio sources should not be a strong contaminant of
the CMB $E$-mode polarization at frequencies above $\simeq
70\,$GHz. However, careful treatment is required for CMB $B$-mode
polarization observations if the tensor-to-scalar ratio, $r$, is
$\ll 10^{-2}$. If the lensing $B$-mode contribution is removed, the
contamination by point sources remains the main nuisance.
The problem can be addressed in several different ways.

Firstly, it is important to improve our ability to detect sources at
the map level (in both intensity and polarization) using blind and
non-blind techniques. The current approach selects those sources
detected in intensity to mask out any possible related contamination in
the $Q$ and $U$ maps. This is a very conservative approach that, given
the typical compact source detection levels in CMB experiments, could
result in the masking out of a significant fraction of the sky,
particularly at lower frequencies where the beam sizes are
large. Moreover, heavily masked maps can have implications for the
fidelity of diffuse component separation methods as applied to the
data.

Alternatively, an attempt can be made to assess whether polarized
sources are present in the data given the list of detections in
intensity. Such a search is made by studying a map of polarized
intensity, $P$, and searching at all of the candidate positions.
However, the statistical properties of the $P$ map can not be
approximated by a Gaussian distribution, following instead a Rayleigh
distribution. Thus, the usual thresholding approach based on
variance-levels of the signal are not applicable, and the assessment
must define significance levels that account for the local statistics
in the vicinity of the putative source location. This is the method
followed in this paper, and previously by \Planck\
\citep{planck2013_pccs,planck2015_pccs}. Such an approach is much less
conservative than masking all the sources detected in intensity, but
an important caveat applies: it is a non-blind approach and sources
with flux densities just below the detection level in intensity but
with very large polarization fractions can be missed. Therefore, blind
searches in the $P$ maps would be desirable, and perhaps even extended to 
$E$ and $B$ maps given that some component separation methods work
directly with these quantities. 

Secondly, we can seek to improve the source detection process using as
a proxy deep catalogues of compact sources detected with high
resolution telescopes or interferometers in the microwave and sub-mm
frequency range to either mask or enable non-blind searches in the CMB data sets.
We can motivate this as follows.

As mentioned in Sect.~\ref{subsec:xtragal}, unresolved extragalactic
sources contribute both shot noise (Poisson) and clustered power to
the CMB maps. There is a substantial difference between the
contributions of the two main populations, radio sources and dusty
galaxies. The former, that dominate the point source temperature
fluctuations up to 100--150\,GHz, have a low clustering power, so that
their power spectrum is simply white noise
($C_\ell = \hbox{constant}$). The amplitude of the Poisson power
spectrum is
\begin{equation}\label{eq:Poisson}
C_\ell={\mathlarger\int}_{\!\!\!0}^{S_d} {dN\over dS}\, S^2\, dS ,
\end{equation}\label{eq:flutt}
where $dN(S)/dS$ are the differential number counts per steradian of
sources weaker that the detection limit $S_d$. The slope $\beta$ of
the radio source $dN(S)/dS$ ($dN(S)/dS\propto S^{-\beta}$) is always
$<3$, implying that the maximum contribution comes from sources just
below the detection limit. Thus the radio source contamination can be
mitigated, at least in some sky areas, by masking sources detected by
large area ground based surveys, such as the Australia Telescope 20
GHz survey, covering the full southern sky down to flux densities in
the range 40--100\,mJy. Over areas of $\sim 1,000\,\hbox{deg}^2$ one
can take advantage of the deep SPT (at 95 and $\simeq 150$\,GHz) and
ACT (at $\simeq 150$\,GHz) surveys. 

Moreover, in the future, significant progress will be made possible by
the Stage-IV CMB (CMB-S4) surveys \citep{Abazajian2016} which aim to
map most of the sky to sensitivities of $\sim 1\,\mu$K$\cdot$arcmin.
The counts based on the combination of the SPT and ACT surveys with
those from \textit{Planck}
\citep{PlanckCollaborationXIIIstat_prop2011, Planck_counts2013}, that
extend them to bright flux densities, tightly constrain the amplitude
of the power spectrum, making possible a correction of the CMB
contamination at the power spectrum level.

However, it should also be noted that a large fraction of the radio
sources that can be detect in CMB experiments are variable.  The fact
that a source was detected with a high resolution instrument at any
given time does not imply that it is contributing to the signal as
measured by a CMB experiment measured at a different epoch. 
Conversely, some sources might be much brighter 
in the space mission data than when observed from the ground.
One way of
addressing this problem is to plan follow-up observations of sources
with high resolution instruments in such a way that these observations
are performed near simultaneously with the CMB observations. Such a
strategy was already undertaken by \Planck, and involved the
simultaneous observation of radio sources using the Australian Compact
Array. Something similar could be envisaged for future CMB
experiments, although ground based observatories are often heavily
over-subscribed. This has led some CMB projects to invest in the
contruction of dedicated instruments to perform simultaneous follow-up
observations of compact sources, e.g., the Very Small Array at Teide
Observatory.

The situation is substantially different in the case of unresolved
dusty galaxies. The main contribution to their power spectrum comes
from faint, high surface density, high-$z$ galaxies.  In fact, the
amplitude of the Poisson power spectrum can be determined quite
accurately exploiting \textit{Planck} plus ground based surveys: the
SCUBA-2 counts at 350\,GHz \cite[][and references therein]{Hsu2016,
  Geach2017}, SPT and ACT counts at $\simeq 150$\,GHz and
$\simeq 200$\,GHz, extended to $\mu$Jy levels by ALMA surveys
\cite[][and references therein]{Carniani2015, Aravena2016}.  But their
power spectrum is dominated by clustering for $\ell \simlt 2000$
\citep{PlanckCollaborationXXX2014,DeZotti2015}. \textit{Planck} has
allowed measurements of the power spectrum of unresolved point sources
in the multipole range dominated by clustering at 217 GHz, 353 GHz,
and 545 GHz, with some preliminary estimate also at 100\,GHz
\citep{PlanckCollaboration2011CIB,PlanckCollaborationXXX2014,Mak2016}. \textit{Herschel}
\citep{Viero2013}, ACT \citep{Dunkley2013} and SPT
\citep{Reichardt2012} measurements have extended the determinations to
smaller angular scales.  It should be mentioned that despite the low
level of polarization of individual dusty galaxies, the actual
polarization of the CIB emission does not average out in variance
since this measures fluctuations in power.

As in the case of the temperature power spectrum, if indeed dusty
galaxies do not contribute much, the radio source contamination can be
mitigated exploiting deep, large area surveys. Furthermore, the counts in
polarization that \core\ will provide should allow accurate modelling of
the contamination at the power spectrum level. 

\subsection{Further improvements}
\label{subsec:improvements}

The work done for this paper shows the difficulty of reaching precision and accuracy
for the detection of inflationary $B$ modes if $r \simeq $ few $\times 10^{-3}$. A big risk 
for a future space mission is to detect an excess of $B$-mode power on large scales, without
the capability to ascertain its cosmological origin. Such a false detection would reproduce the 
uncertainty that followed the original BICEP2 announcement with ground-based observations.

To avoid this, the future space mission must be designed not only for the capability to reduce
foreground contamination to allow for the detection of primordial $B$-modes. It must also 
be designed to provide means of evaluating the level of residual foreground emission
in $B$-mode maps and spectra.

To this effect, the capability of comparing the results obtained on different patches of sky, 
with different band subsets, with different component separation methods is key. 

Component separation to detect $r \simeq 10^{-3}$ has been shown to be hard. We now
identify paths for further improvement of the component separation, and for the characterization
of errors for the three main methods used in this work.

\paragraph{Complementary observations:}

If the model assumed by {\tt Commander} is accurate enough to precisely hold
over a large range of frequencies, then additional observations might help.
We anticipate that by the \core\ launch-date
external surveys of polarized sources will have accurately measured
their spectral properties, and that better constraints on the spectral index
of the diffuse synchrotron polarization emission over the full sky
will have been achieved by C-BASS \citep{Irfan2015} and QUIJOTE
\citep{Rubino-Martin2012} and follow-up surveys. \core\ would be the 
most sensitive experiment to precisely map dust emission in the relevant
frequency range, although an experiment such as PIXIE \citep{Pixie2016}, with 600 absolutely-calibrated frequency channels from 30 to 6000\,GHz, could bring 
additional information on large angular scales to better model large-scale galactic foreground emission.

\paragraph{Commander:}

In this simulation, to alleviate the issue discussed in Sec.~\ref{subsec:degrade},
 we have applied {\tt Commander} on maps generated directly at 
$N_{\rm side}=16$. The analysis of real sky data will require to find
a way to mitigate the impact of variations of emission laws within sky pixels 
and along the line-of-sight. This difficulty, which has been bypassed 
in the present work, will have to be addressed in the future.

As a first step, we must investigate ways to extend the {\tt Commander} parametric
fit for $B$-modes to smaller pixels and larger multipole ranges, e.g.
$2\leq \ell \leq 200$, by optimizing the algorithm in some way to
relax the actual large computational costs. We must also find a way to validate within the analysis the 
parameterization that is assumed for each of the foreground emissions.

\paragraph{SMICA:}

The {\tt SMICA} method relies on the effective dimension of the
foreground subspace in polarization (Sect.~\ref{subsec:smica_method}),
which is a strong prior assumption, to extract the cosmological
signal. In the current {\tt SMICA} algorithm, the dimension of the
foreground subspace is fixed to an ad-hoc value, although this is
likely to vary both over the sky and the angular scales probed,
depending on the relevance of each of the foregrounds with respect to
the noise. A possible improvement would be to follow the approach
adopted by the {\tt GNILC} component separation method, developed for
intensity in \cite{Remazeilles2011b} and
\cite{Planck_PIP_XLVIII}. Here, any prior assumption on the number of
foregrounds is relaxed and the effective dimension of the foreground
subspace is estimated locally both over the sky and over physical
scales by using wavelet decomposition of the sky maps and thresholding
with respect to the local foreground signal-to-noise ratio.

\paragraph{NILC:}

In the present paper, a standard {\tt NILC} algorithm has been used.
There has been no attempt to optimize spectral windows, nor to optimize the rejection of
foreground contamination, rather than minimizing the total variance of the reconstructed maps.
Such improvements may prove useful to reduce the foreground contamination in the reconstructed maps, and possibly to 
find ways of evaluating residual emission. Further investigations are left to future work on this topic.


\section{Conclusions}
\label{sec:conclusions}

In this paper, we have considered whether the \core\ satellite
mission, as proposed to ESA's M5 call, allows the removal of
astrophysical foregrounds at the accuracy required to measure the
power spectrum of primordial CMB $B$-modes on both the reionization
and recombination scales for inflationary models with tensor-to-scalar
ratios over a range of interesting values, an optical depth to
reionization, $\tau=0.055$ \citep{Planck_lowl_2016} and the other
$\Lambda$CDM model cosmological parameters defined by the \Planck\
best-fit values \citep{planck2015_overview}.

In particular, three independent component separation methods of
either a parametric ({\tt Commander}) or blind ({\tt SMICA} and {\tt
  NILC}) nature were applied to a set of simulations consistent with
observations of the sky by \core.  These simulations include
foregrounds covering a range of complexity as modelled by the PSM, but
at least including contributions due to synchrotron and dust with
variable spectral indices, with additional components due to AME and
point-sources considered in the most challenging cases.  The impact of
gravitational lensing effects has also been tested by comparing
results either without the addition of a lensing $B$-mode
contribution, or including them at both the predicted level or after
an idealized 60\% delensing has been applied.

We have determined that, even at the current stage of development more
than one decade before the potential launch of the \core\ satellite,
each of the three component separation methods are independently able
to reconstruct the cosmological $B$-mode signal, without bias, for
tensor-to-scalar values $r \gtrsim 5\times 10^{-3}$. Specifically, in
the absence (or perfect control) of lensing \core\ allows the
determination of a tensor-to-scalar ratio at the $10^{-2}$ or
$5\times 10^{-3}$ level at a significance exceeding $10\sigma$.
However, when lensing effects are included, the significance of
detection at the latter amplitude is lowered to about $4\sigma$, with,
for example, the {\tt Commander} estimate yielding a value of
$r=\left(5.4 \pm 1.5\right)\times 10^{-3}$.  Nevertheless, one of the
methods -- {\tt Commander} -- operating in the low-$\ell$ regime only
suggests that a value of $r$ as low as ${r=2.5\times 10^{-3}}$ can be
determined without bias and with $\sim 8\sigma$ significance in the
absence of lensing.  Therefore, it appears that, at the very least,
observations by \core\ would allow constraints to be imposed on the
energy-scale of inflation for the Starobinsky $R^2$ inflation model,
for which $r\sim 4.2\times10^{-3}$
\citep{Starobinsky1980,Mukhanov1981,Starobinski1983}.

Importantly, however, we have demonstrated that when the
tensor-to-scalar ratio is decreased to a critical value of
$r \sim 10^{-3}$, then the spectral complexity of the foregrounds is
sufficient to bias the measurement of the primordial CMB $B$-mode
power spectrum, both on reionization scales ($\ell < 12$) in the case
of parametric component separation techniques, and over all angular
scales when utilizing blind methods, even in the absence of lensing
effects.  What is evident is that, in order to measure $r=10^{-3}$,
foreground residuals after component separation have to be modelled
and marginalized over in the likelihood estimation, in which case
\core\ is able to correct for the bias and measure the
tensor-to-scalar ratio with an accuracy
$\sigma(r=10^{-3})\sim 2 \times 10^{-4}$, in the absence of lensing
effects. However, the presence of lensing $B$-modes, together with the
addition of further foreground contributions due to AME and
point-sources, decreases the sensitivity to $r$ to
$\sigma(r=10^{-3})\sim 5 \times 10^{-4}$.
We note that \cite{Alonso2017} have claimed similar uncertainties on
$r$ for a high-resolution ground-based CMB-Stage IV experiment,
although the authors do not include a contribution from extra-galactic
sources.
In addition, and perhaps unexpectedly, even if delensing of the
$B$-mode signal is achieved at the 60\% level, the uncertainty on the
tensor-to-scalar ratio remains at this level, despite the fact that
the cosmic variance contribution from the lensing $B$-modes must have
been reduced by a corresponding amount. The uncertainty is therefore
dominated by foreground residuals rather than lensing cosmic variance.
In this case, the use of external information on the spectral index
variation of the diffuse synchrotron emission and the polarized
point-source contribution should increase the significance of the
detection by \core\ to $\sigma(r=10^{-3}) \sim 3 \times 10^{-4}$,
after foreground cleaning and 60\% delensing
(Sect.~\ref{subsec:delens}).

We have identified different sources of potential bias in the
reconstruction of the primordial CMB $B$-mode signal: (i) incorrect
foreground modelling; (ii) inadequate frequency range or lack of
frequency channels sampling it; and (iii) systematic bias due to
averaging foreground spectral indices through pixelization. While
these sources of bias likely have no significant impact on the
reconstruction of CMB temperature and $E$-mode polarization, they
become critical for the reconstruction of the faint primordial CMB
$B$-mode polarization especially if $r\sim 10^{-3}$.  Despite the
absence of spectral mismatch between the foreground model and the
data, and a perfect scaling of the foregrounds in each pixel area
through direct $N_{\rm side}=16$ simulations, we have found that the
spectral variations of the synchrotron spectral index are not
accurately reconstructed, even though the fit of the total sky
emission in polarization can be considered satisfactory on the basis
of low chi-square statistics. This, ultimately, generates some bias on
the estimate of a tensor-to-scalar ratio at the level of
$r=10^{-3}$. However, fixing the synchrotron spectral indices and the
dust temperatures to their input values and fitting the dust spectral
index only with {\tt Commander} helps to remove the bias on the
reconstructed CMB $B$-mode power spectrum on reionization scales, and
reduces the overall uncertainty on $r$ after foreground cleaning and
delensing (right panel of Fig.~\ref{Fig:delensing_vs_foregrounds}).
Therefore, \core\ could benefit from the addition of low-frequency
channels below $60$\,GHz. This could be achieved in part by including
information from full-sky polarization surveys at lower frequencies,
such as C-BASS \citep{Irfan2015} and QUIJOTE
\citep{Rubino-Martin2012}.  A counterpoint to this, however, is that
we would then need accurate modelling of diffuse AME polarization,
which can no longer be ignored as a contaminant to low-$r$ CMB
$B$-modes at such frequencies.

The high resolution of the \core\ frequency channels also allows the
determination of the CMB $E$-mode polarization signal 
up to relatively high multipoles. 
This, in principle, will
then allow the accurate reconstruction of the lensing potential,
$\phi$, using well-known quadratic lensing estimators \citep{Hu2002}
as applied to the reconstructed CMB $E$-mode polarization
map. 
Subsequently, proper delensing of the CMB $B$-mode power spectra could
reduce the lensing contribution to the cosmic variance by $\sim 60$\%,
enabling the detection of the recombination bump for a level of $r$ 
that is out of reach for lower-resolution
CMB polarization experiments such as \textit{LiteBIRD} \citep{Litebird2016} or
\textit{PIXIE} \citep{Pixie2016}. 

It should be noted that systematic effects (calibration errors, beam
asymmetries, bandpass mismatch, etc) will have an impact on component
separation results and on the accuracy of foreground removal.
\cite{2010MNRAS.401.1602D} demonstrated that small calibration errors
(less than $1$\%) can perturb the weighting of the frequency maps
implemented in the {\tt NILC} component separation method, resulting
in a partial cancellation of the reconstructed CMB signal, especially
for the high signal-to-noise measurements. Beam asymmetries can also
be regarded as an imperfect calibration for {\tt NILC}, and also
affect the inferred {\tt NILC} weighting similarly. Given the high
sensitivity required to measure the primordial $B$-modes at a level of
$r\sim10^{-3}$, an accurate calibration of the instrument becomes
essential. Detector bandpass mismatch can also impact the results of
parametric fitting methods, such as {\tt Commander}, due to the
modication of the foreground spectra integrated over the bandpass from
one detector to another. The control of such systematics effects is
discussed in a companion paper \citep{ECO_systematics}. However,
ideally the effects of foregrounds and systematics actually should be
controlled collectively rather than independently. This will be
investigated in future work.

In summary, we have performed a detailed component separation and
likelihood analysis, going beyond a simple Fisher approach, to
reconstruct the exact shape of the primordial CMB $B$-mode power
spectrum as might be measured by \core\ in order to study both
sensitivities and biases to the tensor-to-scalar ratio measurements
after foreground cleaning. In Sect.~\ref{sec:fisher}, we have compared
our component separation results with a forecasting method that
performs an effective statistical averaging of CMB and noise
realisations and foreground uncertainties, and we have found
consistency between both approaches when analysing data with a
tensor-to-scalar ratio of $r=10^{-3}$ in the presence of foregrounds and
lensing. 
Although additional refinements are required, 
the component separation results seem promising and we can speculate
that a detection of $B$-mode polarisation corresponding to $r\sim 10^{-3}$ will be achievable by a \core-like mission with a potential launch-date
around 2030.


\acknowledgments{The research leading to these results has received funding from the ERC Starting Consolidator Grant (no.~307209). Some of the results in this paper have been derived using the \healpix\ package \citep{Gorski2005}.  We acknowledge the use of the PSM package \citep{Delabrouille2013} , developed by the
  \Planck\ working group on component separation, for making the
  simulations used in this work. We acknowledge the use of the
  Ulysses cluster at SISSA. This research was 
  partially supported by the RADIOFOREGROUNDS project, funded by the
  European Commission's H2020 Research Infrastructures under the
  Grant Agreement 687312, and the INDARK INFN Initiative. GDZ acknowledges
  support by ASI/INAF agreement no.~2014-024-R.1. R.F.-C., E.M.-G., and P.V. acknowledge support from the Spanish MINECO project ESP2015-70646-C2-1-R (cofinanced with EU FEDER funds), Consolider-Ingenio 2010 project CSD2010-00064 and from the CSIC ``Proyecto Intramural Especial'' project 201550E091. JGN acknowledges financial support from the Spanish MINECO for a 'Ramon y Cajal' fellowship (RYC-2013-13256) and the I+D 2015 project AYA2015-65887-P (MINECO/FEDER). CJM is supported by an FCT Research Professorship, contract reference IF/00064/2012, funded by FCT/MCTES (Portugal) and POPH/FSE.}

\appendix


\section{Component Separation Methods}
\label{sec:methods_appendix}

\subsection{Bayesian Parametric Fitting}
\label{subsec:commander_method}

A physical model is fit to a set of observations within a Bayesian
parametric framework using the 
{\tt Commander} algorithm \citep{Eriksen2008}.
The sky model, $\bdm(\nu,p)$, as fitted to the \core\ sky maps,
$\bdd(\nu,p)$, at each frequency, $\nu$, and for each pixel, $p$, is
parametrized by
\bea
\label{eq:fit}
\bdm(\nu,p) &=& a(\nu)\,\bds^{cmb}(p) \cr
            &+& \left({\nu\over \nu_0^{s}}\right)^{\beta_s(p)}\, \bds^{sync}(p) \cr
            &+& \left({\nu\over\nu_0^{d}}\right)^{\beta_d(p)}{B_{\nu}\left(T_d(p)\right)}\,\bds^{dust}(p) \cr
            &+& \bdn(\nu,p),
\eea 
where $\bds^{cmb}(p)$ is the amplitude of the CMB ${Q,U}$ polarization
anisotropies, $\bds^{sync}(p)$ is the amplitude of the polarized
synchrotron ${Q,U}$ radiation, $\bds^{dust}(p)$ is the amplitude of
the polarized thermal dust ${Q,U}$ radiation, $\bdn(\nu,p)$ is the
instrumental noise for \core\ in the ${Q,U}$ Stokes parameters, $a(\nu)$ is
the frequency spectrum of the CMB, $\beta_s(p)$ is the synchrotron
spectral index, $\beta_d(p)$ is the thermal dust spectral index, and
$T_d(p)$ is the dust temperature.  Here, we fix $a(\nu)$ to be constant
and equal to unity in thermodynamic temperature units. In addition to
the amplitudes and spectral indices of the components, the $E$- and
$B$-mode angular power spectra
$C_\ell=\left\{C_\ell^{EE},C_\ell^{BB}\right\}$ of the CMB are fitted
in a self-consistent way with the amplitude and spectral index maps
using the Gibbs sampling scheme detailed later in the
section. Considering both  the $Q$ and $U$ Stokes parameters in the
analysis, there are $2\times 19 = 38$ channels from \core, while
the parametric model consists of $2\times 3+3+2 = 11$ parameters in
total to be fitted to the data,
\bea
\bds = \left(\bds^{cmb},\bds^{dust},\bds^{sync}\right), \boldsymbol{\beta} = \left(\beta_d,T_d,\beta_s\right), C_\ell = \left\{C_\ell^{EE},C_\ell^{BB}\right\} = \langle\vert\bds_{\ell m}^{cmb}\vert^2\rangle.
\eea
As shown in \cite{Remazeilles2016}, the performance of the {\tt
  Commander} component separation method relies on the accurate
parametrization of the spectral properties of the foregrounds, and
small foreground modelling errors can strongly bias an estimate of a
low tensor-to-scalar ratio. In this work, we fit a power-law for
synchrotron and a single modified blackbody for thermal dust,
thereby avoiding any mismatch between the data and the model, in
order to characterize the uncertainty on the tensor-to-scalar ratio
that is only due to foreground residuals after component
separation. It should be noted that we have verified that omitting AME
in the parametric model has no impact for {\tt Commander} on the
estimate of the tensor-to-scalar ratio, so that we do not attempt
to fit for its contribution. We have determined that this absence of impact on the
results is due to the frequency coverage of \core. The CNM spectrum of
AME steepens so sharply at frequencies $ \gtrsim 50$\,GHz that AME
polarization becomes insignificant in the lowest frequency band,
$60$\,GHz, of the instrument.

Component separation is achieved by computing the joint CMB-foreground
posterior distribution, which, according to Bayes' theorem, is given
by
 \bea\label{eq:posterior}
P\left(\bds,\boldsymbol{\beta},C_\ell \big| \bdd \right) \propto \mathcal{L}\left(\bdd \big| \bds,\boldsymbol{\beta},C_\ell\right)P\left(\bds,\boldsymbol{\beta},C_\ell\right),
\eea 
where $P\left(\bds,\boldsymbol{\beta},C_\ell\right)$ is the prior
distribution of the parameters. Here, we do not set any prior on
$C_\ell$ and on the amplitudes of the components but we assume
Gaussian priors for the foreground spectral parameters,
$P\left(\beta_s\right)\sim\mathcal{N}(-3,0.1)$,
$P\left(\beta_d\right)\sim\mathcal{N}(1.6,0.3)$, and
$P\left(T_d)\sim\mathcal{N}(19.4\,{\rm K},1.5\,{\rm K}\right)$. The
Gaussian priors are multiplied by a Jeffreys ignorance prior in order
to suppress the prior volume of the likelihood space for non-linear
parameters \citep{Jeffreys1946,Eriksen2008}. By assuming the noise to
be Gaussian and uncorrelated across frequencies, the likelihood is then
\bea
\mathcal{L}\left(\bdd \big| \bds,\boldsymbol{\beta},C_\ell\right) \propto \exp\left\{{-1\over 2} \sum_\nu \left(\bdd(\nu,p) - \bdm(\nu,p)\right)^t \tN^{-1}\left(\bdd(\nu,p) - \bdm(\nu,p)\right) \right\},
\eea
where $\tN$ is the noise covariance matrix in pixel space.

The Bayesian parametric fitting approach enables the end-to-end propagation of the foreground uncertainties to the
reconstructed CMB power spectrum and then to an estimate of the
tensor-to-scalar ratio. The algorithm also provides a map of the
chi-square goodness-of-fit, measuring the mismatch between the model
and the data in each pixel,
\bea
\label{eq:chi2}
\chi^2 (p) = \sum_\nu \left({\bdd(\nu,p) - \bdm(\nu,p)\over \sigma_\nu(p)}\right)^2,
\eea
where $\sigma_\nu(p)$ is the noise pixel $p$. The chi-square map
provides useful feedback on the fidelity of the foreground modelling as well as a
possible criterion to improve masking a posteriori, by discarding the pixels
where the chi-square statistic is too high in successive iterations of
the component separation. However, in this paper we use a simpler
masking approach, as described in Sect.~\ref{subsec:parametric}.

Direct computation of the joint CMB-foreground posterior distribution
Eq.~(\ref{eq:posterior}) is not feasible as it would require the
distribution to be mapped out over a multidimensional grid, the
size of which grows exponentially with the number of parameters. A more
tractable approach is to map out the joint CMB-foreground posterior
distribution by Gibbs sampling
\citep{Wandelt2004,Eriksen2004,Eriksen2008}, i.e., each parameter is
sampled alternately from iterative conditional probabilities,
according to the following Gibbs sampling scheme,
\bea
\label{eq:gibbs}
\widehat{\bds}^{~(i+1)} &\leftarrow& P\left(\widehat{\bds} \,\big|\, \widehat{C}_\ell^{~(i)},\boldsymbol{\widehat{\beta}}^{~(i)},\bdd\right),\cr
\boldsymbol{\widehat{\beta}}^{~(i+1)} &\leftarrow& P\left(\boldsymbol{\widehat{\beta}} \,\big|\, \widehat{\bds}^{~(i+1)}, \bdd\right),\cr
\widehat{C}_\ell^{~(i+1)} &\leftarrow& P\left(\widehat{C}_\ell \,\big|\, \widehat{\bds}^{~(i+1)}\right),
\eea 
where $(i)$ denotes the samples at the $i^{th}$ iteration in the Markov
chain. It has been demonstrated mathematically that as the number of iterations
approaches infinity then the Gibbs scheme converges to the sampling with
the full joint CMB-foreground posterior distribution, after some initial
burn-in \citep{Wandelt2004}. Here, we use $2000$ iterations in the
Markov chain but discard the first $500$ Gibbs samples (burn-in) to
ensure that we have reached convergence. 

The amplitudes of the components are sampled by the conditional
Gaussian distribution
\bea
P\left(\bds \big| C_\ell,\bdd\right) &\propto& P\left(\bdd \big| \bds,C_\ell\right)P\left(\bds \big| C_\ell\right)\cr
&\propto& e^{(-1/2)\left(\bdd - \bds\right)^T\tN^{-1}\left(\bdd - \bds\right)}e^{(-1/2)\bds^T\tS^{-1}\bds}\cr
&\propto& e^{(-1/2)\left(\bds - \widehat{\bds}\right)^T\left(\tS^{-1}+\tN^{-1}\right)\left(\bds - \widehat{\bds}\right)},
\eea
where $\tS$ and $\tN$ are the CMB and noise covariance matrices in
pixel space, and $\widehat{\bds}$ is the Wiener filter solution:
\bea
\widehat{\bds} = \left(\tS^{-1}+\tN^{-1}\right)^{-1}\tN^{-1} \bdd.
\eea
In practice, the Wiener estimate is computed by solving the following
equation with the conjugate gradient method,
\bea\left(\tS^{-1}+\tN^{-1}\right)\widehat{\bds} = \tN^{-1} \bdd + \tS^{-1/2}w_0 + \tN^{-1/2}w_1,
\eea
where $w_0,w_1 \sim \mathcal N(0,1)$.
By assuming the CMB fluctuations to be Gaussian and isotropic, the CMB
power spectrum is thus sampled from the conditional distribution,
\bea
P\left(C_\ell \big| \bds\right) \propto {e^{-{\left(2\ell +1\right)\over 2C_\ell}\left({1\over 2l+1} \sum_{m=-\ell}^\ell \vert \bds_{\ell m}^{\rm cmb} \vert^2\right)} \over C_\ell^{(2\ell+1)/2}},
\eea
which corresponds to an Inverse-Gamma distribution when interpreted as a
 function of $C_\ell$.\footnote{In the case of polarization, the conditional distribution actually corresponds to a mix of an Inverse-Wishart distribution (the correlated T and E-mode parts) and an Inverse-Gamma distribution (the B-mode part).} By writing the pseudo-power spectrum of the CMB
 map as
\bea
\widehat{C}_\ell = {1\over 2l+1} \sum_{m=-\ell}^\ell \vert \bds_{\ell m}^{cmb} \vert^2,
\eea
the $C_\ell$ sampling distribution can be recast as
\bea\label{eq:cl_distribution}
\ln  P\left(C_\ell \big| \bds\right) = \ln  P\left(C_\ell \big| \widehat{C}_\ell\right) = {(2\ell+1)\over 2}\left[ \ln\left({\widehat{C}_\ell\over C_\ell}\right) - {\widehat{C}_\ell\over C_\ell} + 1\right].
\eea
For the sampling of the foreground spectral indices,
$\boldsymbol{\beta}$, we refer to \cite{Eriksen2008} for more details.

The posterior distribution of the CMB $E$- and $B$-mode power spectra
can then be estimated by using the Blackwell-Rao approximation
\citep{Chu2005},
\bea\label{eq:cl_posterior}
P\left(C_\ell \big| \bdd\right) &=& \int d\bds P\left(C_\ell,s \big| \bdd\right),\cr
&=& \int d\bds P\left(C_\ell \big| \bds \right)P\left(\bds \big| \bdd\right),\cr
&=& \int D\widehat{C}_\ell P\left(C_\ell \big| \widehat{C}_\ell \right)P\left(\widehat{C}_\ell \big| \bdd\right),\cr
&=& {1\over N_G} \sum_i P\left(C_\ell \big| \widehat{C}_\ell^{(i)}\right),
\eea
where the sum runs over $N_{G}$ Gibbs samples $\widehat{C}_\ell^{(i)}$, and $P\left(C_\ell \big|
  \widehat{C}_\ell^{(i)}\right)$ is given by
Eq.~(\ref{eq:cl_distribution}). The Blackwell-Rao estimate becomes an
exact approximation of the posterior $C_\ell$ distribution as the
number of Gibbs samples increases \citep{Chu2005,Eriksen2008}.

In summary, the {\tt Commander} algorithm enables component
separation, power spectrum estimation, and cosmological parameter
estimation (see Sect.~\ref{subsec:parametric}) in a self-consistent
way using MCMC Gibbs sampling and
the Blackwell-Rao approximation. 

Note that in Sect.~\ref{sec:compsep_results}, where we integrate low-
and high-$\ell$ results
for an optimal estimation of $r$ based on a larger range of
multipoles, we multiply the posteriors for the cosmological parameters
obtained from {\tt Commander}, using a Blackwell-Rao approximation and
non-Gaussian likelihood, with those obtained from either {\tt NILC} or
{\tt SMICA}, using a Gaussian likelihood. This is because the latter
are not suited to the Blackwell-Rao approach.

\subsection{Internal Linear Combination in Needlet space}
\label{subsec:nilc_method}

As an alternative approach for the removal of foregrounds present in
the sky maps, we implement a blind Internal Linear Combination (ILC)
method
\citep{Tegmark1996,Bennett2003,Tegmark2003,Eriksen_ilc2004}. This
method only attempts to reconstruct the CMB signal, without using any prior
information about foregrounds. It is based on two specific
assumptions. Firstly, that the CMB is frequency independent
in thermodynamic unit,  and secondly, that the CMB is uncorrelated
with foreground signals. The ILC method
then estimates the CMB, ${\widehat S}$, as a weighted linear combination of
the set of input multi-frequency sky maps such that the variance of the estimate is
minimum, with unit response to the flat CMB frequency spectrum,
\begin{eqnarray} 
\widehat S = w^{T} X = \frac{\displaystyle{a^{T} \widehat R^{-1}}}{\displaystyle{a^{T} \widehat R^{-1} a}} X =\frac{\displaystyle{a^{T} \widehat R^{-1}}}{\displaystyle{a^{T} \widehat R^{-1} a}} \left(a^{ } S + F + N\right),
\label{equ:ilc} 
\end{eqnarray}
where $X$ is the vector of frequency maps, $a$ the constant
frequency spectrum of the CMB signal $S$, $F$ the total foreground
signal, $N$ the instrumental noise for the different frequency
channels, and ${\widehat R}$ the frequency-frequency covariance
matrix. The first condition guarantees minimum contamination by
foregrounds and instrumental noise whereas the second condition
guarantees that the CMB signal is conserved without bias.
The presence of the foregrounds induces
correlated errors across frequencies, so that the ILC weights adjust
themselves to minimise the foreground residuals present in the the weighted
linear combination.
However, in reality the weights result from a
trade-off between minimising the foregrounds and minimising the
instrumental noise contribution in the reconstructed CMB map.

However, there are also some drawbacks to the method. As discussed by
a number of authors \citep{saha2006blind, souradeep2006angular,
  hinshaw2007three, saha2008cmb, van2009lecture, delabrouille2009full,
  saha2011foreground}, the component of interest (CMB) and the
foreground signals must be uncorrelated for proper ILC performance. On
finite data sets, this can only be approximately true, and empirical
correlations between the CMB and foregrounds generate a bias in the
reconstructed CMB on large angular scales. In addition, as shown by
\citep{dick2010impact}, the ILC method tends to amplify calibration
errors such that the reconstructed CMB map exhibits significantly
lower variance than the true sky with strong suppresssion of CMB
features, in particular in the high signal-to-noise regime.

The ILC method can be straightforwardly implemented in either real
(pixel) space or in harmonic space. Thus, sets of ILC weights can
either be computed for different regions of the sky or for different
angular scales, respectively, which allows for variations of the data
covariance matrix in either space.  However, the ILC in harmonic space
does not take into account the fact that noise can be a significant
source of CMB measurement error in at high Galactic latitude, while
foreground signals are more important at low Galactic
latitude. Conversely, the ILC in pixel space does not take into
account the fact that the noise dominates on small angular scales,
while diffuse Galactic foreground emission dominates on large angular
scales.

In order to overcome this problem, we implement the ILC on a frame of
spherical wavelets called needlets \citep{Narcowich2006}, a
component-separation approach that we now refer to as the Needlet Internal Linear
Combination ({\tt NILC}) method. This technique has already been applied broadly
in CMB data analysis \citep{delabrouille2009full, Remazeilles2011,
  Remazeilles2011b, Basak2012, Basak2013, Remazeilles2013}.  The
needlets enable localized filtering in both pixel space and harmonic
space because they have compact support in the harmonic domain, while
still being very well localized in the pixel domain
\citep{Narcowich2006,marinucci2008spherical}. The needlet
decomposition allows the ILC weights to vary both smoothly on large angular
scales and rapidly on small angular scales, which is not possible
by sub-dividing the sky into different areas prior to any processing.

The needlet windows, $h^{j}_{l}$, in harmonic space that we use in our
analysis are defined as follows,
\begin{eqnarray} 
h^{j}_{l} = \left\{
\begin{array}{rl} 
\cos\left[\left(\frac{l^{j}_{peak}-l}{l^{j}_{peak}-l^{j}_{min}}\right)
\frac{\pi}{2}\right]& \text{for } l^{j}_{min} \le l < l^{j}_{peak},\\ 
\\
1\hspace{0.5in} & \text{for } l = l_{peak},\\
\\
\cos\left[\left(\frac{l-l^{j}_{peak}}{l^{j}_{max}-l^{j}_{peak}}\right)
\frac{\pi}{2}\right]& \text{for } l^{j}_{peak} < l \le l^{j}_{max} 
\end{array} \right. 
\end{eqnarray}
In terms of $h_{l}^{j}$, the spherical needlets are defined as
\begin{eqnarray}
\Psi_{j k}(\hat n)=\sqrt{\lambda_{jk}}\sum _{l=l^{j}_{min}}^{l^{j}_{\max }}\sum_{m=-l}^l 
h_{l}^{j}\,Y_{l m}^{*}(\hat n)\,Y_{l m}(\hat\xi_{jk}),
\end{eqnarray}
where the $\{\xi_{jk}\}$ denote a set of cubature points on the sphere for
scale $j$. In practice, we identify these points with the pixel
centres of the \healpix\ pixelization scheme \citep{Gorski2005}. Each
index $k$ corresponds to a particular \healpix\ pixel, at a resolution
parameter $N_{\mathrm{side}}(j)$ specific to that scale $j$. The
cubature weights $\lambda_{jk}$ are inversely proportional to the
number $N_{j}$ of pixels used for the needlet decomposition,
i.e., $\lambda_{jk}=\frac{\displaystyle{4\pi}}{\displaystyle{N_{j}}}$.

\begin{figure}[tbp]
\centering
\includegraphics[width=0.5\textwidth]{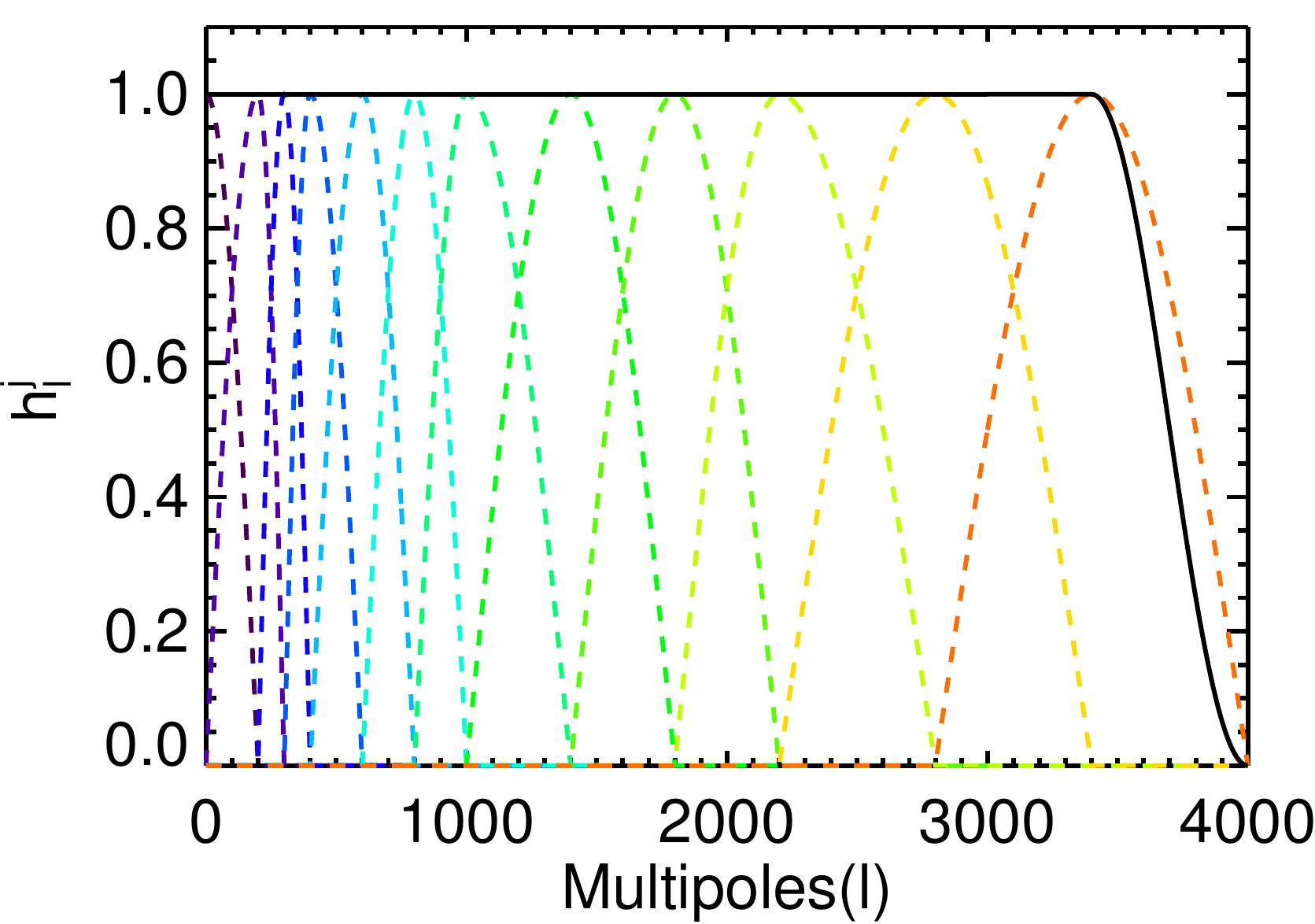}
\caption{Needlet windows used in the {\tt NILC} component separation
  (dashed lines) applied to the simulated \core\ data set in this
  paper.  The solid black line shows the normalization of the needlet
  bands, i.e., the total transfer function applied to the original map
  after needlet decomposition and synthesis of the CMB map from the
  needlet coefficients.}
\label{fig:needlet-bands-core} 
\end{figure}

Given a set of needlet functions, any sky map of a spin-$0$ field
$X(\hat n)$ (such as the CMB temperature anisotropy, or the $E$- and $B$-modes
of CMB polarization) on the sphere can be expressed as
\begin{eqnarray}
X(\hat n) =\sum _{l=0}^{l_{\max }}\sum_{m=-l}^lX_{lm}Y_{lm}(\hat n)=\sum _{j}^{}\sum _{k}^{}\beta^{X}_{j k}\Psi_{j k}(\hat n),
\end{eqnarray}
where the needlet coefficients, $\beta^{X}_{j k}$, of the sky map are denoted as
\begin{eqnarray}
\beta^{X}_{j k}=\left <X,\Psi_{j k}\right>=\sqrt{\lambda_{j k}} \sum _{l=0}^{l_{\max
  }}\sum_{m=-l}^l h_l^j\,X_{l m}\,\,Y_{l m}(\xi _{j k}).
\end{eqnarray}
For each scale $j$, the ILC filter has a compact support between the
multipoles $l^{j}_{min}$ and $l^{j}_{max}$ with a peak at
$l^{j}_{peak}$ (Fig. \ref{fig:needlet-bands-core}) The needlet
coefficients, $\beta^{X}_{jk}$, are computed on the \healpix\ grid
points, $\xi_{j k}$, with a resolution parameter, \nside\, equal to
the smallest power of $2$ larger than $l^{j}_{max}/2$.

Once an estimate of the CMB sky has been evaluated, the corresponding 
CMB angular power spectrum can be evaluated using a a pseudo-$C_{\ell}$ estimator
\citep{Hivon2002,Chon2004,Szapudi2005}. This method is computationally
much faster than maximum likelihood and provides optimal results at
intermediate to high $\ell$s. In order to compute the covariance on
our measurement of angular power spectrum, we have followed the method
described in \cite{Tristram2005}. 

In this paper, although the {\tt NILC} weights are computed from
full-mission sky maps, the impact of instrumental noise residuals on
the measurement of the angular power spectra is avoided by producing
independent CMB maps for the two half-mission data sets. These maps
are obtained by applying the full mission {\tt NILC} weights to the
half-mission sky maps. Each data point of the angular power spectra is
then obtained from the average of all possible cross-half-mission
angular power spectra.

Suppose we have $N$ measurements (one per
half-mission) of CMB fields such that the residuals of noise in these
measurements are statistically independent from each other. 
The estimator of angular power spectra with $(2l+1)$ weights for binning is given by
\begin{eqnarray}
\widehat{C}^{XY}_{b}=\frac{\displaystyle{\sum^{l_{max}^{b}}_{l=l_{min}^{b}}(2l+1)\widehat{C}^{XY}_{l}}}{\displaystyle{\sum^{l_{max}^{b}}_{l=l_{min}^{b}}(2l+1)}} \quad \widehat{C}^{XY}_{l}=\left(M^{XY}\right)^{-1}_{ll^{\prime}}\widehat{D}^{XY}_{l^{\prime}} \quad  X,Y=\left\{T,E,B\right\},
\end{eqnarray}
where $l_{min}^{b}$ and $l_{max}^{b}$ are respectively the minimum and maximum values of the multipole for the b-th bin, and $\widehat{D}^{XY}_{l}$ is the average of all possible cross-half-mission spectra for the recovered CMB fields after applying the mask under consideration:
\begin{eqnarray}
\widehat{D}^{XY}_{l}=\frac{1}{N(N-1)}\sum^{N}_{I=1}\sum^{N}_{J=1}\widehat{D}^{XY,IJ}_{l}\epsilon_{IJ} \qquad \epsilon_{IJ}=\left(1-\delta_{IJ}\right) \quad  I,J=1,...,N.
\end{eqnarray} 
The corresponding coupling matrix $M^{XY}_{ll^{\prime}}$, in terms of
the angular power spectrum $W^{XY}_{l}$ of the mask, is given by
\begin{eqnarray}
M^{XY}_{ll^{\prime}}=\mathlarger{\frac{2l^{\prime}+1}{4\pi}\mathlarger{\sum}_{l^{\prime\prime}}\left(2l^{\prime\prime}+1\right)W^{XY}_{l^{\prime\prime}}\,\,\wg{l}{l^{\prime}}{l^{\prime\prime}}{0}{0}{0}^2} \quad W^{XY}_{l}=\frac{1}{2l+1}\sum^{m}_{l=-m}W^{X}_{lm}W^{Y*}_{lm}.\enspace\enspace
\end{eqnarray} 
The covariance $\Xi^{XY,XY}_{ll^{\prime}}$ of $\widehat{C}^{XY}_{l}$ is by definition:
\begin{eqnarray}
\Xi^{XY,XY}_{ll^{\prime}}=\left<\left(\widehat{C}^{XY}_{l}-\left<\widehat{C}^{XY}_{l}\right>\right)\left(\widehat{C}^{XY}_{l_{\prime}}-\left<\widehat{C}^{XY}_{l_{\prime}}\right>\right)\right>=\left(M^{XY}\right)^{-1}_{ll_{1}}\Sigma^{XY,XY}_{l_{1}l_{2}}\left(M^{XY}\right)^{-1}_{l_{2}l^{\prime}},\enspace\enspace\enspace
\end{eqnarray} 
where $\Sigma^{XY,XY}_{l_{1}l_{2}}$ is the covariance of masked angular spectra $\widehat{D}^{XY}_{l}$:
\begin{eqnarray}
\Sigma^{XY,XY}_{l_{1}l_{2}}=\sum^{N}_{I,J,K,L=1}\frac{M^{\left(2\right)}_{l_{1}l_{2}}\left(W^{XX,YY}\right)C^{XI,XK}_{l_{1}}C^{YI,YK}_{l_{2}}+M^{\left(2\right)}_{l_{1}l_{2}}\left(W^{XY,XY}\right)C^{XI,YL}_{l_{1}}C^{XK,YJ}_{l_{2}}}{2l_{2}+1}\,\epsilon_{IJ}\epsilon_{KL}.\nonumber
\end{eqnarray}
The coupling matrix
$M^{(2)}_{ll^{\prime}}$, in
terms of the angular power spectrum, $W^{XY,X^{\prime}Y^{\prime}}$, of the product of the masks for the fields $X$ and $Y$ is given by
\begin{eqnarray}
M^{(2)}_{ll^{\prime}}\left(W^{XY,X^{\prime}Y^{\prime}}\right)=\mathlarger{\frac{2l^{\prime}+1}{4\pi}\mathlarger{\sum}_{l^{\prime\prime}}\left(2l^{\prime\prime}+1\right)W^{XY,X^{\prime}Y^{\prime}}_{l^{\prime\prime}}\wg{l}{l^{\prime}}{l^{\prime\prime}}{0}{0}{0}^2},
\end{eqnarray}
\begin{eqnarray}
W^{XY,X^{\prime}Y^{\prime}}_{l}=\frac{1}{2l+1}\sum^{m}_{l=-m}W^{(2)XY}_{lm}W^{(2)X^{\prime}Y^{\prime}*}_{lm}.
\end{eqnarray}
The covariance of the estimator of the binned angular power spectra, $\widehat{C}^{XY}_{b}$, is then given by
\begin{eqnarray}
\widehat{\Xi}^{XY,XY}_{bb^{\prime}}=\frac{\displaystyle{\sum^{l_{max}^{b}}_{l=l_{min}^{b}}\sum^{l_{max}^{b^{\prime}}}_{l=l_{min}^{b^{\prime}}}(2l+1)(2l^{\prime}+1)\widehat{\Xi}^{XY,XY}_{ll^{\prime}}}}{\displaystyle{\left(\sum^{l_{max}^{b}}_{l=l_{min}^{b}}(2l+1)\right)\left(\sum^{l_{max}^{b^{\prime}}}_{l=l_{min}^{b^{\prime}}}(2l+1)\right)}}.
\end{eqnarray}

\subsection{Spectral Matching}
\label{subsec:smica_method}

The use of the Spectral Matching Independent Component Analysis ({\tt
  SMICA}) method \citep{2003MNRAS.346.1089D} assumes that the total
sky emission observed across a set of frequency bands, in any pixel
$p$, or any harmonic mode $(\ell,m)$ is a noisy sum of components,
that can be written, in the most general form, as,
\begin{equation}
\mathbf{x}  =  \sum_{\rm comp} \mathbf{c} + \mathbf{n}.
\label{eq:mixture-model}
\end{equation}
where the sum runs over all components of the sky emission, and where the
contribution of each component to the total observed sky emission is
fully described by a vector $\mathbf{c}$ that represents its emission
in the set of all observed frequency bands. For each pixel or
harmonic mode, all of $\mathbf{x}$, the various components
$\mathbf{c}$ and $\mathbf{n}$ are vectors of dimension 
$n_{\nu}$.

While no assumption has been made so far about the properties of the
components, such a decomposition is of particular interest when the
components are \emph{independent}. Indeed, together with the
assumption of independence between the sky components and noise,
this guarantees that the multivariate power spectrum of $\mathbf{x}$ is the
sum of the multivariate power spectra of the components and of the
noise, with no cross terms.
\begin{equation}
\tX_\ell  =   \sum_{\rm comp}  \tC_\ell  +  \tN_\ell,
\label{eq:full-data-spectrum1}
\end{equation}
where the sum runs over independent sky components, and for each
$\ell$, each of $\tX_\ell$, $\tN_\ell$  and of all of the $\tC_\ell$
is an $n_{\nu} \times n_{\nu}$ 
covariance matrix. 

The independence of components depends on their physical origin. CMB
emission is independent from Galactic foreground emission, however,
the spinning dust emission in our Galaxy can not be assumed to be
independent from the thermal dust emission.

We now turn to modeling further the emission of each physical
component, $\mathbf{c}$,  as a linear mixture of $n \leq n_{\nu}$
independent, unphysical, templates.
\begin{equation}
\mathbf{c}  =  \tA \, \mathbf{s},
\label{eq:multidimensional_comp}
\end{equation}
where $\tA$ is an $n_{\nu} \times n$ 
matrix, and $\mathbf{s}$ a set of
$n$ templates that describe component $\mathbf{c}$ (hereafter denoted
as the \emph{sources} of component $\mathbf{c}$). Such a decomposition
is always possible independently of the nature and of the physical
properties of the component: in the worst case scenario, the emission
of a component can be modeled using as sources the $n = n_{\nu}$
templates of emission of the component in all the bands of
observation. The matrix $\tA$ is then the identity matrix. However,
this decomposition into sources is more interesting when $n < n_{\nu}$. 
We denote as the \emph{dimension of a component} the minimum
number of sources required to represent with sufficient accuracy\footnote{So
  that the difference between the real emission of the component and
  its model with $n<n_{\nu}$ 
  sources is not detectable given the
  observational noise.} its emission in all of the observed bands. 
The extreme case of $n=1$ corresponds to a component that scales
perfectly with frequency, a model valid for the CMB emission. For
example, Galactic ISM emission, considered as one single component,
typically requires $n>1$ templates that may or may not be identified
with emission from specific processes such as free-free or
synchrotron.

The multivariate power spectrum $\tC_\ell$ of an $n$--dimensional
component can be written as
\begin{equation}
\tC_\ell  =   \tA \, \tS_\ell \, \tA^t,
\label{eq:full-data-spectrum2}
\end{equation}
where $\tS_\ell$ is the multivariate power spectrum of the sources of
component (for each $\ell$, an $n \times n$ matrix). The diagonal
elements of $\tS_\ell$ represent the elements of the auto-spectra and
the off-diagonal elements the cross-spectra of the sources
$\mathbf{s}$. For a one-dimensional component, $\tS_\ell$ is for each
$\ell$ a $1 \times 1$ matrix, that represents the value of the power
spectrum of that particular component for multipole $\ell$.

The idea of the {\tt SMICA} method is to adjust a \emph{model} of the
covariance matrices of the various independent components and of the
noise to achieve the best match, in the maximum likelihood sense, to the
\emph{observed} covariance matrices of the data.
Observed covariances are estimated from the observations using
spectral band averages of the form
\begin{equation}
\widehat \tX_q  =   \frac 1 {{\mathcal{N}}_q} \sum_{\ell=\ell_{\rm min}(q)}^{\ell_{\rm max}(q)}  \sum_{m=-\ell}^\ell \mathbf{x}^t  \mathbf{x},
\label{eq:full-data-spectrum3}
\end{equation}
where $q$ indexes the spectral band, and ${{\mathcal {N}}_q}$ is the
number of modes $(\ell,m)$ in the band. In practice, one must take
into account the fact that the original maps are in general at a
different resolution, and that a mask must be applied to select only
regions where the Galactic emission is low and simple enough to be
accurately modeled as a component of sufficiently low dimensionality for the
spectral fit to be possible. 

The model that we adjust comprises in general three sky emission
components, and noise
\begin{itemize}
\item CMB, modeled as a 1-dimensional component and specified by its
  frequency dependence;
\item a multi-dimensional `catch-all' component that models the
  emission of the Galactic ISM as well as the diffuse background of
  extra-galactic sources;
\item an additional 1-dimensional component that models the emission
  of strong extra-galactic sources;
\item a noise component for which the covariance matrix is fixed to
  its exact value.
\end{itemize}

For the data, at each frequency a $B$-mode map is generated from the
$Q$ and $U$ Stokes parameter data. A spherical harmonic
transform is computed up to $\ell_{\mathrm{max}}=6000$, and then
the corresponding $a^{B}_{\ell m}$ are convolved with a Gaussian beam of
20 arcminutes FWHM, before synthesising the  $B$-mode map.
These maps are then masked before we finally compute the
weighted covariance matrices that {\tt SMICA} will adjust to the
model.

The statistical properties of the noise are assumed to be known a
priori (i.e., the term $\tN_\ell$ in Eq.~\ref{eq:full-data-spectrum1}
is known). Alternatively, a parametric model can be assumed for a
joint estimation of sky model and noise model parameters. Generically,
the set of model parameters $\theta$ that we assume comprises the set
of band-average auto and cross spectra of all sources of sky emission,
and the elements of the `mixing matrix' $\tA_{\rm fg}$ of the  `catch
all' component of foreground sky emission. The adjustment criterion is
to maximize the likelihood of the model given the observed
band-averaged covariance of the observations. The best fit estimated parameter set $\widehat \theta$
is obtained as :
\begin{equation}
\widehat \theta  =  \arg \min \left [   \phi(\theta) \right ]
\end{equation}
where
\begin{equation}
 \phi(\theta)    =
   \sum_{q=1}^Q
   {\mathcal N}_q
   \
   D\left(\widehat \tX_q \, ,  \sum_{\rm comp}  \tC_q(\theta)  +  \tN_q(\theta) \right) 
\end{equation}
and where $D(\cdot,\cdot)$ is a measure of divergence between two positive
$n\times n$ matrices defined by
\begin{equation}
   \label{eq:kullR}
   D(R_1,R_2)
   =
   \trace \left( R_1R_2\inv \right) - \log\det (R_1R_2\inv) - n
   .
\end{equation}
We refer the reader to the {\tt SMICA} publications
\citep{2003MNRAS.346.1089D, Cardoso2008} for further details about the
method.


\section{Forecasts for the \core\ component separation problem}
\label{sec:fisher}

We describe in this section a complementary approach to the component
separation methods applied on single CMB and noise realizations as
presented in the main sections of this article. The framework, named
\textsc{xForecast}~(\cite{Stompor2016}), is an extension of the
\textsc{CMB4cast}\footnote{\protect Publicly accessible at
  \url{http://portal.nersc.gov/project/mp107/index.html}} method described in \cite{Errard2011,Errard2012,Errard2016}. 

\textsc{xForecast} optimizes a CMB- and noise-averaged spectral
likelihood, therefore providing an estimate of the ensemble-averaged
spectral parameters, the statistical and systematic foreground
residuals, and the likelihood for the tensor-to-scalar ratio, $r$.
Whereas \textsc{CMB4cast} allows the
estimation of the impact of statistical residuals on $\sigma(r)$,
\textsc{xForecast} derives the statistically-meaningful performance of
\core\ in the specific case of a parametric component separation
approach, in particular wih respect to the bias on the estimation of
$r$.

\subsection{Formalism}
\label{ssec:fisher_methodology}

Similarly to the formalism used by {\tt Commander} (see
Sect.~\ref{subsec:parametric}), we use the parametric maximum-likelihood approach as introduced in, e.g., \cite{Brandt1994,Eriksen2006,Stompor2009}. At a sky pixel $p$, the
measured amplitudes at all frequencies are concatenated in a data
vector $d$, such that
\begin{eqnarray}
	\centering
		d(\nu, p) = \mathbf{A}(\nu, p)\,s^{\rm true}(p) + n(\nu, p)
	\label{eq:data_modeling}
\end{eqnarray}
where 
\begin{itemize}
	\item $\mathbf{A}$ is the mixing matrix, which contains the scaling laws of all sky components (CMB,
          foregrounds). Under the parametric formalism, we assume that
          the mixing matrix $\mathbf{A}$ can be parametrized by a set
          of spectral parameters $\beta$:
	\begin{eqnarray}
	\centering
		\mathbf{A} \equiv \mathbf{A}(\beta).
	\end{eqnarray}
	\item $s^{\rm true}(p)$ contains the true amplitudes of the sky
          signals, scaled at a reference frequency;
	\item $n(\nu, p)$ is the instrumental noise, assumed white in this analysis.
\end{itemize}
We will not write the $\nu$ argument in most of the equations below.
Given Eq.~\ref{eq:data_modeling}, the component separation and
cosmological analysis is performed in three steps:
\begin{itemize}
	\item \textbf{the estimation of the mixing matrix or,
            equivalently, the estimation of the spectral parameters.}
          This is achieved through the optimization of a spectral
          likelihood, $\mathcal{L}_{spec}(\beta)$, as detailed
          in~\cite{Stompor2009}. In order to estimate the
          statistically-averaged performance of the component
          separation for a given instrumental configuration,
          \cite{Stompor2016} propose a spectral likelihood averaged
          over a statistical ensemble of the noise realizations,
\begin{eqnarray}
\langle {\cal L}_{spec} \rangle =  -{\rm tr}\, \sum_p \bigg\{(\mathbf{N}(p)^{-1} - \mathbf{P}(p)) \Big( \mathbf{\hat d}(p) \mathbf{\hat d}(p)^t+ \mathbf{N}(p)\Big)\bigg\}.
\label{eqn:likeAvFinal}
\end{eqnarray}
where frequency-frequency $\mathbf{N}$ is the noise covariance matrix
and $\mathbf{\hat d}$ is the noiseless sky signal
i.e., $\mathbf{A}(p)\,s(p)$.
In Eq.~\ref{eqn:likeAvFinal}, the dependence on the spectral
parameters is confined to the projection operator, $\mathbf{P}(p)$, 
 \begin{eqnarray}
\mathbf{P}(p) \equiv \mathbf{N}(p)^{-1} - \mathbf{N}(p)^{-1}\mathbf{A}(p)\left(\mathbf{A}(p)^t\mathbf{N}(p)^{-1}\mathbf{A}(p)\right)^{-1} \mathbf{A}(p)^t \mathbf{N}(p)^{-1}.
\label{eqn:projDef}
 \end{eqnarray}
$\langle {\cal L}_{spec} \rangle$ can be maximized very efficiently
numerically, given that the number of unknown spectral parameters is
usually limited and that one can capitalize on the analytical
derivatives of the likelihood. \textsc{xForecast}, in the same manner
as \textsc{CMB4cast}, based on
\cite{Errard2011}, uses a semi-analytical expression for the
covariance of the error bars on spectral indices,
$\mathbf{\Sigma}(\beta)$. This gives a computationally efficient way
of estimating the statistical foregrounds residuals.

\begin{figure*}
\centering
	\includegraphics[width=16cm]{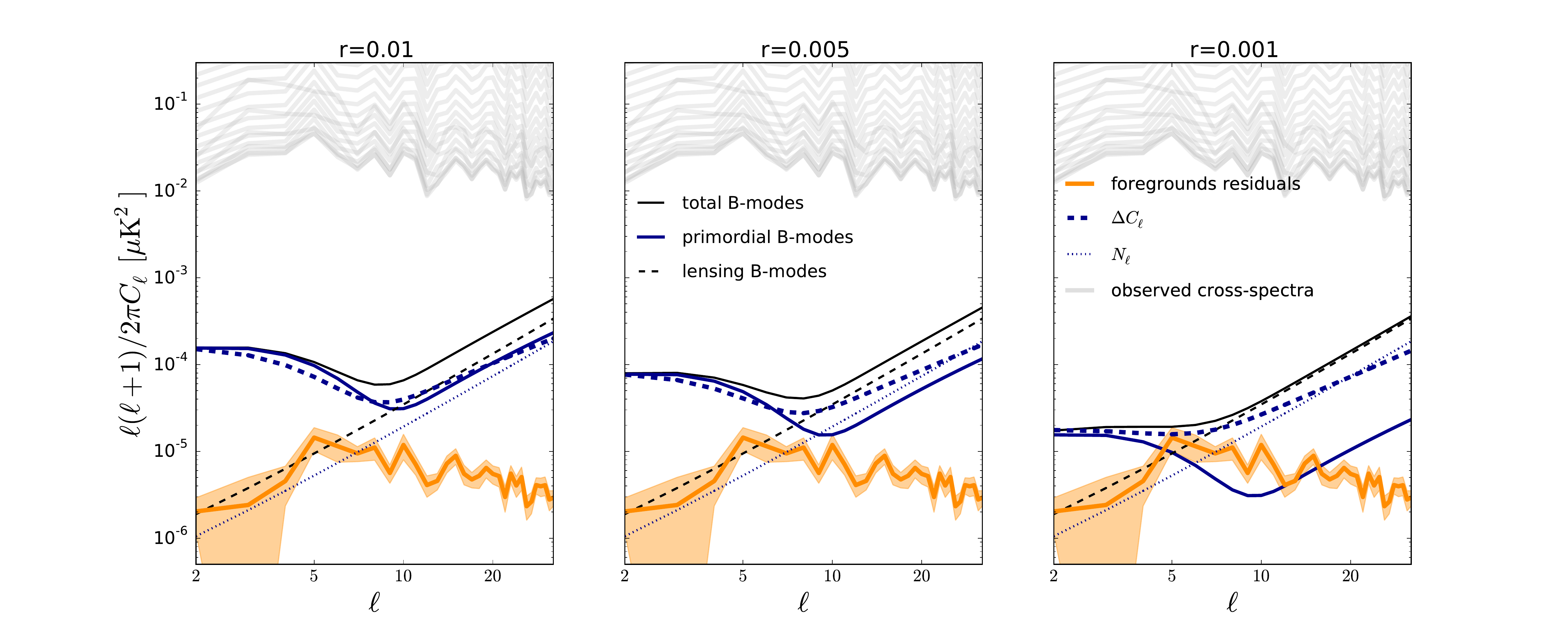}
	\caption{Performance of
          \core\ regarding the level of foregrounds residuals, in the
          case of $r=10^{-2}$ (left), $5\times 10^{-3}$ (middle) and $10^{-3}$
          (right). Results as produced by
          \textsc{xForecast}~\citep{Stompor2016}.}
	\label{fig:xForecast_power_spectra}
\end{figure*}
	
	\item \textbf{the ``inversion'' of Eq.~\ref{eq:data_modeling}
            with the estimated $\mathbf{A}$,} in order to disentangle
          sky components and obtain estimates of the sky signals
          $\tilde{s}$,
	\begin{eqnarray}
		\centering
			\tilde{s}(p) &=&  (\mathbf{A}(p)^t\mathbf{N}(p)^{-1}\mathbf{A}(p))^{-1}\mathbf{A}(p)^t\mathbf{N}(p)^{-1}\mathbf{\hat d}(p) \\
			&\equiv& \mathbf{W}_p(\beta)\mathbf{\hat d}(p).
		\label{eq:normal_equation}
	\end{eqnarray}
	From Eq.~\ref{eq:normal_equation}, it should be evident that the noise
        variance, $\sigma_{\rm CMB}$, associated with the recovered
        CMB map is given by
	\begin{eqnarray}
		\centering
			\sigma^2_{\rm CMB} \equiv \left[  \left(\mathbf{A}^T\mathbf{N^{-1}}\mathbf{A}\right)^{-1} \right]_{\rm CMB \times CMB}
		\label{eq:noise_variance_CMB_map}
	\end{eqnarray}
	The noiseless foreground residuals are then given by
	\begin{eqnarray}
		\mathbf{r}(p) = \mathbf{\tilde{s}}(p) - \mathbf{s}^{\rm true}(p) = \mathbf{W}_p(\beta)\mathbf{\hat d}(p) - \mathbf{s}^{\rm true}(p).
		\label{eqn:resPixDefGen}
	\end{eqnarray}
	Eq.~\ref{eqn:resPixDefGen} can be rewritten and specialized for the CMB component residual,
        \begin{eqnarray}
            \mathbf{r}(p)^{\rm cmb} = \sum_k \mathbf{W}^{0 k}_p(\beta)\mathbf{\hat f}(p)^{\left(k\right)} \equiv \sum_k \mathbf{W}^{0 k}_p(\beta)\mathbf{F}_{p k} 
        \label{eqn:resCMBpix}
        \end{eqnarray}
        which does not contain CMB signal. $\mathbf{F}$ is  a
        foreground matrix, and the $k$th column defines the total
        foreground contribution to the $k$th frequency channel.
      	\cite{Stompor2016} perform a Taylor expansion of the residuals
        with respect to the scaling parameters around the
        maximum-likelihood values, $\tilde{\beta}$,
        \begin{eqnarray}
        \begin{array}{l l l}
        \mathbf{r}^{\rm cmb}(p)(\beta)  & \simeq & {\displaystyle \sum_k \mathbf{W}^{0 k}_p(\tilde \beta)\mathbf{F}_{pk}
        + \sum_{k, \beta} \,\delta \beta \left.\frac{\partial \mathbf{W}^{0k}_p}{\partial \beta}\right|_{\tilde \beta}\mathbf{F}_{pk}}\\
        &+& {\displaystyle \sum_{k, \beta,\beta'} \,\delta \beta \delta \beta' \left.\frac{\partial^2 \mathbf{W}^{0k}_p}{\partial \beta\partial \beta'}\right|_{\tilde \beta}\mathbf{F}_{pk}}.
        \end{array}
        \label{eqn:resExp}
        \end{eqnarray}
        Evaluating $\langle
        \mathbf{r}^{\rm cmb\ \dagger}_{\ell m} \mathbf{r}^{\rm
          cmb}_{\ell' m'}\rangle_{m,m'}$ in harmonic space leads to an analytical
        expression for residuals power spectrum which can be found in
        \cite{Stompor2016}. Schematically, 
	\begin{eqnarray}
		\centering
		{C}^{\rm res}_{\ell} = {C}^{\rm stat.\ res}_{\ell}\left(\mathbf{\Sigma}(\tilde \beta), s(p)^{\rm true}\right) + {C}^{\rm syst.\ res}_{\ell}\left( \tilde \beta, \beta^{\rm true}, s(p)^{\rm true}\right)
		\label{eqn:Clres_schematic_expr}
	\end{eqnarray}	
	The first term corresponds to the statistical residuals,
        generated by the finite error bar on the spectral indices,
        $\mathbf{\Sigma}$. Note that this term is the only one used in the
        \textsc{CMB4cast} framework \citep{Errard2011,Errard2016}. The second term corresponds
        to the systematic residuals, sourced by the mismatch between
        the fitted mixing matrix, $\mathbf{A}(\tilde \beta)$, and the
        true mixing matrix used to generate the sky simulations.
	
	\item \textbf{Optimization of a CMB+noise-averaged
            cosmological likelihood}. \cite{Stompor2016} start
          from a standard Gaussian likelihood, which accounts only for
          noise, CMB signal and  statistical foreground residuals in
          the recovered CMB map, $\tilde{s}(p)$. Assuming that they all
          are Gaussian with the total covariance given by
          $\mathbf{C}$, one can write
\begin{eqnarray}
	-2\,\ln {\cal L}_{\rm cosmo} = \mathbf{a}^t \mathbf{C}^{-1} \mathbf{a} + \ln \det \mathbf{C},
	\label{eqn:parLike}
\end{eqnarray}
where $\mathbf{a}$ is a harmonic representation of the map obtained
after the component separation procedure. In addition to the CMB
signal, it can include the noise as well as the statistical and
systematic residuals. Similarly to~\cite{Errard2011}, this latter contributionis
ignored in the assumed data covariance matrix, $\mathbf{C}$. 
The cosmological likelihood averaged over the instrumental noise and
CMB signal realizations is given by (\cite{Stompor2016}),
\begin{eqnarray}
\langle -2\,\ln {\cal L}_{\rm cosmo} \rangle = {\rm tr} \, \mathbf{C}^{-1} \mathbf{E} + \ln \det \mathbf{C},
\label{eqn:parLikeAv}
\end{eqnarray}
where $\mathbf{E} \equiv \langle \mathbf{a}\mathbf{a}^t\rangle$ is the correlation matrix of the data. 

\end{itemize}

\begin{figure}
\centering
		\includegraphics[width=8cm]{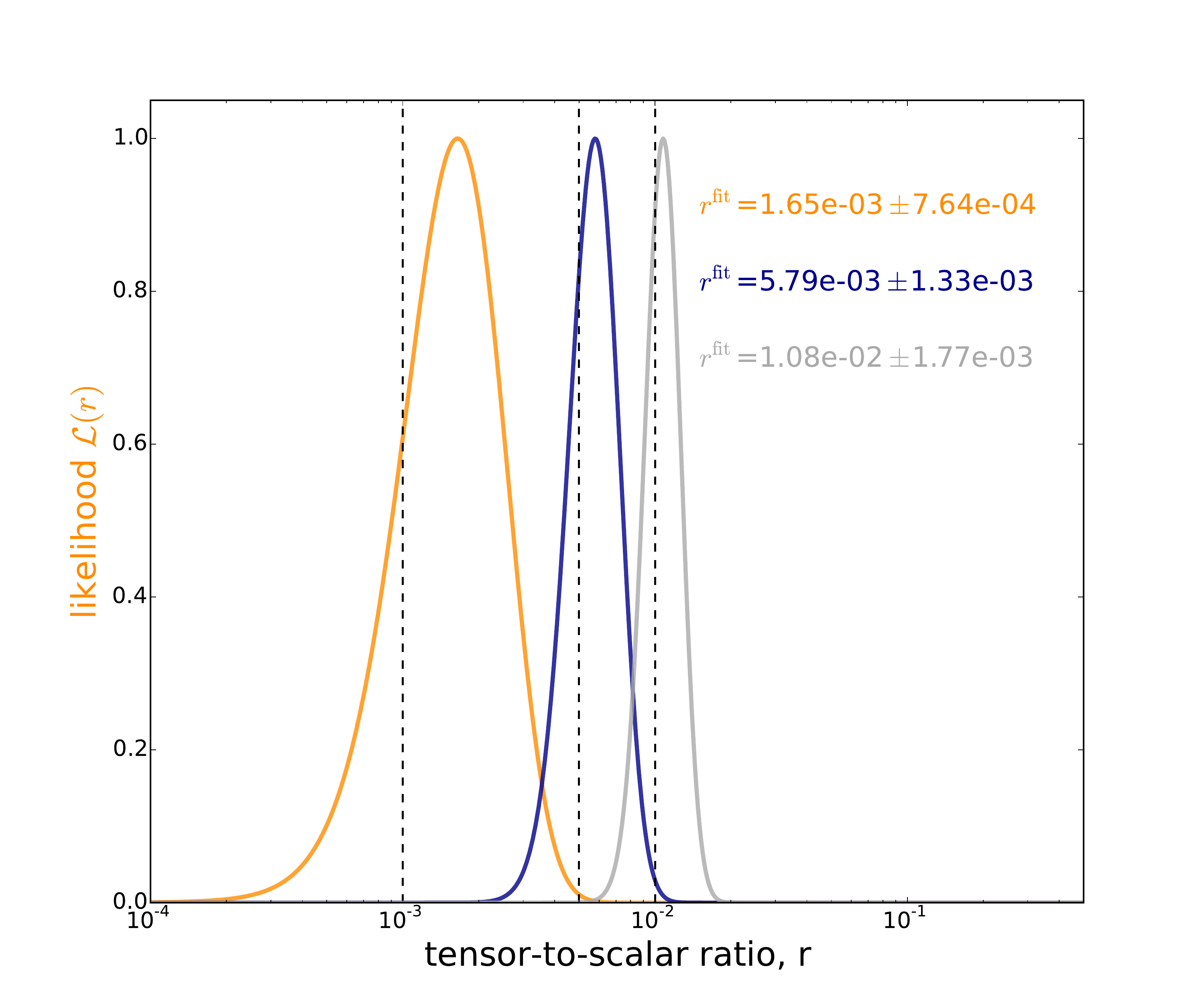}
	\caption{Propagation of the foreground residuals depicted in
          Fig~\ref{fig:xForecast_power_spectra} in CMB+noise-averaged
          cosmological likelihoods, obtained for different input
          values of the tensor-to-scalar ratios: $r=10^{-3}$, $5\times 10^{-3}$
          and $10^{-2}$. Note that these ensemble-averaged likelihoods can be interpreted as the statistical distributions of the fitted $r$ values obtained for a given CMB and noise realization. }
	\label{fig:xForecast_likelihood}
\end{figure}

\subsection{Application of \textsc{xForecast} to \core\ simulations}

In the exercise of cleaning the foregrounds in the \core\ simulated
sky maps, we simply consider two diffuse polarized astrophysical
foregrounds: dust and synchrotron. Similarly to Eq.~\ref{eq:fit}, these are assumed to follow a
grey-body and power-law spectra as,
\begin{eqnarray}
	\centering
		A_{\rm sync}(\nu, \nu_{\rm ref}) \equiv \left(\frac{\nu}{\nu_{\rm ref}}\right)^{\beta_s},
	\label{eq:As_def}
\end{eqnarray}
where the reference frequency $\nu_{\rm ref}=150$\;GHz. We consider a
modified grey-body emission law for the dust,
\begin{eqnarray}
	\centering
		A_{\rm dust}(\nu, \nu_{\rm ref}) \equiv \left( \frac{\nu}{\nu_{\rm ref}}\right)^{\beta_d+1}\frac{e^{\frac{h\nu_{\rm ref}}{k\,T_d}}-1}{e^{\frac{h\nu}{kT_d}} -1 },
	\label{eq:Ad_def}
\end{eqnarray}
These expressions are used to build the mixing matrix, $\mathbf{A}$,
involved in the spectral likelihood, cf. Eq.~\ref{eqn:likeAvFinal}.

After the numerical optimization of the spectral likelihood, Eq.~\ref{eqn:likeAvFinal}, we evaluate the level of statistical and systematic residuals, Eq.~\ref{eqn:Clres_schematic_expr}, and finally look at the cosmological likelihood, Eq.~\ref{eqn:parLikeAv}, for three values of the tensor-to-scalar ratio, $r=10^{-2}$, $r=5\times 10^{-3}$ and $r=10^{-3}$.
\textsc{xForecast} hence gives complementary results to the ones presented in the previous sections as it gives the ensemble-averaged level of foregrounds residuals as well as the averaged distribution of recovered $r$.

Results obtained in the case of the \core\ baseline after foreground cleaning, but without any delensing, are shown in Figs.~\ref{fig:xForecast_power_spectra} and~\ref{fig:xForecast_likelihood}. 
It should be noted that the bias on $r$ is strongly related to both
the complexity of the simulated foregrounds and the fidelity of the
chosen parametrization, Eqs.~\ref{eq:As_def},
\ref{eq:Ad_def}. In addition, the uncertainty on the measurement of $r$
depends primarily on the noise variance, the level of the statistical foreground
residuals (sourced by the uncertainty on the estimation of the spectral
parameters), and the lensing variance. The latter can potentially
be improved on by delensing the cleaned $B$-modes map.


\bibliographystyle{plainnat}
\bibliography{ECO_foregrounds}


\end{document}